\documentclass[]{raa}           
\usepackage{graphicx,times}
\usepackage{natbib}
\usepackage{amssymb,amsmath}
\bibpunct{(}{)}{;}{a}{}{,}

\usepackage[a4paper=true,pagebackref=true]{hyperref}%,dvipdfm=true
\usepackage{longtable}
\usepackage{footnote}
\usepackage{threeparttable}
\newcommand{\degree}{^\circ}
\hypersetup{pdftitle = The title of my PDF, pdfauthor = My name, pdfsubject= The subject, pdfkeywords = keyword1 keyword2 keyword3} 
\hypersetup{colorlinks = true, linkcolor = green, anchorcolor = red, citecolor = blue, filecolor = red, pagecolor = red, urlcolor = red}

\begin{document}

   \title{In Search for Infall Motion in molecular clumps II: HCO$^+$ (1-0) and HCN (1-0) Observations toward a Sub-sample of Infall Candidates
%$^*$
%\footnotetext{\small $*$ Supported by the National Natural Science Foundation of China.}
}

   \volnopage{Vol.0 (20xx) No.0, 000--000}      %%preserved for Editor. DOn't remove!
   \setcounter{page}{1}          %%starting page, preserved for Editor. DOn't remove!

   \author{Yang Yang
      \inst{1,2}
   \and Zhibo Jiang 
      \inst{1,2,3}
   \and Zhiwei Chen
      \inst{1}
   \and Shaobo Zhang
      \inst{1}
   \and Shuling Yu
      \inst{1,2}
   \and Yiping Ao
      \inst{1}
   }

   \institute{Purple Mountain Observatory, Chinese Academy of Sciences, Nanjing, Jiangsu 210008, People's Republic of China; {\it zbjiang@pmo.ac.cn}\\
        \and
             University of Science and Technology of China, Chinese Academy of Sciences, Hefei, Anhui 230026, People's Republic of China\\
        \and
             China Three Gorges University, Yichang, Hubei 443002, People's Republic of China\\
\vs\no
   {\small Received~~20xx month day; accepted~~20xx~~month day}}

\abstract{Gravitational accretion accumulates the original mass, and this process is crucial for us to understand the initial phases of star formation. Using the specific infall profiles in optically thick and thin lines, we searched the clumps with infall motion from the Milky Way Imaging Scroll Painting (MWISP) CO data in previous work. In this study, we selected 133 sources of them as a sub-sample for further research and identification. The excitation temperatures of these sources are between 7.0 and 38.5 K, while the H$_2$ column densities are between $10^{21}$ and $10^{23}$ cm$^{-2}$. We have observed optically thick lines HCO$^+$ (1-0) and HCN (1-0) using the DLH 13.7-m telescope, and found 56 sources of them with blue profile and no red profile in these two lines, which are likely to have infall motions, with the detection rate of 42\%. It suggests that using CO data to restrict sample can effectively improve the infall detection rate. Among these confirmed infall sources, there are 43 associated with Class 0/I young stellar objects (YSOs), and 13 are not. These 13 sources are probably associated with the sources in earlier evolutionary stage. By comparison, the confirmed sources which are associated with Class 0/I YSOs have higher excitation temperatures and column densities, while the other sources are colder and have lower column densities. Most infall velocities of the sources we confirmed are between 10$^{-1}$ to 10$^0$ km s$^{-1}$, which is consistent with previous studies.
}

   \authorrunning{Y. Yang et al.}            %author_head in even pages
   \titlerunning{In Search for Infall Motion in molecular clumps II}  % title_head in odd pages
   \maketitle

%________________________________________________ sections below
% 
\section{Introduction}           %% first-level sections will be auto-capitalized
\label{sect:intro}

The study of star formation shows that stars form through inside-out gravitation collapse in dense molecular clumps (\citealt{Shu+etal+1987}), while the accretion flow is maintained in the subsequent evolutionary processes to feed the forming stars. Gravitational infall mainly occurs in Class 0/I young stellar objects (YSOs) (\citealt{Bachiller+1996}). These sources are at the beginning stages of evolution, and their emissions are dominated by the infalling envelopes (e.g. \citealt{Allen+etal+2004}). After Class II stage, the infall motion is no longer dominates. Therefore, the research on infall motion helps us to better understand the physical process of the initial stage of star formation.
 
According to the model of infalling clumps (e.g. \citealt{Leung+Brown+1977}), a pair of optically thick and thin lines can be used as a tracer for infall motion. The optically thick line will self-absorb when passing through the clumps. When the gas falls into the core, line will produce a Doppler shift along the line of sight. This causes the self-absorption of the optically thick line to shift slightly to the red. In such a case, the line produces a double-peaked profile with the blue peak stronger than the red one. If the self-absorption is weak, the red peak may not be so obvious, and the line will show a blue peak profile with a red shoulder, or a single-peaked profile with the peak skewed to the blue. This kind of feature is commonly referred to as blue profile. Meanwhile, the optically thin line shows a single-peaked profile, with the peak between the double-peaks of the optically thick line. If the optically thick line shows a single peak, the peak of optically thin line is slightly skewed to the red in comparison. The velocity of optically thin line can be used to indicate the central radial velocity. Therefore, optically thin line is a good way to distinguish between self-absorption and multiple components. In recent years, using this method, a number of researches have been conducted on the infall motion (e.g. \citealt{He+etal+2015, Calahan+etal+2018, Saral+etal+2018}), but most of them have focused on known star-forming regions. If we are able to obtain an unbiased large-scale infall sample in the Milky Way, we may estimate the number of clumps with infall motion, which will help us to infer the lifetime of infall. 

Instead of looking for infall candidates in known star-forming regions, we use the Milky Way Imaging Scroll Painting (MWISP) data by DLH 13.7-m telescope to search the infall candidates in the Galactic disk. The MWISP project provides $^{12}$CO (1-0) and its isotopes $^{13}$CO (1-0) and C$^{18}$O (1-0). These three lines can be used as a start to search for infall motions in the molecular clumps. Using the combinations of $^{12}$CO and $^{13}$CO, $^{13}$CO and C$^{18}$O, about 2,200 candidates were preliminarily identified by blue profiles (\citealt{Jiang+etal+{2020, in prep}}). We further selected 133 candidates of them for follow-up study. Compared with CO, HCO$^+$ and HCN are generally optically thick lines, and can trace denser regions, which are believed to be more effective tracers of infall profile (e.g. \citealt{Vasyunina+etal+2011, Peretto+etal+2013, Yuan+etal+2018, Zhang+etal+2018}).
 In this paper, using the DLH 13.7-m telescope, we carried out HCO$^+$ (1-0) and HCN (1-0) lines observations toward our sub-sample. Based on the observations of these two lines and optically thin line C$^{18}$O (1-0), we further identified our sources, and discussed the physical properties and infall velocities of the confirmed infall sources.

In Section 2, we briefly introduce our source selection criteria and the observations. The results and discussion based on CO data, and HCO$^+$ and HCN lines analysis are given in Section 3. Finally, the summary in Section 4.

\section{The Sample and Observations}
\label{sect:Obs}

The MWISP project is an on-going, unbiased CO survey along the Galactic plane in the range of $l=[-10\degree, 250\degree]$, $b=[-5\overset{\degree}{.}2, 5\overset{\degree}{.}2]$, observing $^{12}$CO (1-0), $^{13}$CO (1-0), and C$^{18}$O (1-0) lines simultaneously (\citealt{Su+etal+2019}). At present, the project has completed about 80\%. Using the combinations of $^{12}$CO and $^{13}$CO, $^{13}$CO and C$^{18}$O, about 2,200 candidates showing blue profiles were identified by machine search and manual check up to the end of 2017 (\citealt{Jiang+etal+{2020, in prep}}). As the first step to confirm these candidates, we selected 133 of them with significant blue profiles in $^{12}$CO, and C$^{18}$O intensities greater than 1 K as a sub-sample for further study. Table~\ref{Tab:tab1} lists the selected sources, where the kinematic distances are calculated by the model from \cite{Reid+etal+2014}, and if the sources have kinematic distance ambiguity (KDA), the nearer one is chosen.

%________________________________________ Table 1: Basic parameters

\begin{table}
\begin{center}
  \caption[]{ The Derived Clump Parameters.}\label{Tab:tab1} 
%Please Capitalize the First Letter of Each Notional Word in table's caption
 \begin{tabular}{ccccccc}
  \hline\noalign{\smallskip}
Source & RA & Dec & V$_{LSR}$ & Distance & T$_{ex}$($^{12}$CO) & log($N($H$_2))$ \\
Name & (J2000) & (J2000) & (km s$^{-1}$) & (kpc) & (K) & (cm$^{-2}$) \\
  \hline\noalign{\smallskip}
%  \endhead\noalign{\smallskip}
% $^*$
G002.97+4.22 & 17:36:36.4 & -24:11:31 & 19.7 & 4.88  & 9.5  & 21.20 \\
G012.34-0.09$^\dagger$ &  18:12:53.7 & -18:17:01 & 34.2  & 3.35  & 14.6  & 22.25 \\
G012.72+0.69$^\dagger$ &  18:10:46.9 & -17:34:15 & 17.3  & 1.94  & 21.8  & 22.19 \\
G012.77-0.19$^\dagger$ &  18:14:08.2 & -17:57:04 & 35.6  & 3.38  & 33.0  & 23.13 \\
G012.82-0.19$^\dagger$ &  18:14:13.3 & -17:54:53 & 34.9  & 3.33  & 38.5  & 23.61 \\
G012.87-0.20$^\dagger$ &  18:14:22.1 & -17:52:02 & 35.5  & 3.36  & 25.5  & 23.01 \\
G012.88-0.24$^\dagger$ &  18:14:32.4 & -17:52:48 & 35.8  & 3.38  & 22.3  & 23.24 \\
G012.96-0.23$^\dagger$ &  18:14:39.6 & -17:48:36 & 35.2  & 3.33  & 17.1  & 23.07 \\
G013.79-0.23$^\dagger$ &  18:16:19.5 & -17:04:39 & 38.0  & 3.39  & 17.9  & 22.63 \\
G013.97-0.15$^\dagger$ &  18:16:22.0 & -16:53:02 & 40.2  & 3.50  & 19.9  & 22.65 \\
  \hline\noalign{\smallskip}
\end{tabular}\\
\end{center}
\tablecomments{0.86\textwidth}{Columns are (from left to right) the source name, its equatorial coordinate, its local standard of rest (LSR) velocity, its heliocentric distance, its excitation temperature, and its H$_2$ column density. $^\dagger$ Source are associated with Class 0/I YSOs. The full table is available in the Appendix B.1.}
\end{table}

HCO$^+$ and HCN trace the dense regions in molecular clouds, and they are generally optically thick lines in these regions. Their line profiles can provide some information of gas motions in dense clumps (e.g. \citealt{Smith+etal+2012}). From May 14 to 25, 2018, we carried out the follow-up single-pointing observations (project code: 18A004) toward the 133 selected sources. The front-end of telescope is the 3$\times$3 beam sideband-separating Superconducting Spectroscopic Array Receiver (SSAR) (\citealt{Shan+etal+2012}), and we used the spectrum from the beam 2 of the array. The back-end is a set of 18 Fast Fourier Transform Spectrometers (FFTSs), which can analyze signals from the upper and lower sidebands(\citealt{Su+etal+2019}). The frequencies of HCO$^+$ (1-0) and HCN (1-0) lines are about 89.2 GHz and 88.6 GHz, respectively. The angular resolution of the telescope is about 62$^{\prime\prime}$, and the main beam efficiency is about 56.8\% at these frequencies (\citealt{Gong+etal+2018}). Since each sideband is 1 GHz, we can observe them at the same time. We have observed a total of 102 targets of these two lines. The data of the remaining 31 targets were obtained from the Millimeter Wave Radio Astronomy Database\footnote{\url{http://www.radioast.nsdc.cn/}} (among them, fourteen targets lack HCN data). The data were reduced using the CLASS of the GILDAS package\footnote{\url{http://www.iram.fr/IRAMFR/GILDAS/}}. The baselines of the spectra have been first-order fitted, and the fitting regions are $V_{LSR}\,\pm$ 40 km s$^{-1}$ excepted the emission regions. Since the observed lines have different frequencies, their velocity resolutions are also different. We smoothed all the spectra to a spectrometer resolution of 0.2 km s$^{-1}$. The on-source integration time is at least 7 minutes for each source, and the typical system temperatures ($T_{sys}$) are between 150 to 250 K. For some sources with low altitude or high $T_{sys}$, we repeated observations until the RMS values of these lines are less than 0.1 K at a resolution of 0.2 km s$^{-1}$. The total observing time is about 53 hours for this project.

\section{Results and Discussion}
\label{sect:res}

\subsection{Properties of the Sample}

\subsubsection{Excitation Temperature and Column Density}

\begin{figure}[h]
  \begin{minipage}[t]{0.495\linewidth}
  \centering
   \includegraphics[width=80mm]{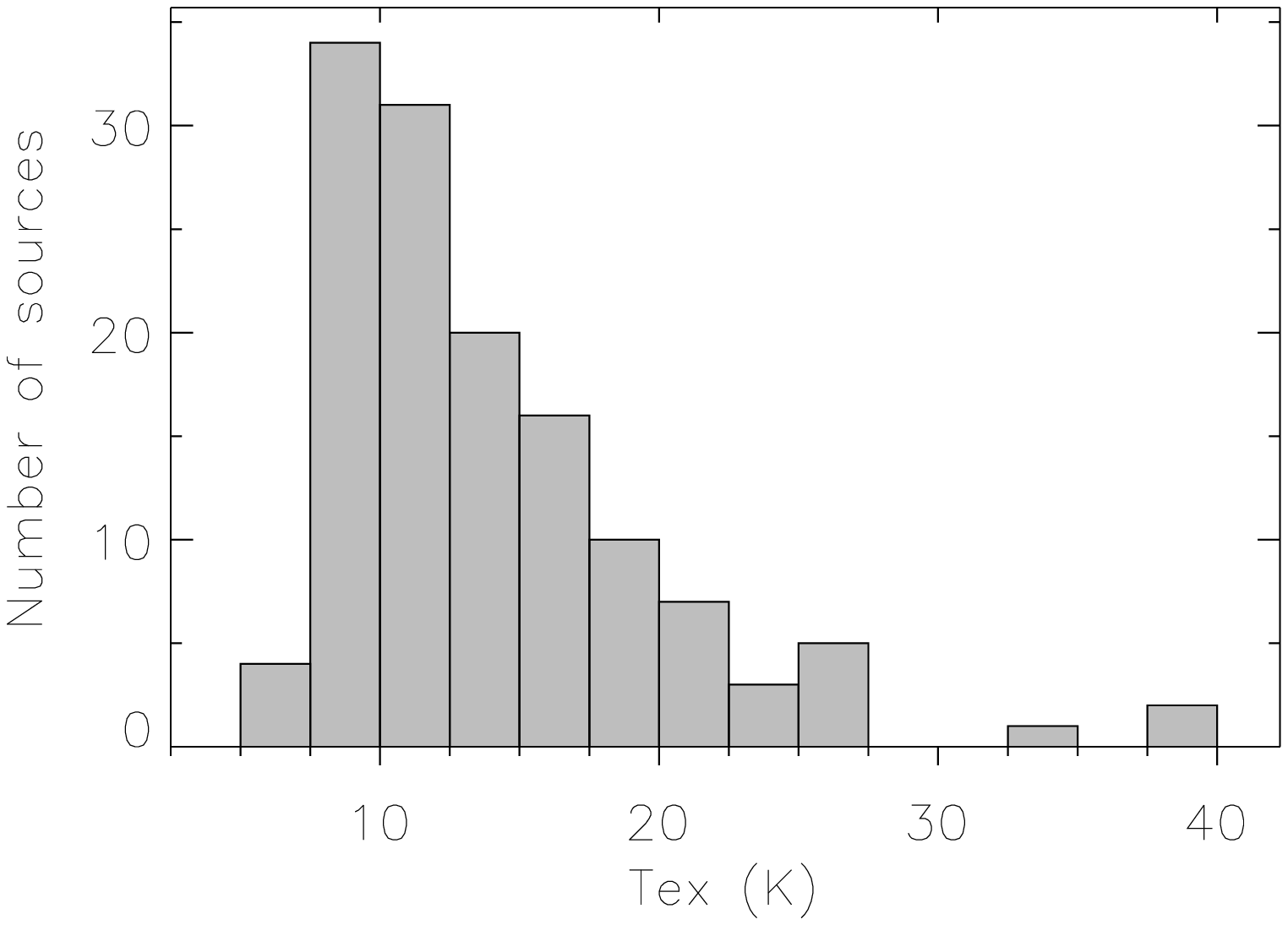}
%   \caption{{\small Amplitudes evolution of . . .} }
  \end{minipage}%
%  \label{Fig:fig2}
  \begin{minipage}[t]{0.495\textwidth}
  \centering
   \includegraphics[width=80mm]{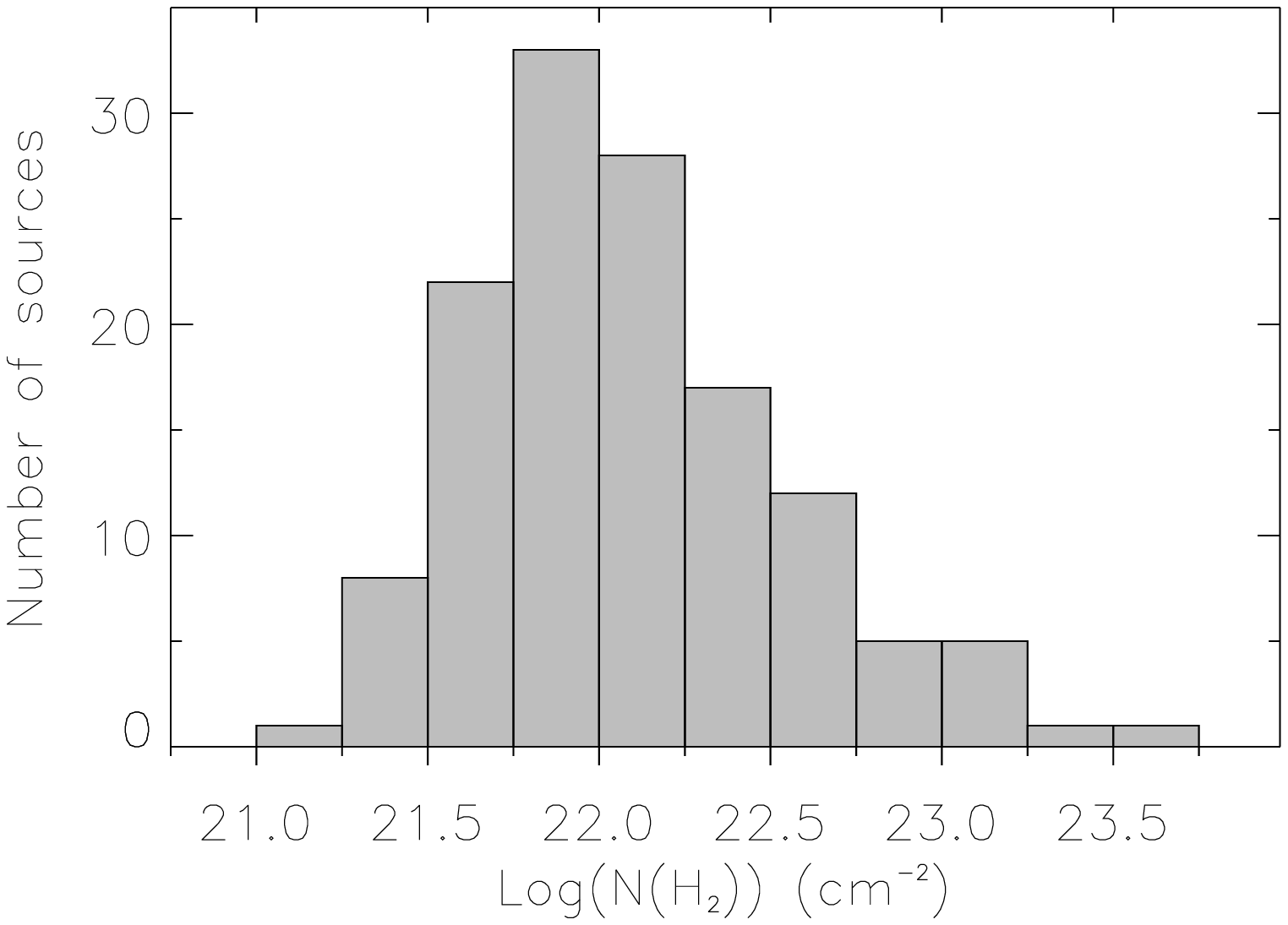}
%  \caption{{\small Amplitude variation of AN Lyn.}}
  \end{minipage}%
  \caption{{\small Distributions of the excitation temperatures and H$_2$ column densities of 133 sources. Left panel: The excitation temperature distribution. Right panel: The H$_2$ column density distribution derived from C$^{18}$O (1-0).
}}
  \label{Fig:fig1}
\end{figure}

CO lines are widely used to estimate the excitation temperature and column density of molecular cloud because of their high stability, high abundance, wide distribution and easy measurement. In this paper, we use the optically thick line $^{12}$CO to calculate the excitation temperature of the selected sources. Since there are self-absorption in $^{12}$CO of our sources, we use Gaussian fitting of the waist and foot portions of the profiles to obtain the expected peak intensity values. We fit each line three times and then select the one with maximum peak value. The peak values derived by this method are sensitive to the choice of the fitting portions. After carefully adjusting the fitting portions, we found that the errors are less than 20\% in most cases (\citealt{Jiang+etal+{2020, in prep}}). The excitation temperature $T_{ex}$ can be estimated by the peak intensity value $T(^{12}$CO$)$:
\begin{equation}
  T_{ex}=\frac{h\nu}{k}\left\{ \ln \left[ 1
  +\left(\frac{kT(^{12}{\rm CO})}{h\nu}+\frac{1}{e^{h\nu/kT_{bg}}-1}\right) ^{-1}\right]\right\} ^{-1}
\label{eq:LebsequeI}
\end{equation}
Where $h\nu/k = 5.53$ for $^{12}$CO (1-0), and $T_{bg}$ is cosmic background radiation temperature, i.e. 2.7 K. The uncertainty of $T_{ex}$ may be mainly due to our previous choice of $^{12}$CO fitting portion. This makes it difficult to give the error values. We can only estimate that the errors of $T_{ex}$ are less than 20\% in most cases. The left panel of Figure~\ref{Fig:fig1} shows the distribution of $T_{ex}$. As can be seen from the figure, the excitation temperature deviates from the normal distribution. The median value of our sources is 12.2 K. About 98\% of the sources have an excitation temperature of less than 30 K, and 38 of them have $T_{ex}$ $<$ 10 K, which suggests that our sources are mostly cold clumps. The $T_{ex}$ of remaining three sources G012.77-0.19, G012.82-0.19, and G081.72+0.57 are greater than 30 K. These sources may be some hotter molecular clumps.

In addition, we use the C$^{18}$O column density to estimate the H$_2$ column density (\citealt{Sato+etal+1994}):
\begin{equation}
  N({\rm C^{18}O})=2.24\times10^{14}\frac{T_{ex}\tau({\rm C^{18}O})\Delta v({\rm C^{18}O})}{1-e^{-h\nu/kT_{ex}}}{\rm cm^{-2}}
\label{eq:LebsequeI}
\end{equation}
where $h\nu/k = 5.27$ for C$^{18}$O (1-0), $\tau($C$^{18}$O$)$ is the optical depth and $\Delta v($C$^{18}$O$)$ is the FWHM of line profile. The abundance ratio we assume here is $N($H$_2)/N($C$^{18}$O$)\approx7\times10^6$ (\citealt{Warin+etal+1996, Castets+langer+1995}). The uncertainty of $N($H$_2)$ may be more difficult to determine due to the uncertainties of $T_{ex}$, the local thermal equilibrium assumption, and the abundance ratio of $N($H$_2)/N($C$^{18}$O$)$. These factors may bring several times of the error to $N($H$_2)$ values. As shown in the right panel of Figure~\ref{Fig:fig1}, the H$_2$ column densities of our sources are between $10^{21}$ and $10^{23}$ cm$^{-2}$, with the median value of $1.03\times10^{22}$ cm$^{-2}$. Although the x-axis is in logarithmic scale, the H$_2$ column density still seems to deviate from the normal distribution at the high-density end. It may be a power-law distribution at the high-density end.

%----------------------------------------------------- Fig 1:

\subsubsection{Association with Infrared Point Sources}

In order to better study the star formation status of these 133 sources, we check the association between the infrared point sources and our selected sources. If the infrared point sources located within a radius of 1$^{\prime}$ of our sources, we believe that this source is associated with the infrared point sources. There are 23 sources associated with IRAS point sources (\citealt{Helou+Walker+1988}), and 6 of them are associated with ultracompact HII (UC HII) regions from the criteria by \cite{Wood+Churchwell+1989}. After excluding the foreground stars, almost all sources are associated with WISE point sources (\citealt{Wright+etal+2010}), and 71 of them are associated with Class 0/I YSOs according to the criteria given by \cite{Koenig+etal+2012}. For our sources in the Galactic latitudes $|b| \leq 1 \degree$ and longitudes $l = 0 - 65\degree$, Spitzer GLIMPSE (\citealt{Benjamin+etal+2003, Churchwell+etal+2009}) and MIPSGAL (\citealt{Gutermuth+Heyer+2015}) surveys can also be used to estimate the evolutionary stages empirically. Sixty-four of our sources are associated with GLIMPSE and MIPSGAL point sources. \cite{Gutermuth+etal+2008} gave the selected criteria for Class 0/I YSOs. According to this criteria, sixty sources are associated with Class 0/I YSOs. Combined with these infrared point source tables, a total of 94 sources are associated with Class 0/I YSOs, and 6 of them are associated as UC HII regions. The remaining 39 sources are not associated with Class 0/I YSOs. These sources are probably in the early infall stage. The median values of $T_{ex}$ for the sources that are associated with Class 0/I YSOs and not associated with Class 0/I YSOs are 13.3 and 10.5 K, while the corresponding mean values of $N($H$_2)$ are $1.43\times10^{22}$ and $6.79\times10^{21}$ cm$^{-2}$, respectively. Both distributions display that the sources associated with Class 0/I YSOs have slightly higher excitation temperatures and column densities.

\subsection{HCO$^+$/ HCN Blue-profile Sources}

About 80\% of our sources show HCO$^+$ (1-0) emissions. Among them, the line profiles of 54 sources are double-peaked profiles, and 50 sources are single-peaked profiles. The other three sources show complex line profiles, with three or more peaks, e.g. G045.45+0.05, and we have not classified these line profiles. The hyperfine structure of HCN (1-0) produces complex line profiles. In this paper, we only consider the main peak. Based on the observations, we have found that 82 sources show HCN emissions, with the detection rate of 69\%. There are 34 sources with double-peaked profile and 45 sources with single-peaked profile. The remaining sources show complex features in their main peaks, and we have not classified their profiles.

%----------------------------------------------------- Fig 2:
\begin{figure}[h]
  \begin{minipage}[t]{0.495\linewidth}
  \centering
   \includegraphics[width=80mm]{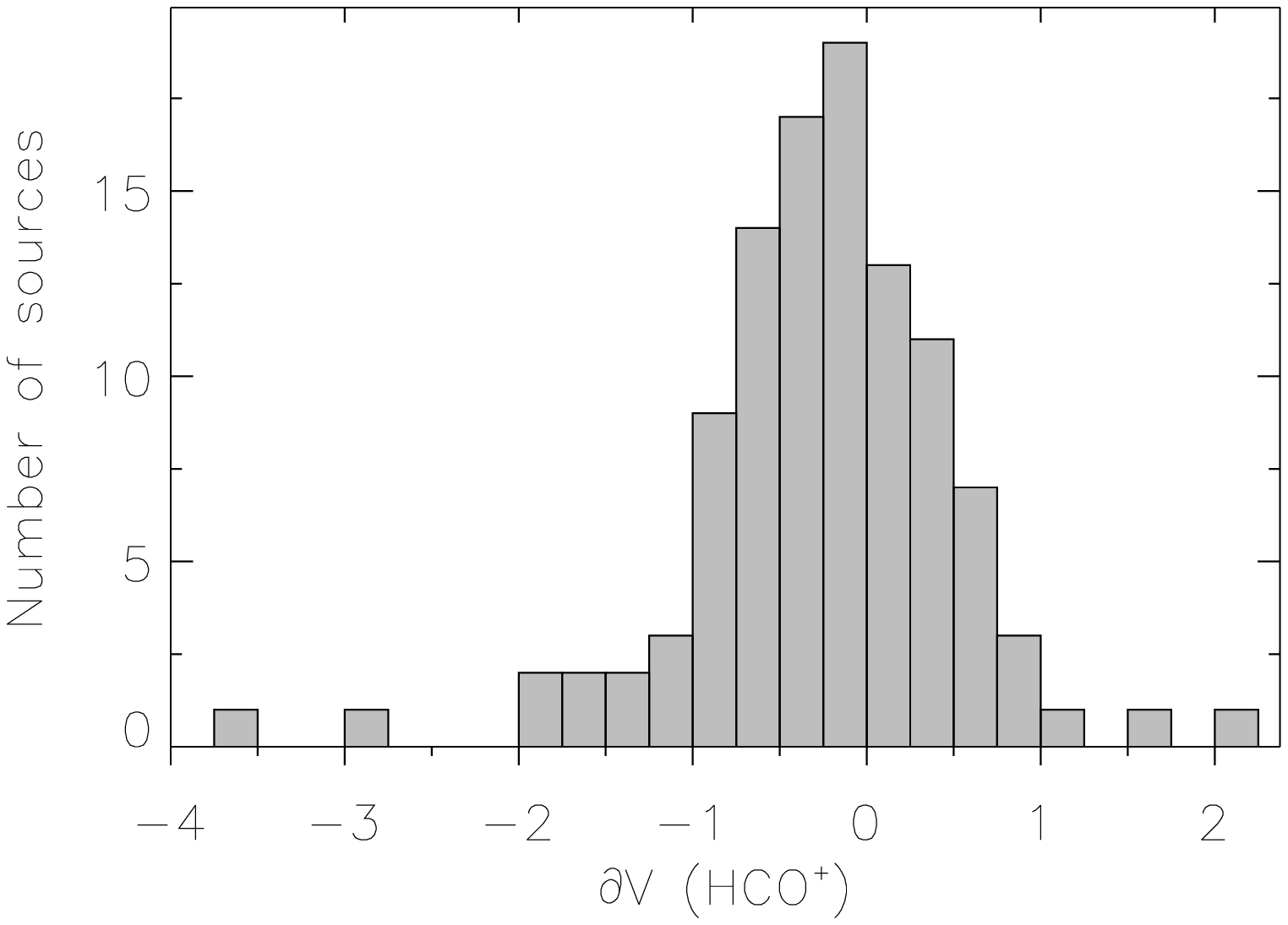}
%   \caption{{\small Amplitudes evolution of . . .} }
  \end{minipage}%
  \begin{minipage}[t]{0.495\textwidth}
  \centering
   \includegraphics[width=80mm]{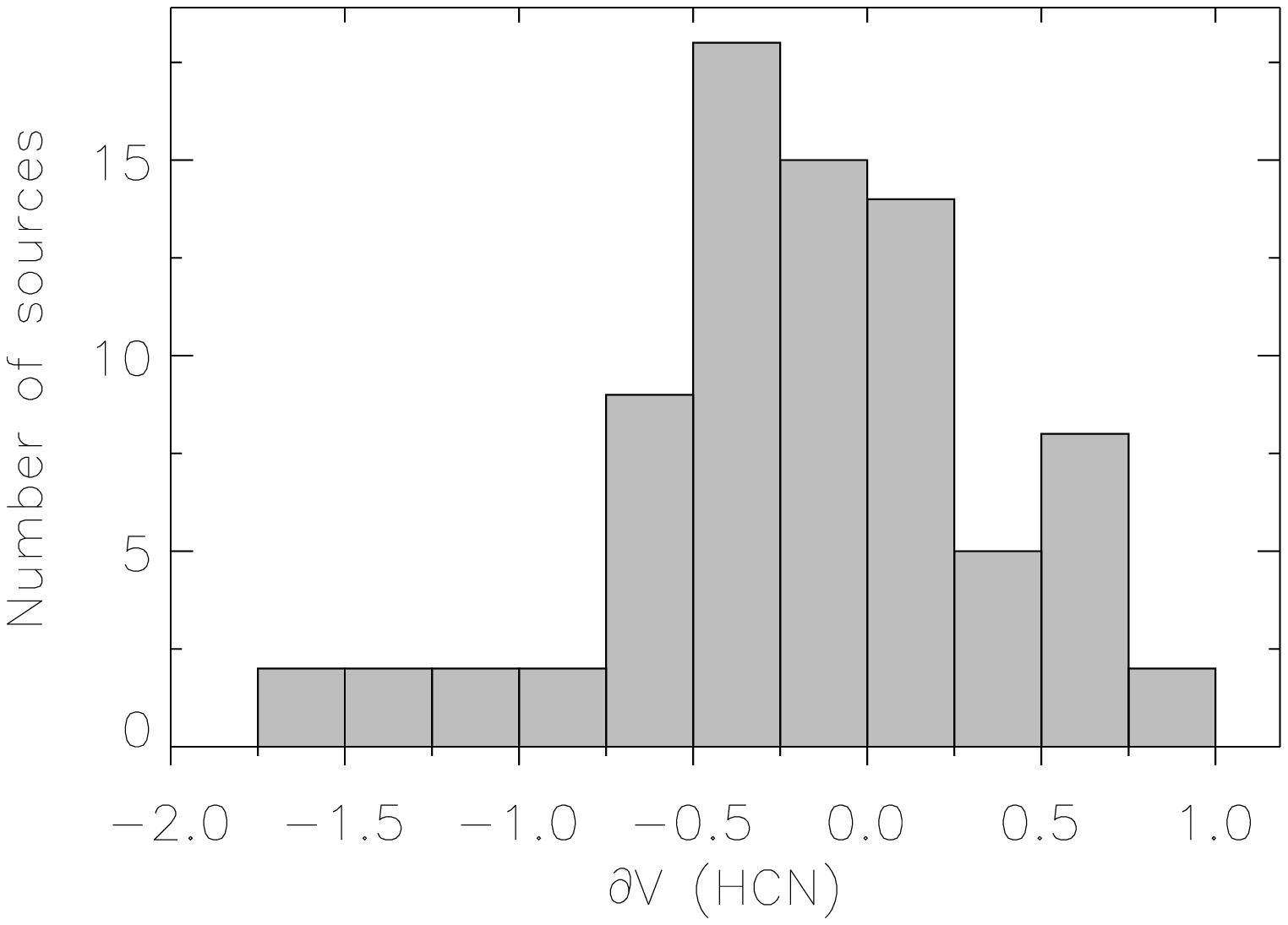}
%  \caption{{\small Amplitude variation of AN Lyn.}}
  \end{minipage}%
  \caption{{\small The dimensionless parameter $\delta V$ of HCO$^+$ (1-0) and HCN (1-0) of 133 sources. Left panel: The $\delta V$ distribution of HCO$^+$ (1-0) of our sources. Right panel: The $\delta V$ distribution of HCN (1-0) of our sources. }}
  \label{Fig:fig2}
\end{figure}

%________________________________________ Table 2: Observation parameters

\begin{table}
\begin{center}
  \caption[]{ The Derived Line Parameters and Profiles of Sample.}\label{Tab:tab2} 
%Please Capitalize the First Letter of Each Notional Word in table's caption
 \begin{tabular}{cccccccc}
  \hline\noalign{\smallskip}
Source & V$_{thick}$ & V$_{thick}$ & V$_{thin}$ & $\Delta$ V & $\delta$ V & $\delta$ V & Profile \\
Name & HCO$^+$ (1-0) & HCN (1-0) & C$^{18}$O (1-0) & C$^{18}$O (1-0) & HCO$^+$ (1-0) & HCN (1-0) &  \\
  & (km s$^{-1}$) & (km s$^{-1}$) & (km s$^{-1}$) & (km s$^{-1}$) &  &  &  \\  \hline\noalign{\smallskip}
% $^*$
G002.97+4.22 & - & 19.25(0.19) & 19.67(0.02) & 0.24(0.01) & - & -1.75(0.79) & N,B \\
G012.34-0.09 & - & - & 34.34(0.05) & 2.51(0.11) & - & - & N,N \\
G012.72+0.69 & 16.70(0.09) & 17.26(0.26) & 17.28(0.04) & 1.66(0.08) & -0.35(0.05) & -0.01(0.15) & B,N $^{\alpha}$ \\
G012.77-0.19 & 33.42(0.20) & 35.16(0.20) & 35.78(0.02) & 3.90(0.05) & -0.61(0.05) & -0.16(0.05) & B,N \\
G012.82-0.19 & 34.69(0.04) & 34.68(0.06) & 34.94(0.01) & 4.52(0.02) & -0.06(0.01) & -0.06(0.01) & N,N \\
G012.87-0.20 & 34.90(0.37) & 34.32(0.20) & 35.42(0.03) & 4.41(0.06) & -0.12(0.08) & -0.25(0.05) & N,B \\
G012.88-0.24 & 38.17(0.03) & No data & 35.90(0.01) & 3.94(0.03) & 0.58(0.01) & No data & R \\
G012.96-0.23 & 32.20(0.14) & No data & 35.23(0.02) & 4.87(0.05) & -0.62(0.03) & No data & B $^{\alpha,\beta}$ \\
G013.79-0.23 & 38.81(0.06) & 38.59(0.11) & 38.01(0.02) & 2.30(0.05) & 0.35(0.03) & 0.25(0.05) & R,R \\
G013.97-0.15 & 37.97(0.12) & 37.82(0.13) & 39.66(0.03) & 3.18(0.07) & -0.53(0.04) & -0.58(0.04) & B,B $^{\alpha}$ \\
  \hline\noalign{\smallskip}
\end{tabular}\\
\end{center}
\tablecomments{1.0\textwidth}{Columns are (from left to right) the source name, peak velocity of HCO$^+$ (1-0), peak velocity of HCN (1-0), peak velocity of C$^{18}$O (1-0), FWHM of C$^{18}$O (1-0), asymmetry of HCO$^+$ (1-0), asymmetry of HCN (1-0), and profile of HCO$^+$ (1-0) and HCN (1-0). The values in parentheses give the uncertainties. The HCO$^+$ (1-0) and HCN (1-0) profiles are evaluated: B denotes blue profile, R denotes red profile, and N denotes neither blue or red or no emission. $^{\alpha,\beta,\gamma}$ indicates a source which are associated with the ATLASGAL compact sources, the ATLASGAL cold high-mass clumps, and the BGPS sources. The full table is available in the Appendix B.2.}
\end{table}

%----------------------------------------------------- Fig 3:
\begin{figure}[h]
  \centering
   \includegraphics[width=7.5cm, angle=0]{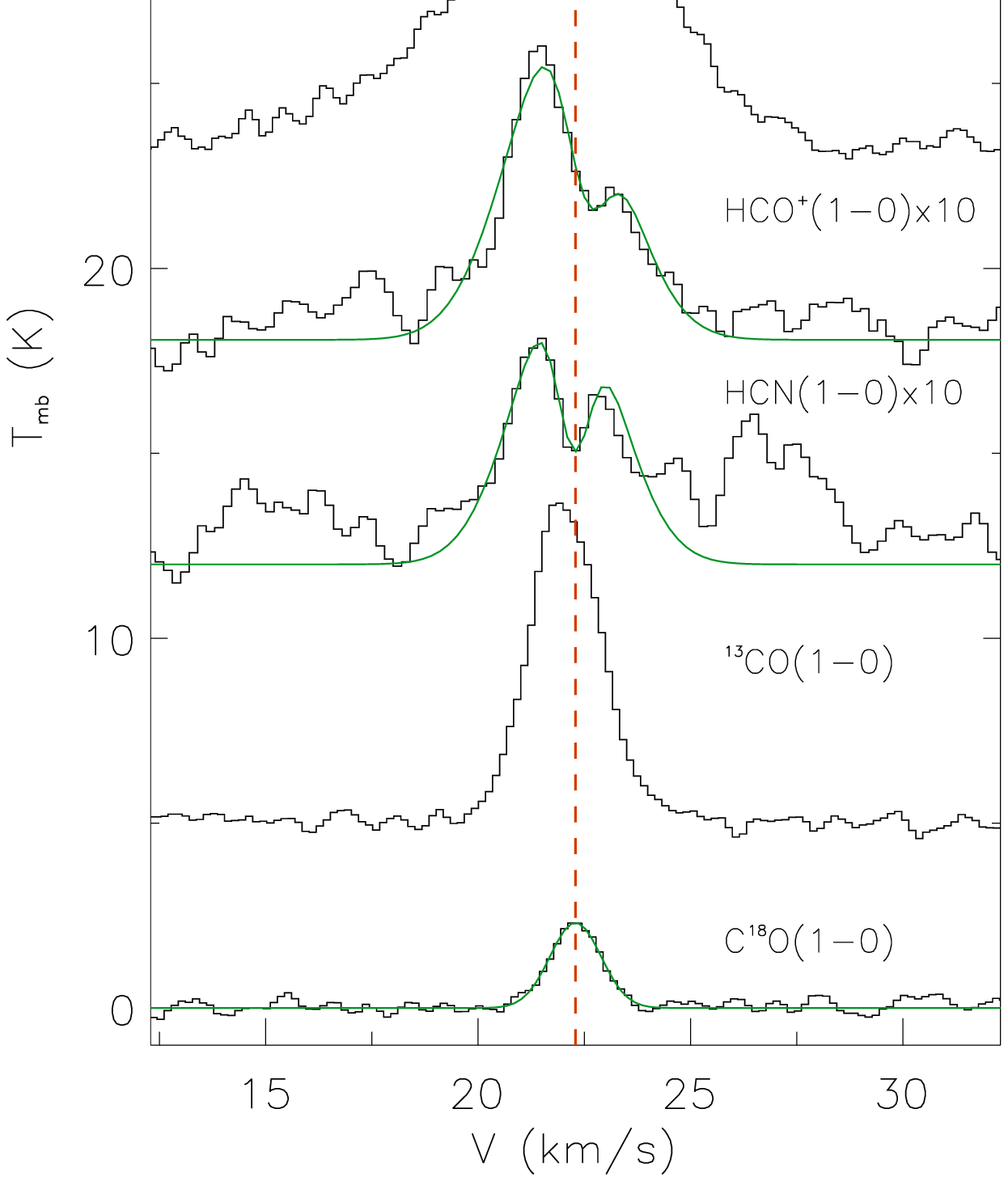}
  \caption{{\small Example of the confirmed infall sources G053.12+0.08. The lines from bottom to top are C$^{18}$O (1-0), $^{13}$CO (1-0), HCN (1-0), HCO$^+$ (1-0) and $^{12}$CO (1-0), respectively. The green lines are the results of Gaussian fitting of C$^{18}$O (1-0), HCO$^+$ (1-0) and HCN (1-0) lines, and the dashed red line indicates the central radial velocity of C$^{18}$O (1-0).  
}}
  \label{Fig:fig3}
\end{figure}

Using the dimensionless parameter $\delta V$ defined by \cite{Mardones+etal+1997}, we can further quantify the asymmetry of HCO$^+$ and HCN lines:
\begin{equation}
  \delta V = \frac{V_{thick} - V_{thin}}{\Delta V_{thin}}
\label{eq:eq1}
\end{equation}
where $V_{thick}$ and $V_{thin}$ are the peak velocities of optically thick and thin lines, and $\Delta V_{thin}$ is the FWHM of the optically thin line. The $V_{thick}$ values are taken from the brightest emission peak positions of HCO$^+$ (1-0) and HCN (1-0) lines, and the $V_{thin}$ and FWHM are taken from the Gaussian fitting of C$^{18}$O (1-0). The dimensionless parameters $\delta V$ of HCO$^+$ and HCN lines are given in Table~\ref{Tab:tab2}, and the statistical distributions are shown in Figure~\ref{Fig:fig2}. Following the criteria of \cite{Mardones+etal+1997}, i.e. sources with $\delta V < -0.25$ are considered to have significant blue profiles, sources with $\delta V > 0.25$ are considered to have significant red profiles, and sources with $-0.25 < \delta V < 0.25$ have neither asymmetry profiles, fifty-two sources show blue profiles for HCO$^+$ line (thirty-one of them show double-peaked profiles), while twenty-five sources show red profiles for this line. For the HCN line, thirty-six sources show blue profiles (fifteen of them show double-peaked profiles) and fifteen sources show red profiles. The line profiles of the remaining sources do not show asymmetry. The profile asymmetries are also given in Table~\ref{Tab:tab2}. Compared to the HCN lines, HCO$^+$ have a higher detection rate and show more line profile asymmetries, suggesting that this line is better for tracing infall motion.

We identify that clumps with infall profile evidence in at least one optically thick line, and no expansion evidence are infall candidates (\citealt{Gregersen+etal+1997}). Therefore, according to the HCO$^+$ and HCN observations given in our study, there are 56 sources meet this criterion and we consider them as confirmed infall sources with higher confidence. Figure~\ref{Fig:fig3} shows an example of them, which has significant infall profiles on both HCO$^+$ and HCN lines. The detection rate of the confirmed sources is about 42\%. This relatively high detection rate suggests that it is effective to restrict sample with CO data from the MWISP project.
 Among these confirmed sources, forty-three sources are associated with Class 0/I YSOs (three of them are associated with UC HII regions), accounting for 77\% of all confirmed sources. The remaining 13 sources are not associated with Class 0/I YSOs.

\subsection{Physical Properties of the Confirmed Sources}

\subsubsection{Physical Properties}

%----------------------------------------------------- Fig 4:
\begin{figure}[h]
  \begin{minipage}[t]{0.47\linewidth}
  \centering
   \includegraphics[width=62mm,angle=90]{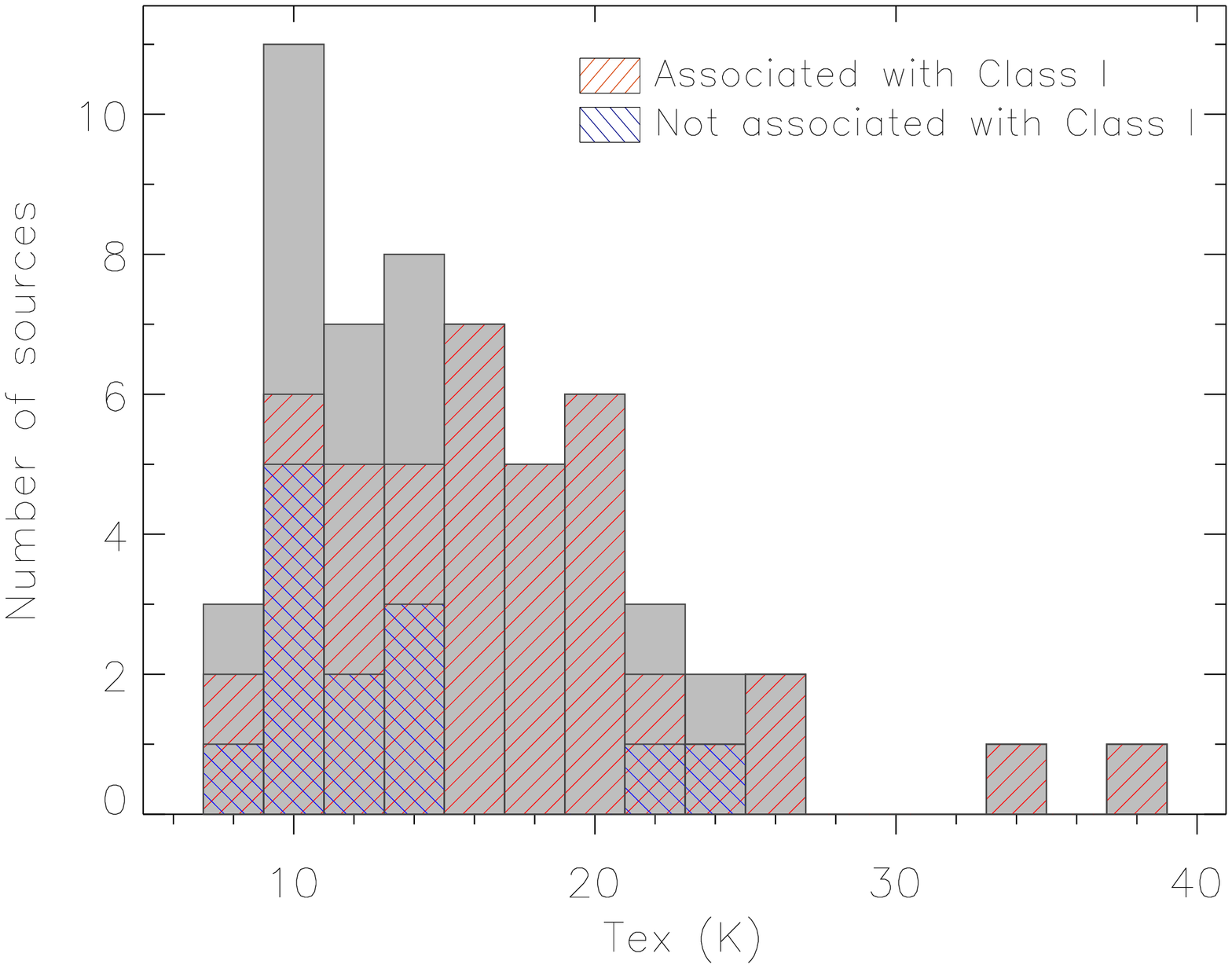}
%   \caption{{\small Amplitudes evolution of . . .} }
  \end{minipage}%
%  \label{Fig:fig2}
  \begin{minipage}[t]{0.495\textwidth}
  \centering
   \includegraphics[width=87mm,angle=0]{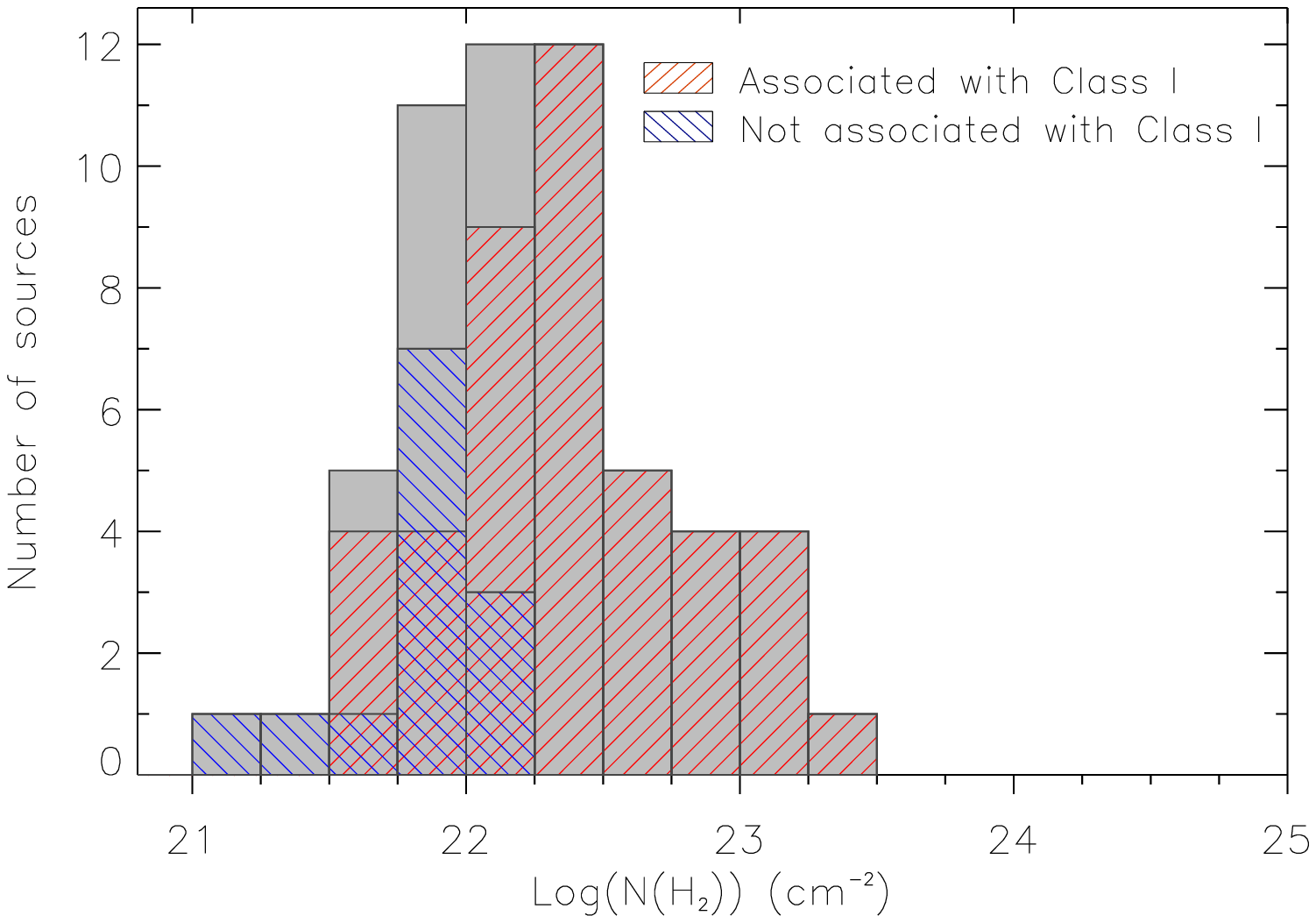}
%  \caption{{\small Amplitude variation of AN Lyn.}}
  \end{minipage}%
  \caption{{\small Distributions of the excitation temperatures and H$_2$ column densities for confirmed sources. Left panel: The excitation temperature $T_{ex}$ distribution for all confirmed sources (gray histogram), the confirmed sources associated with Class 0/I YSOs (red histogram) and not associated with Class 0/I YSOs (blue histogram). Right panel: The H$_2$ column densities for all confirmed sources (gray histogram), the confirmed sources are associated with Class 0/I YSOs (red histogram) and not associated with Class 0/I YSOs (blue histogram).
}}
  \label{Fig:fig4}
\end{figure}

Figure~\ref{Fig:fig4} shows the distributions of some physical parameters for the confirmed infall sources with different evolutionary stages. Among them, the red histograms show the distributions of confirmed sources which are associated with Class 0/I YSOs. The blue histograms show the distributions of the confirmed sources which are not associated with Class 0/I YSOs, and these confirmed sources may be in the earlier evolutionary stage. For all confirmed sources, their excitation temperatures are between 8.7 and 33.0 K, with a median value of 14.4 K. The distribution still deviates from the normal distribution, and is not much different from the $T_{ex}$ distribution of 133 sources. The H$_2$ column density shows an approximately normal distribution when the x-axis is a logarithmic scale, with the median value of $1.67\times10^{22}$ cm$^{-2}$. For confirmed sources of different evolutionary stages, the confirmed sources associated with Class 0/I YSOs have significantly higher $T_{ex}$ and H$_2$ column densities. And those sources that are not associated with Class 0/I YSOs are mostly cold, and their H$_2$ column densities are less than the median value of all confirmed sources. 

Among the 56 confirmed sources, six (i.e. G014.25-0.17, G025.82-0.18, G028.20-0.07, G029.60-0.63, G037.05-0.03, and G049.07-0.33) are associated with the APEX Telescope Large Area Survey of the Galaxy (ATLASGAL) cold high-mass clumps (\citealt{Wienen+etal+2012}), of which G025.82-0.18, G028.20-0.07, G029.60-0.63, and G049.07-0.33 are also associated with ATLASGAL compact sources (\citealt{Contreras+etal+2013, Urquhart+etal+2014}). In addition, eight confirmed sources are only associated with the ATLASGAL compact sources. The excitation temperatures of these sources are about ten to twenty Kelvin, and the median value is 17.3 K, which is slightly higher than the median value of all 56 confirmed sources. These sources have relatively high hydrogen column densities, and about 78\% of them have a column density higher than the median value of all confirmed sources. In addition, there are 21 sources associated with the Bolocam Galactic Plane Survey (BGPS) sources (\citealt{Ginsburg+etal+2013}). The coordinates of these BGPS sources are within a radius of 1$^{\prime}$ of our sources, and the central radial velocities differ by less than 3 km s$^{-1}$. The distributions of the physical properties of these sources are similar to the distributions of all confirmed sources, except that their median values of excitation temperature and column density are slightly higher than the confirmed sources.

\subsubsection{Infall Velocities}

%________________________________________ Table 3: Infall velocities

\begin{table}
\begin{center}
  \caption[]{ Infall Velocities of Confirmed Infall Sources with Double-peaked HCO$^+$ (1-0) Profile.}\label{Tab:tab3} 
%Please Capitalize the First Letter of Each Notional Word in table's caption
 \setlength{\tabcolsep}{9mm}{
 \begin{tabular}{cccccc}
  \hline\noalign{\smallskip}
%Source & $V_{in}$ & $\dot{M}$ & Source & $V_{in}$ & $\dot{M}$  \\
Source & $V_{in}$ & Source & $V_{in}$ \\
%Name & (km s$^{-1}$) & ($10^{-3}$ $M_{\odot}$ yr$^{-1}$) & Name & (km s$^{-1}$) & ($10^{-3}$ $M_{\odot}$ yr$^{-1}$) \\  
Name & (km s$^{-1}$) & Name & (km s$^{-1}$) \\
  \hline\noalign{\smallskip}
G012.72+0.69 & 0.24(0.22)  & G053.11+0.09 & 1.03(0.64) \\
G012.87-0.20 & 2.02(1.37)  & G053.12+0.08 & 0.64(0.21) \\
G012.96-0.23 & 11.22(7.04)  & G053.14+0.09 & 0.19(0.13) \\
G013.97-0.15 & 0.49(0.12)  & G079.24+0.53 & 3.92(3.75) \\
G014.02-0.19 & 2.50(0.62)  & G079.71+0.15 & 0.68(0.57) \\
G014.11-0.16 & 1.88(1.54)  & G081.72+0.57 & 5.38(0.21) \\
G014.26-0.17 & 2.44(2.01)  & G081.72+1.28 & 0.91(0.20) \\
G017.09+0.82 & 0.20(0.06)  & G081.90+1.43 & 0.11(0.04) \\
G025.82-0.18 & 8.49(1.51)  & G082.17+0.07 & 0.08(0.02) \\
G026.32-0.07 & 6.71(6.13)  & G085.05-1.25 & 2.29(0.70) \\
G028.20-0.07 & 0.36(0.03)  & G108.99+2.73 & 0.88(0.23) \\
G028.97+3.54 & 2.43(0.32)  & G110.32+2.52 & 0.43(0.25) \\
G029.60-0.63 & 9.84(3.20)  & G110.32+2.54 & 0.01(0.01) \\
G036.02-1.36 & 1.44(1.19)  & G121.31+0.64 & 1.63(0.40) \\
G039.45-1.17 & 0.35(0.19)  & G172.77+2.09 & 0.22(0.11) \\
G053.10+0.11 & 0.49(0.18)  &  &  \\
  \hline\noalign{\smallskip}
\end{tabular}}
\end{center}
\tablecomments{0.9\textwidth}{Columns are (from left to right) the source name, its infall velocity, the source name, and its infall velocity. The values in parentheses give the uncertainties.}
\end{table}

%----------------------------------------------------- Fig 5:
\begin{figure}[h]
  \centering
   \includegraphics[width=8.5cm]{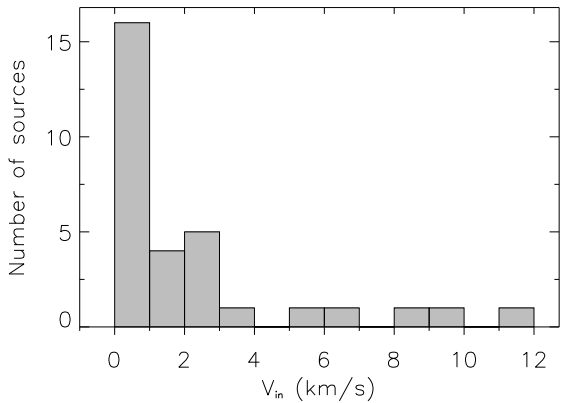}
  \caption{{\small Distribution of infall velocities for 31 confirmed sources with double-peaked profile in HCO$^+$ (1-0).  
}}
  \label{Fig:fig5}
\end{figure}

\cite{Myers+etal+1996} gave a method to estimate the infall velocity of clumps based on their model and the assumption of $V_{in}\ll\Delta V_{thin}(2\ln\tau)^{1/2}$:
\begin{equation}
  V_{in}=\frac{\Delta V_{thin}^{2}}{V_{red}-V_{blue}}\ln\left(\frac{1+e^{(T_{blue}-T_{dip})/T_{dip}}}{1+e^{(T_{red}-T_{dip})/T_{dip}}}\right)
\label{eq:LebsequeI}
\end{equation}
where $V_{red}$ and $V_{blue}$ are the velocities of the red and blue peaks of optically thick line, while $T_{red}$ and $T_{blue}$ are the corresponding peak intensities. $T_{dip}$ is the intensity of self-absorption dip between the two peaks, and $\Delta V_{thin}$ is the same as in Section 3.2. Since the above formula contains the velocities and peak intensities corresponding to the blue and red peaks of the optically thick lines, the confirmed sources with the double-peaked profile can use this formula to calculate the infall velocities. Observations suggest that HCO$^+$ lines show more double-peaked profiles (see Section 3.2) than the HCN lines, so we estimated the infall velocities of 31 confirmed sources with double-peaked HCO$^+$ line profiles. The obtained $V_{in}$ values are listed in Table~\ref{Tab:tab3}, and Figure~\ref{Fig:fig5} shows its distribution. In previous studies, the infall velocities of the host clumps were about 10$^{-1}$ to 10$^{0}$ km s$^{-1}$ (e.g. \citealt{Rygl+etal+2013, Qin+etal+2016}). Most of our results fall within this range, with the median value of 0.91 km s$^{-1}$. However, some of the $V_{in}$ we calculated are lager than expected. For example G012.96-0.23, its calculated infall velocity reached 11.22 km s$^{-1}$, and the error was quite high. We analyzed the above formula and found that the second item on the right has a large influence on the $V_{in}$ value. When the signal-to-noise ratio is not large enough, the noise is likely to cause a large deviation in $T_{dip}$ and even $T_{red}$ values. This will affect the estimation of the infall velocity. Another reason may be that these sources are not satisfy the assumption given by this model. 

We also tried to calculate the mass infall rates of these sources using $V_{in}$, but we lack the mapping data, only the C$^{18}$O integrated intensity data can be used to estimate the sizes of clumps. According to previous studies, for low-mass star formation, the mass infall rates onto the host clumps are about $10^{-5}$ to $10^{-6}$ $M_{\odot}$ yr$^{-1}$ (e.g. \citealt{Palla+Stahler+1993, Whitney+etal+1997, Kirk+etal+2005}). For the massive star formation, the rates are between $10^{-2}$ to $10^{-4}$ $M_{\odot}$ yr$^{-1}$ (e.g. \citealt{Rygl+etal+2013, He+etal+2015}). Unfortunately, the results we obtained are an order of magnitude larger than the values given in these studies. Our estimation of the clump sizes are very simple and may be the cause of excessive mass infall rates. Future higher resolution mapping observations may limit the infall scale, and help us to obtain more reliable mass infall rates.

\section{Summary}
\label{sect:summary}

In the previous work, using the CO data from the MWISP project, approximately 2,200 candidates with infall spectral characteristics were identified by machine search and manual check. We selected 133 sources of them with significant blue profiles in $^{12}$CO, and C$^{18}$O intensities $>$ 1 K as a sub-sample for further study. Using the DLH 13.7-m telescope, we carried out observations of optically thick lines HCO$^+$ (1-0) and HCN (1-0) toward these sources. The results are summarized as follows.

(i) There are 107 sources showing HCO$^+$ emission and 82 sources showing HCN emission, with the detection rates of 80\% and 69\%, respectively. Using the dimensionless parameter $\delta V$ to quantify the line profle asymmetry, we have found that 52 sources show blue profiles and 25 sources show red profiles in the HCO$^+$ lines, while 36 sources show blue profiles and 15 sources show red profiles in the HCN lines. By comparison, HCO$^+$ lines have a higher detection rate and show more line profile asymmetries.

(ii) Taking the observations of HCO$^+$ and HCN into account, we have identified 56 sources, which may be more reliable infall sources. The detection rate reached 42\%, which suggests that our source selection criteria of CO data play an effective role in improving the detection rate of infall motion.

(iii) The excitation temperatures of these 56 confirmed infall sources are between 8.7 and 33.0 K, with a median value of 14.4 K, and the distribution deviates from the normal distribution. When the x-axis is a logarithmic scale, the H$_2$ column density shows a normal distribution, and the median value is $1.67\times10^{22}$ cm$^{-2}$.

(iv) Of these sources, forty-three are associated with Class 0/I YSOs, which have higher excitation temperatures and column densities, accounting for 77\% of the sub-sample. The remaining 13 sources are not associated with Class 0/I YSOs. These confirmed sources may be in the early infall stage.

(v) For the confirmed sources with blue double-peaked profile of HCO$^+$ line, we have estimated their infall velocities. Most sources have infall velocities between 10$^{-1}$ to 10$^0$ km s$^{-1}$, with the median value of 0.91 km s$^{-1}$. %This is consistent to previous studies.

In the following work, mapping observations with higher resolution and sensitivity telescope will help us to obtain spatial details of these confirmed sources, and further estimate the mass infall rates of the host clumps.

%The preliminary photometric results on the $\cdots\cdots$
%are reported along with an introduction to
%the user-compiled IRAF task $\cdots\cdots$

\begin{acknowledgements}

We are grateful to the staff of the Qinghai Radio Observing Station at Delingha for their assistance and support during the observations. This research has made use of the Milky Way Imaging Scroll Painting (MWISP) data, and we are very grateful to the members of the MWISP team of Purple Mountain Observatory for their support. We also have made use of data products from the Wide-field Infrared Survey Explorer, the NASA/ IPAC Infrared Science Archive, and the observations made with the Spitzer Space Telescope.

This work is supported by National Key R\&D Program of China no. 2017YFA0402702, and the National Natural Science Foundation of China (NSFC) under No.10873037, No.11873093, No.11803091, and No.11933011. MWISP project is supported by National Key R\&D Program of China grant no. 2017YFA0402700, and Key Research Program of Frontier Sciences, CAS, grant no. QYZDJ-SSW-SLH047.

\end{acknowledgements}

%\clearpage

\appendix                  %%appendicial material is supported

\section{Observed spectral line profiles of sources}

%----------------------------------------------------- Fig 6:
\begin{figure}[h]
  \begin{minipage}[t]{0.325\linewidth}
  \centering
   \includegraphics[width=55mm]{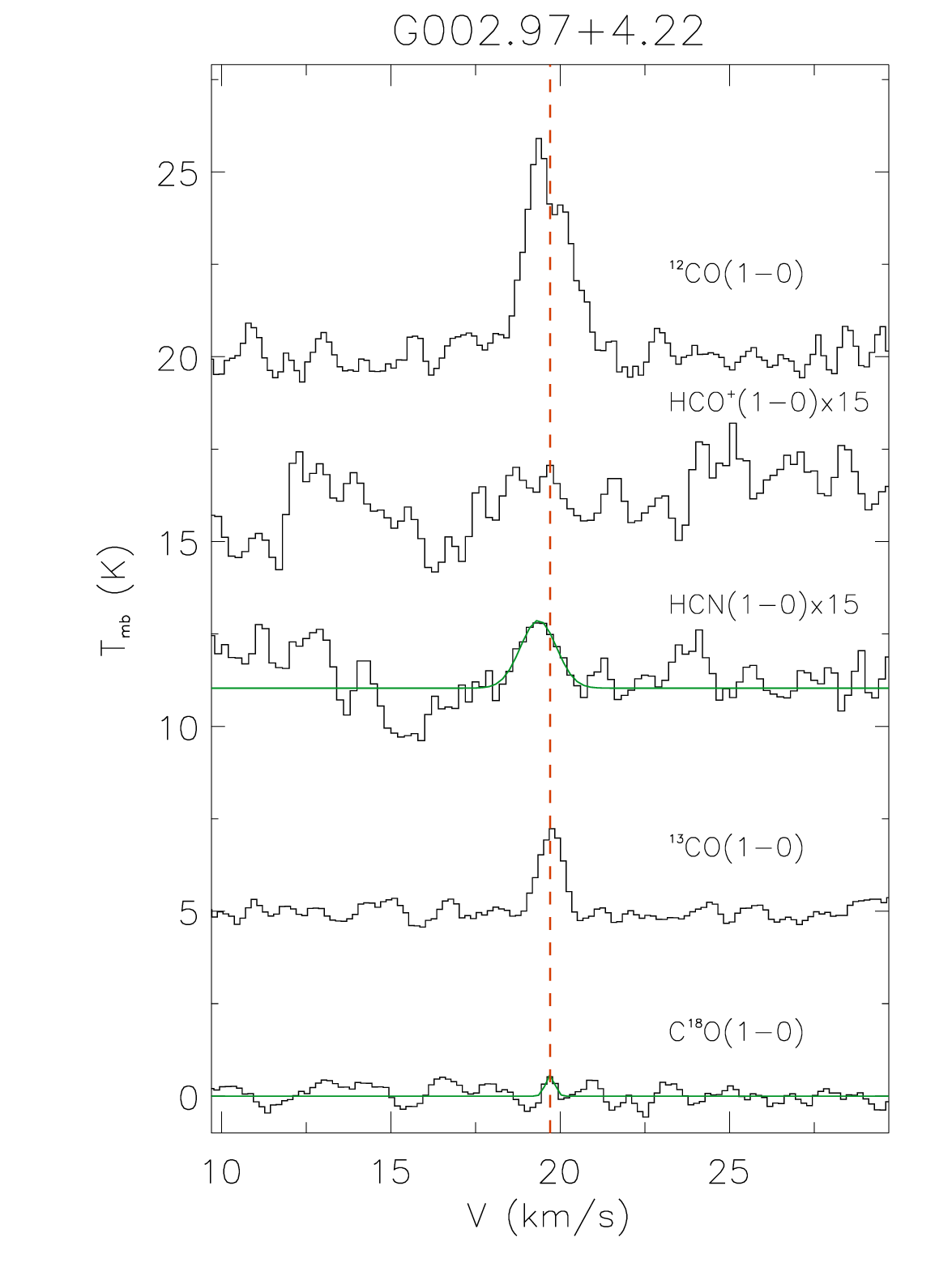}
  \end{minipage}%
  \begin{minipage}[t]{0.325\textwidth}
  \centering
   \includegraphics[width=55mm]{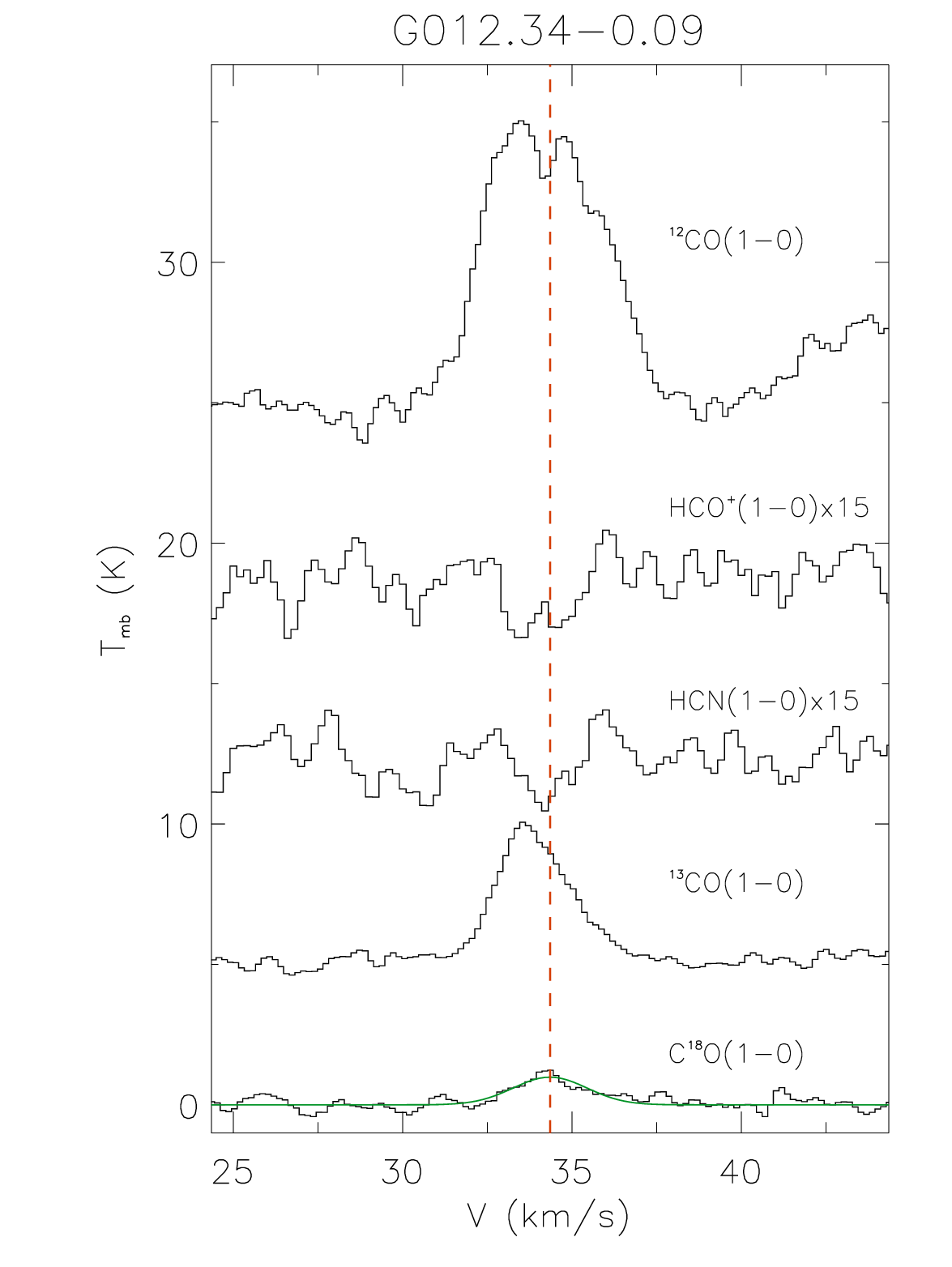}
  \end{minipage}%
  \begin{minipage}[t]{0.325\linewidth}
  \centering
   \includegraphics[width=55mm]{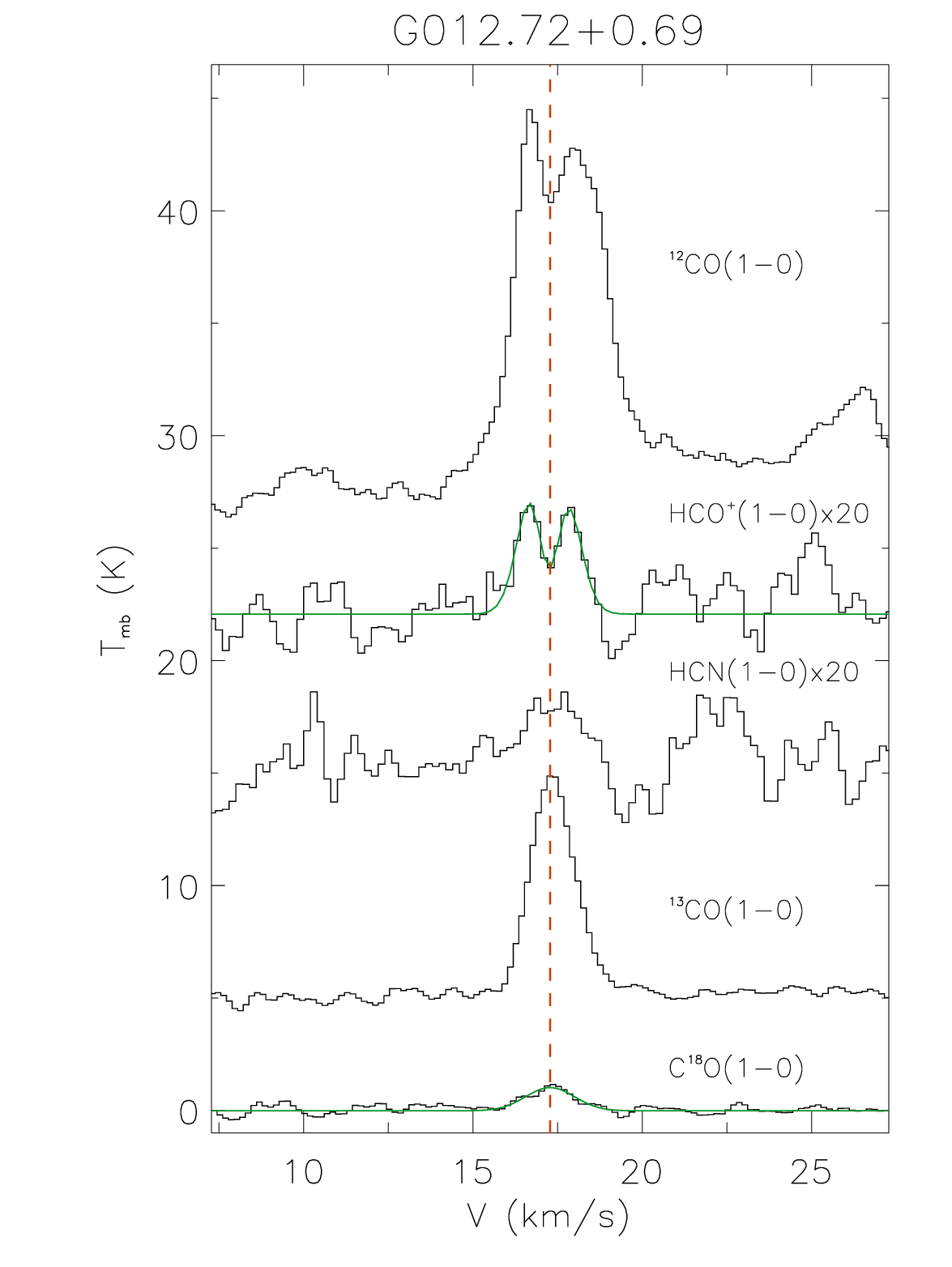}
  \end{minipage}%  
\quad
  \begin{minipage}[t]{0.325\linewidth}
  \centering
   \includegraphics[width=55mm]{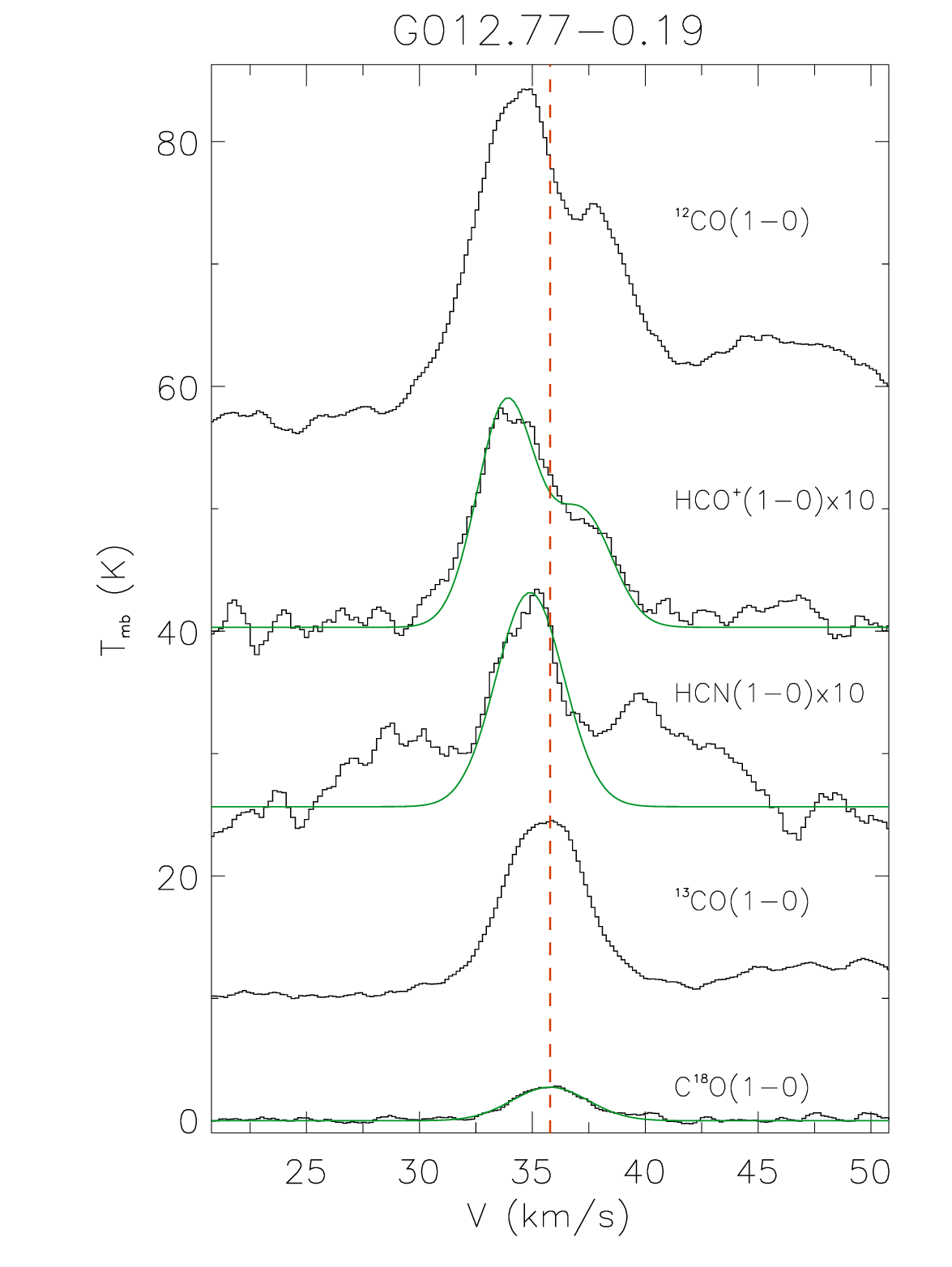}
  \end{minipage}%
  \begin{minipage}[t]{0.325\linewidth}
  \centering
   \includegraphics[width=55mm]{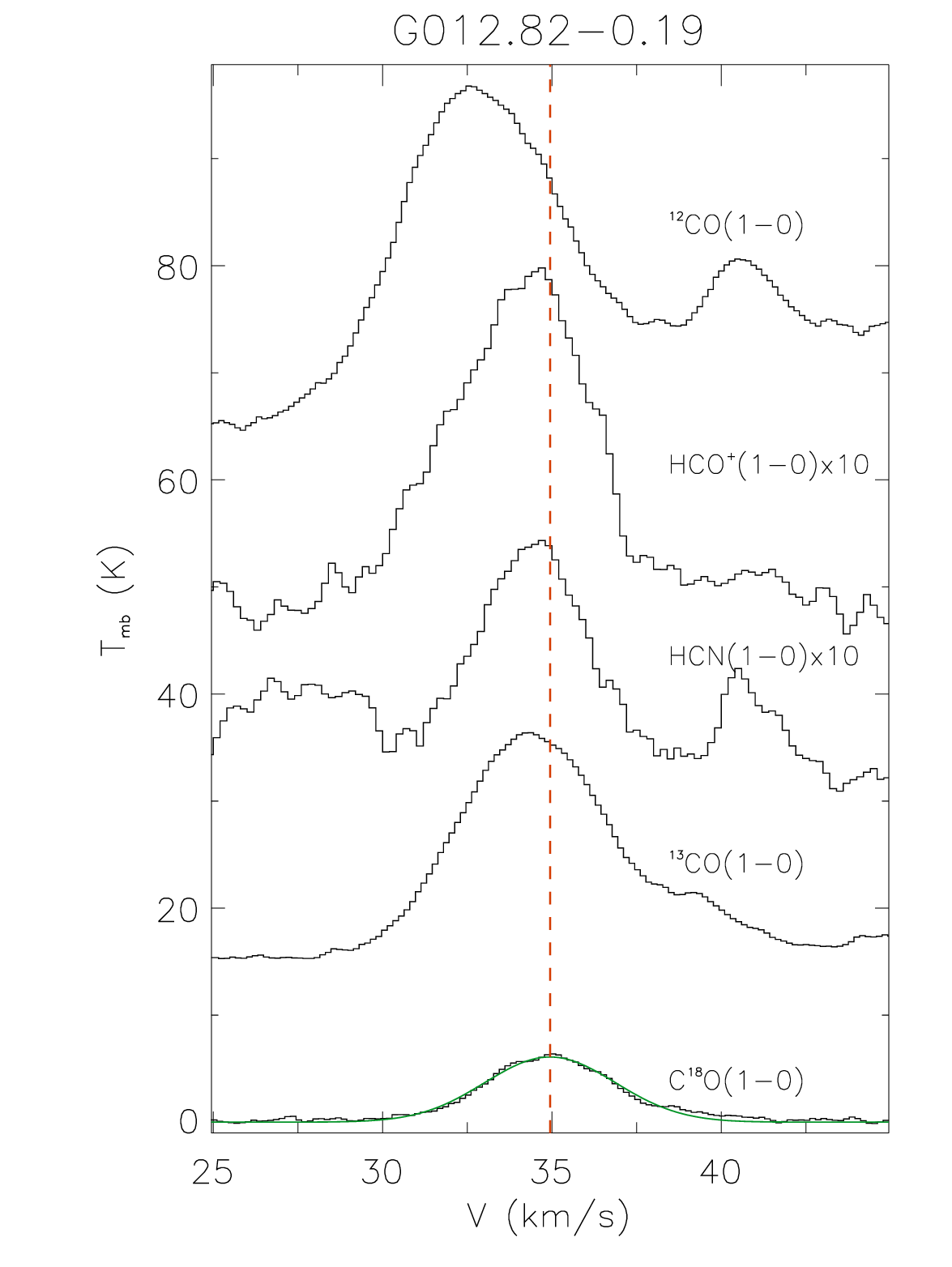}
  \end{minipage}%
  \begin{minipage}[t]{0.325\linewidth}
  \centering
   \includegraphics[width=55mm]{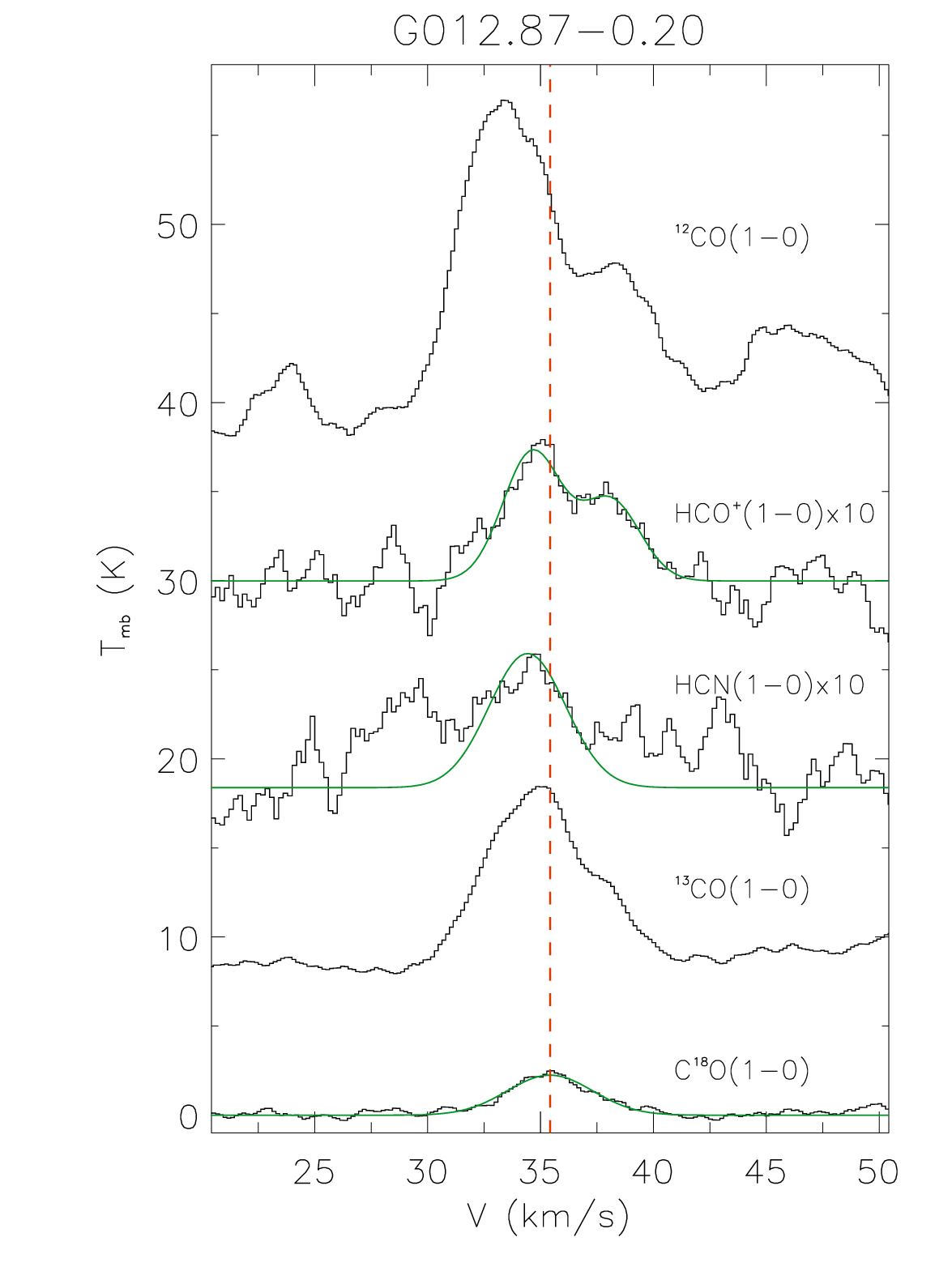}
  \end{minipage}%
\quad
  \begin{minipage}[t]{0.325\linewidth}
  \centering
   \includegraphics[width=55mm]{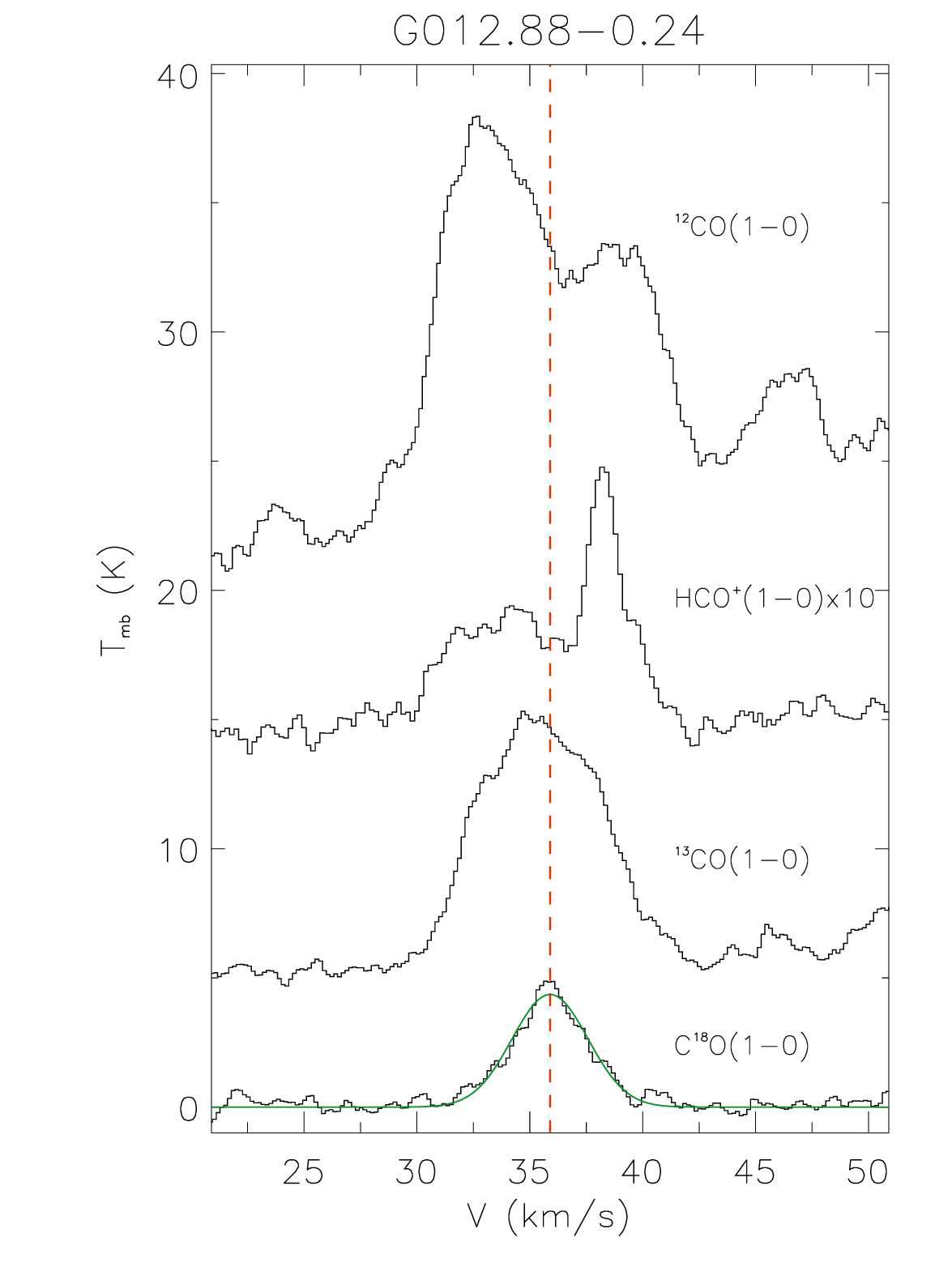}
  \end{minipage}%
  \begin{minipage}[t]{0.325\linewidth}
  \centering
   \includegraphics[width=55mm]{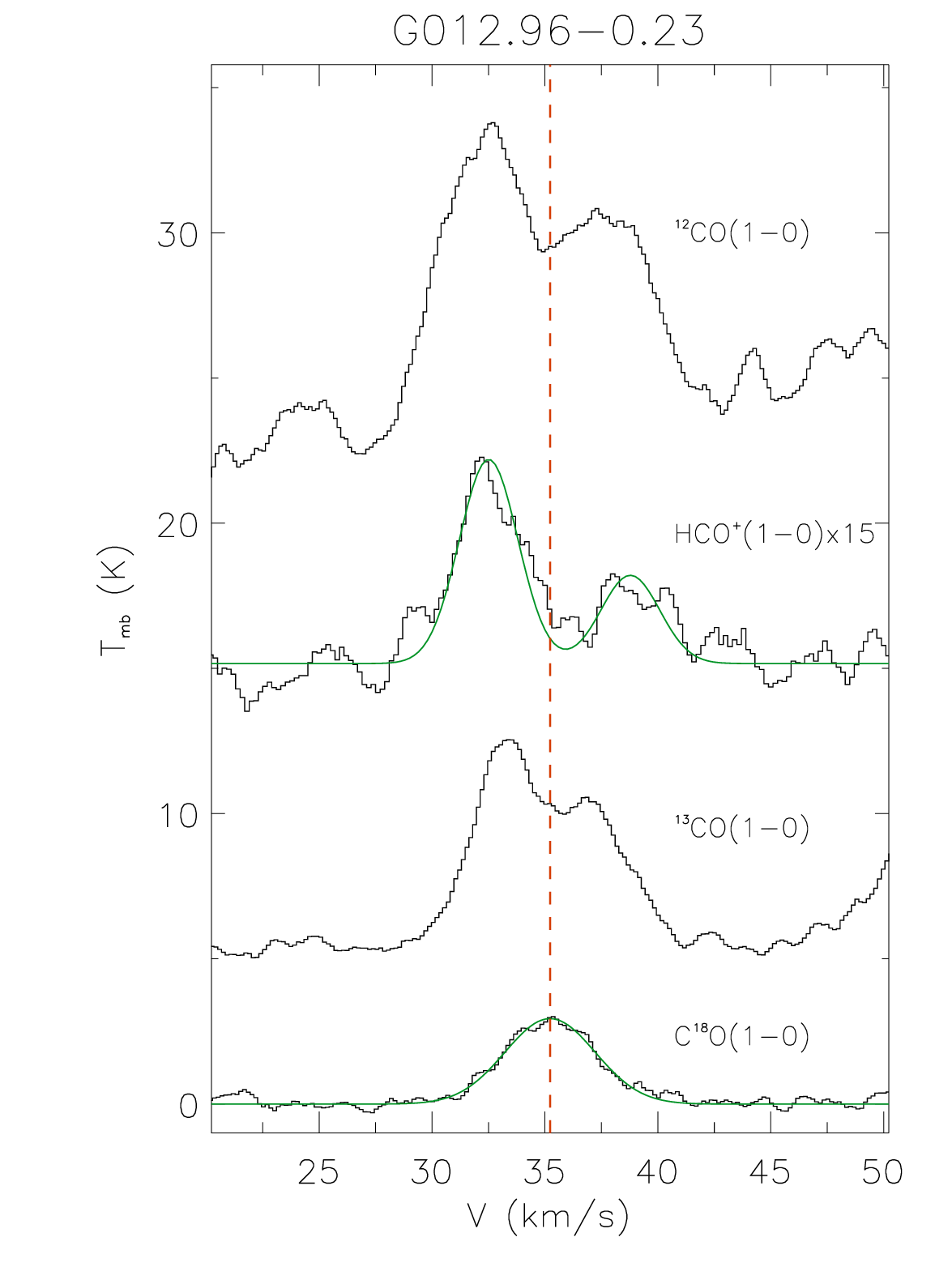}
  \end{minipage}%
  \begin{minipage}[t]{0.325\linewidth}
  \centering
   \includegraphics[width=55mm]{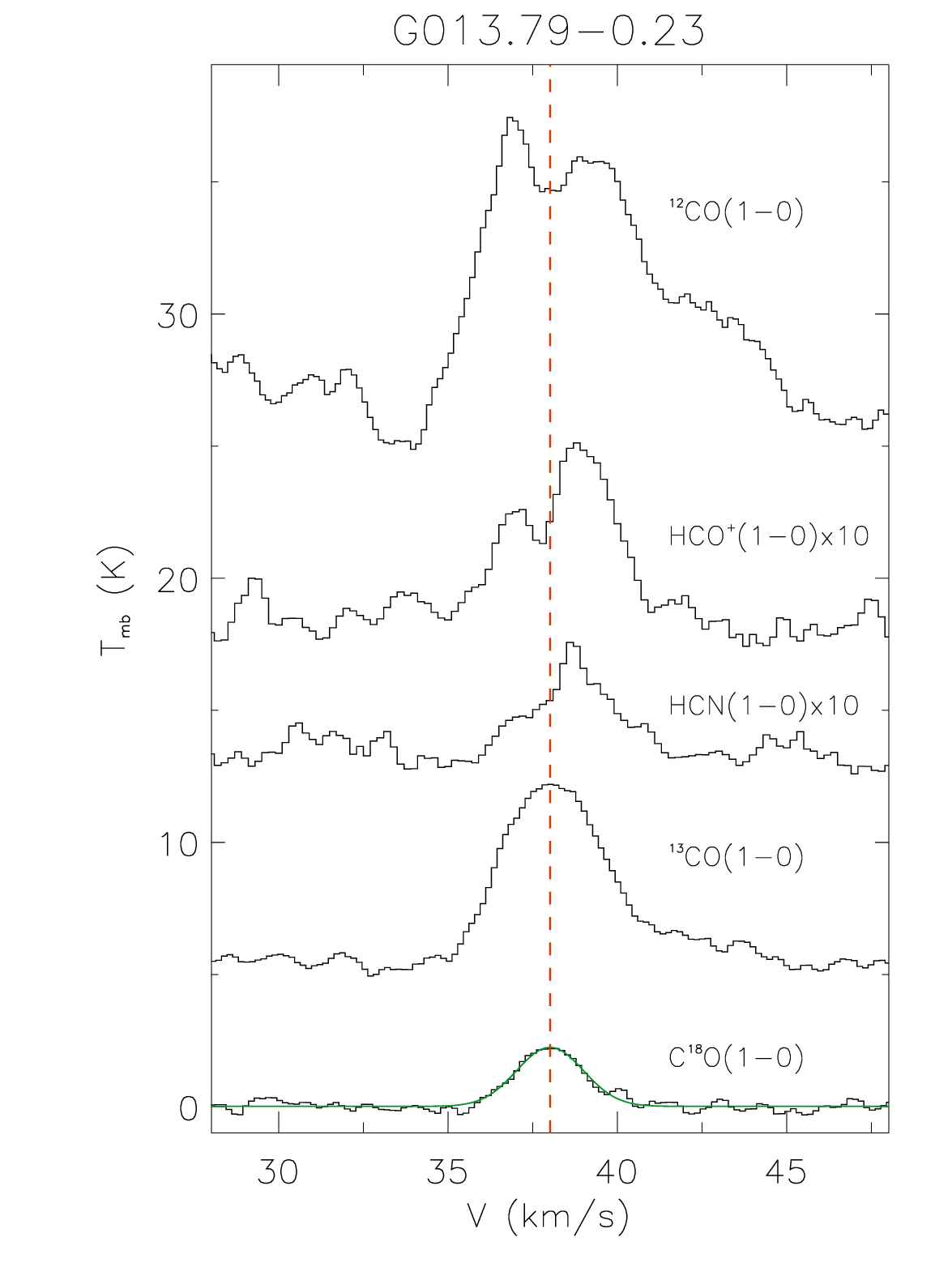}
  \end{minipage}%
  \caption{{\small Line profiles of 133 sources we selected. The lines from bottom to top are C$^{18}$O (1-0), $^{13}$CO (1-0), HCN (1-0) (14 sources lack HCN data), HCO$^+$ (1-0) and $^{12}$CO (1-0), respectively. The dashed red line indicates the central radial velocity of C$^{18}$O (1-0) estimated by Gaussian fitting. For infall candidates, HCO$^+$ (1-0) and HCN (1-0) lines are also Gaussian fitted.}}
  \label{Fig:fig6}
\end{figure}  
  
\begin{figure}[h]
\ContinuedFloat
  \begin{minipage}[t]{0.325\linewidth}
  \centering
   \includegraphics[width=55mm]{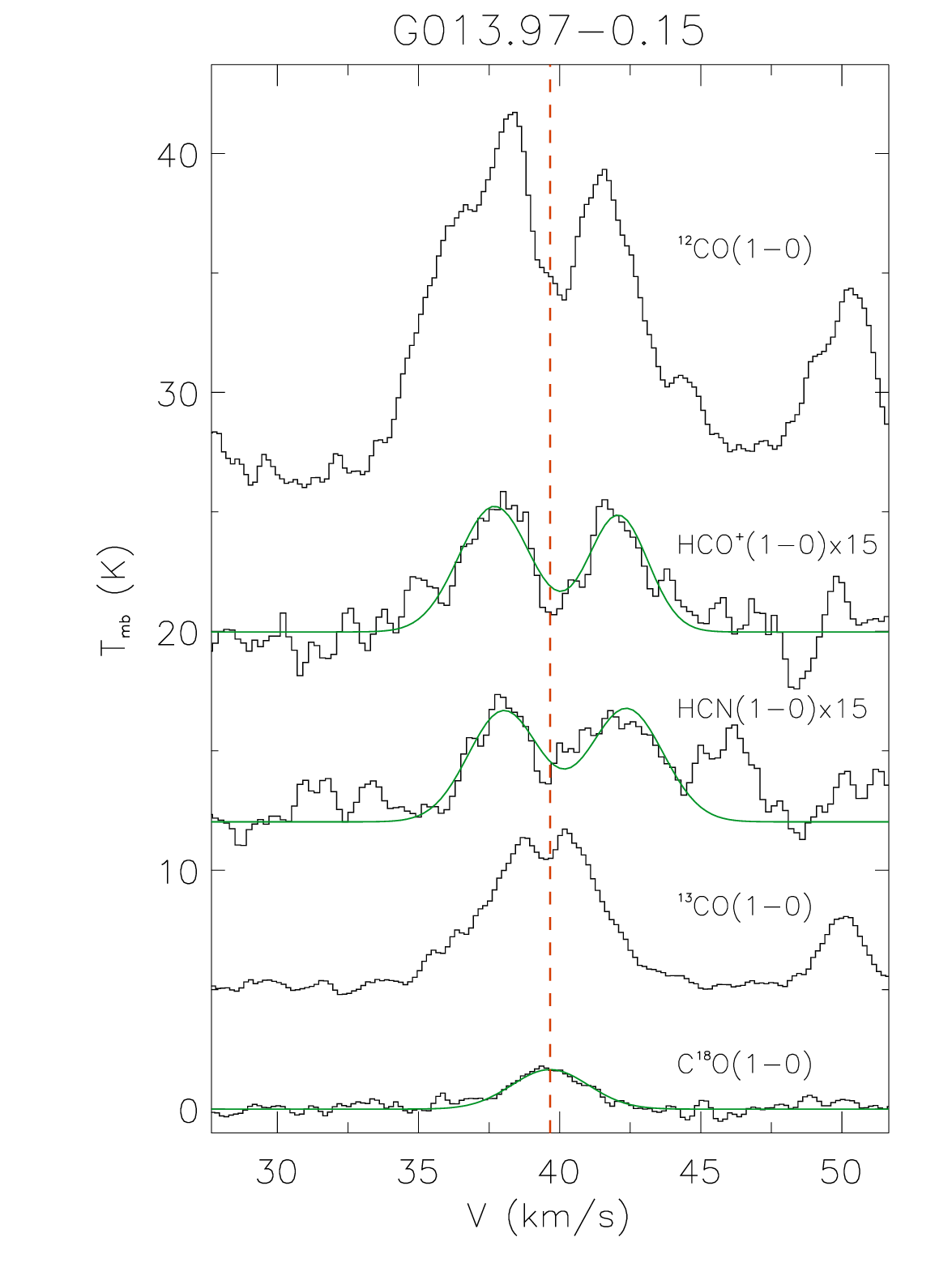}
  \end{minipage}%
  \begin{minipage}[t]{0.325\textwidth}
  \centering
   \includegraphics[width=55mm]{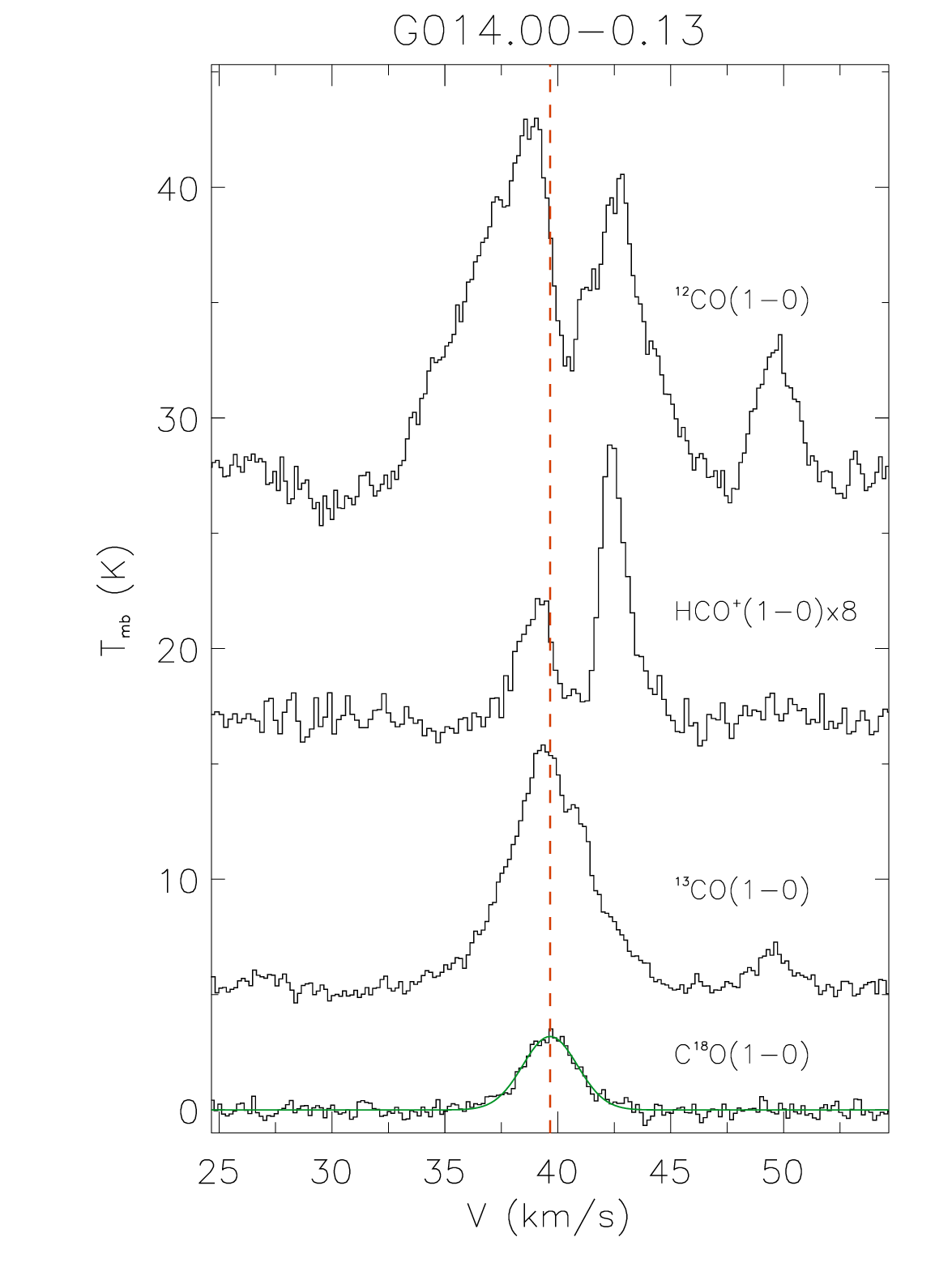}
  \end{minipage}%
  \begin{minipage}[t]{0.325\linewidth}
  \centering
   \includegraphics[width=55mm]{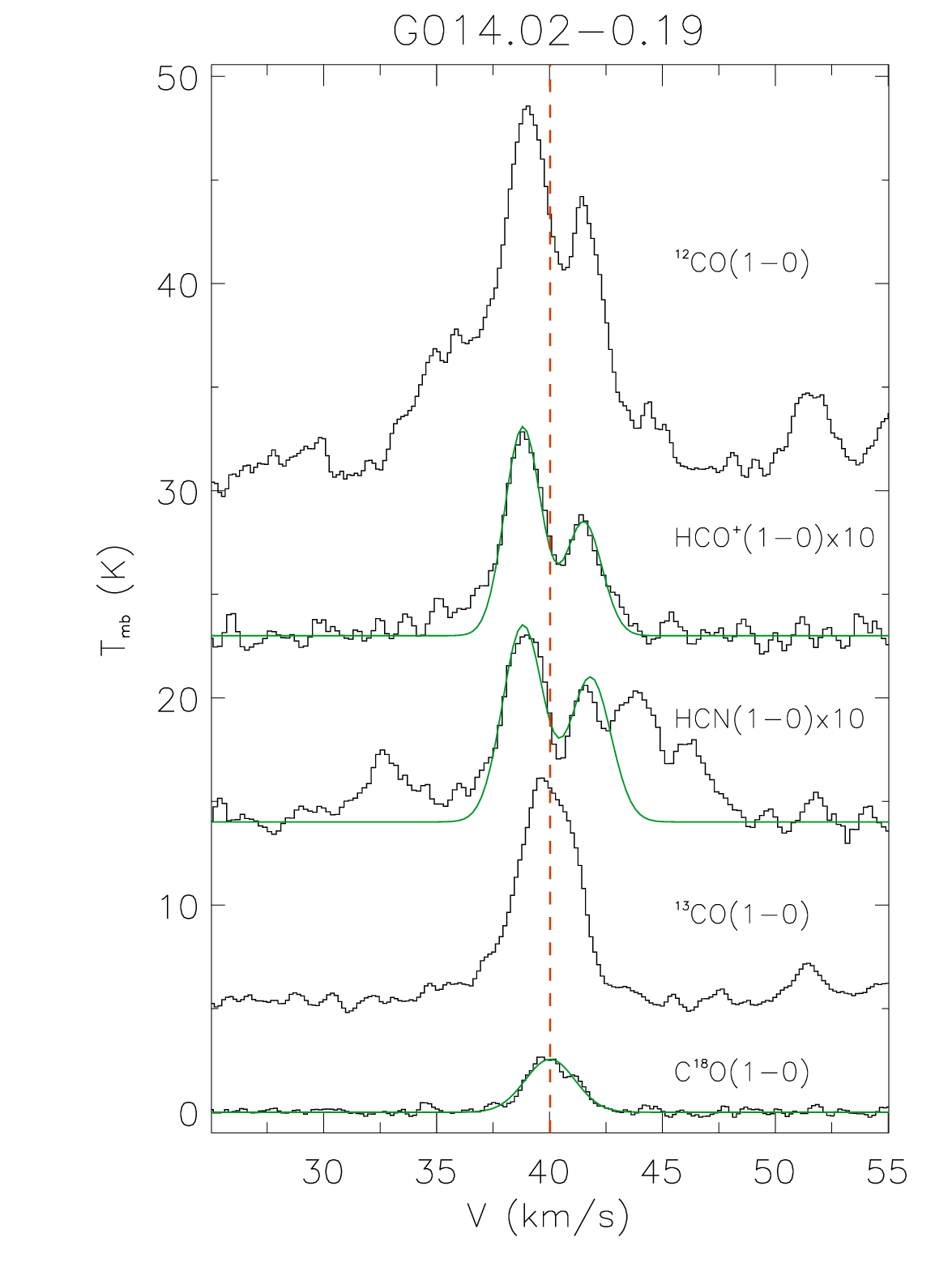}
  \end{minipage}%  
\quad
  \begin{minipage}[t]{0.325\linewidth}
  \centering
   \includegraphics[width=55mm]{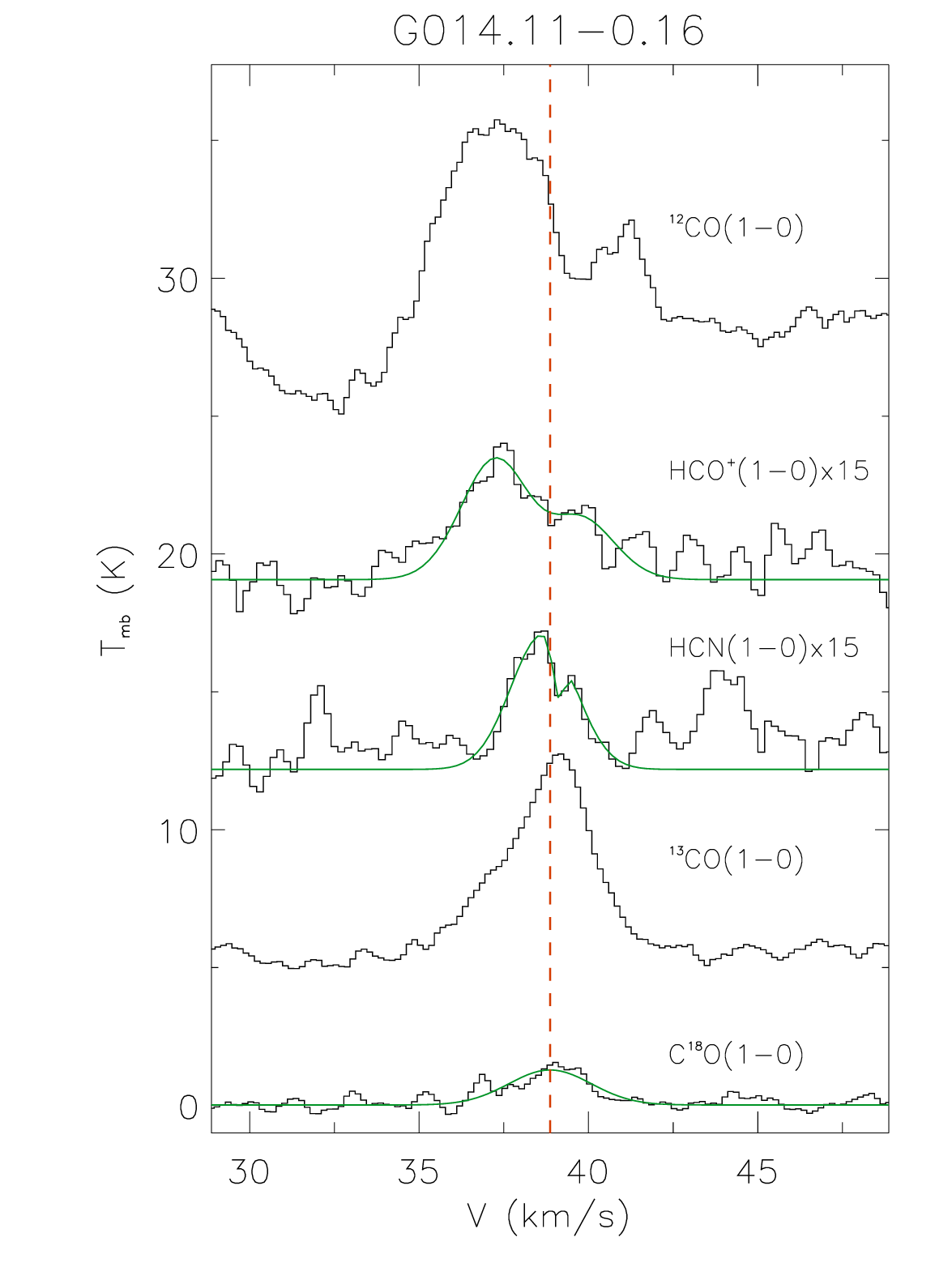}
  \end{minipage}%
  \begin{minipage}[t]{0.325\linewidth}
  \centering
   \includegraphics[width=55mm]{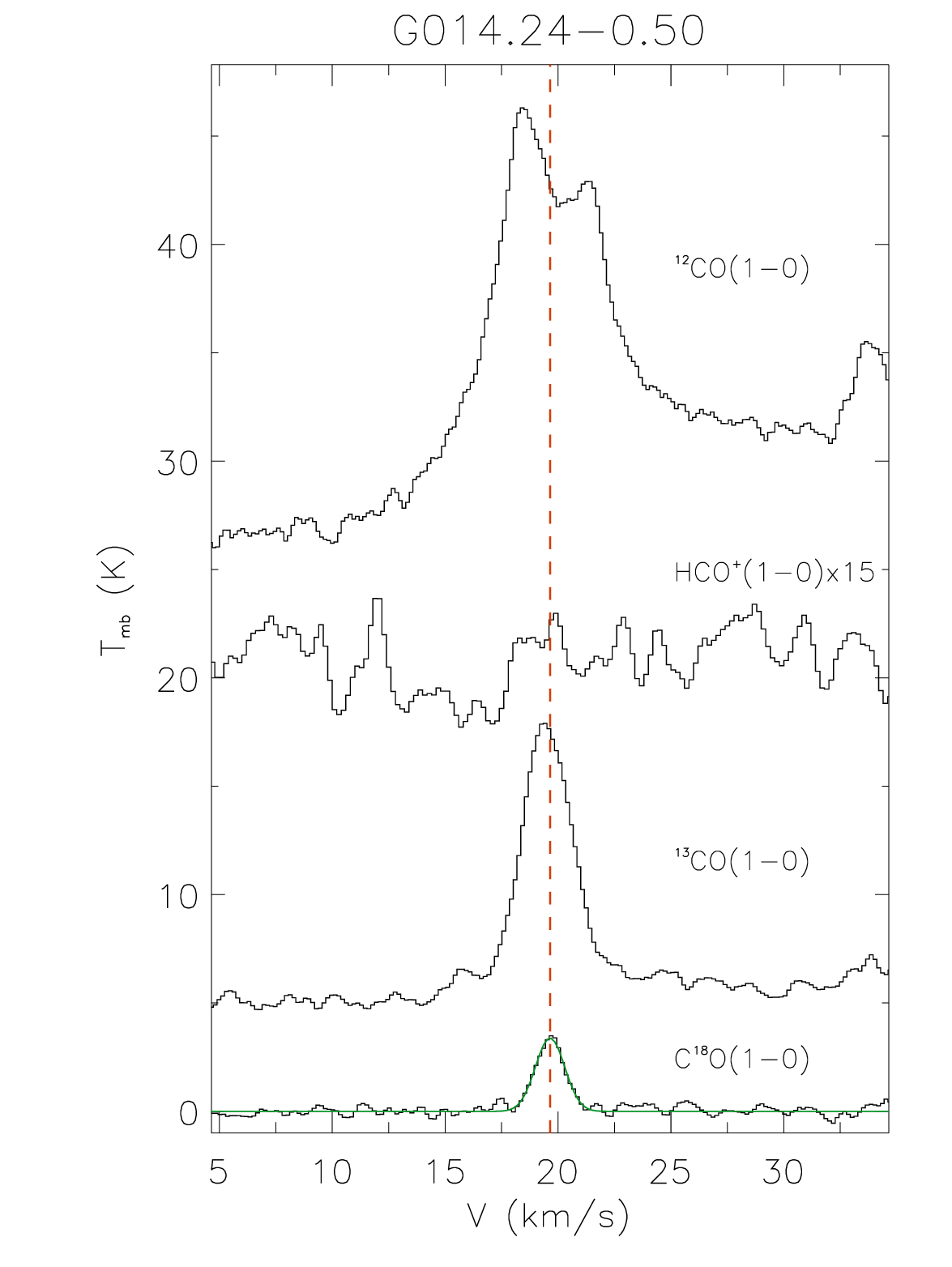}
  \end{minipage}%
  \begin{minipage}[t]{0.325\linewidth}
  \centering
   \includegraphics[width=55mm]{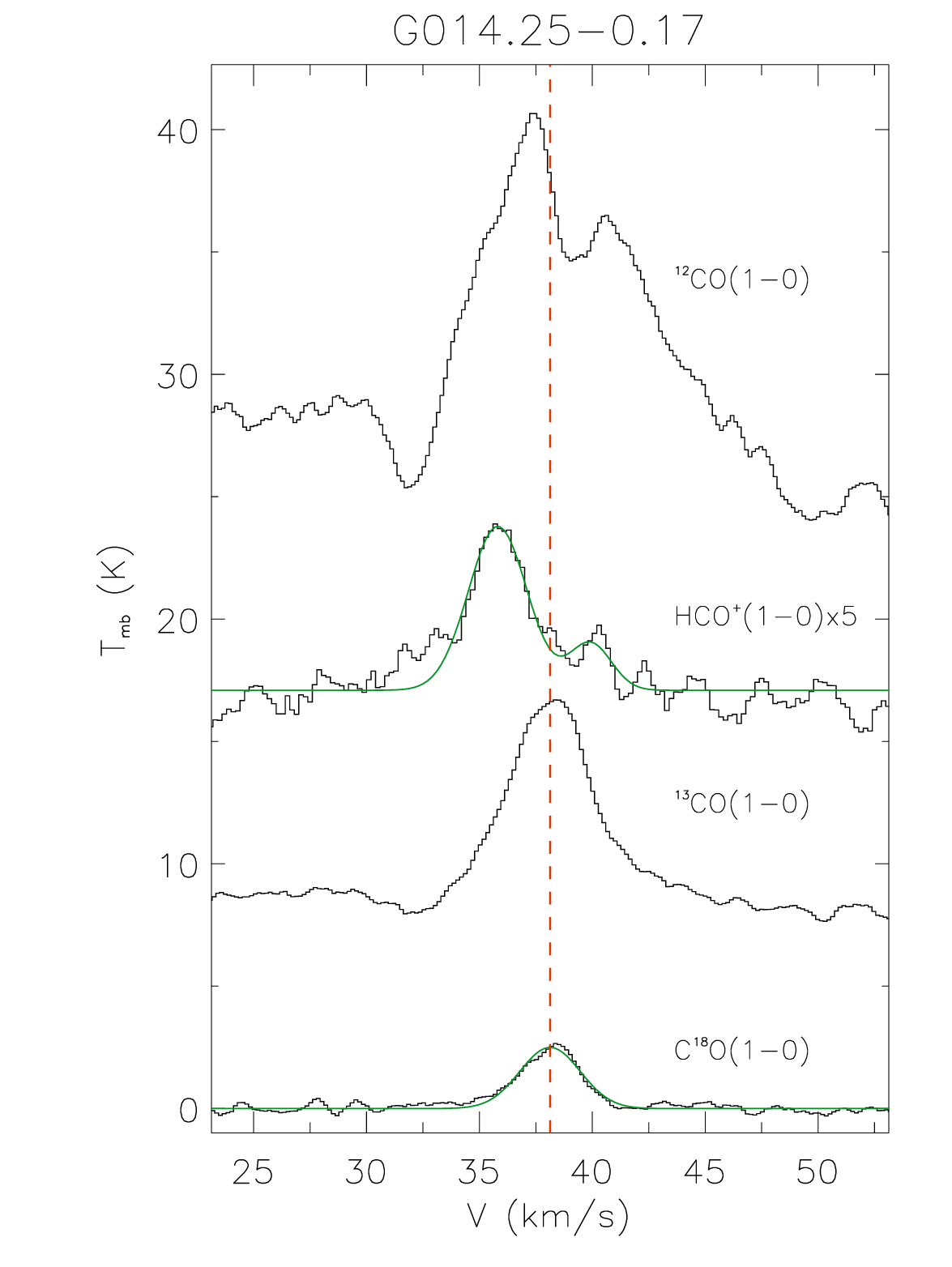}
  \end{minipage}%
\quad
  \begin{minipage}[t]{0.325\linewidth}
  \centering
   \includegraphics[width=55mm]{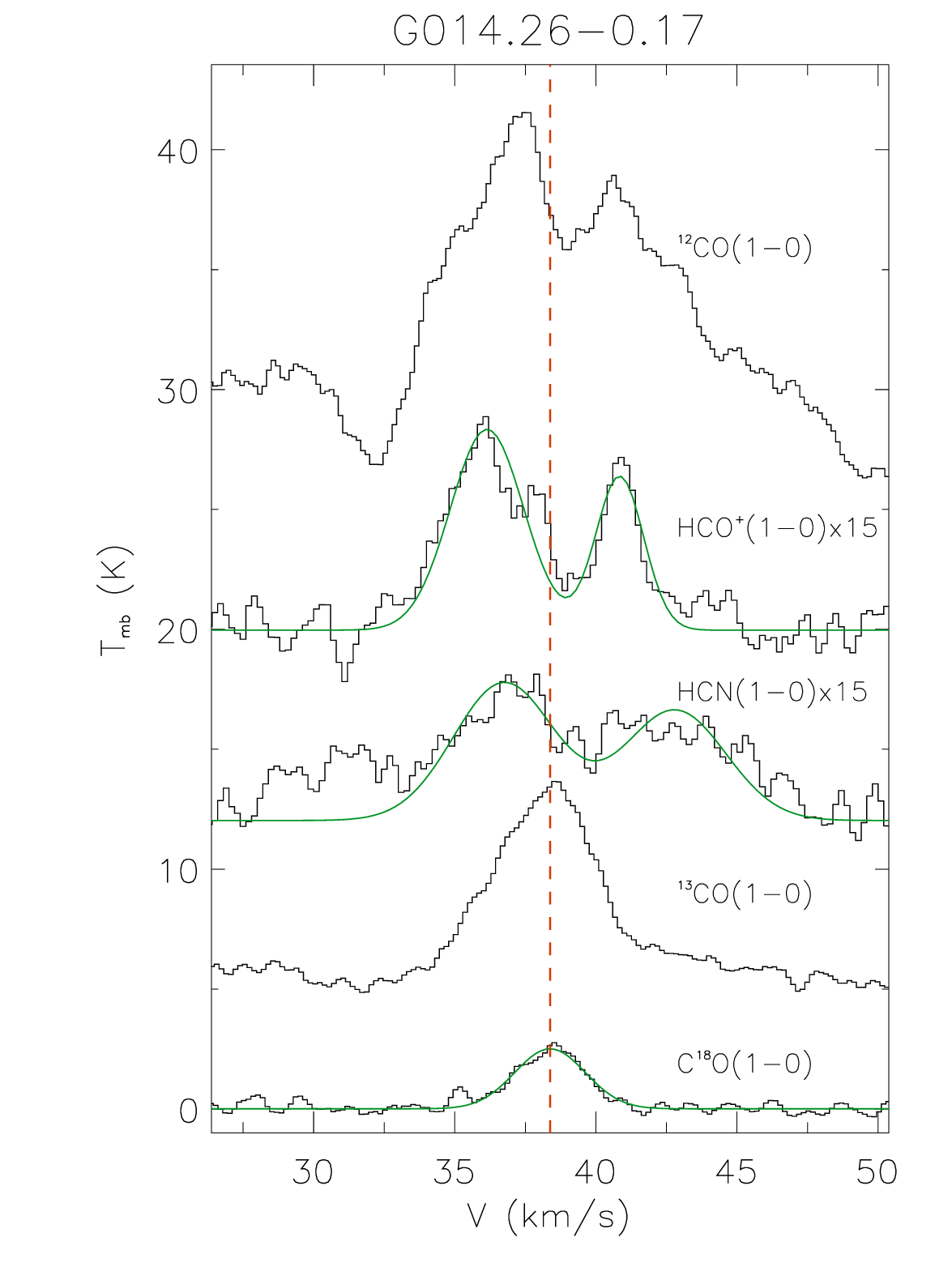}
  \end{minipage}%
  \begin{minipage}[t]{0.325\linewidth}
  \centering
   \includegraphics[width=55mm]{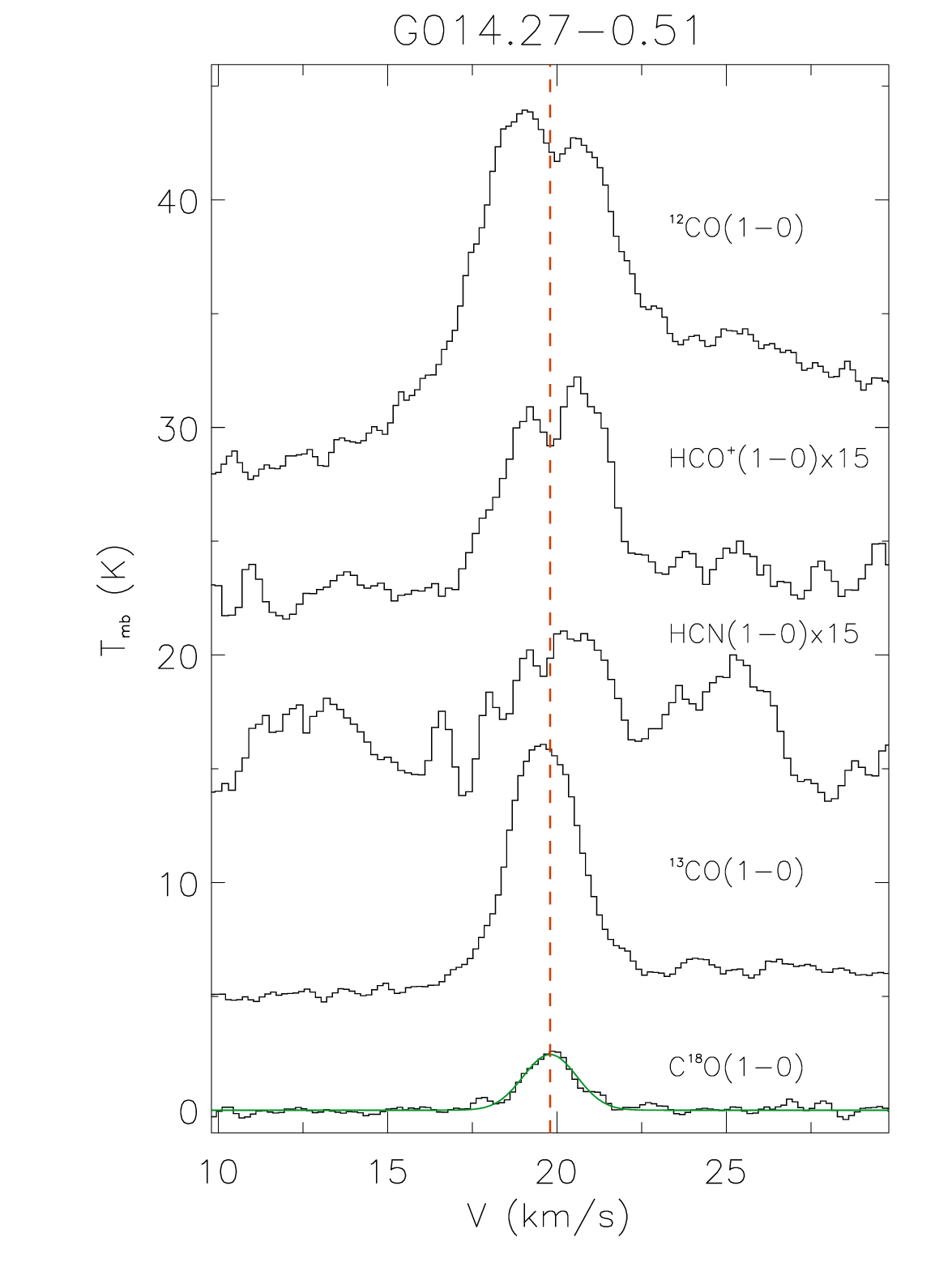}
  \end{minipage}%
  \begin{minipage}[t]{0.325\linewidth}
  \centering
   \includegraphics[width=55mm]{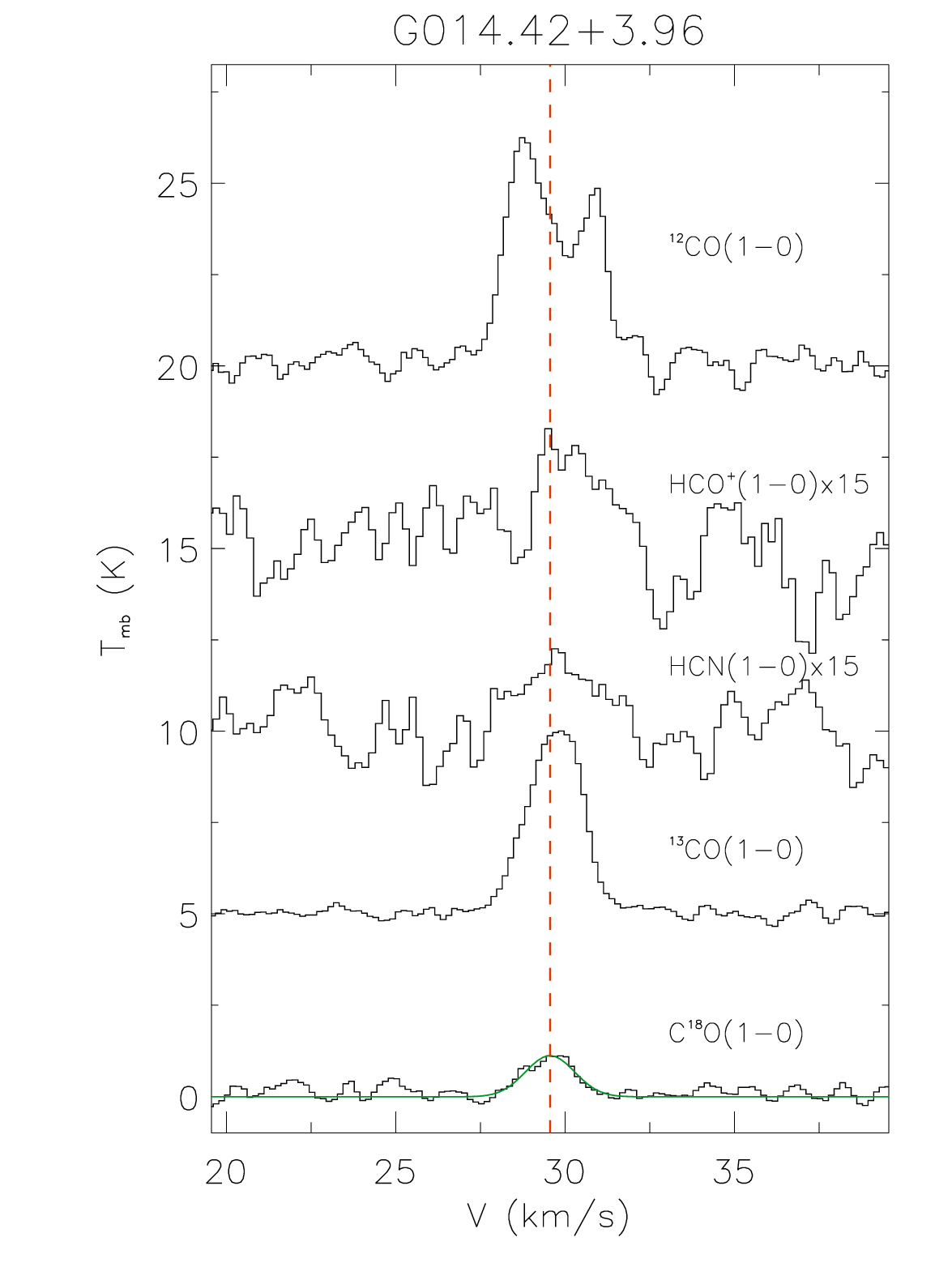}
  \end{minipage}%
  \caption{{\small Line profiles of 133 sources we selected. The lines from bottom to top are C$^{18}$O (1-0), $^{13}$CO (1-0), HCN (1-0) (14 sources lack HCN data), HCO$^+$ (1-0) and $^{12}$CO (1-0), respectively. The dashed red line indicates the central radial velocity of C$^{18}$O (1-0) estimated by Gaussian fitting. For infall candidates, HCO$^+$ (1-0) and HCN (1-0) lines are also Gaussian fitted.}}
  \label{Fig:fig6}
\end{figure} 

\begin{figure}[h]
\ContinuedFloat
  \begin{minipage}[t]{0.325\linewidth}
  \centering
   \includegraphics[width=55mm]{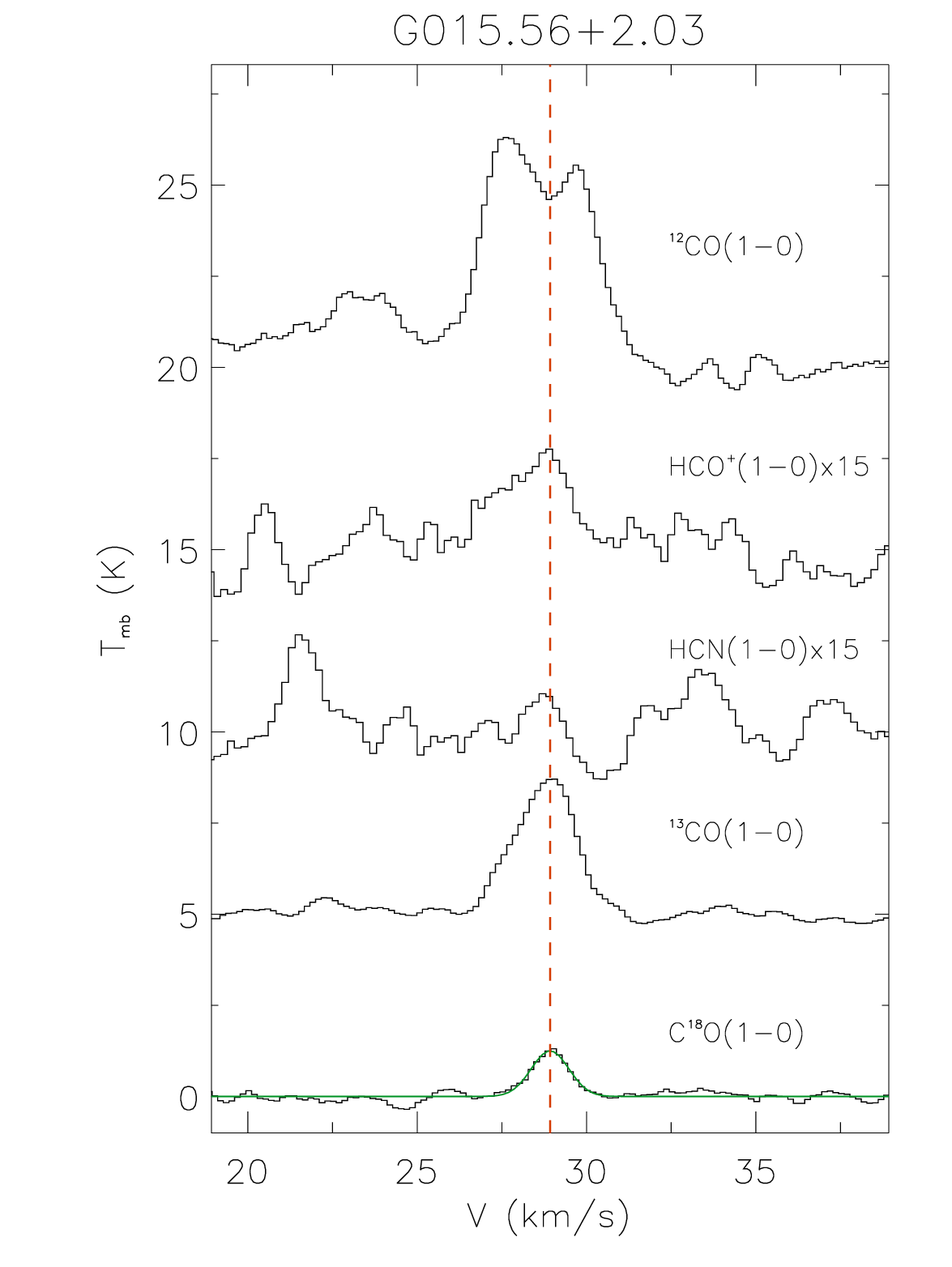}
  \end{minipage}%
  \begin{minipage}[t]{0.325\textwidth}
  \centering
   \includegraphics[width=55mm]{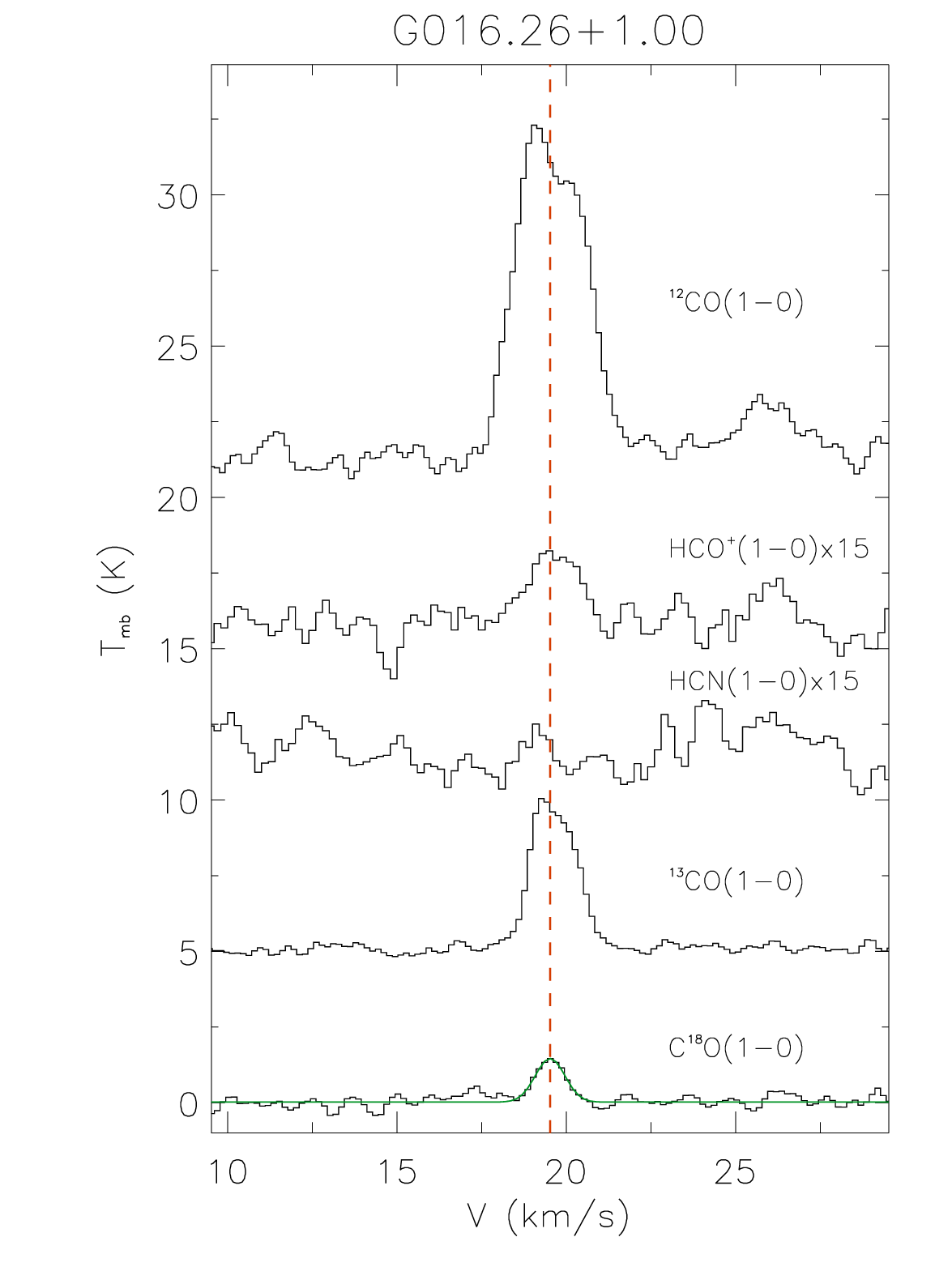}
  \end{minipage}%
  \begin{minipage}[t]{0.325\linewidth}
  \centering
   \includegraphics[width=55mm]{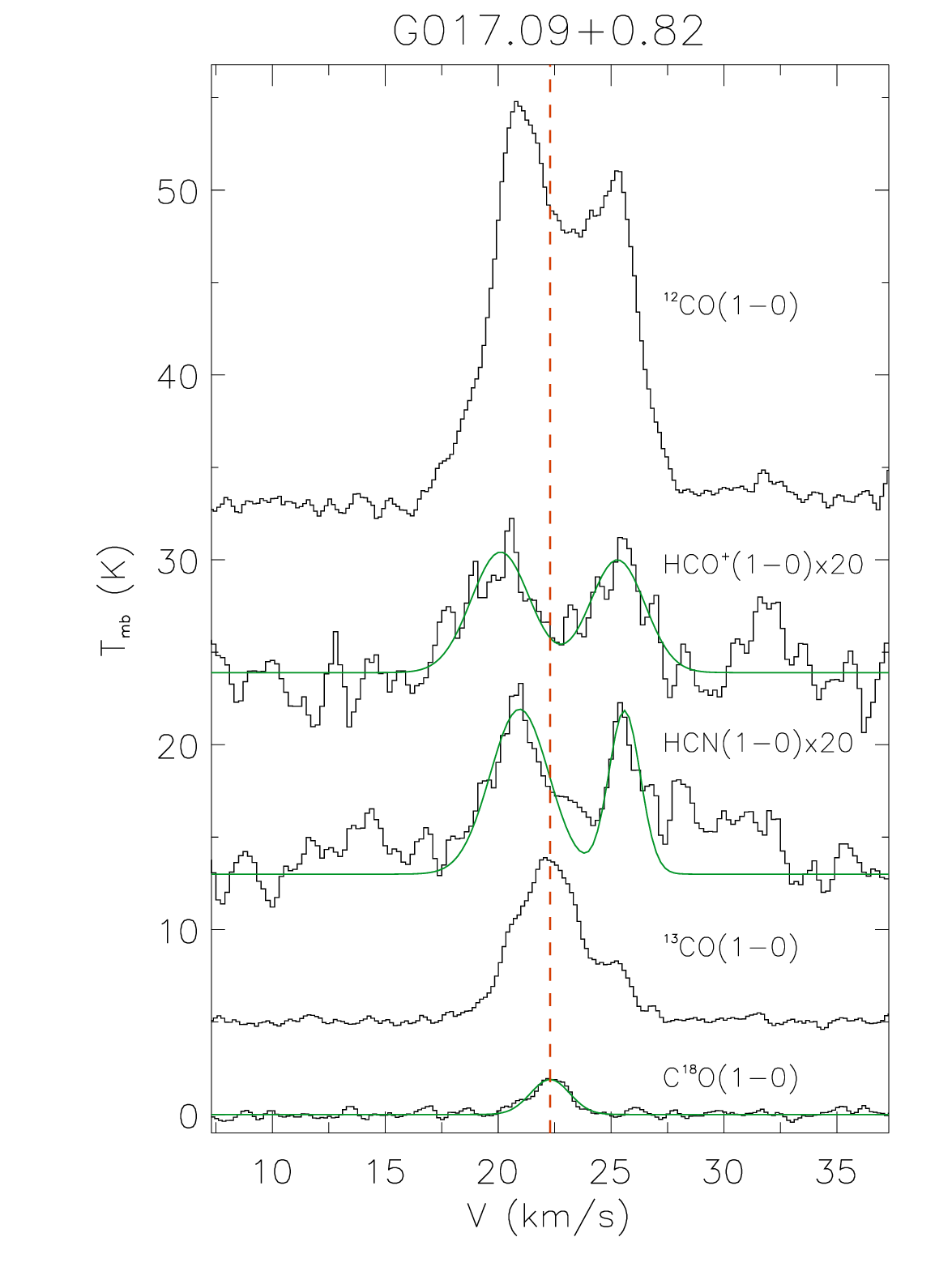}
  \end{minipage}%  
\quad
  \begin{minipage}[t]{0.325\linewidth}
  \centering
   \includegraphics[width=55mm]{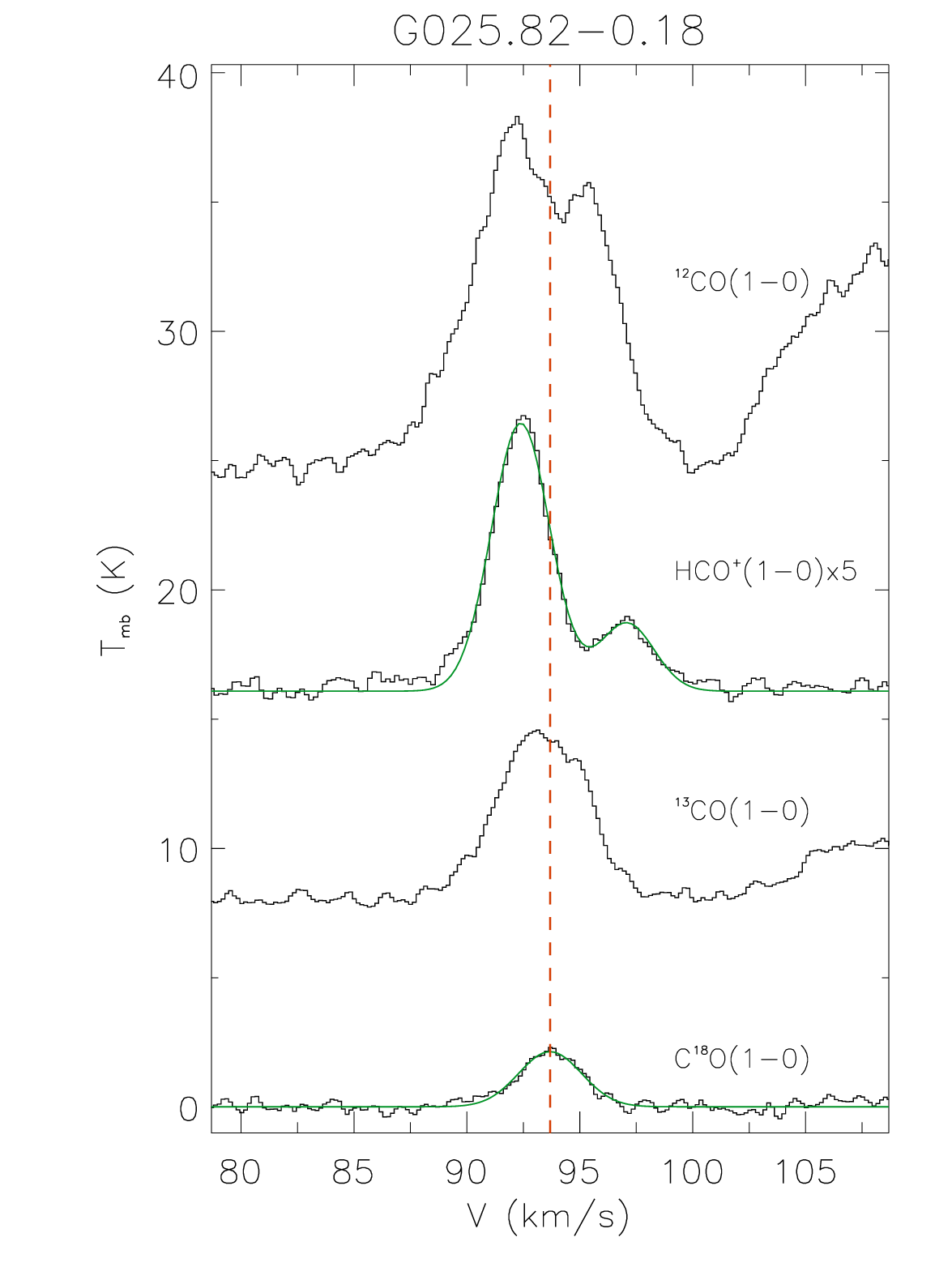}
  \end{minipage}%
  \begin{minipage}[t]{0.325\linewidth}
  \centering
   \includegraphics[width=55mm]{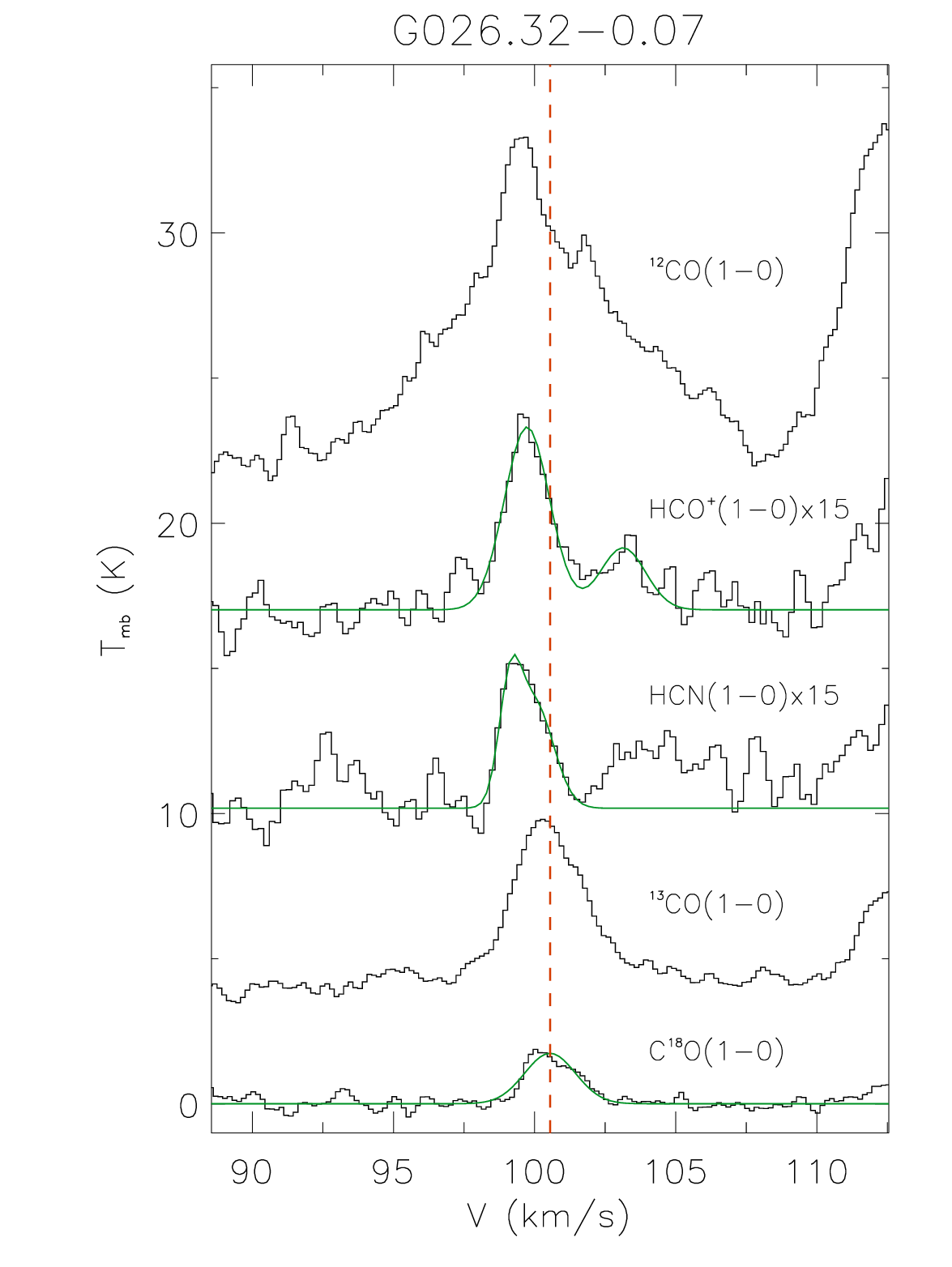}
  \end{minipage}%
  \begin{minipage}[t]{0.325\linewidth}
  \centering
   \includegraphics[width=55mm]{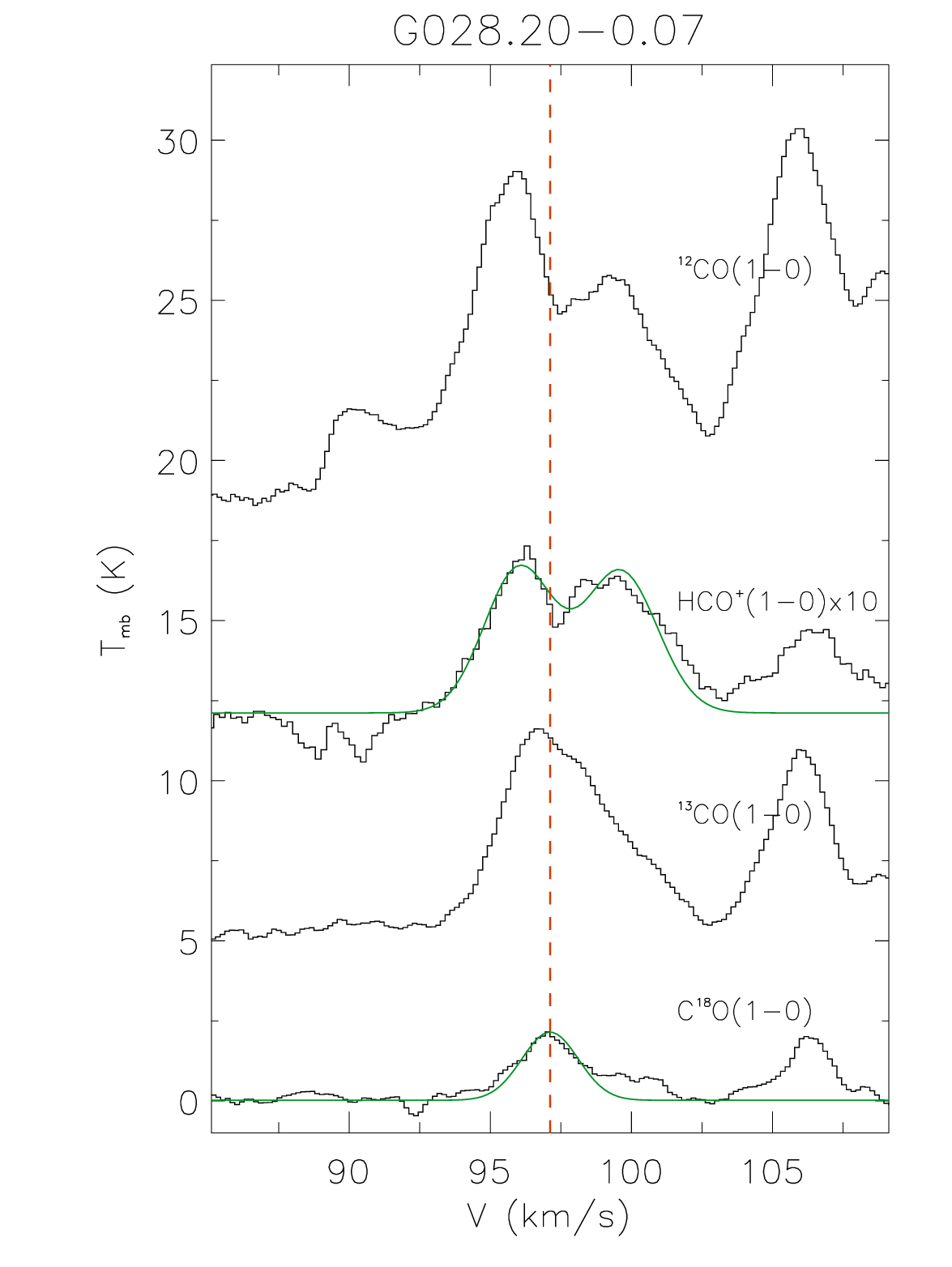}
  \end{minipage}%
\quad
  \begin{minipage}[t]{0.325\linewidth}
  \centering
   \includegraphics[width=55mm]{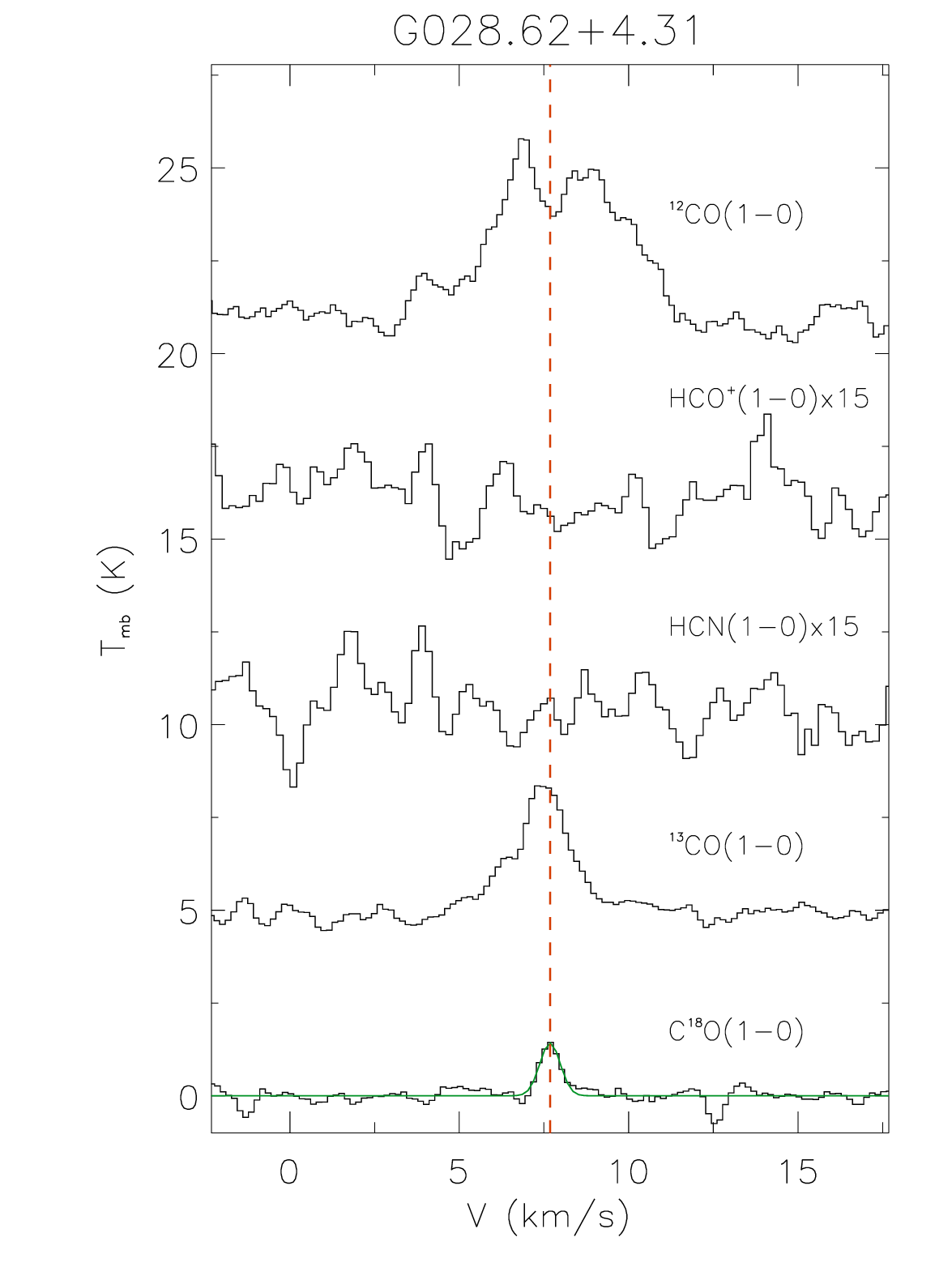}
  \end{minipage}%
  \begin{minipage}[t]{0.325\linewidth}
  \centering
   \includegraphics[width=55mm]{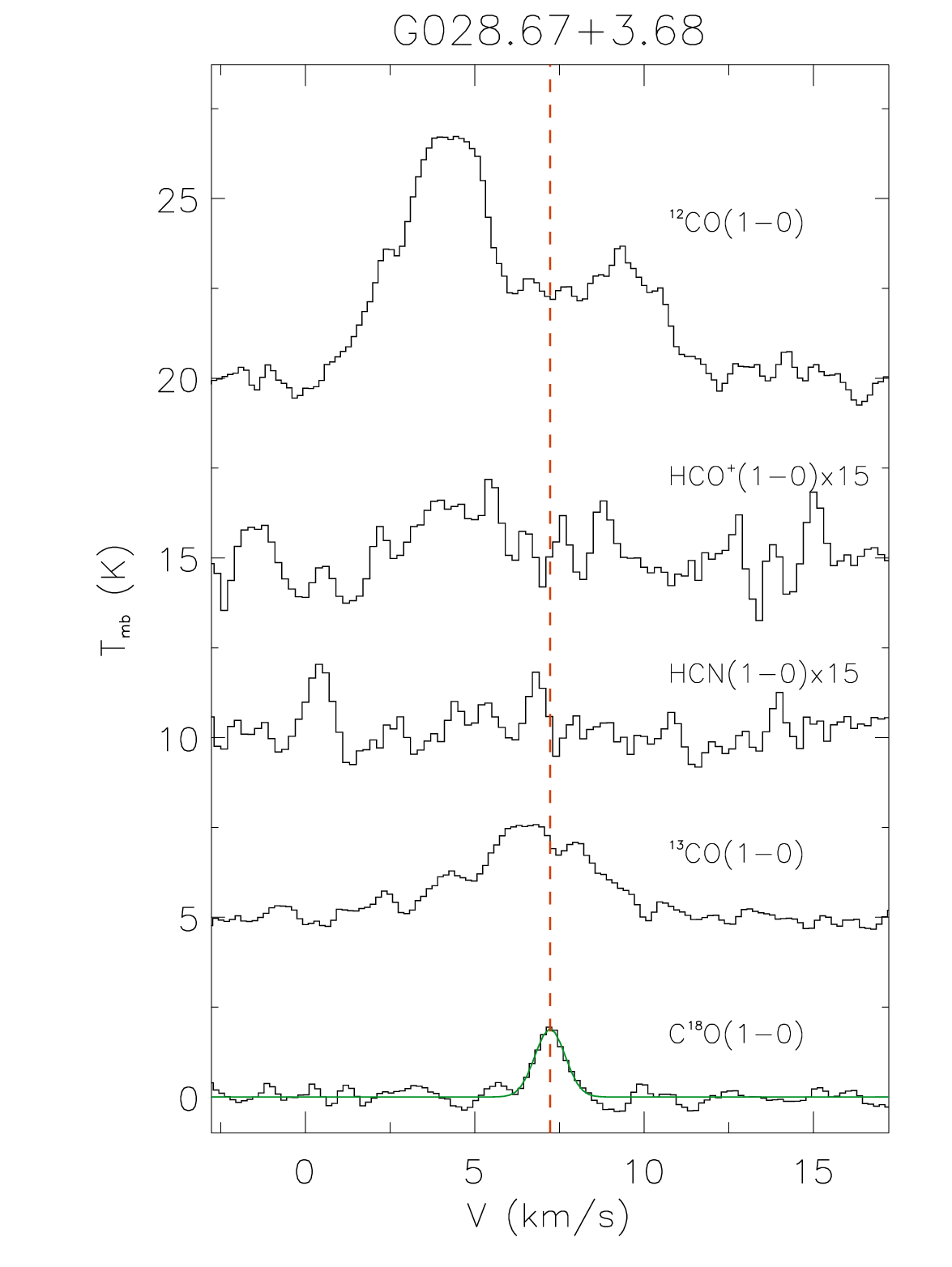}
  \end{minipage}%
  \begin{minipage}[t]{0.325\linewidth}
  \centering
   \includegraphics[width=55mm]{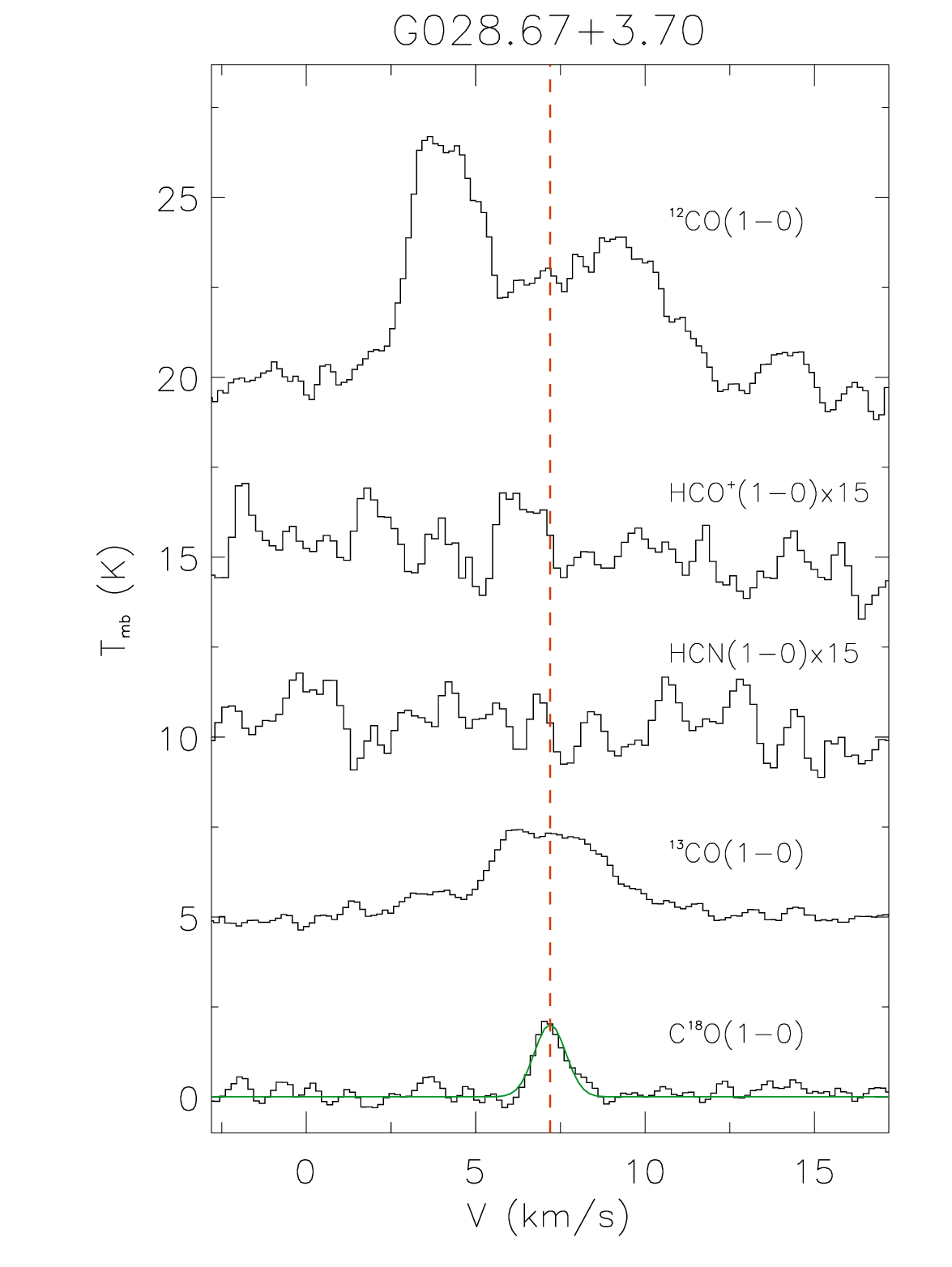}
  \end{minipage}%
  \caption{{\small Line profiles of 133 sources we selected. The lines from bottom to top are C$^{18}$O (1-0), $^{13}$CO (1-0), HCN (1-0) (14 sources lack HCN data), HCO$^+$ (1-0) and $^{12}$CO (1-0), respectively. The dashed red line indicates the central radial velocity of C$^{18}$O (1-0) estimated by Gaussian fitting. For infall candidates, HCO$^+$ (1-0) and HCN (1-0) lines are also Gaussian fitted.}}
  \label{Fig:fig6}
\end{figure} 

\begin{figure}[h]
\ContinuedFloat
  \begin{minipage}[t]{0.325\linewidth}
  \centering
   \includegraphics[width=55mm]{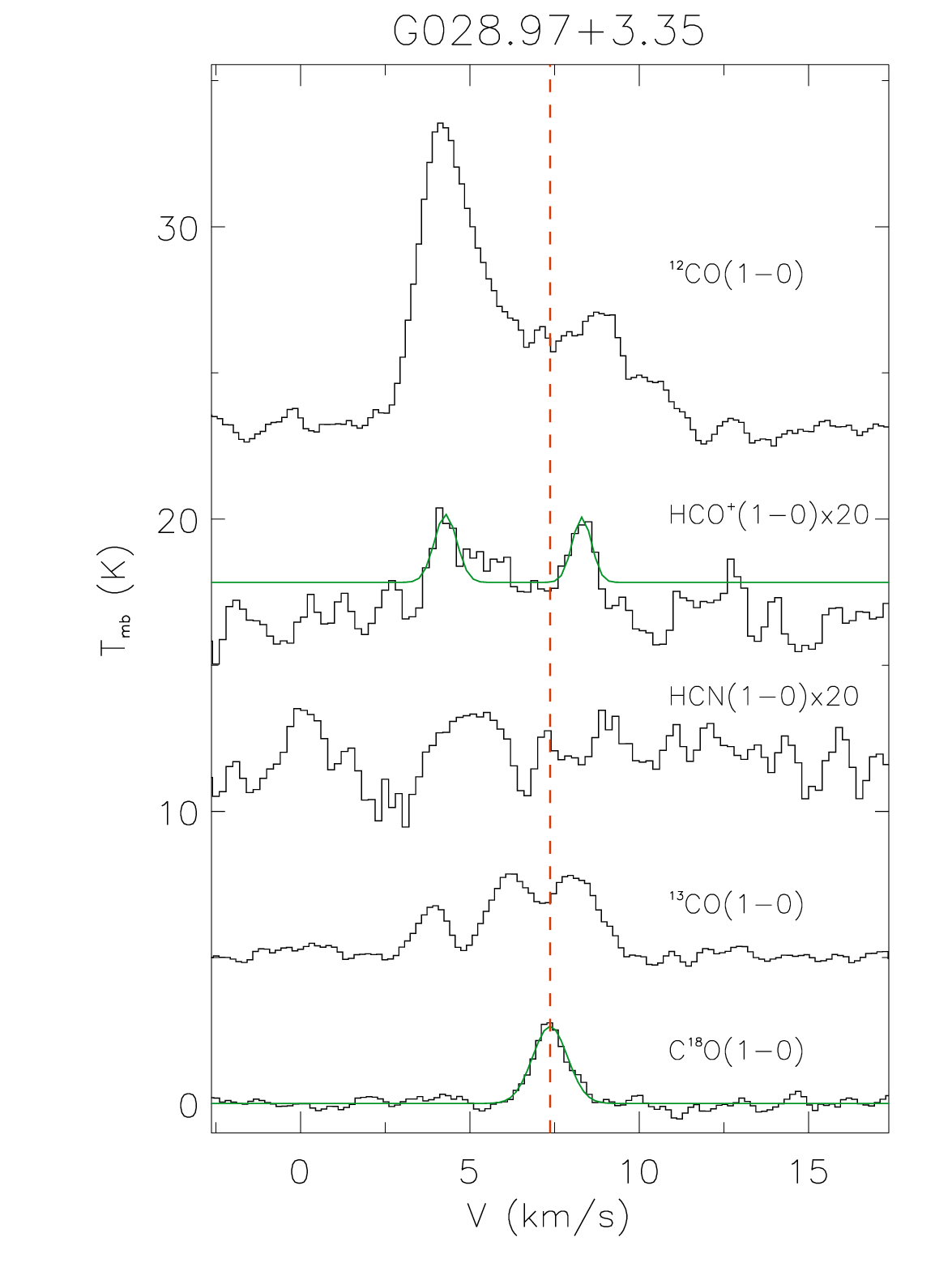}
  \end{minipage}%
  \begin{minipage}[t]{0.325\textwidth}
  \centering
   \includegraphics[width=55mm]{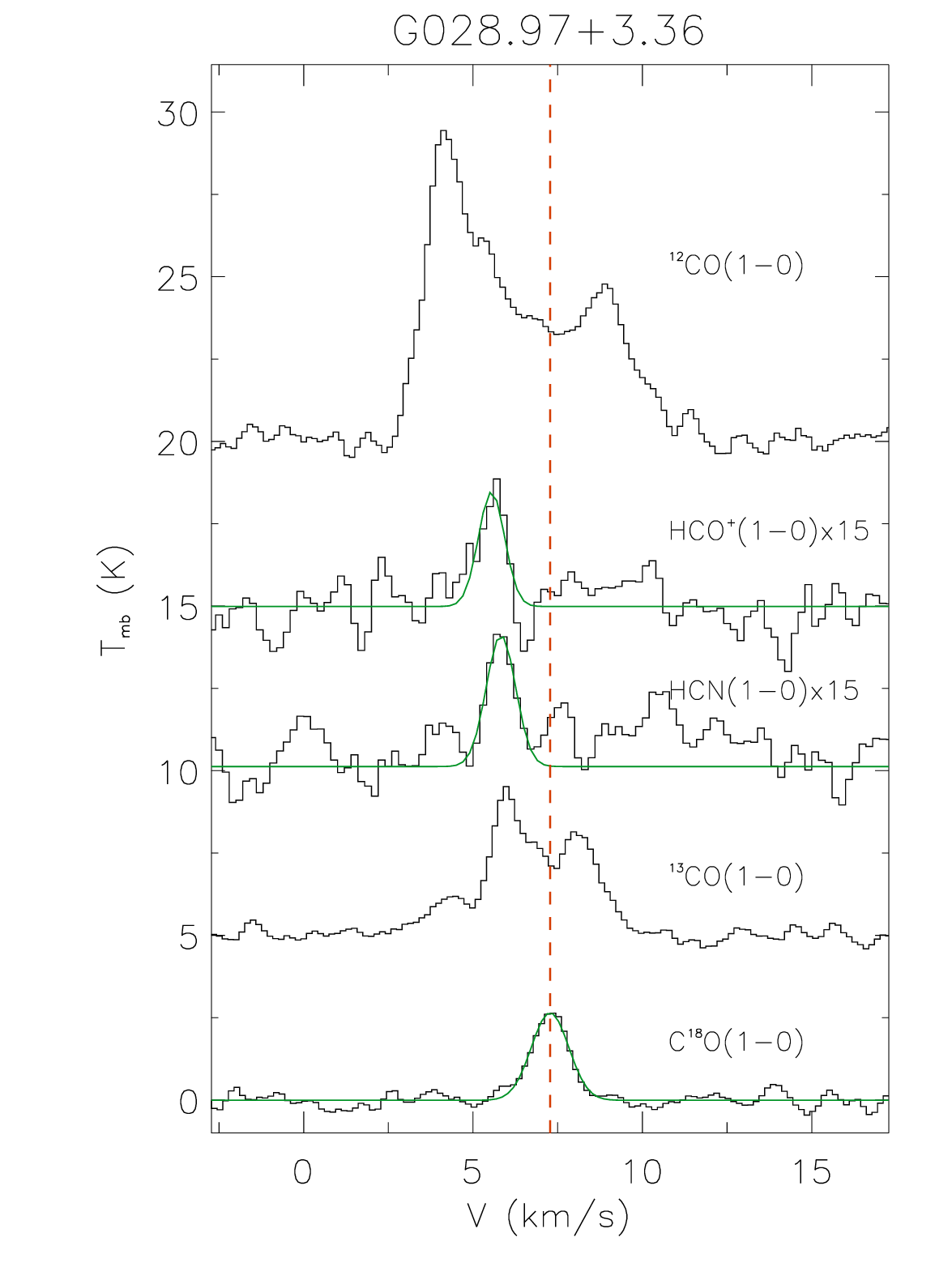}
  \end{minipage}%
  \begin{minipage}[t]{0.325\linewidth}
  \centering
   \includegraphics[width=55mm]{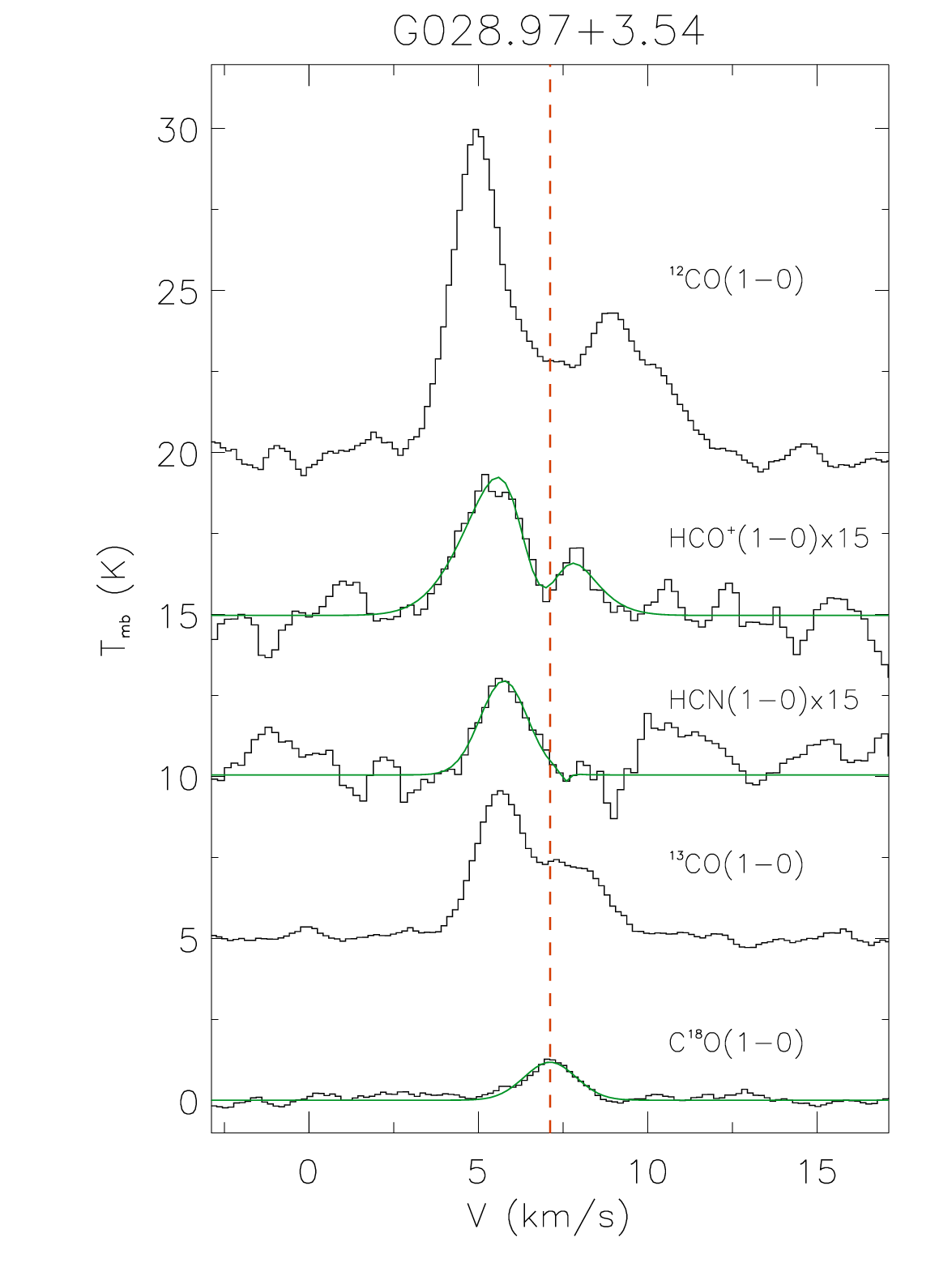}
  \end{minipage}%  
\quad
  \begin{minipage}[t]{0.325\linewidth}
  \centering
   \includegraphics[width=55mm]{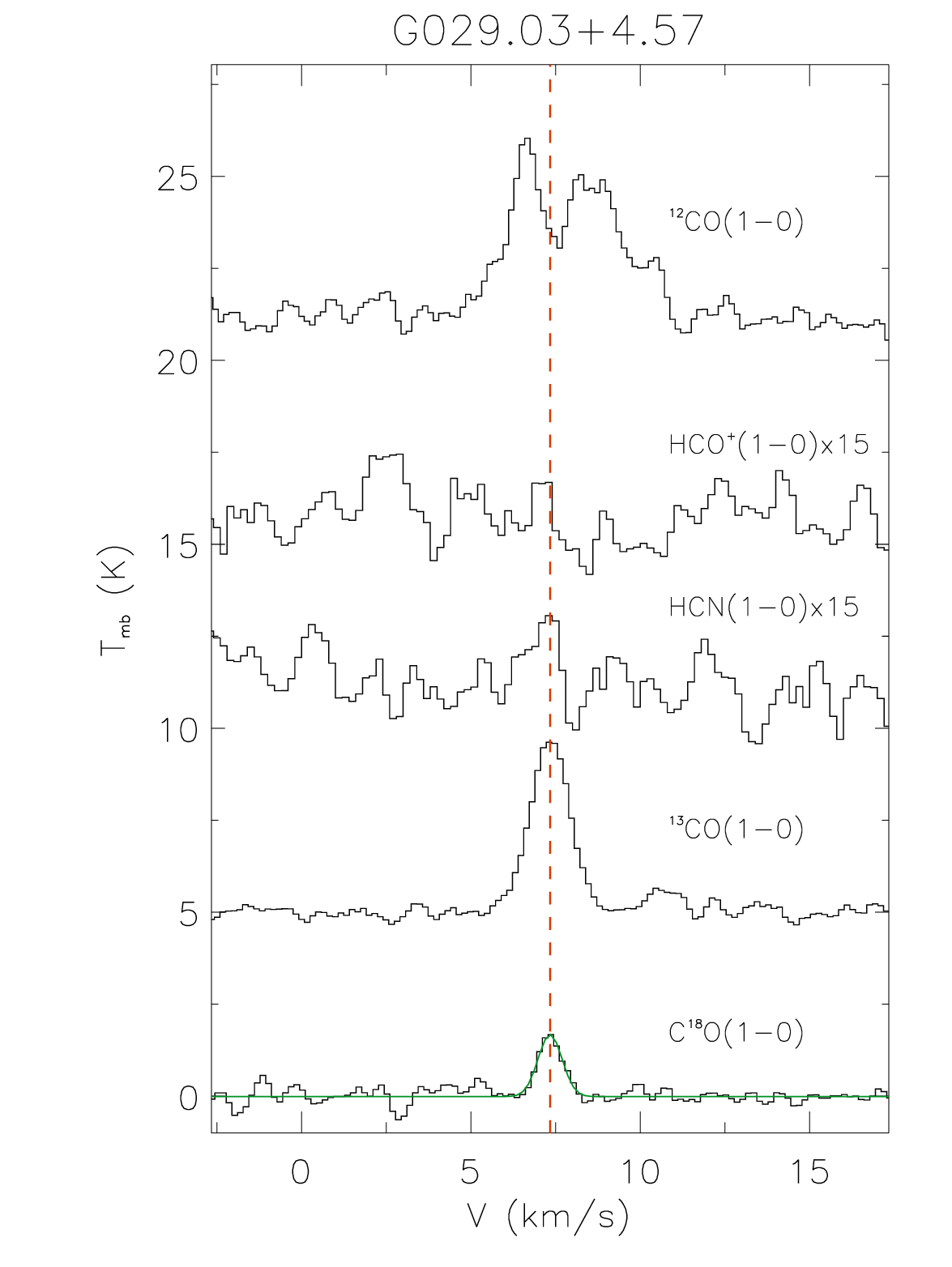}
  \end{minipage}%
  \begin{minipage}[t]{0.325\linewidth}
  \centering
   \includegraphics[width=55mm]{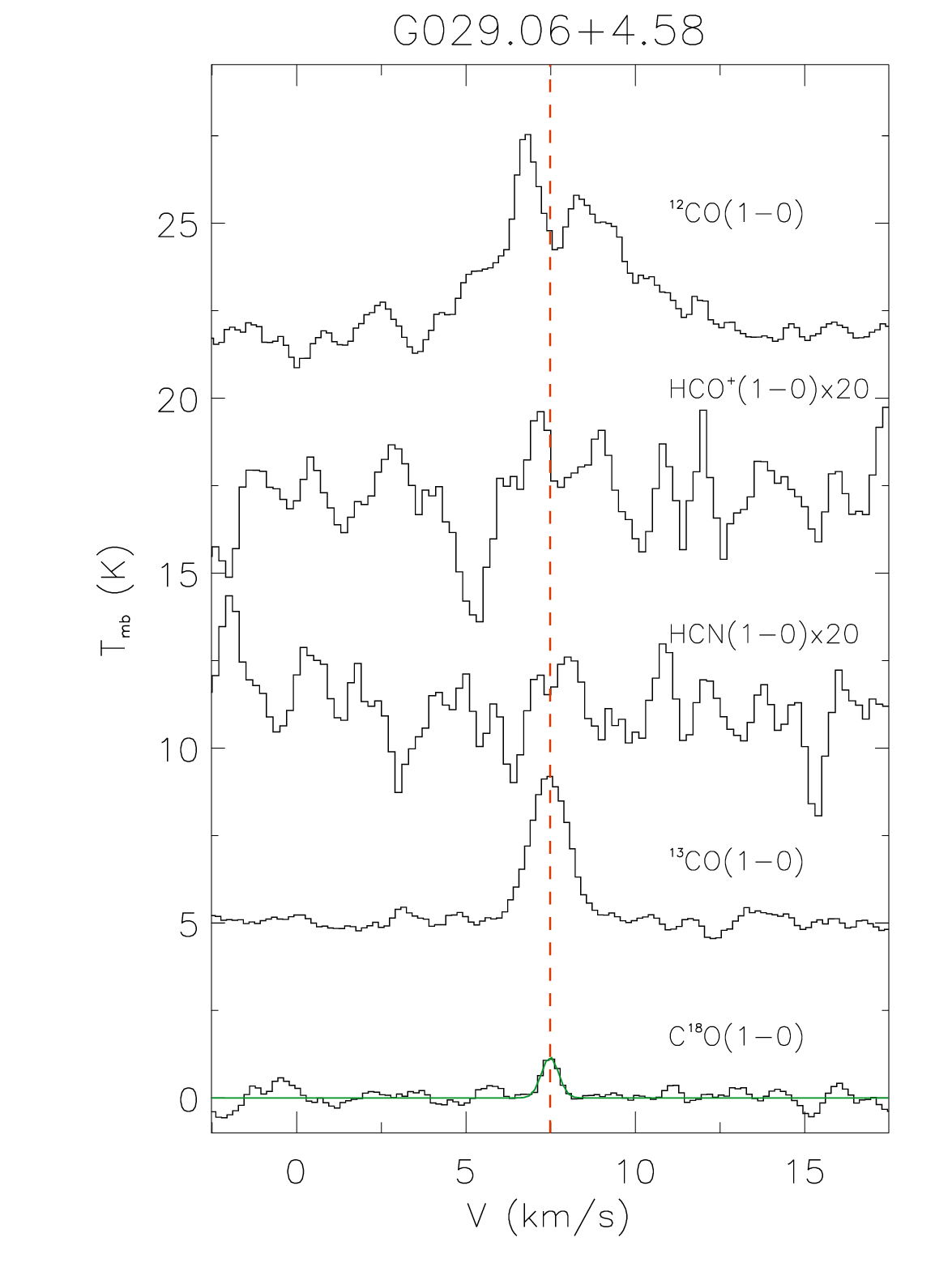}
  \end{minipage}%
  \begin{minipage}[t]{0.325\linewidth}
  \centering
   \includegraphics[width=55mm]{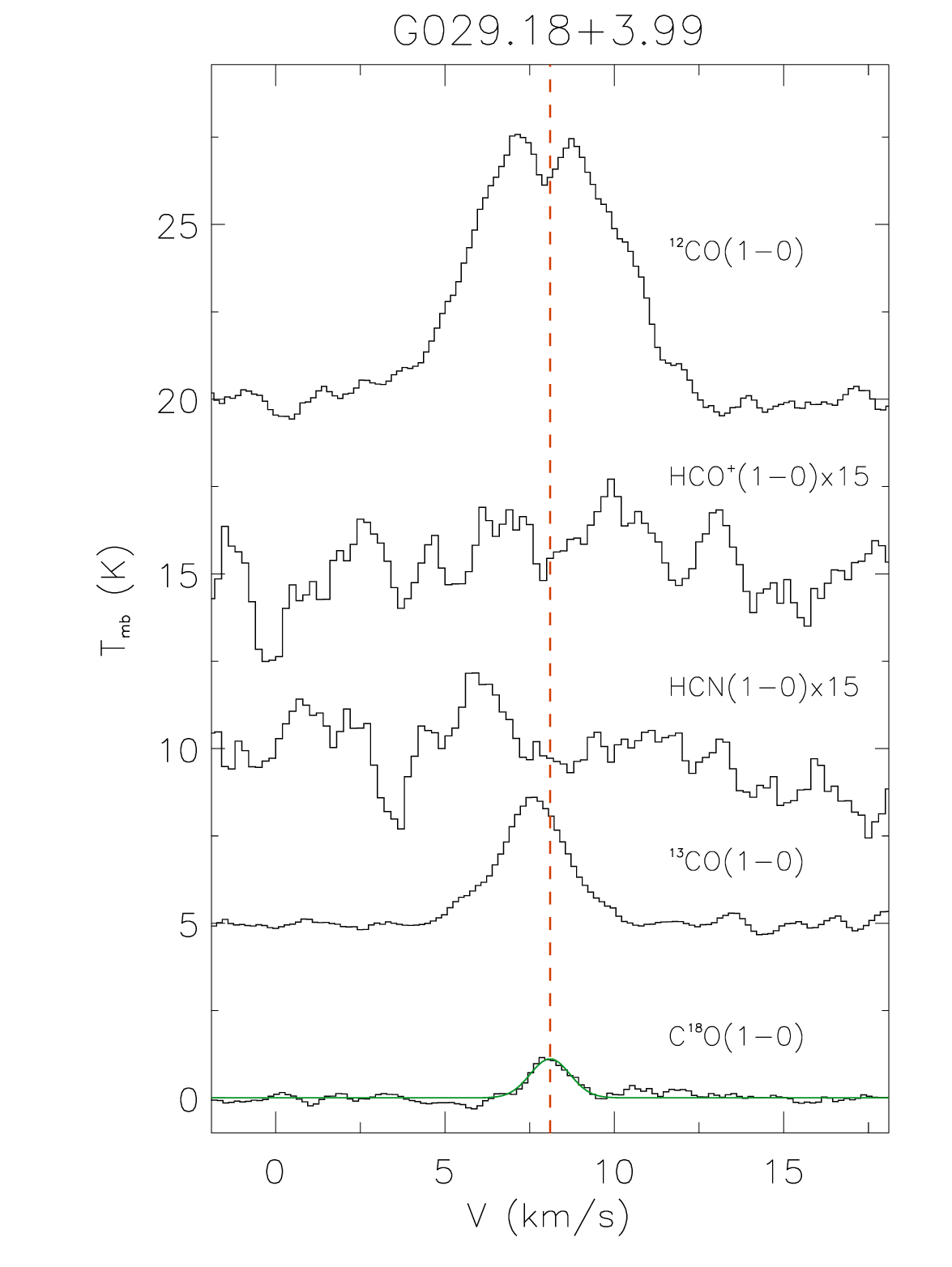}
  \end{minipage}%
\quad
  \begin{minipage}[t]{0.325\linewidth}
  \centering
   \includegraphics[width=55mm]{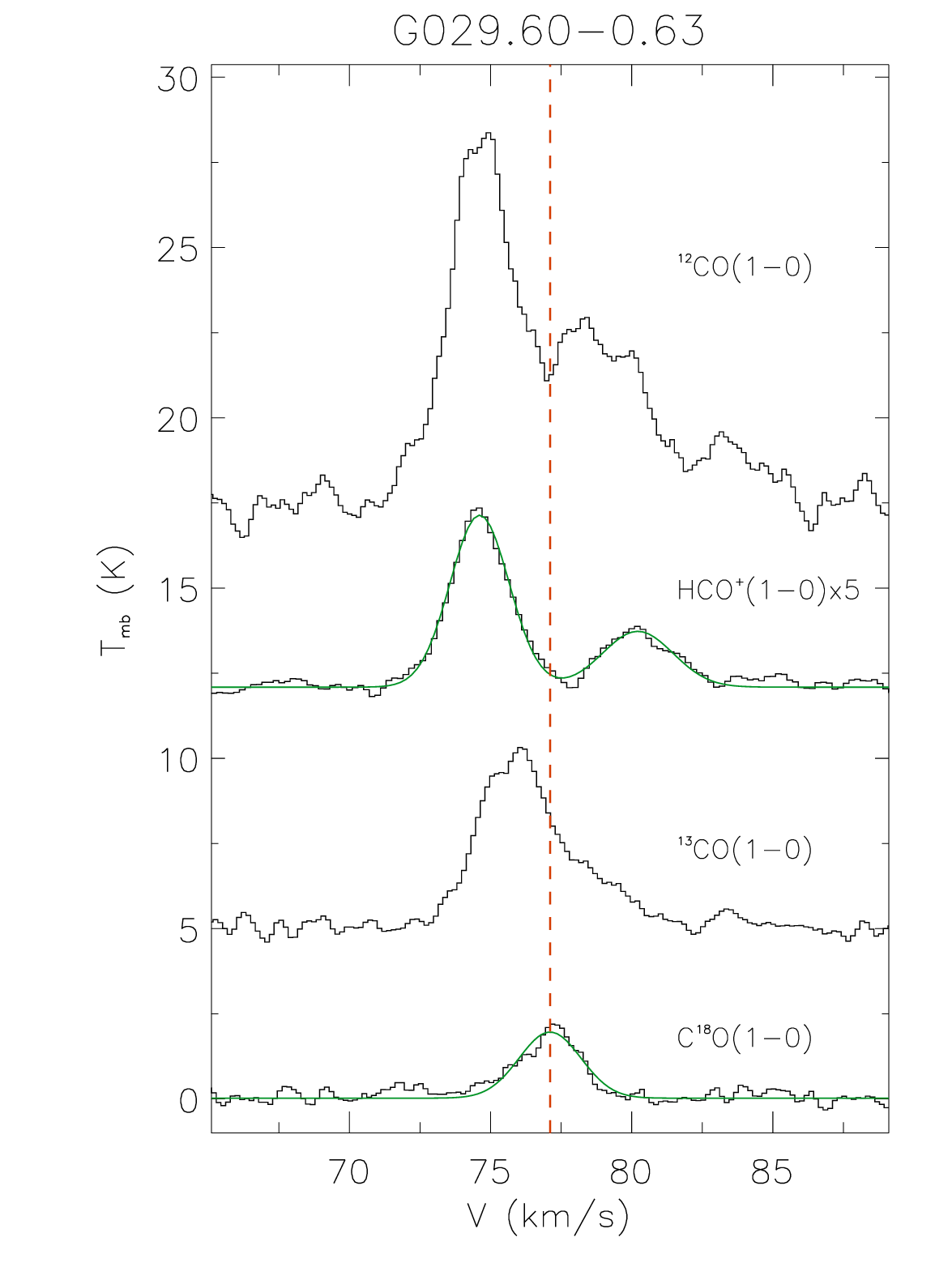}
  \end{minipage}%
  \begin{minipage}[t]{0.325\linewidth}
  \centering
   \includegraphics[width=55mm]{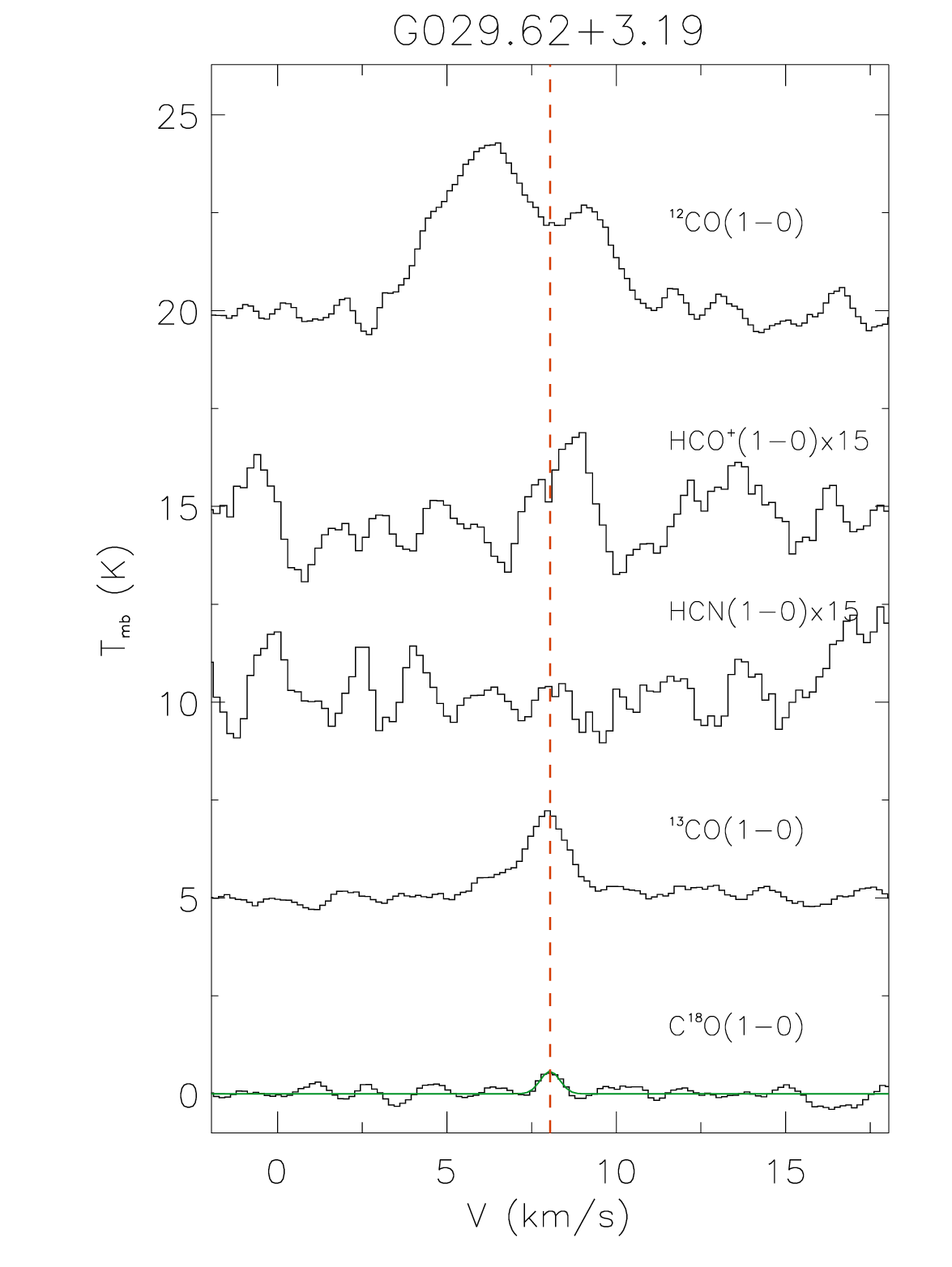}
  \end{minipage}%
  \begin{minipage}[t]{0.325\linewidth}
  \centering
   \includegraphics[width=55mm]{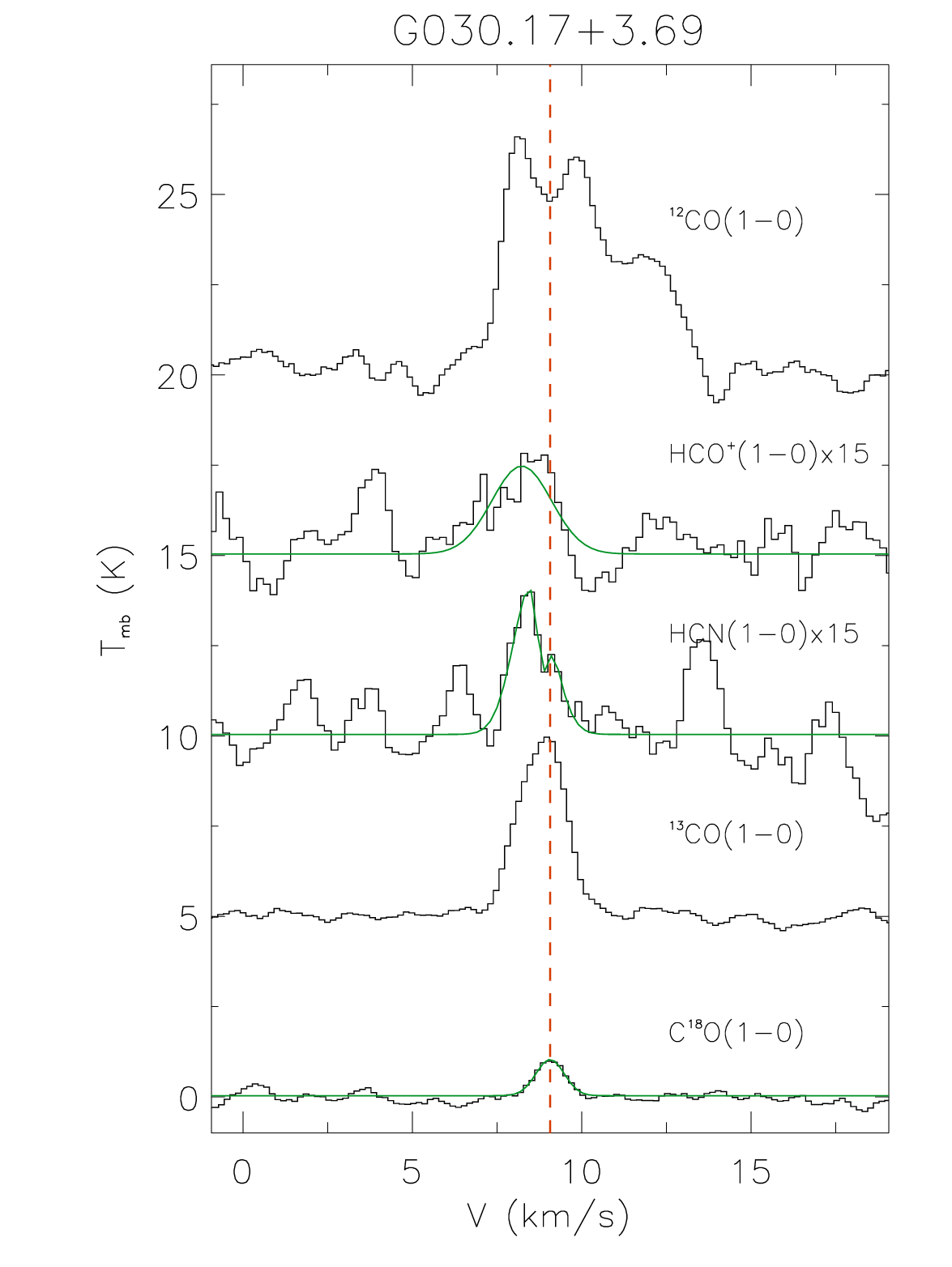}
  \end{minipage}%
  \caption{{\small Line profiles of 133 sources we selected. The lines from bottom to top are C$^{18}$O (1-0), $^{13}$CO (1-0), HCN (1-0) (14 sources lack HCN data), HCO$^+$ (1-0) and $^{12}$CO (1-0), respectively. The dashed red line indicates the central radial velocity of C$^{18}$O (1-0) estimated by Gaussian fitting. For infall candidates, HCO$^+$ (1-0) and HCN (1-0) lines are also Gaussian fitted.}}
  \label{Fig:fig6}
\end{figure} 

\begin{figure}[h]
\ContinuedFloat
  \begin{minipage}[t]{0.325\linewidth}
  \centering
   \includegraphics[width=55mm]{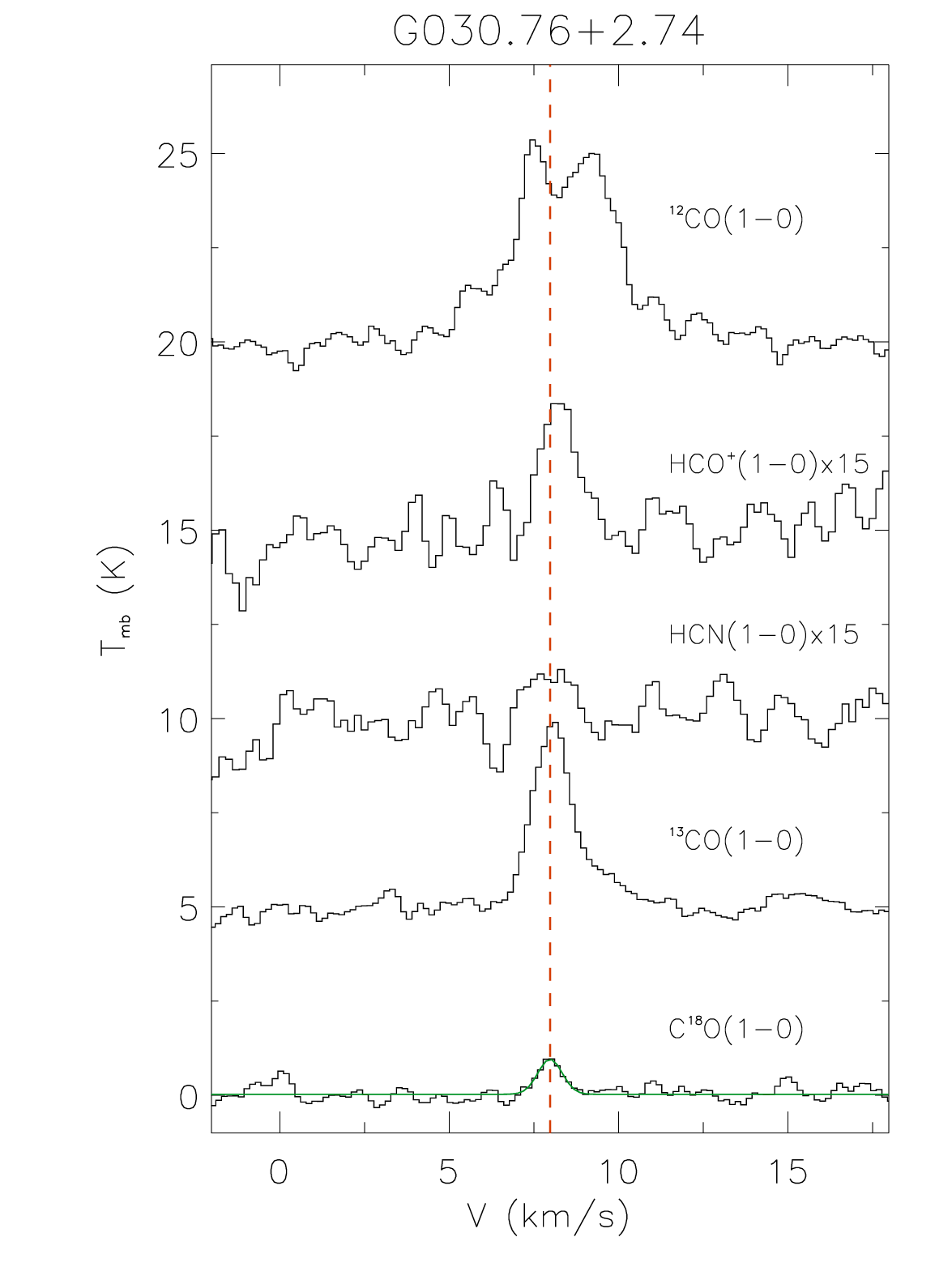}
  \end{minipage}%
  \begin{minipage}[t]{0.325\textwidth}
  \centering
   \includegraphics[width=55mm]{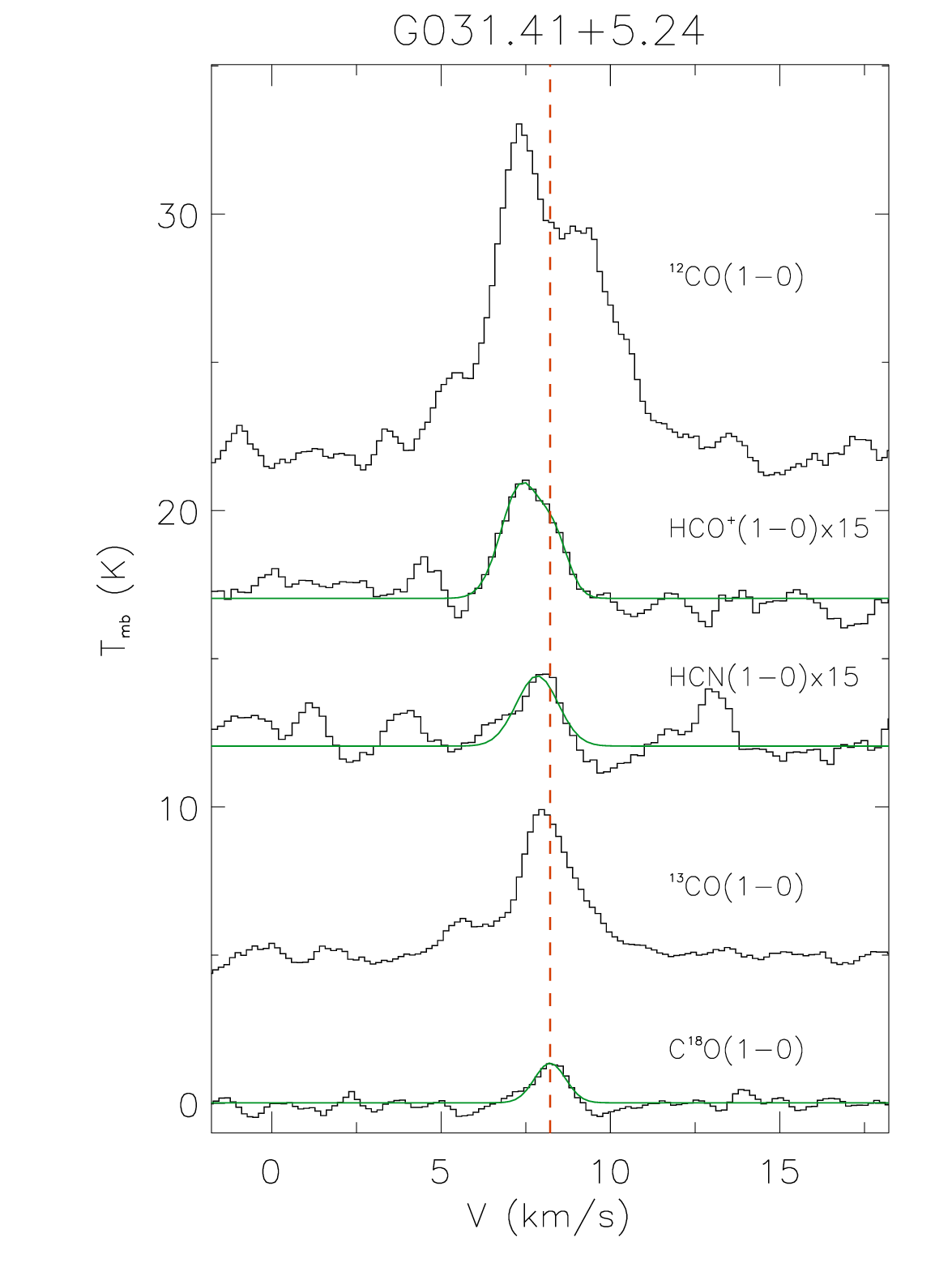}
  \end{minipage}%
  \begin{minipage}[t]{0.325\linewidth}
  \centering
   \includegraphics[width=55mm]{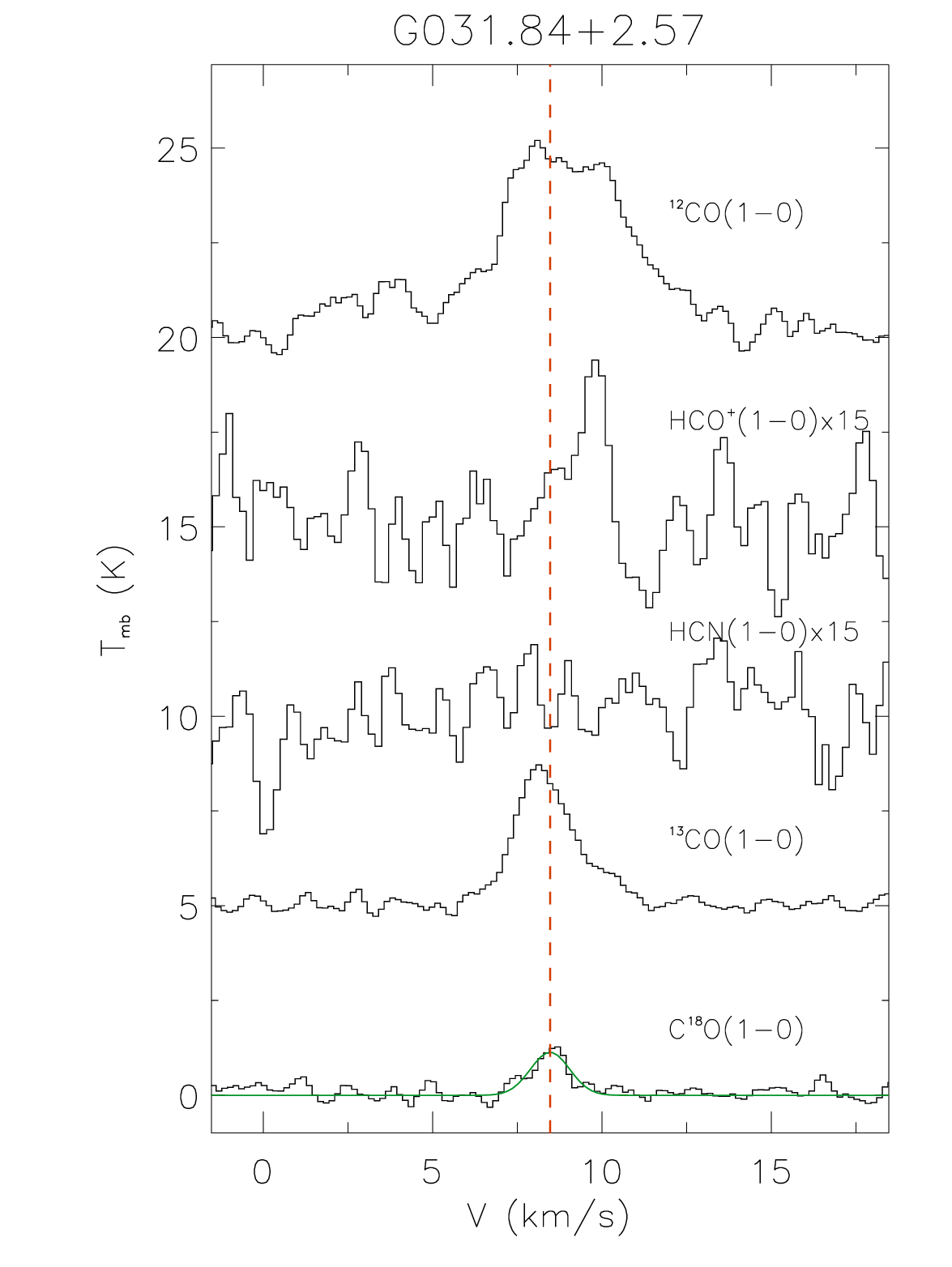}
  \end{minipage}%  
\quad
  \begin{minipage}[t]{0.325\linewidth}
  \centering
   \includegraphics[width=55mm]{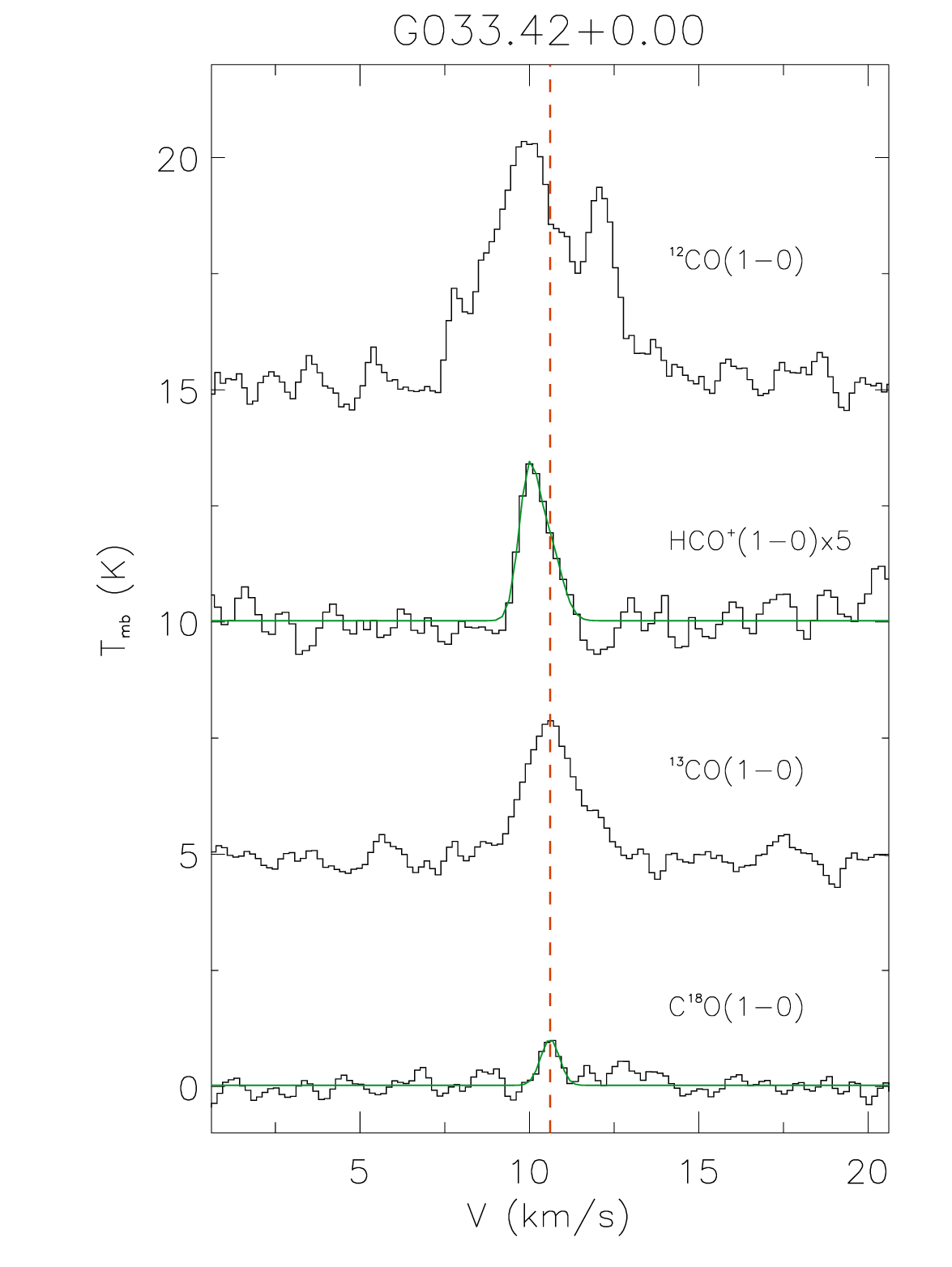}
  \end{minipage}%
  \begin{minipage}[t]{0.325\linewidth}
  \centering
   \includegraphics[width=55mm]{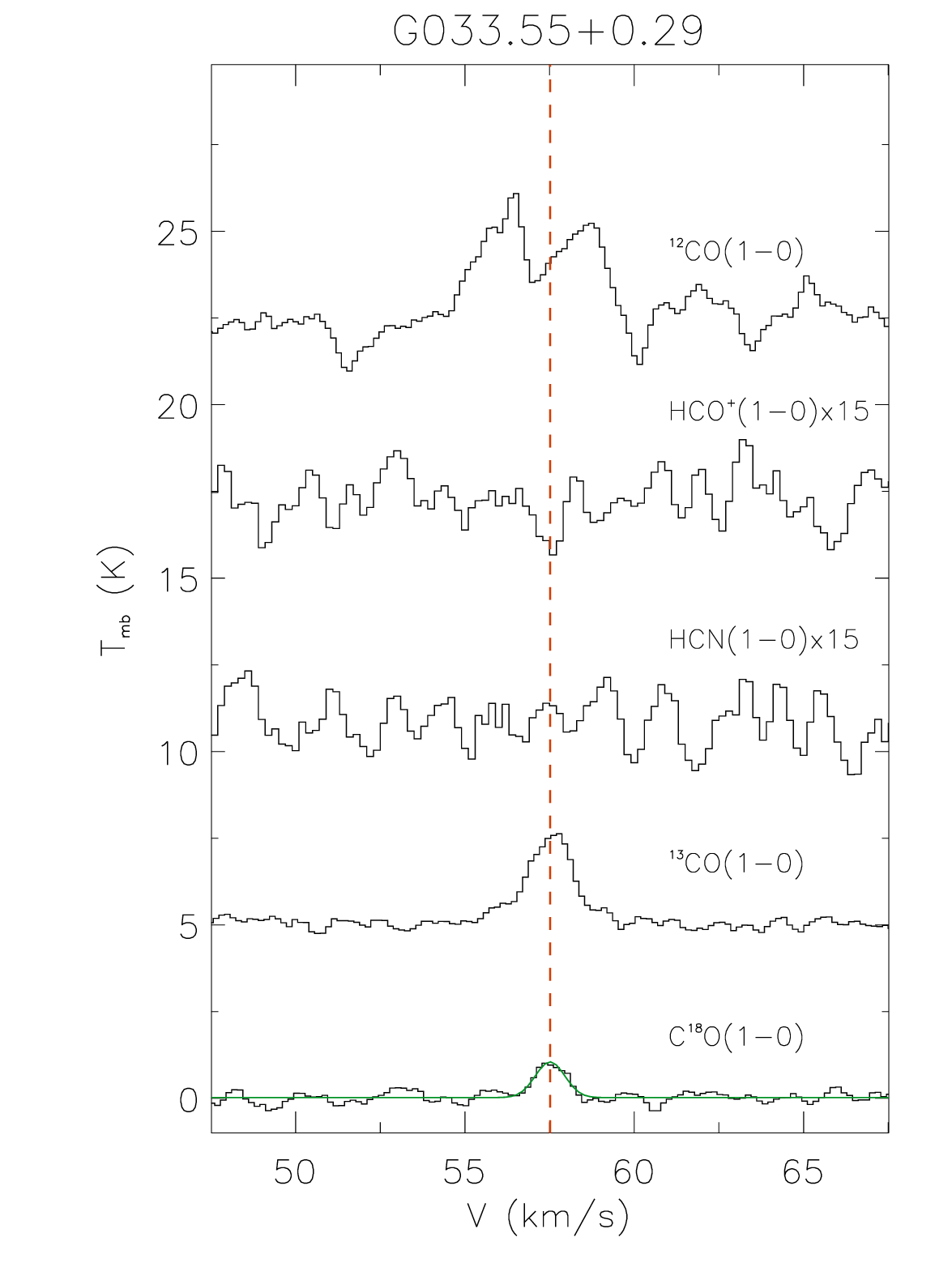}
  \end{minipage}%
  \begin{minipage}[t]{0.325\linewidth}
  \centering
   \includegraphics[width=55mm]{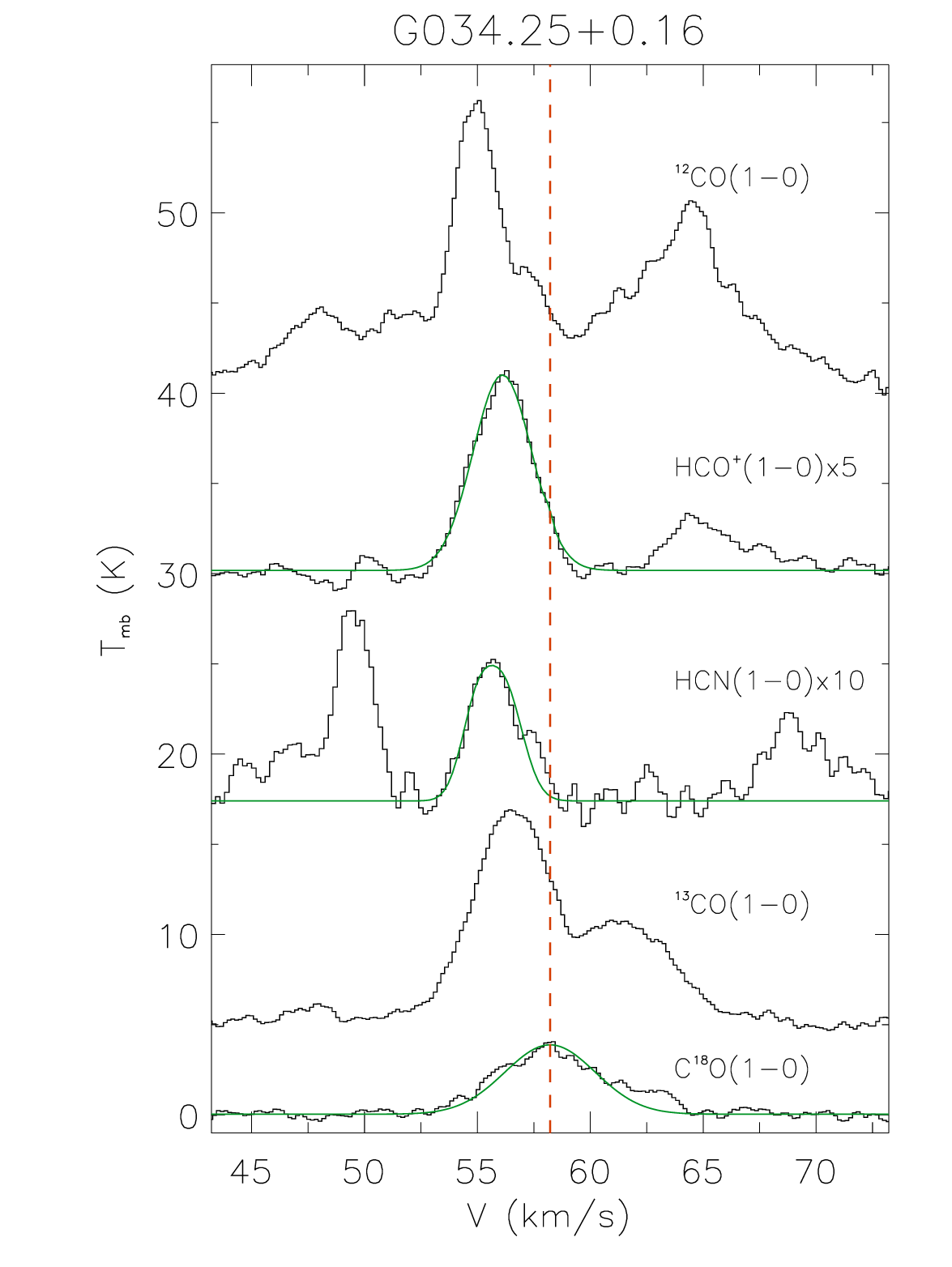}
  \end{minipage}%
\quad
  \begin{minipage}[t]{0.325\linewidth}
  \centering
   \includegraphics[width=55mm]{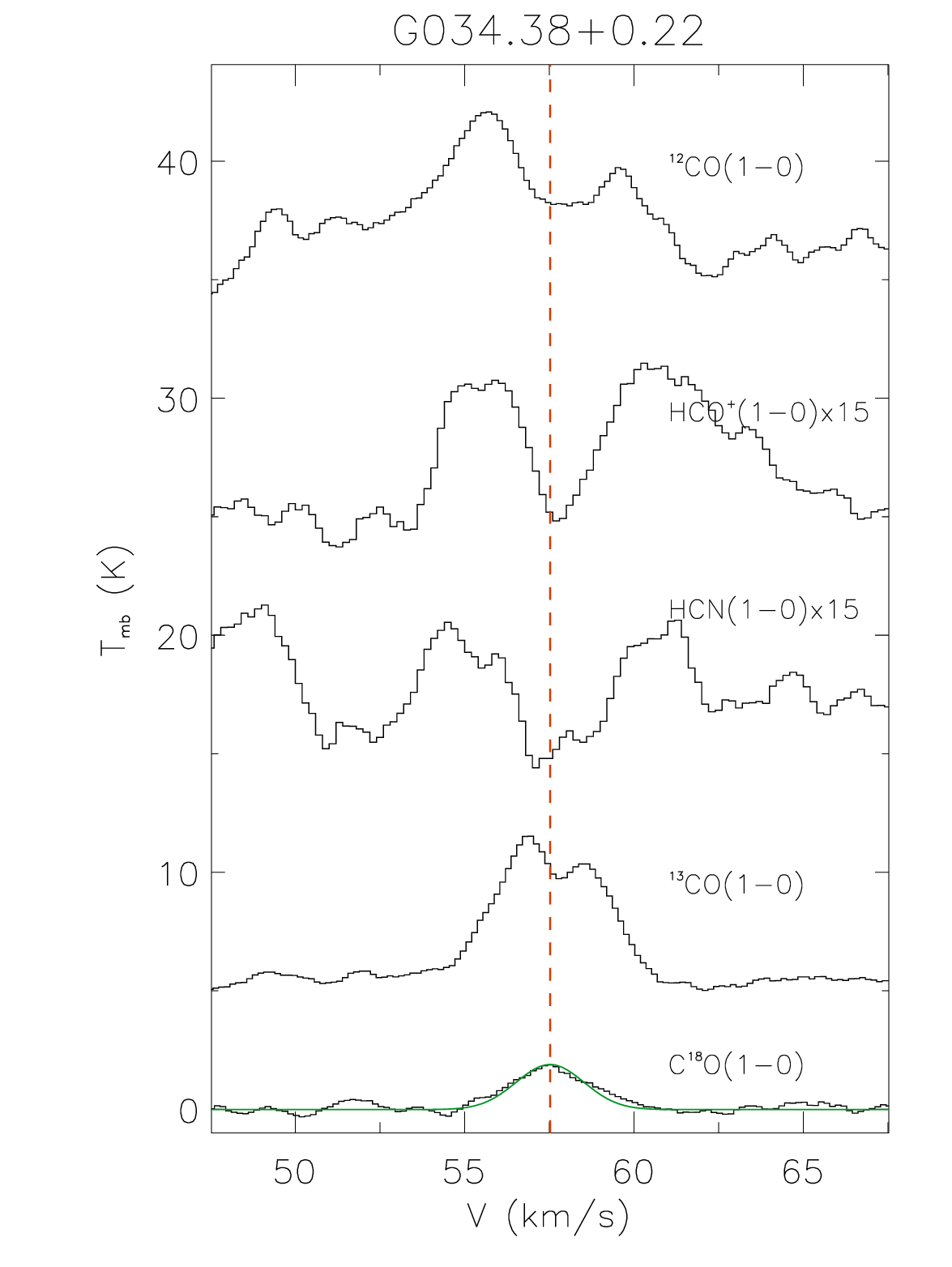}
  \end{minipage}%
  \begin{minipage}[t]{0.325\linewidth}
  \centering
   \includegraphics[width=55mm]{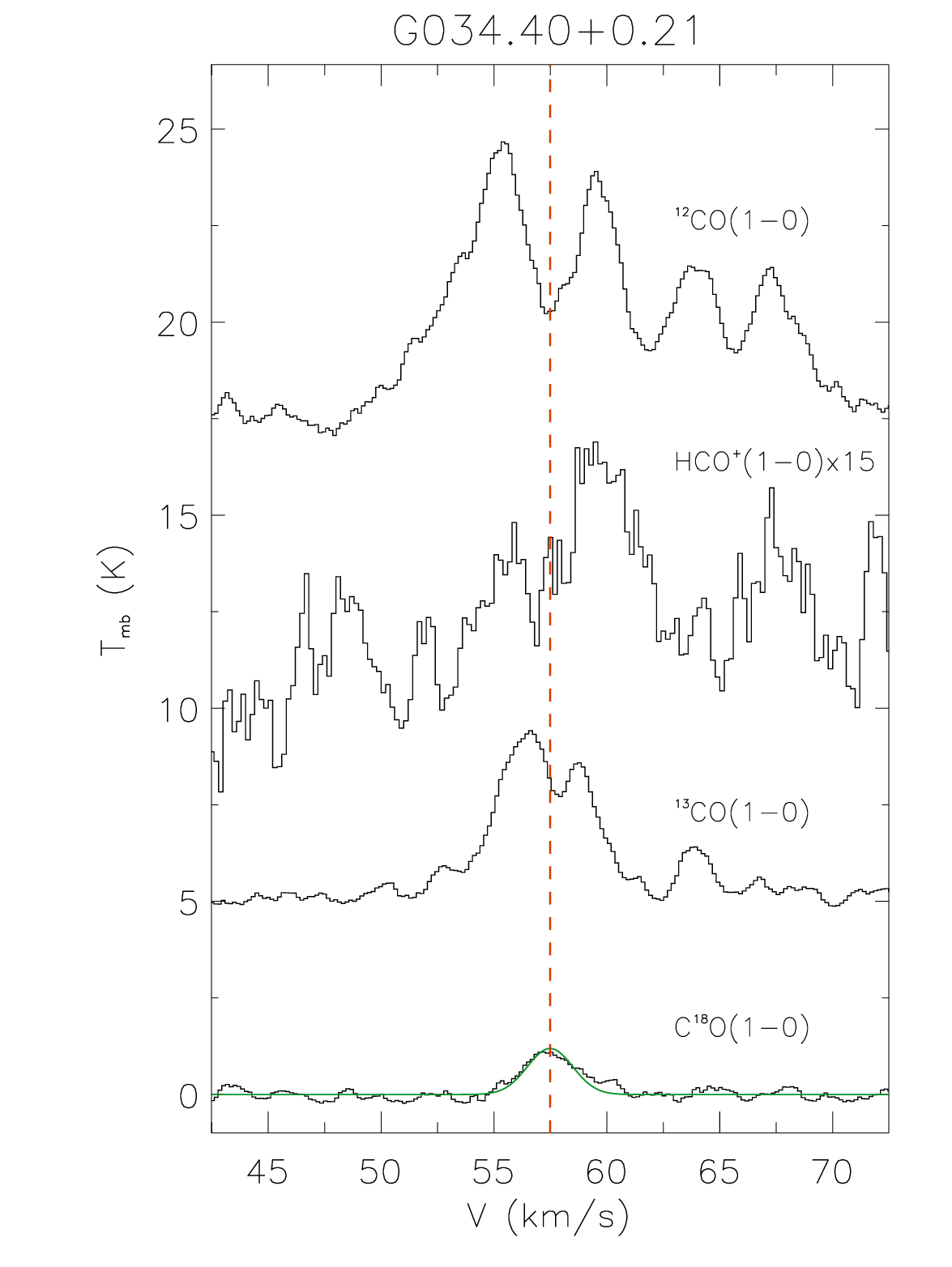}
  \end{minipage}%
  \begin{minipage}[t]{0.325\linewidth}
  \centering
   \includegraphics[width=55mm]{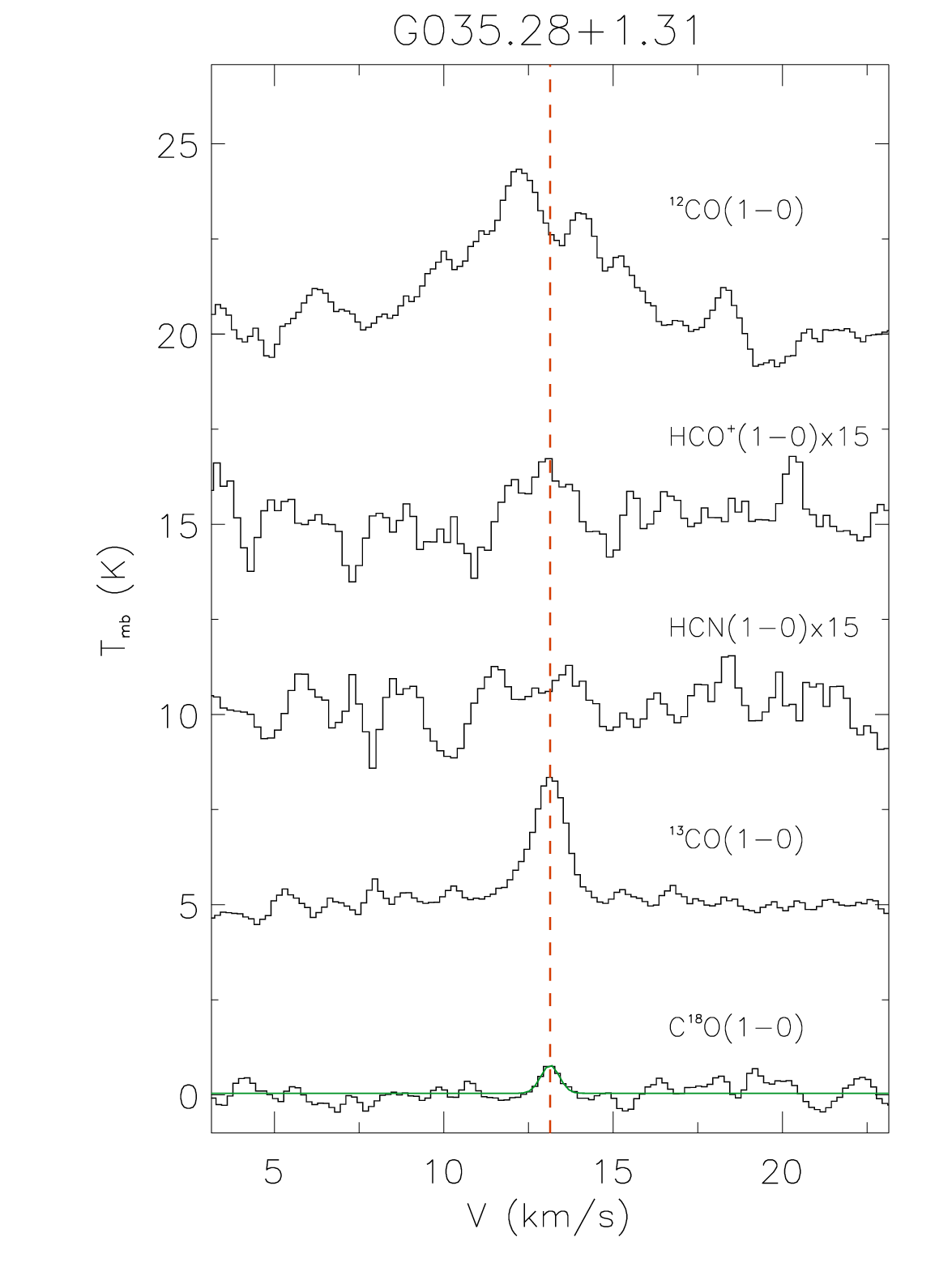}
  \end{minipage}%
  \caption{{\small Line profiles of 133 sources we selected. The lines from bottom to top are C$^{18}$O (1-0), $^{13}$CO (1-0), HCN (1-0) (14 sources lack HCN data), HCO$^+$ (1-0) and $^{12}$CO (1-0), respectively. The dashed red line indicates the central radial velocity of C$^{18}$O (1-0) estimated by Gaussian fitting. For infall candidates, HCO$^+$ (1-0) and HCN (1-0) lines are also Gaussian fitted.}}
  \label{Fig:fig6}
\end{figure} 
  
\begin{figure}[h]
\ContinuedFloat
  \begin{minipage}[t]{0.325\linewidth}
  \centering
   \includegraphics[width=55mm]{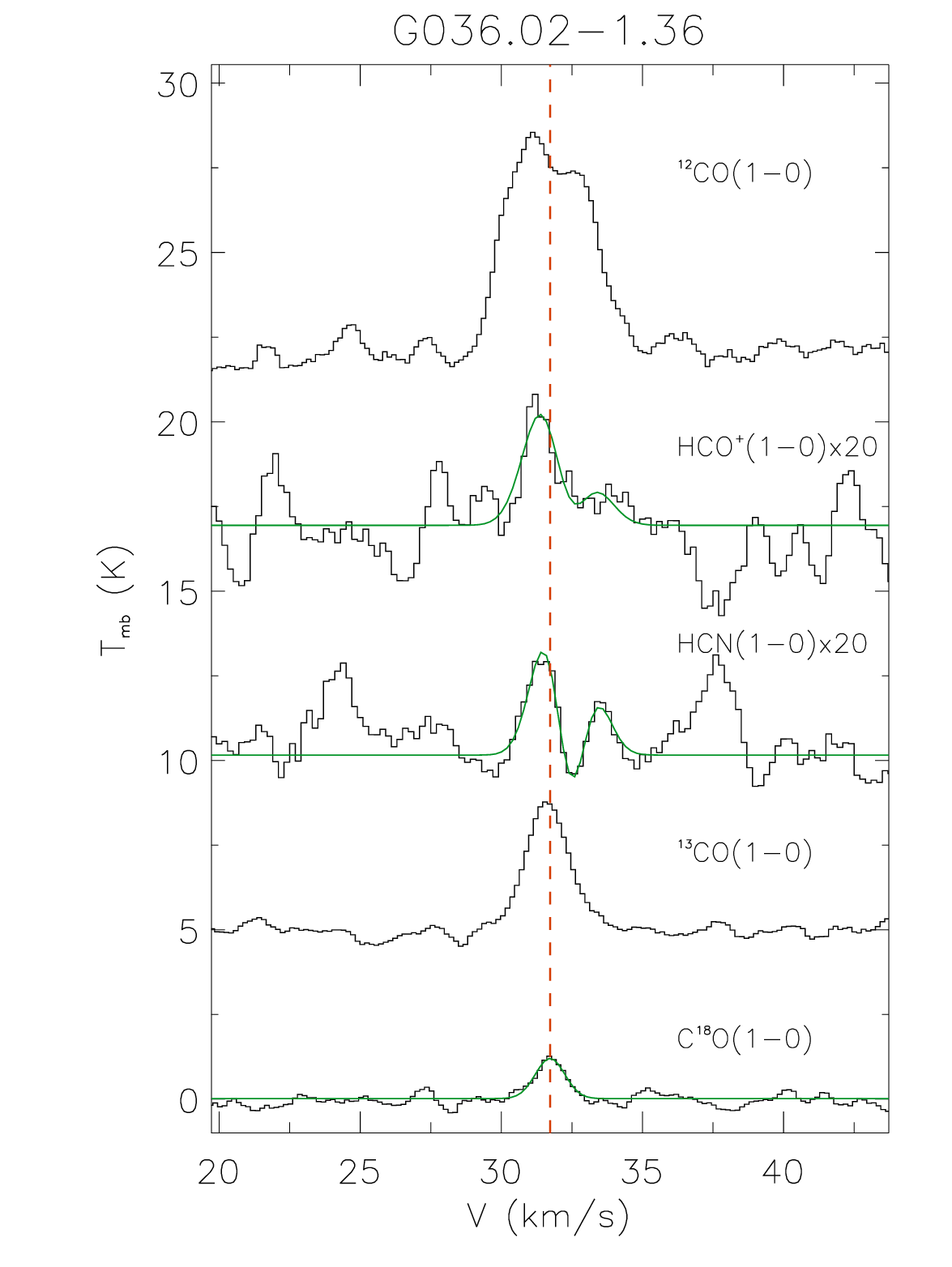}
  \end{minipage}%
  \begin{minipage}[t]{0.325\textwidth}
  \centering
   \includegraphics[width=55mm]{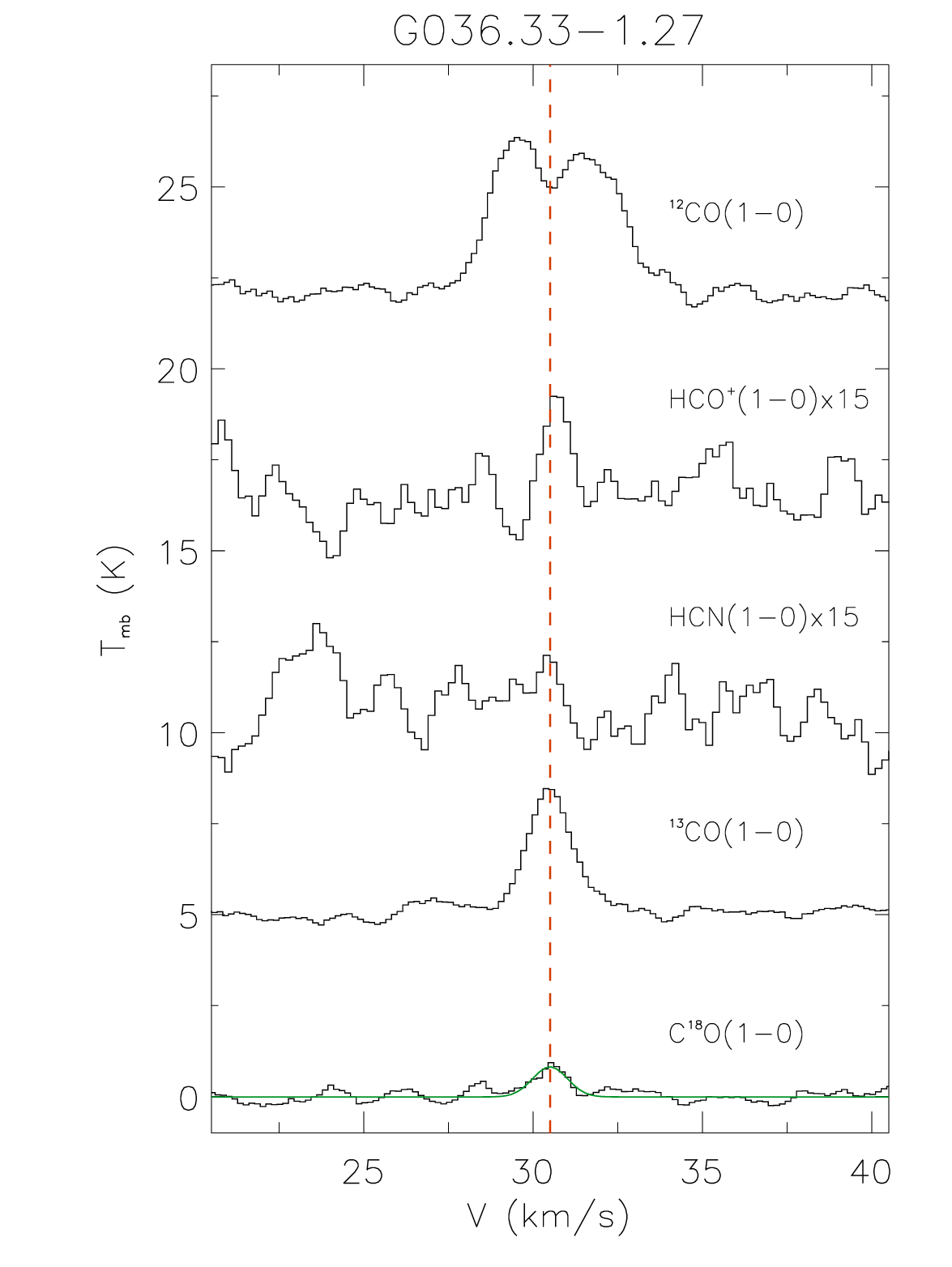}
  \end{minipage}%
  \begin{minipage}[t]{0.325\linewidth}
  \centering
   \includegraphics[width=55mm]{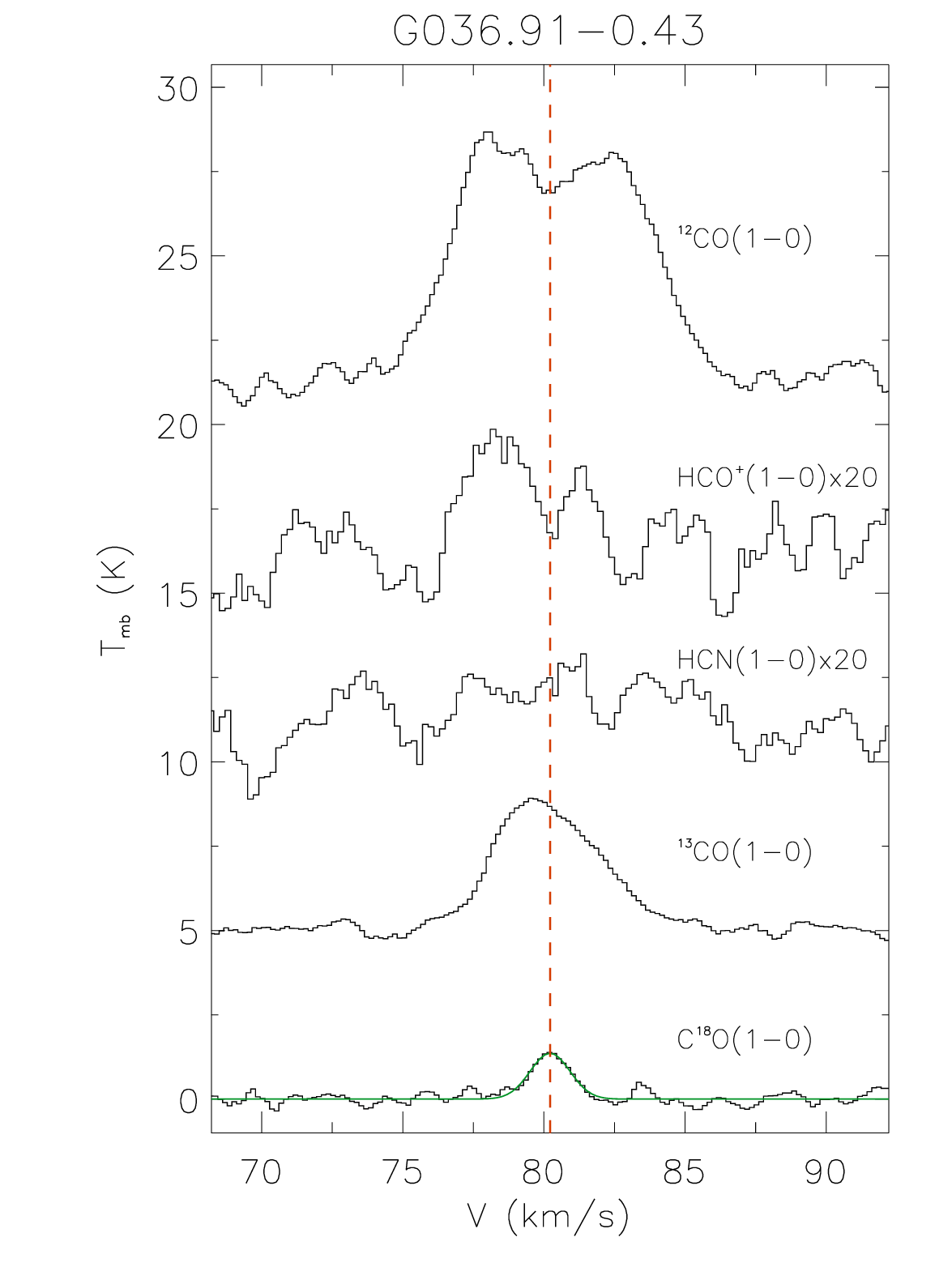}
  \end{minipage}%  
\quad
  \begin{minipage}[t]{0.325\linewidth}
  \centering
   \includegraphics[width=55mm]{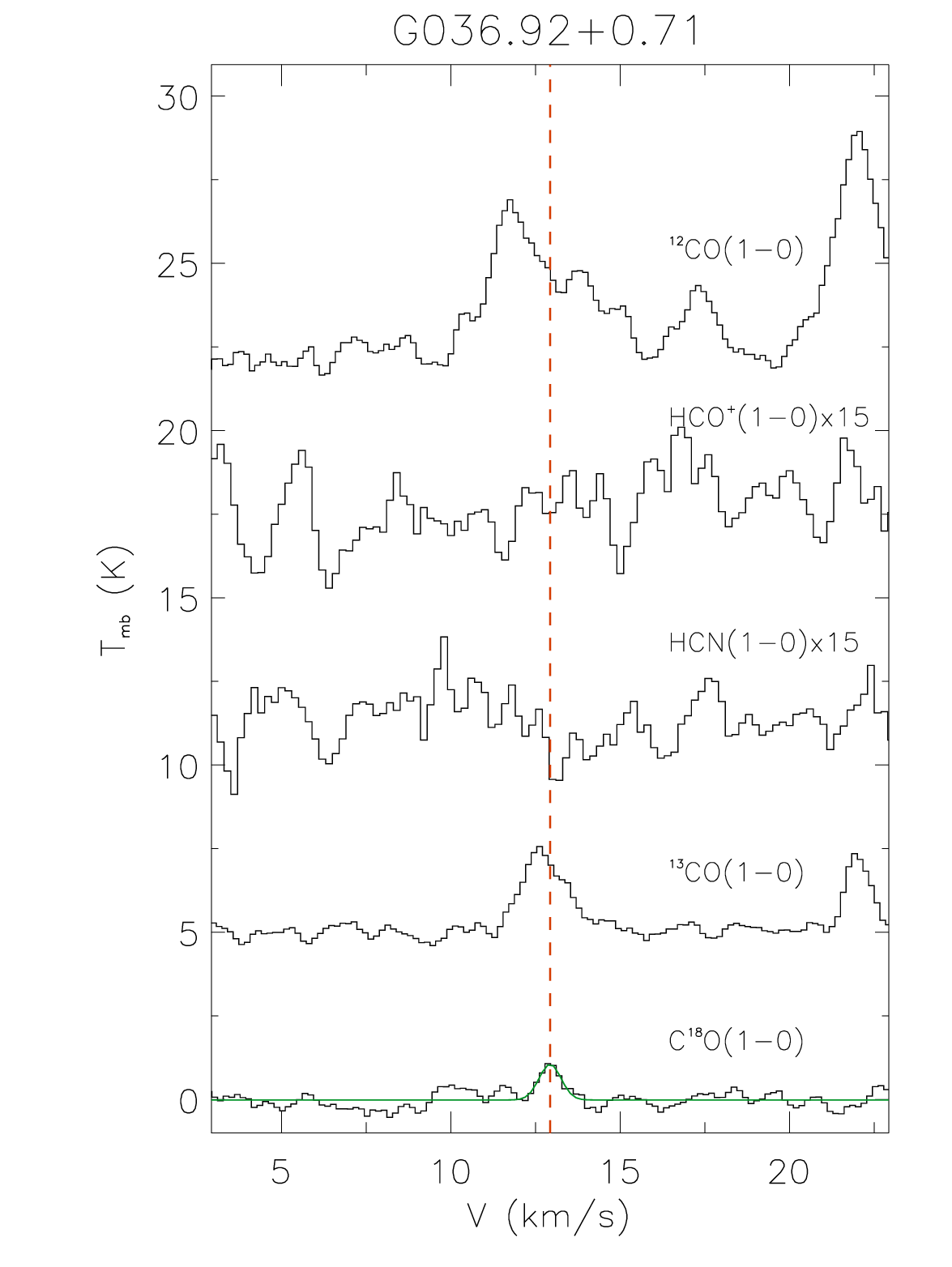}
  \end{minipage}%
  \begin{minipage}[t]{0.325\linewidth}
  \centering
   \includegraphics[width=55mm]{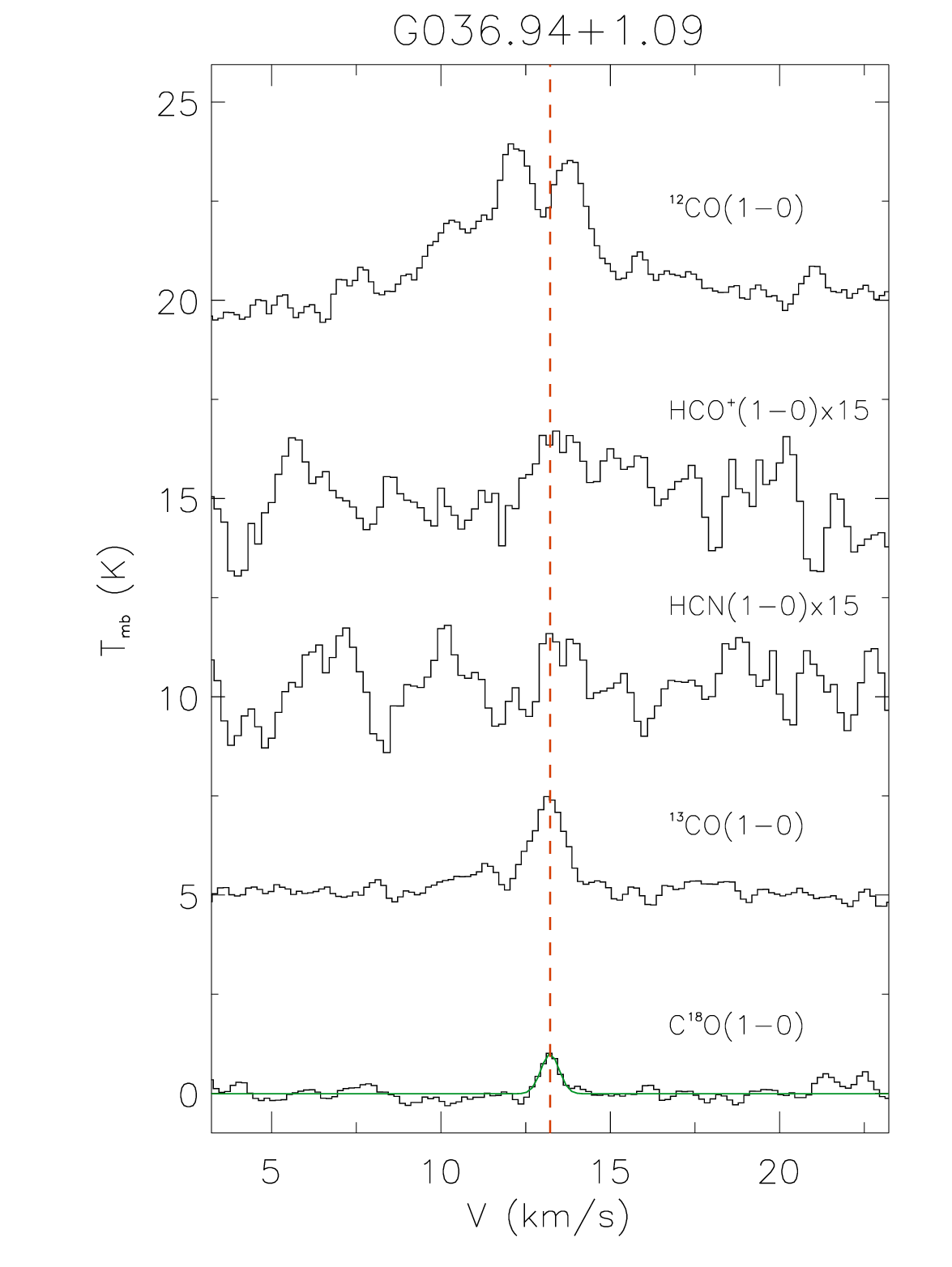}
  \end{minipage}%
  \begin{minipage}[t]{0.325\linewidth}
  \centering
   \includegraphics[width=55mm]{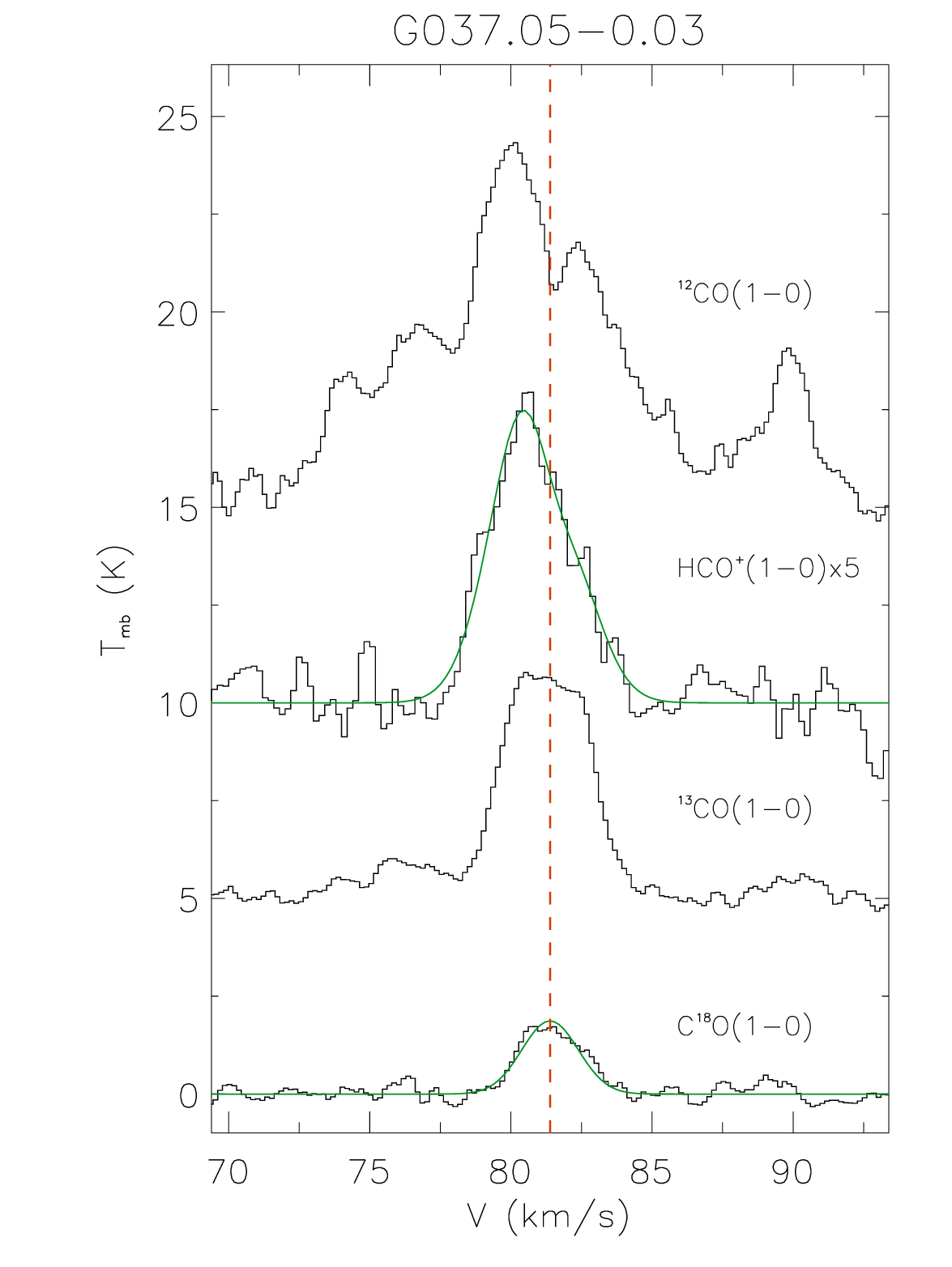}
  \end{minipage}%
\quad
  \begin{minipage}[t]{0.325\linewidth}
  \centering
   \includegraphics[width=55mm]{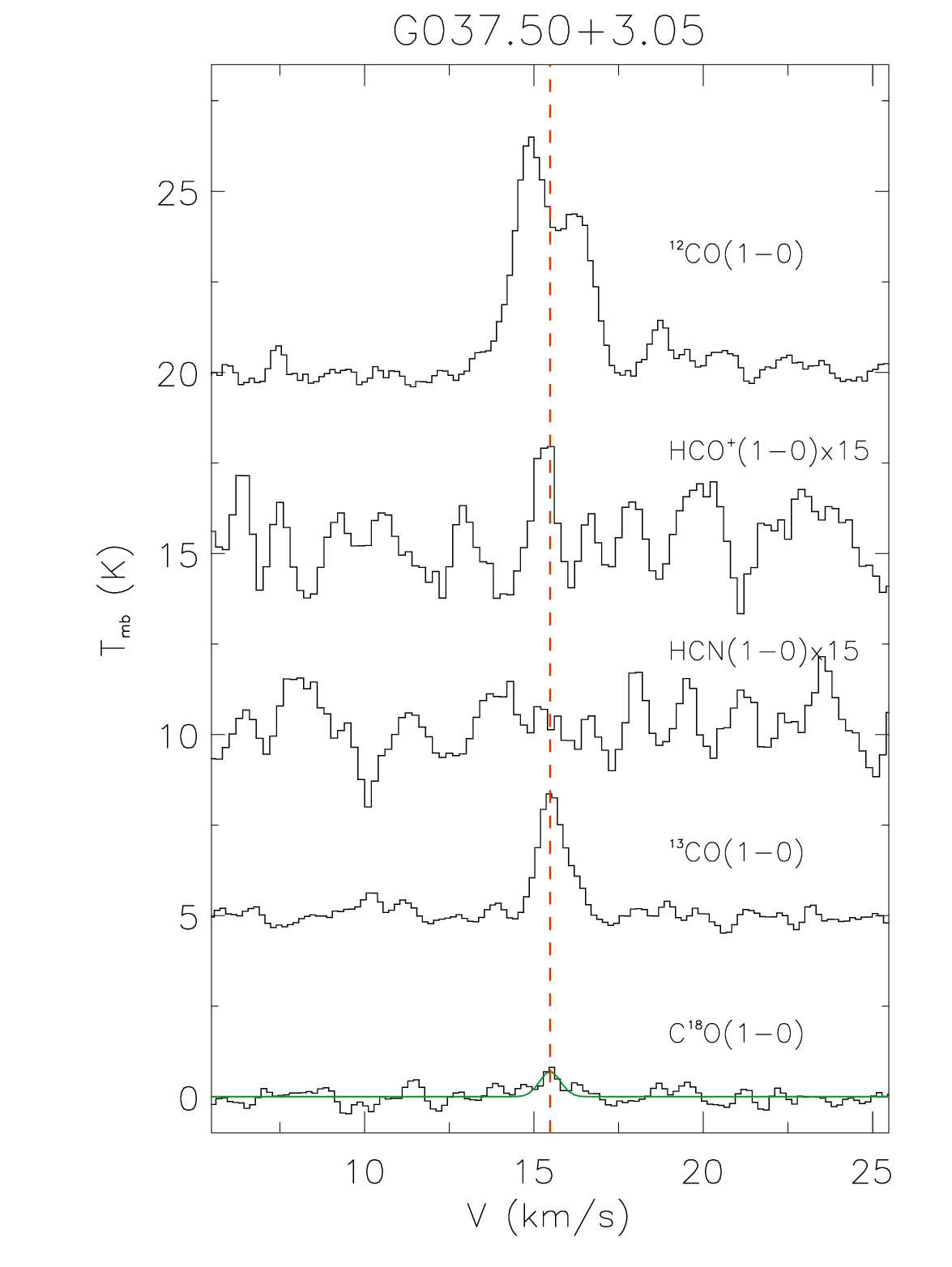}
  \end{minipage}%
  \begin{minipage}[t]{0.325\linewidth}
  \centering
   \includegraphics[width=55mm]{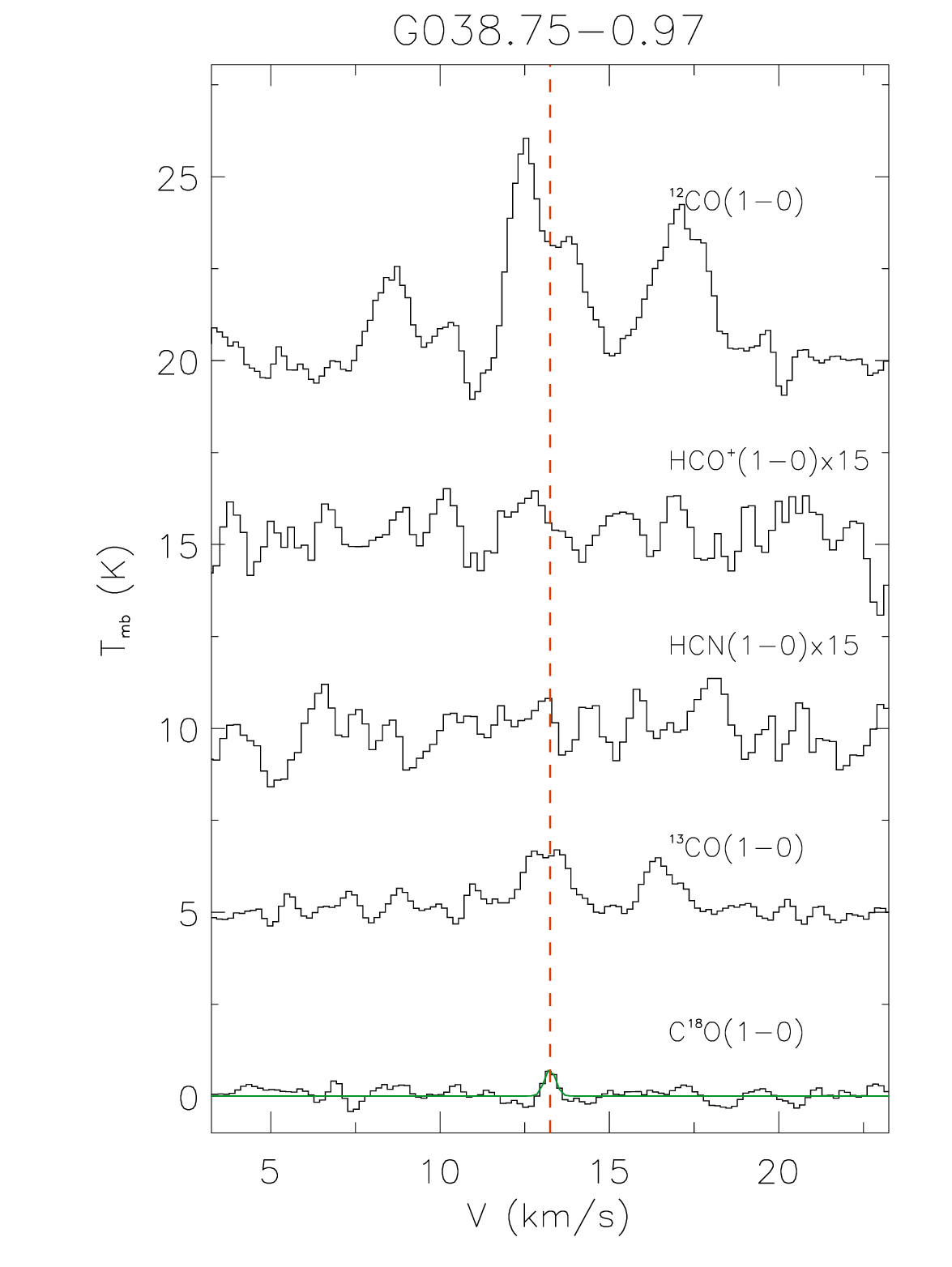}
  \end{minipage}%
  \begin{minipage}[t]{0.325\linewidth}
  \centering
   \includegraphics[width=55mm]{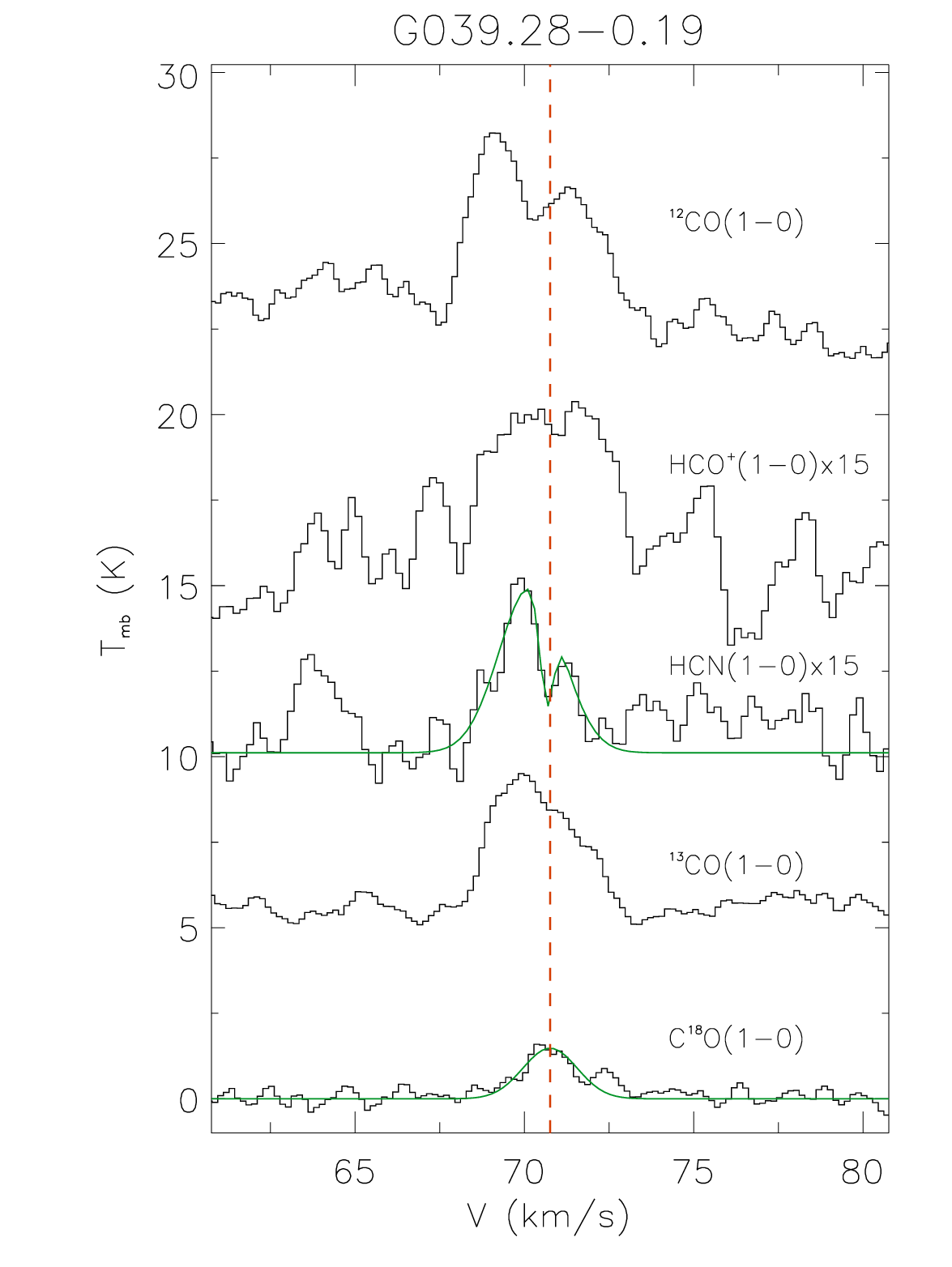}
  \end{minipage}%
  \caption{{\small Line profiles of 133 sources we selected. The lines from bottom to top are C$^{18}$O (1-0), $^{13}$CO (1-0), HCN (1-0) (14 sources lack HCN data), HCO$^+$ (1-0) and $^{12}$CO (1-0), respectively. The dashed red line indicates the central radial velocity of C$^{18}$O (1-0) estimated by Gaussian fitting. For infall candidates, HCO$^+$ (1-0) and HCN (1-0) lines are also Gaussian fitted.}}
  \label{Fig:fig6}
\end{figure} 

\begin{figure}[h]
\ContinuedFloat
  \begin{minipage}[t]{0.325\linewidth}
  \centering
   \includegraphics[width=55mm]{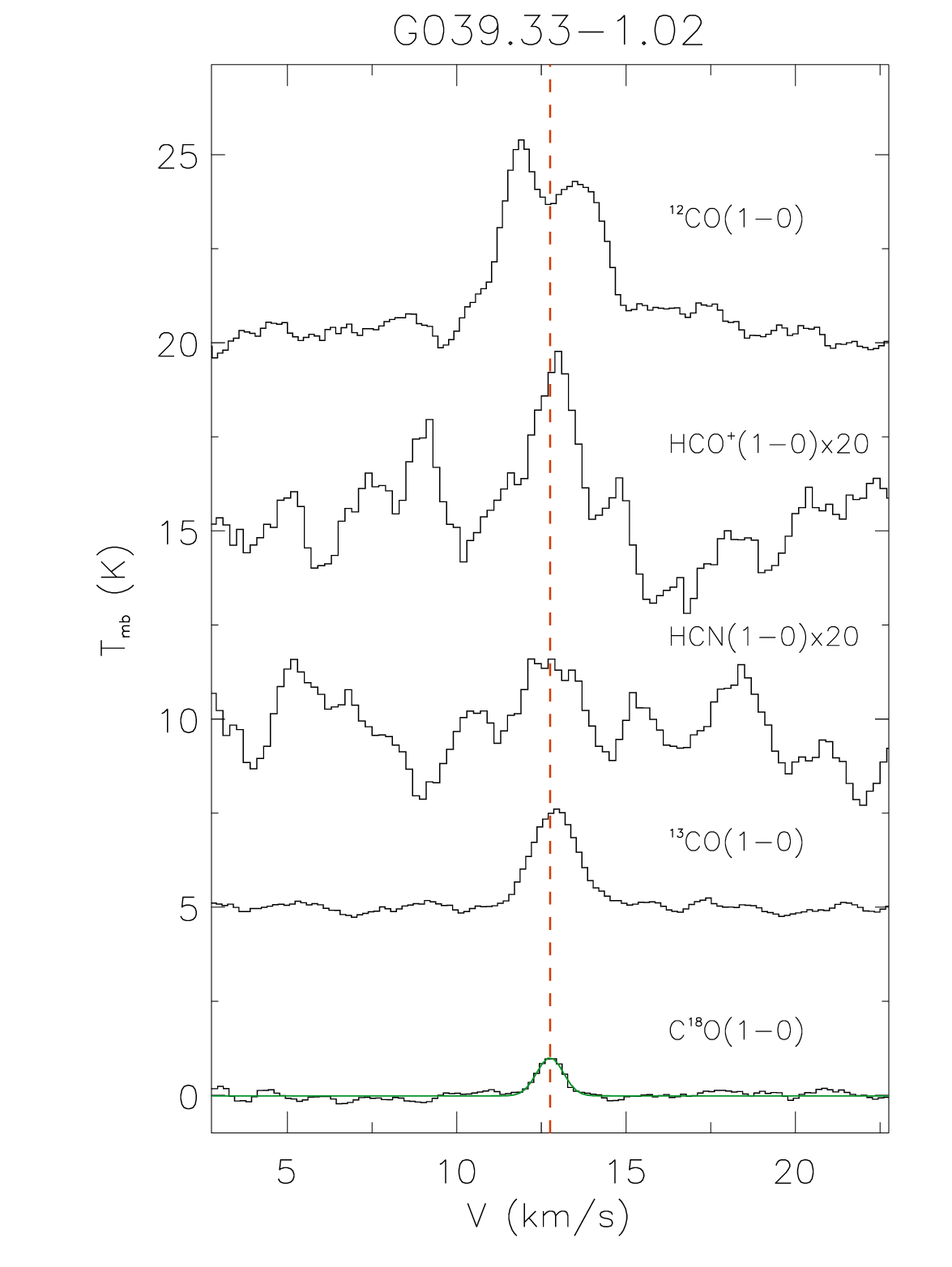}
  \end{minipage}%
  \begin{minipage}[t]{0.325\textwidth}
  \centering
   \includegraphics[width=55mm]{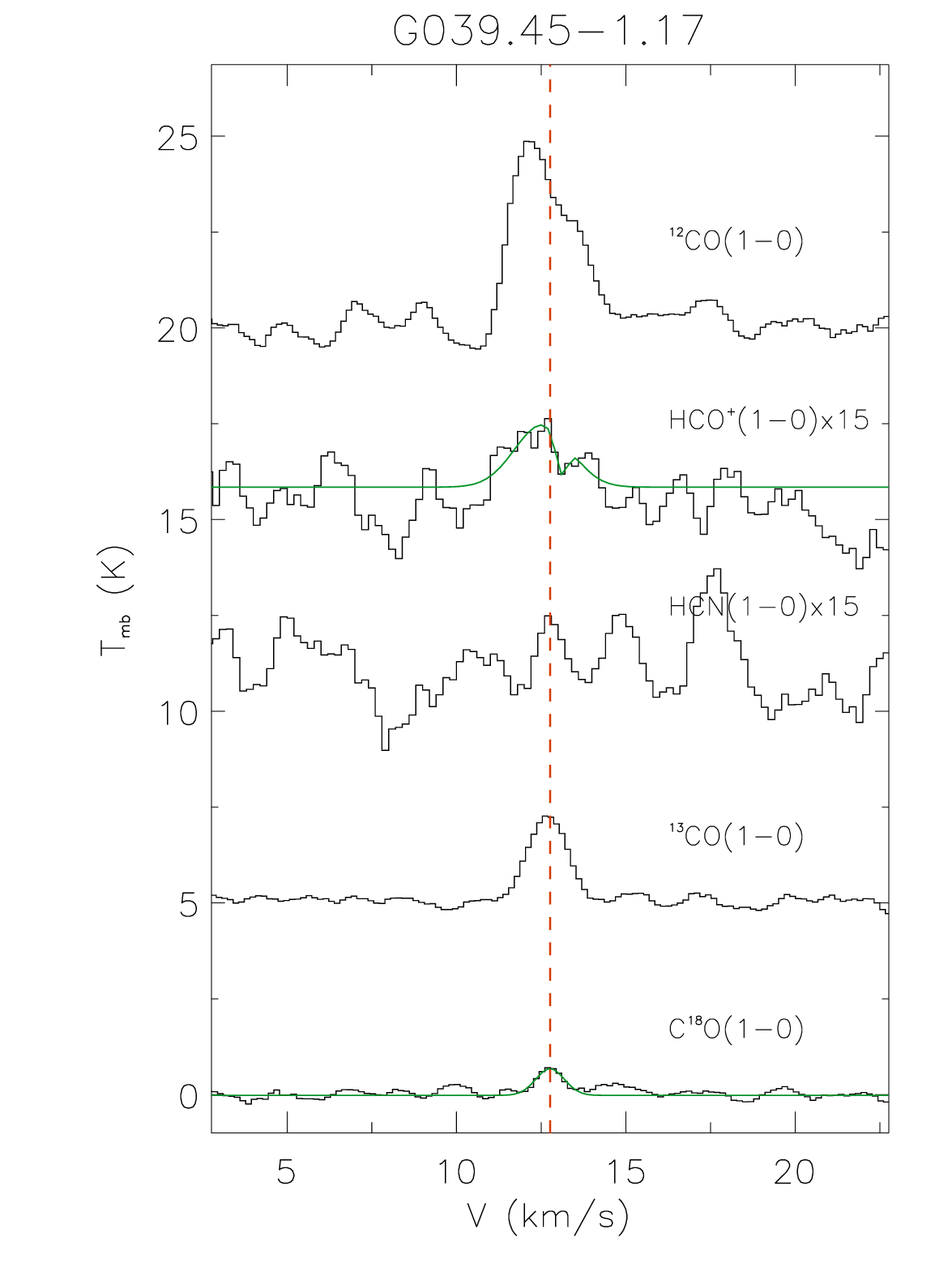}
  \end{minipage}%
  \begin{minipage}[t]{0.325\linewidth}
  \centering
   \includegraphics[width=55mm]{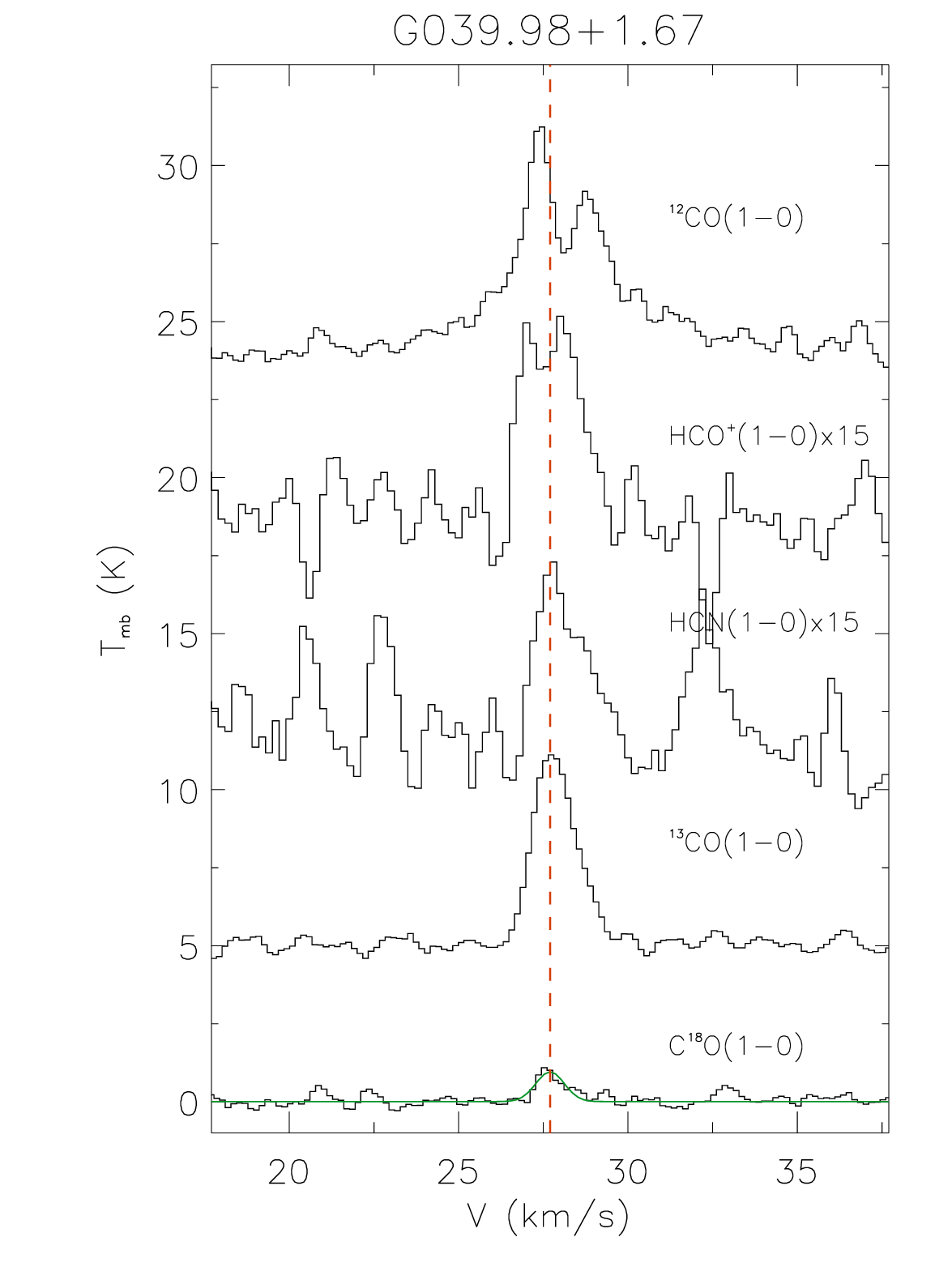}
  \end{minipage}%  
\quad
  \begin{minipage}[t]{0.325\linewidth}
  \centering
   \includegraphics[width=55mm]{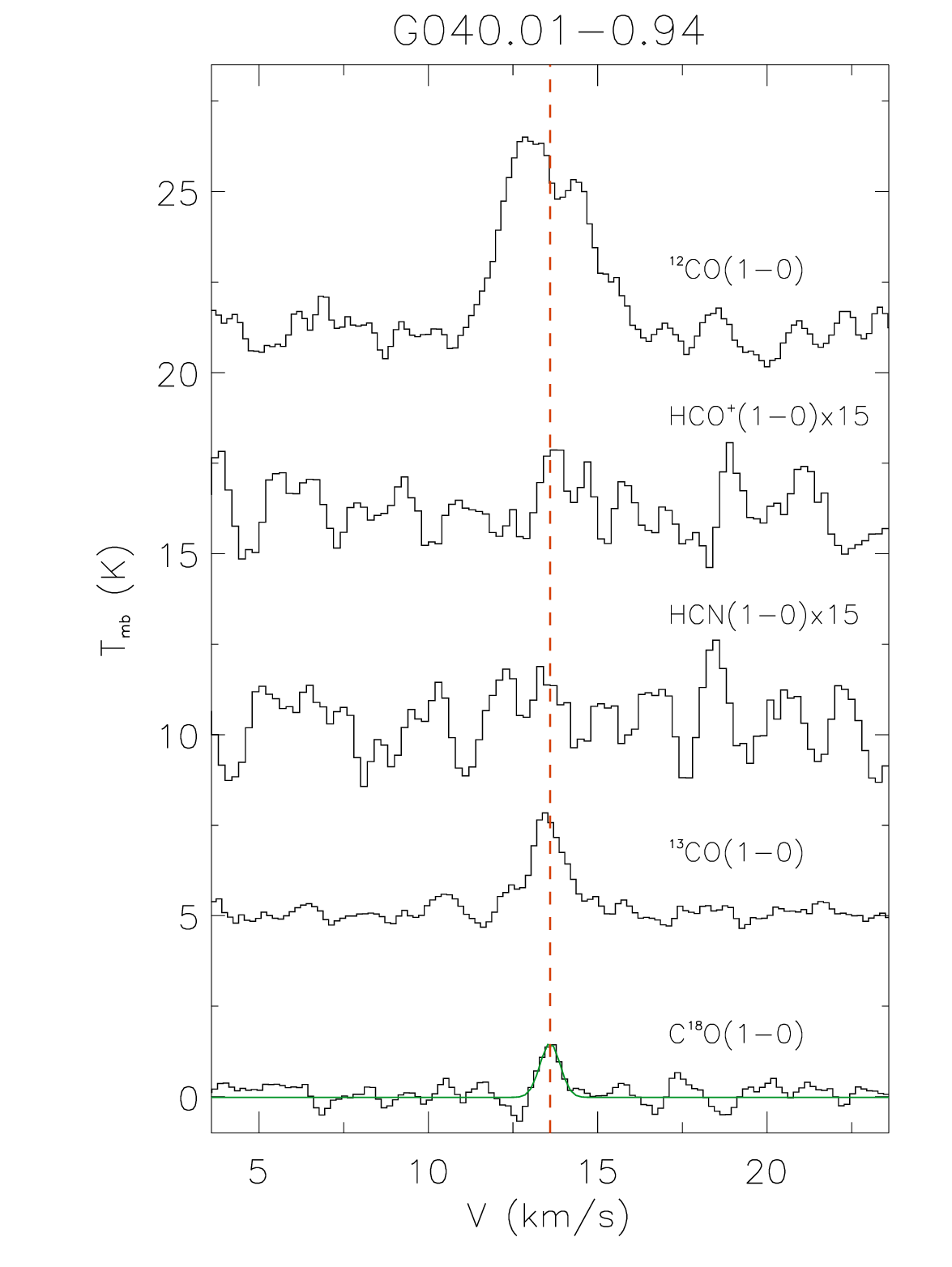}
  \end{minipage}%
  \begin{minipage}[t]{0.325\linewidth}
  \centering
   \includegraphics[width=55mm]{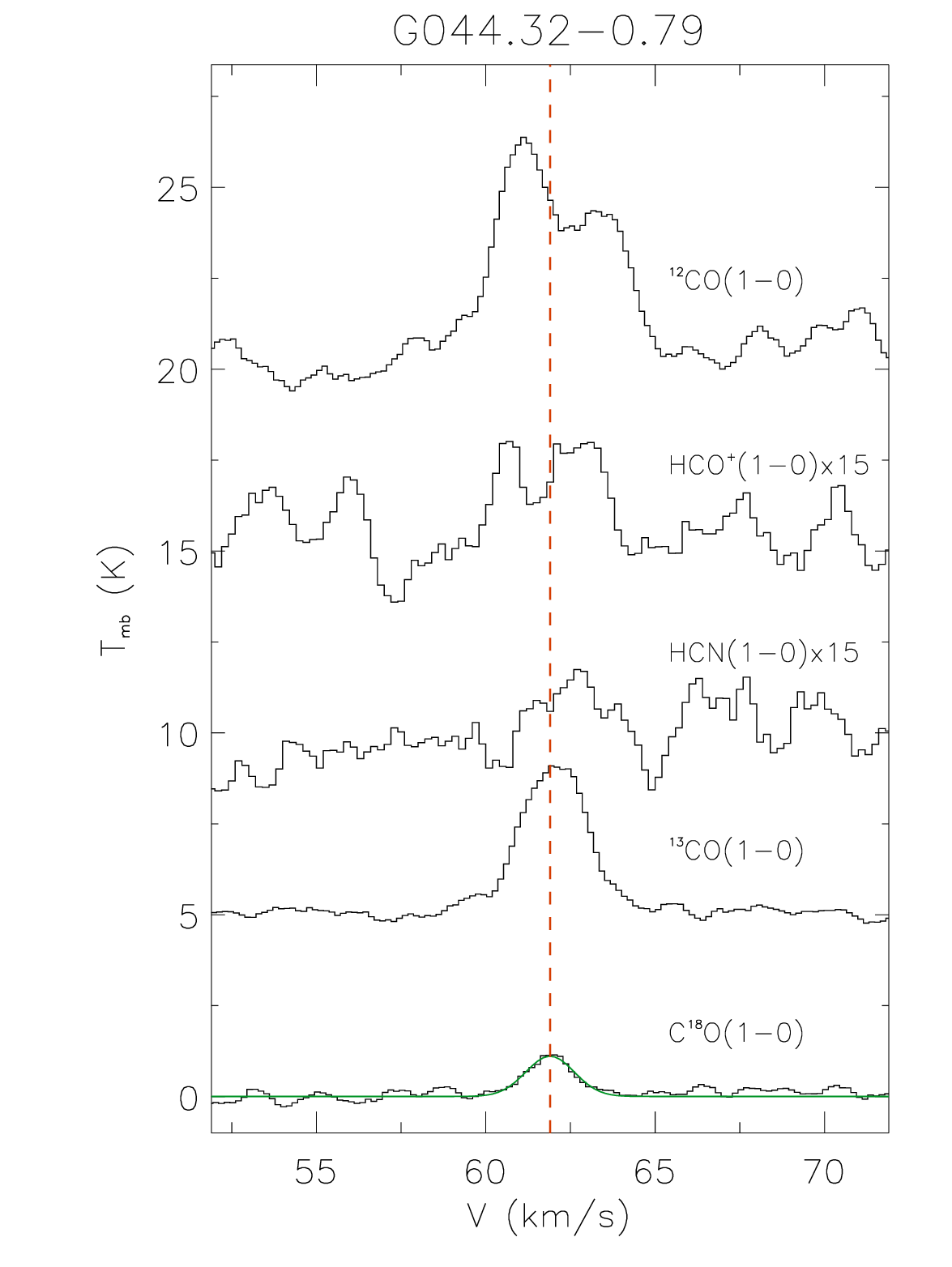}
  \end{minipage}%
  \begin{minipage}[t]{0.325\linewidth}
  \centering
   \includegraphics[width=55mm]{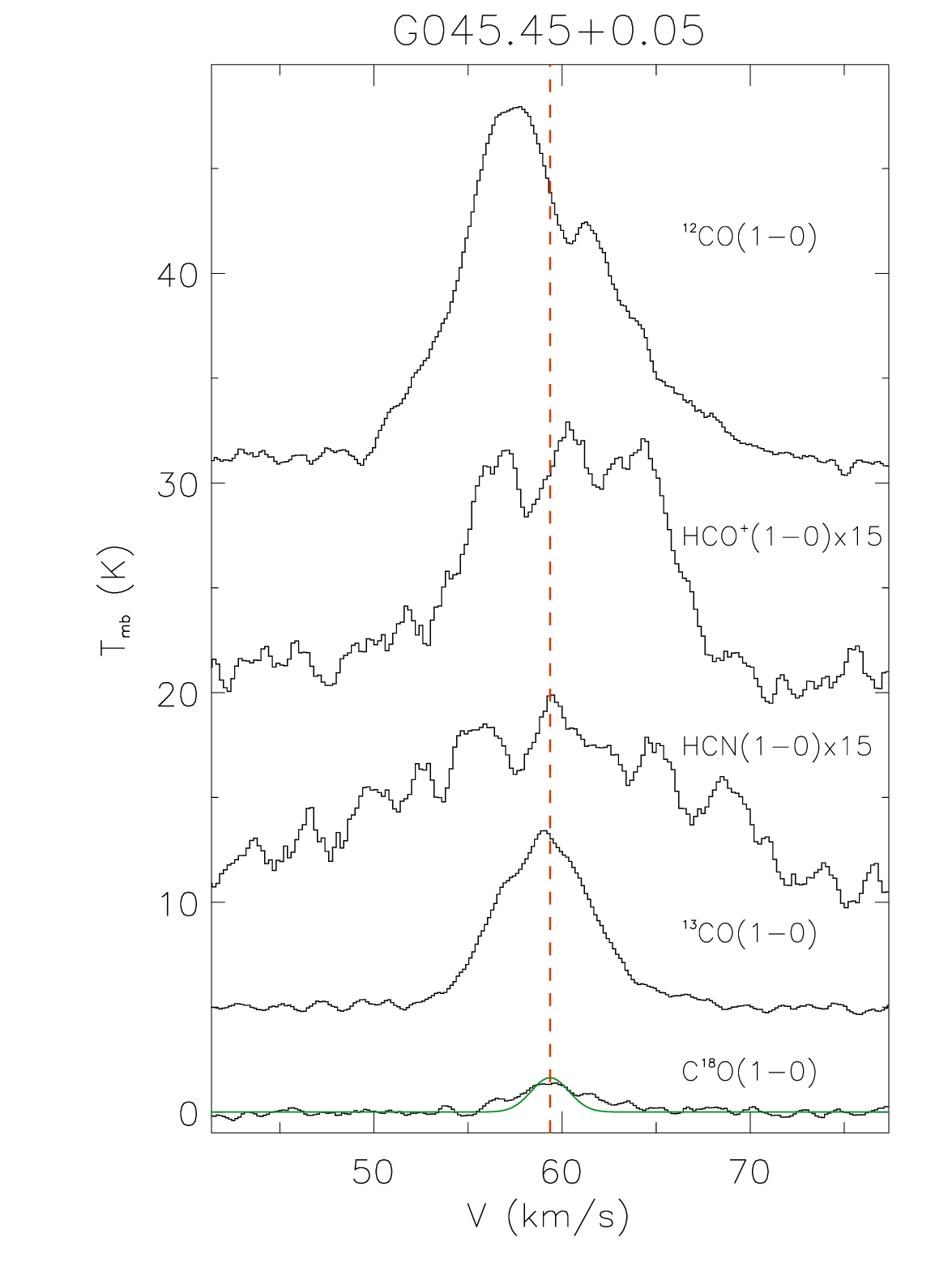}
  \end{minipage}%
\quad
  \begin{minipage}[t]{0.325\linewidth}
  \centering
   \includegraphics[width=55mm]{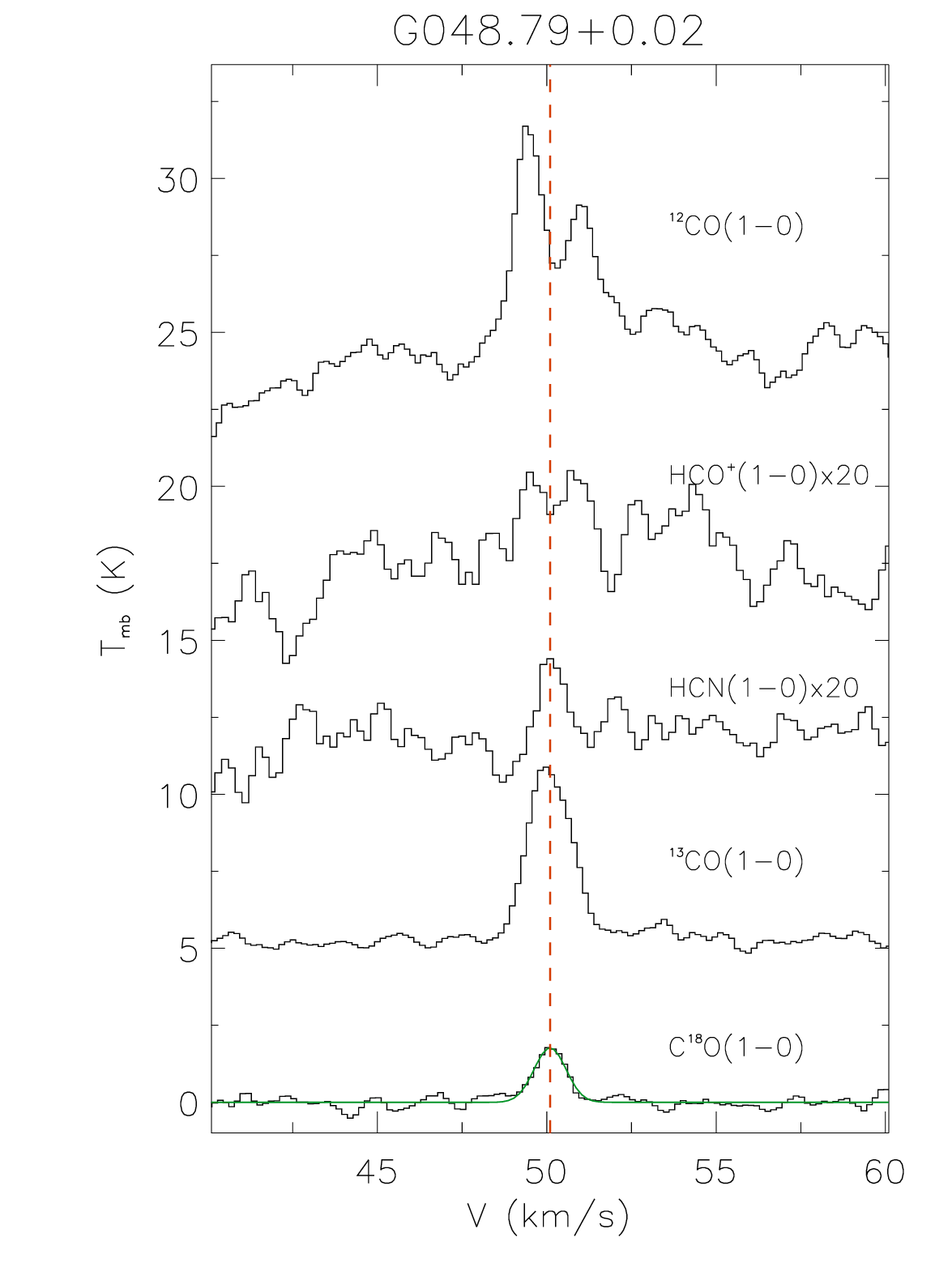}
  \end{minipage}%
  \begin{minipage}[t]{0.325\linewidth}
  \centering
   \includegraphics[width=55mm]{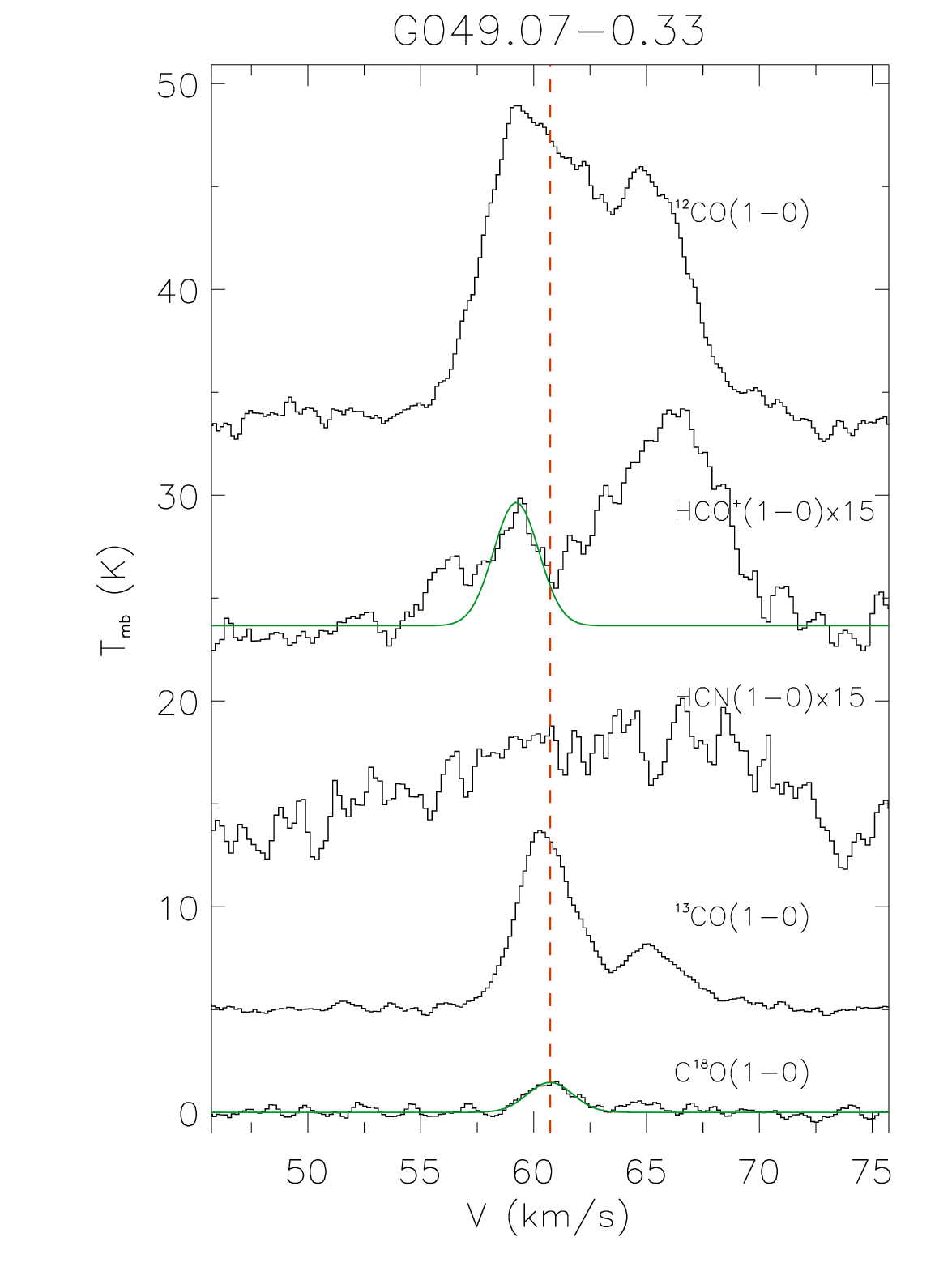}
  \end{minipage}%
  \begin{minipage}[t]{0.325\linewidth}
  \centering
   \includegraphics[width=55mm]{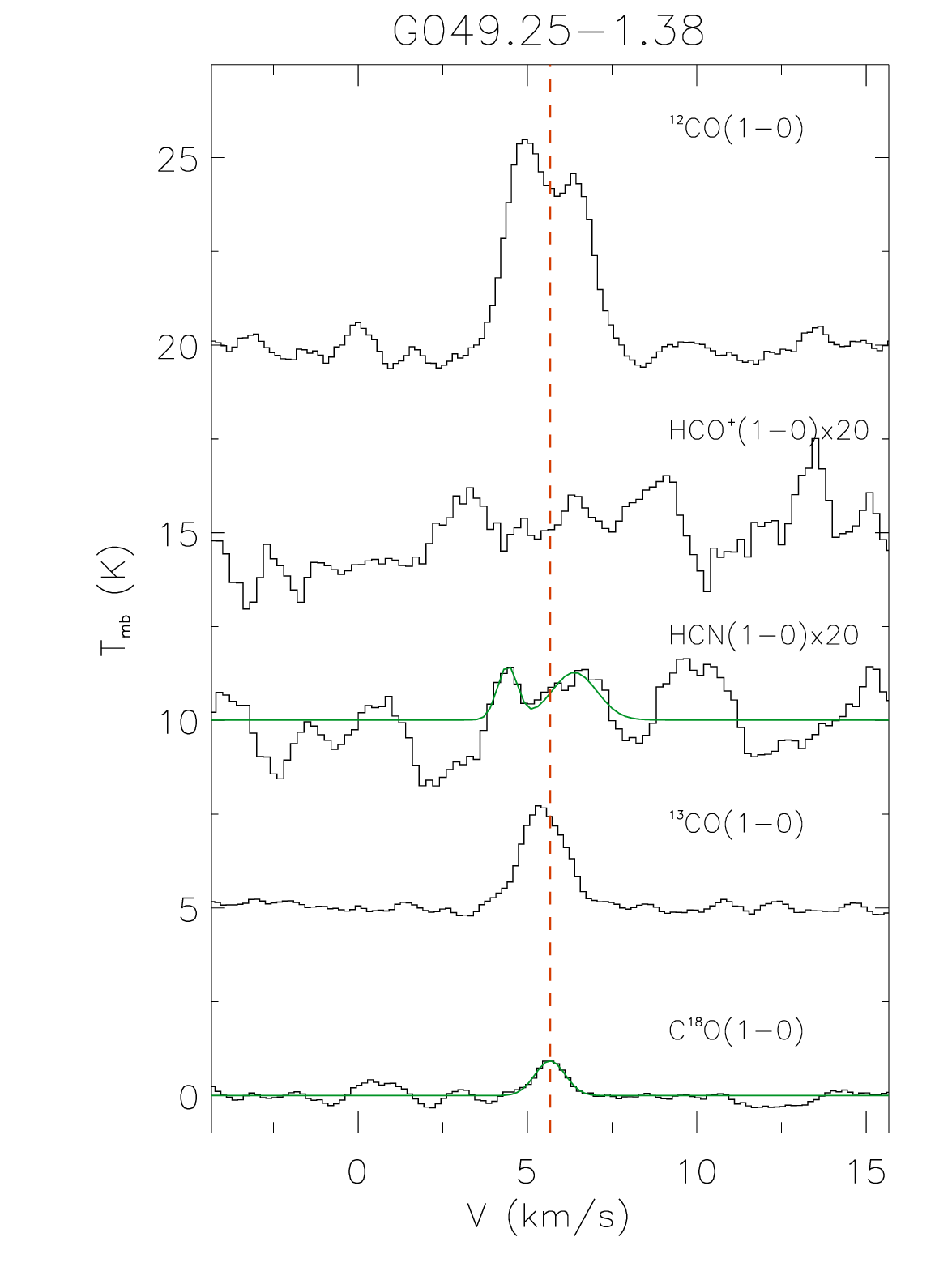}
  \end{minipage}%
  \caption{{\small Line profiles of 133 sources we selected. The lines from bottom to top are C$^{18}$O (1-0), $^{13}$CO (1-0), HCN (1-0) (14 sources lack HCN data), HCO$^+$ (1-0) and $^{12}$CO (1-0), respectively. The dashed red line indicates the central radial velocity of C$^{18}$O (1-0) estimated by Gaussian fitting. For infall candidates, HCO$^+$ (1-0) and HCN (1-0) lines are also Gaussian fitted.}}
  \label{Fig:fig6}
\end{figure} 

\begin{figure}[h]
\ContinuedFloat
  \begin{minipage}[t]{0.325\linewidth}
  \centering
   \includegraphics[width=55mm]{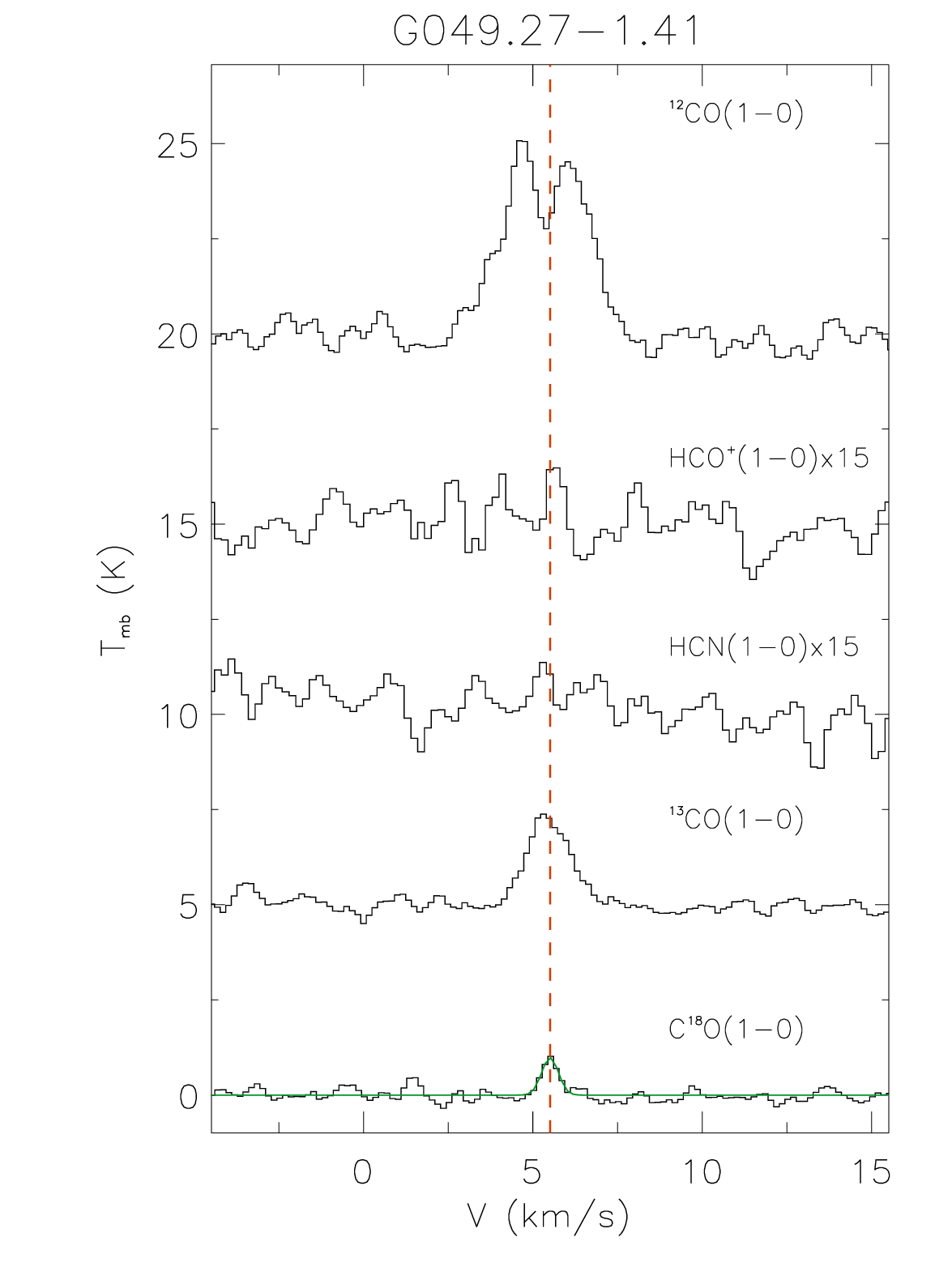}
  \end{minipage}%
  \begin{minipage}[t]{0.325\textwidth}
  \centering
   \includegraphics[width=55mm]{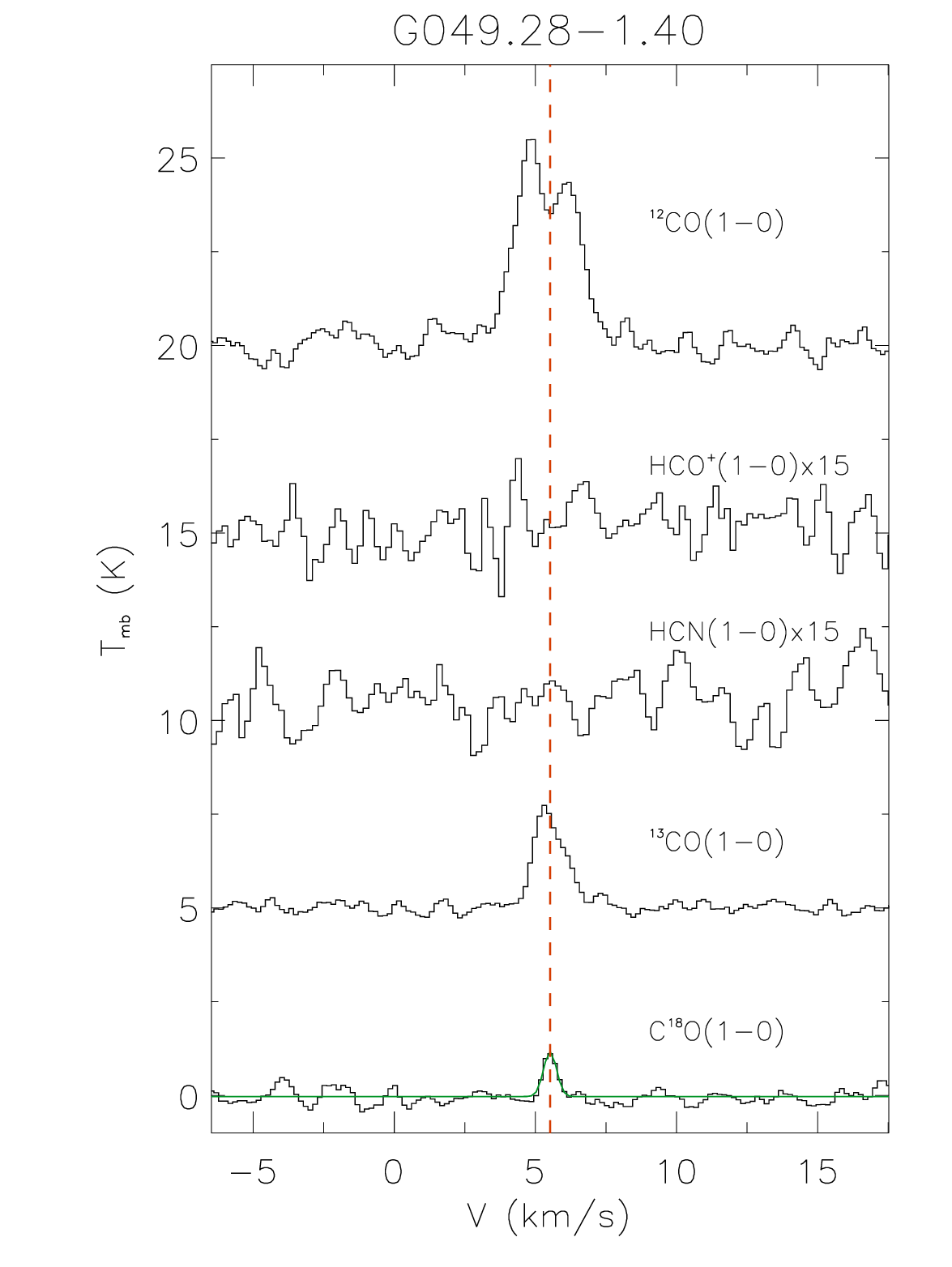}
  \end{minipage}%
  \begin{minipage}[t]{0.325\linewidth}
  \centering
   \includegraphics[width=55mm]{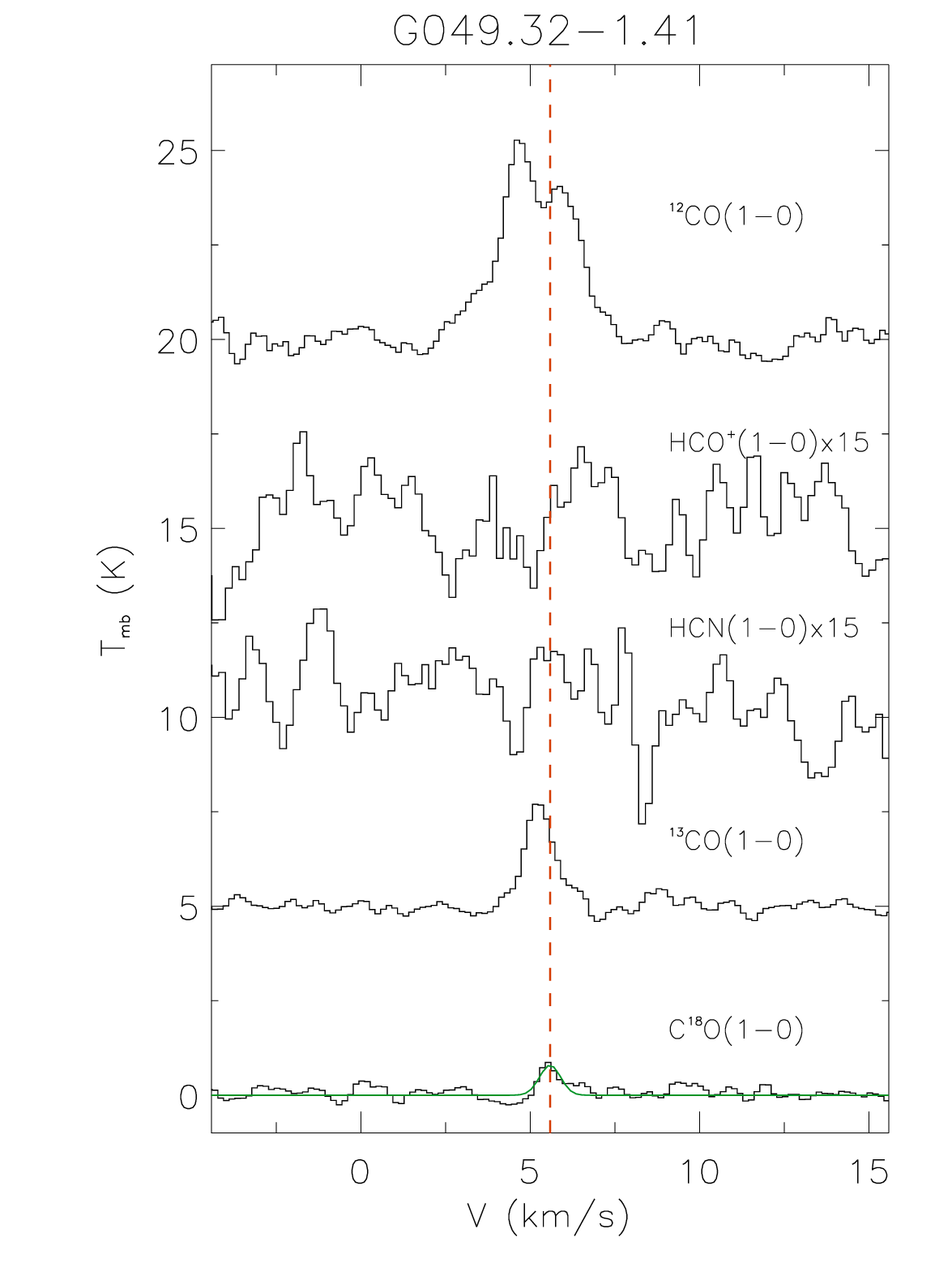}
  \end{minipage}%  
\quad
  \begin{minipage}[t]{0.325\linewidth}
  \centering
   \includegraphics[width=55mm]{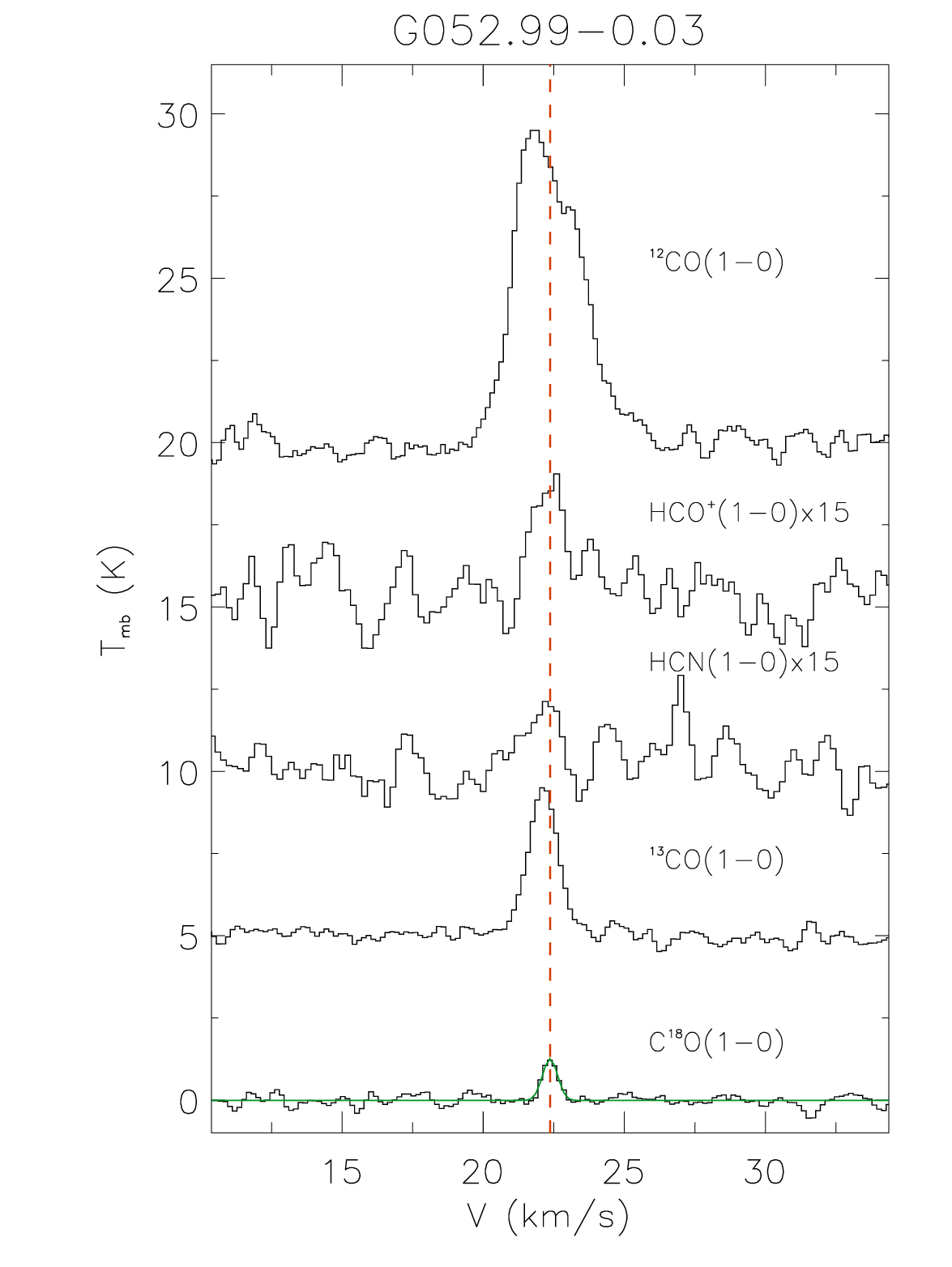}
  \end{minipage}%
  \begin{minipage}[t]{0.325\linewidth}
  \centering
   \includegraphics[width=55mm]{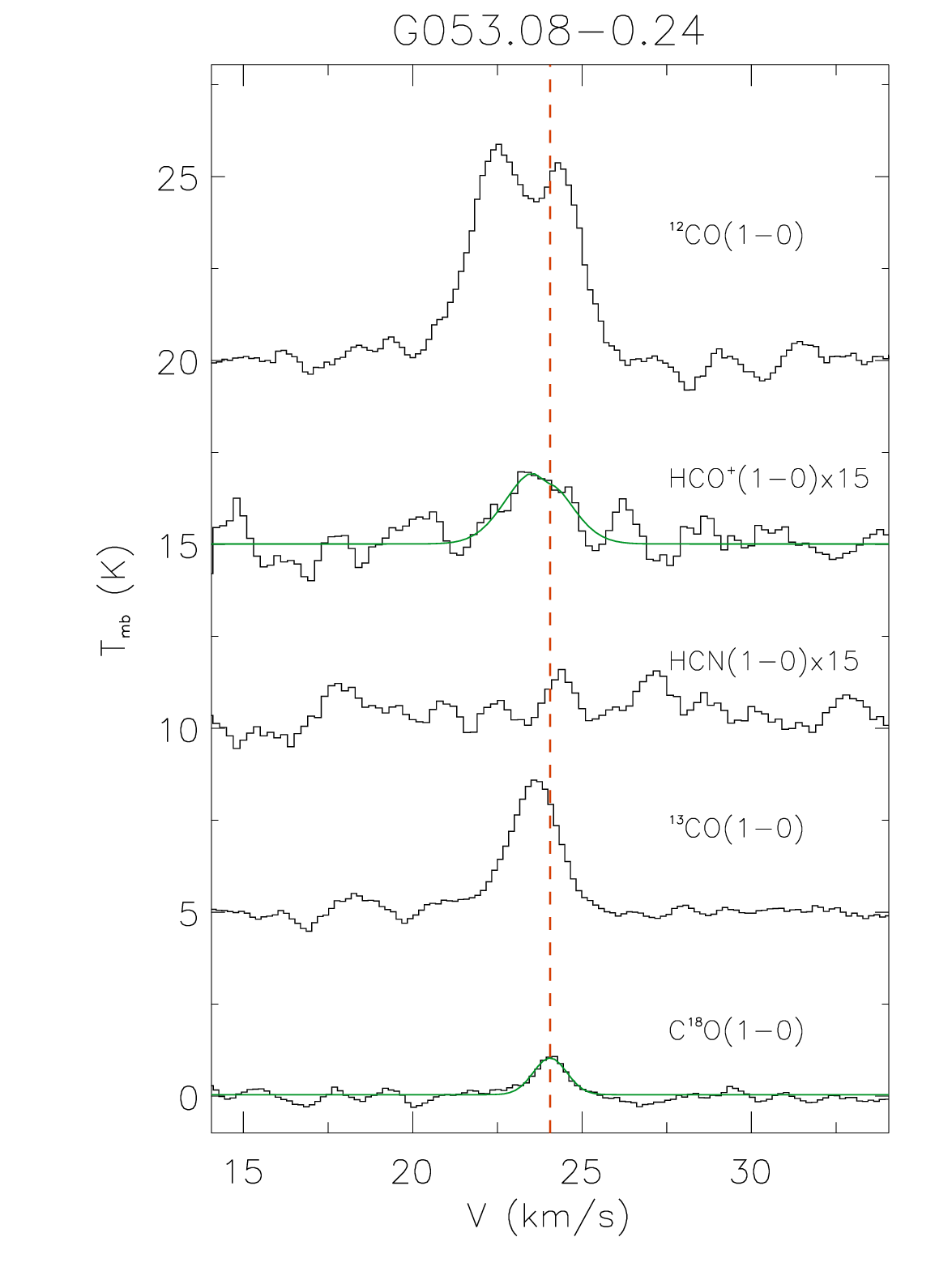}
  \end{minipage}%
  \begin{minipage}[t]{0.325\linewidth}
  \centering
   \includegraphics[width=55mm]{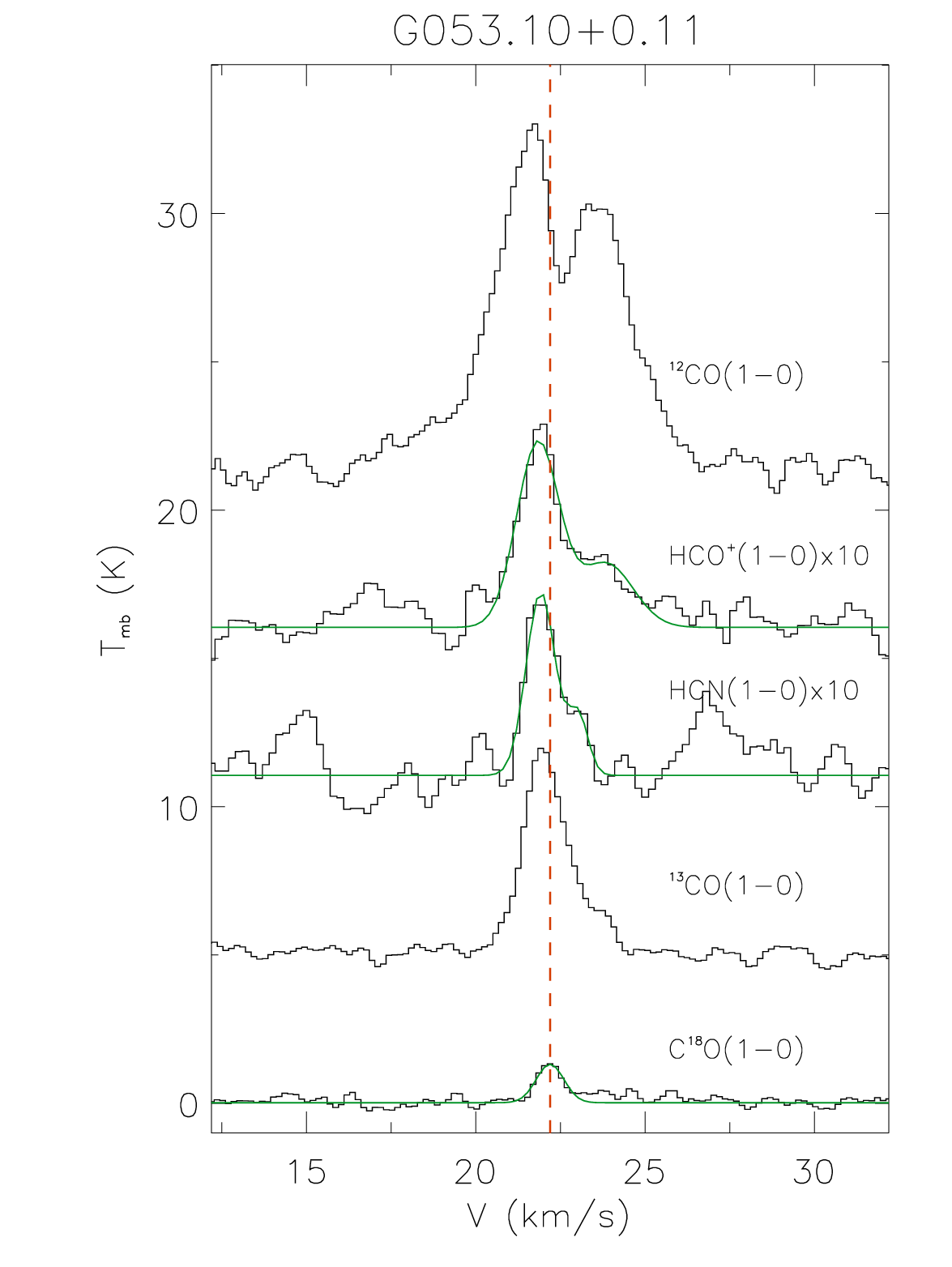}
  \end{minipage}%
\quad
  \begin{minipage}[t]{0.325\linewidth}
  \centering
   \includegraphics[width=55mm]{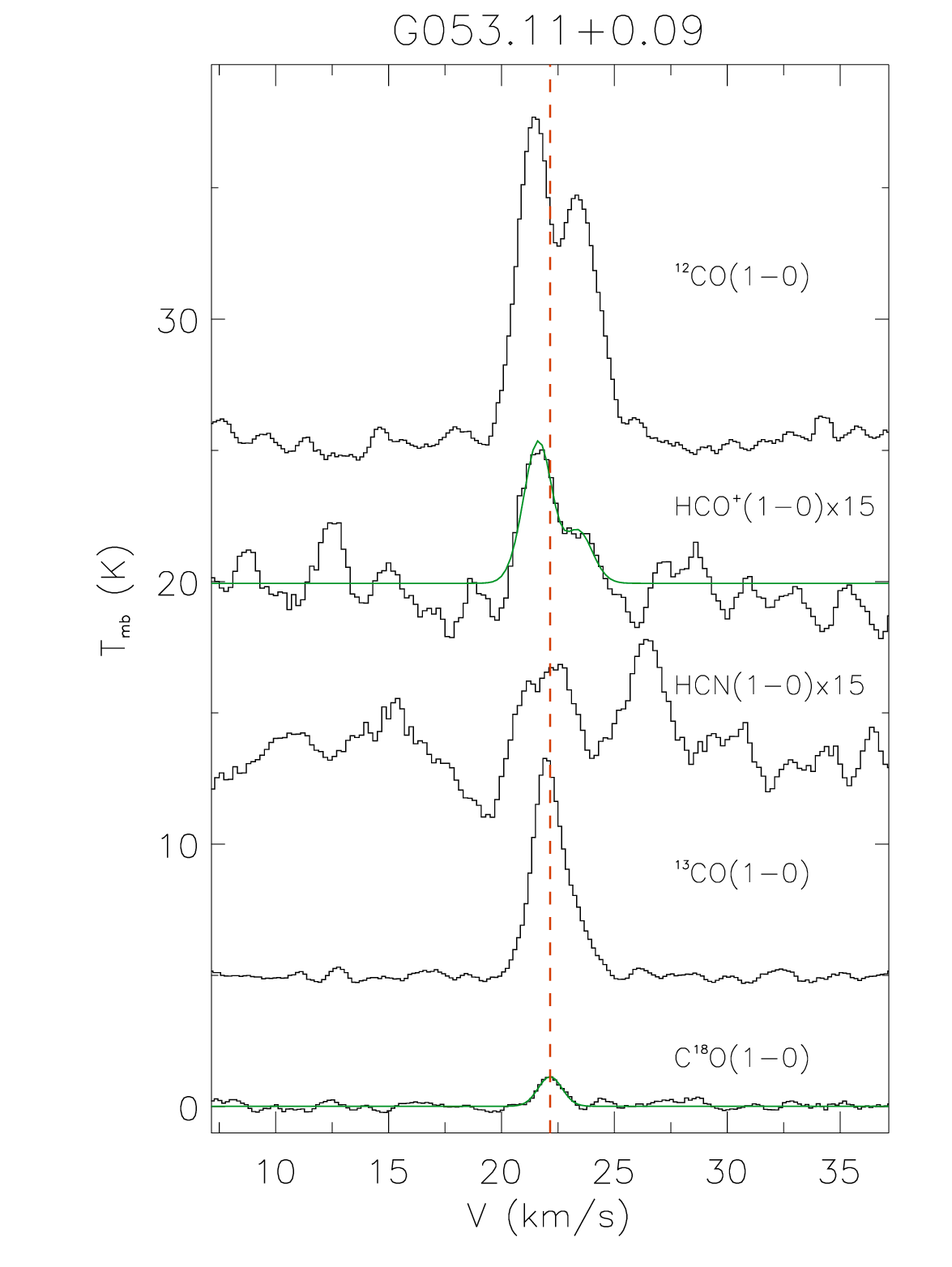}
  \end{minipage}%
  \begin{minipage}[t]{0.325\linewidth}
  \centering
   \includegraphics[width=55mm]{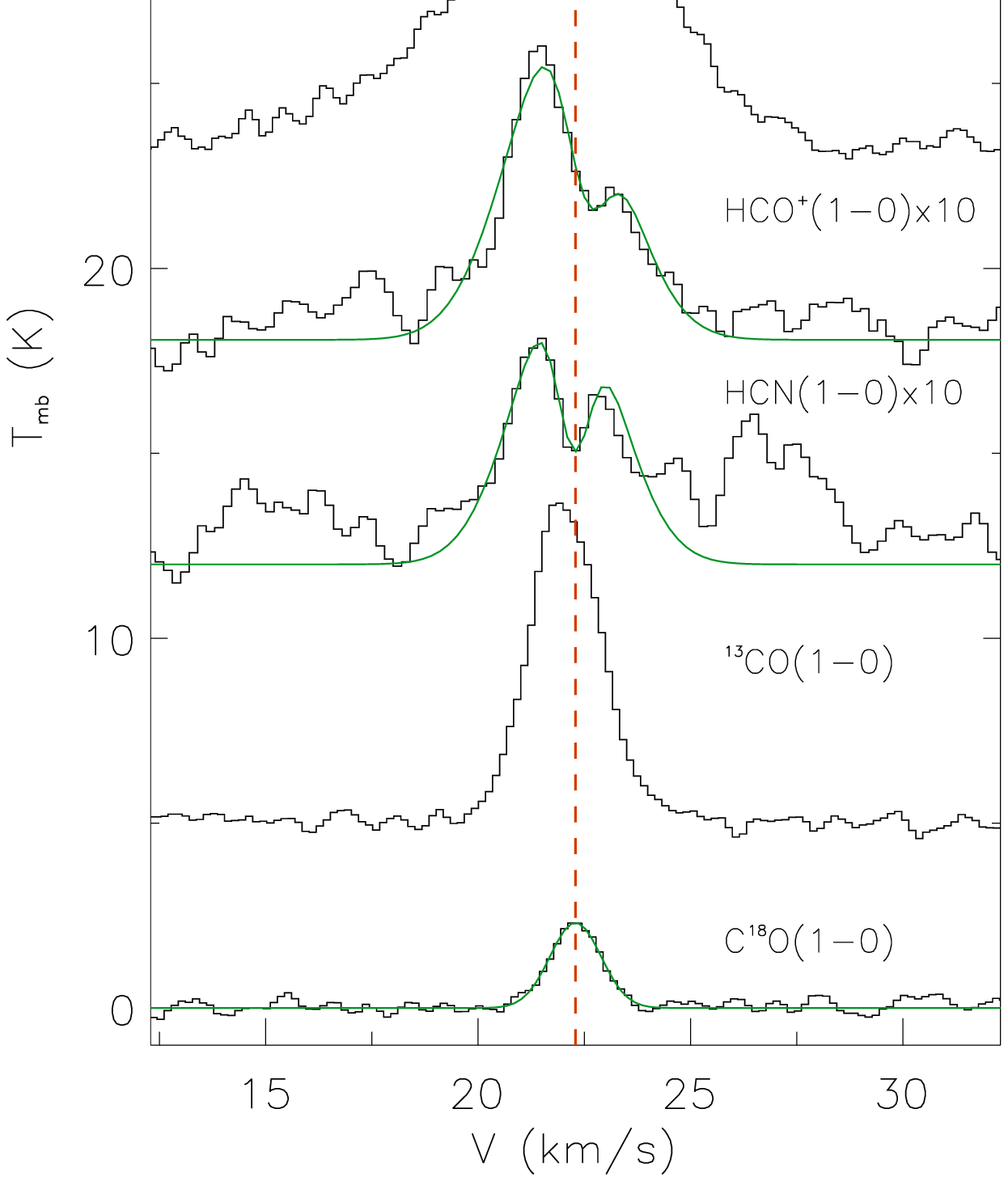}
  \end{minipage}%
  \begin{minipage}[t]{0.325\linewidth}
  \centering
   \includegraphics[width=55mm]{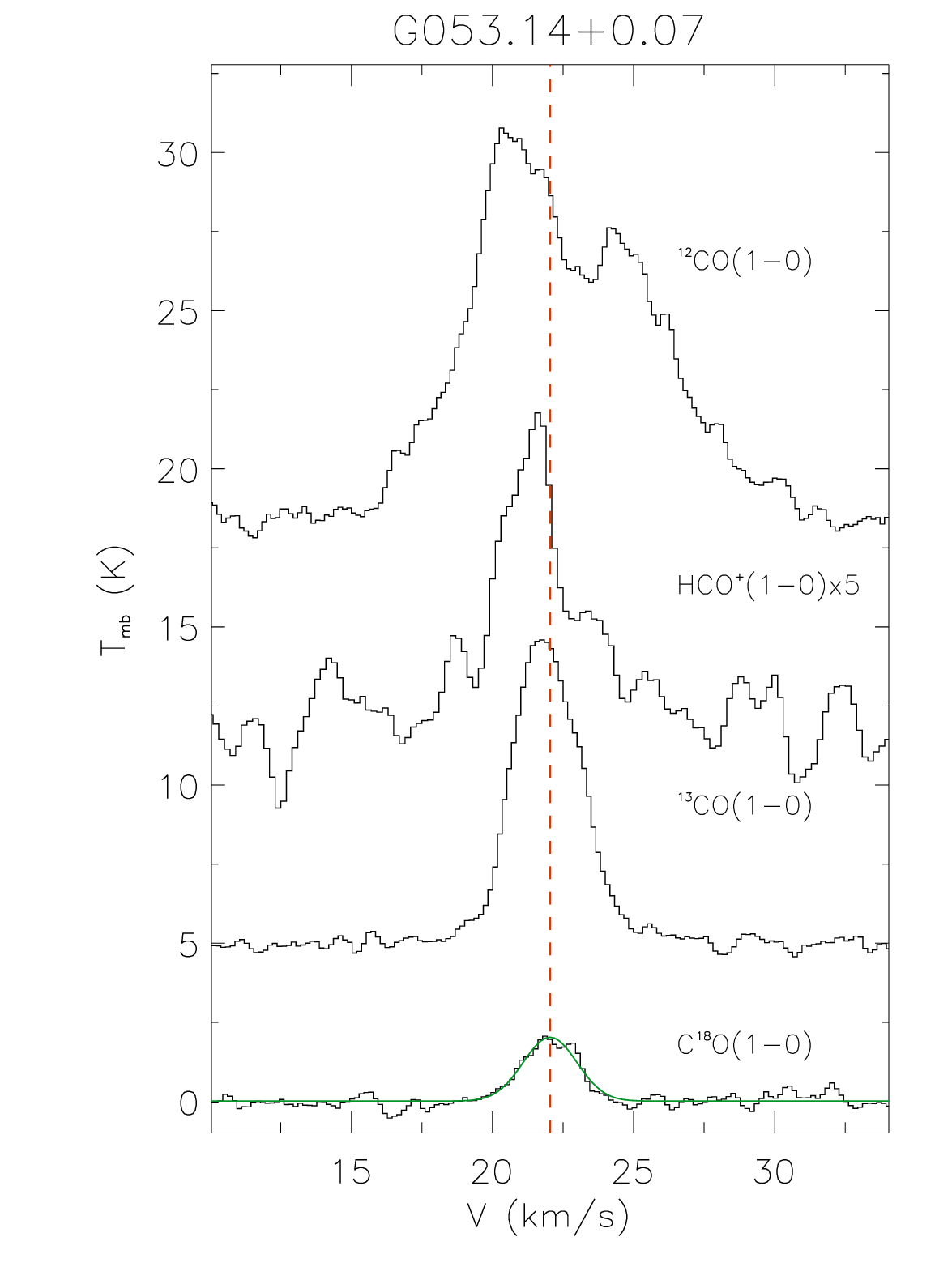}
  \end{minipage}%
  \caption{{\small Line profiles of 133 sources we selected. The lines from bottom to top are C$^{18}$O (1-0), $^{13}$CO (1-0), HCN (1-0) (14 sources lack HCN data), HCO$^+$ (1-0) and $^{12}$CO (1-0), respectively. The dashed red line indicates the central radial velocity of C$^{18}$O (1-0) estimated by Gaussian fitting. For infall candidates, HCO$^+$ (1-0) and HCN (1-0) lines are also Gaussian fitted.}}
  \label{Fig:fig6}
\end{figure} 

\begin{figure}[h]
\ContinuedFloat
  \begin{minipage}[t]{0.325\linewidth}
  \centering
   \includegraphics[width=55mm]{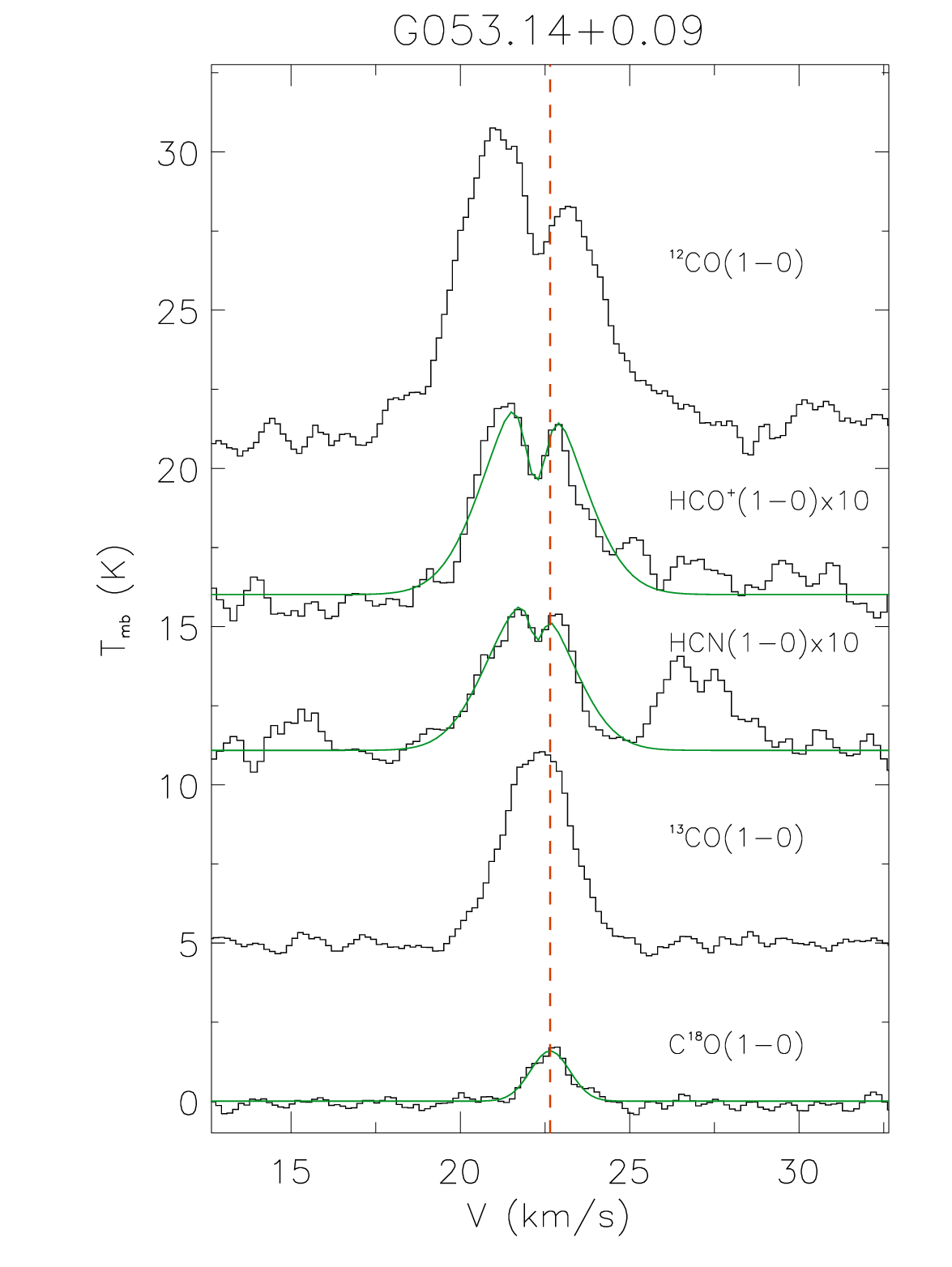}
  \end{minipage}%
  \begin{minipage}[t]{0.325\textwidth}
  \centering
   \includegraphics[width=55mm]{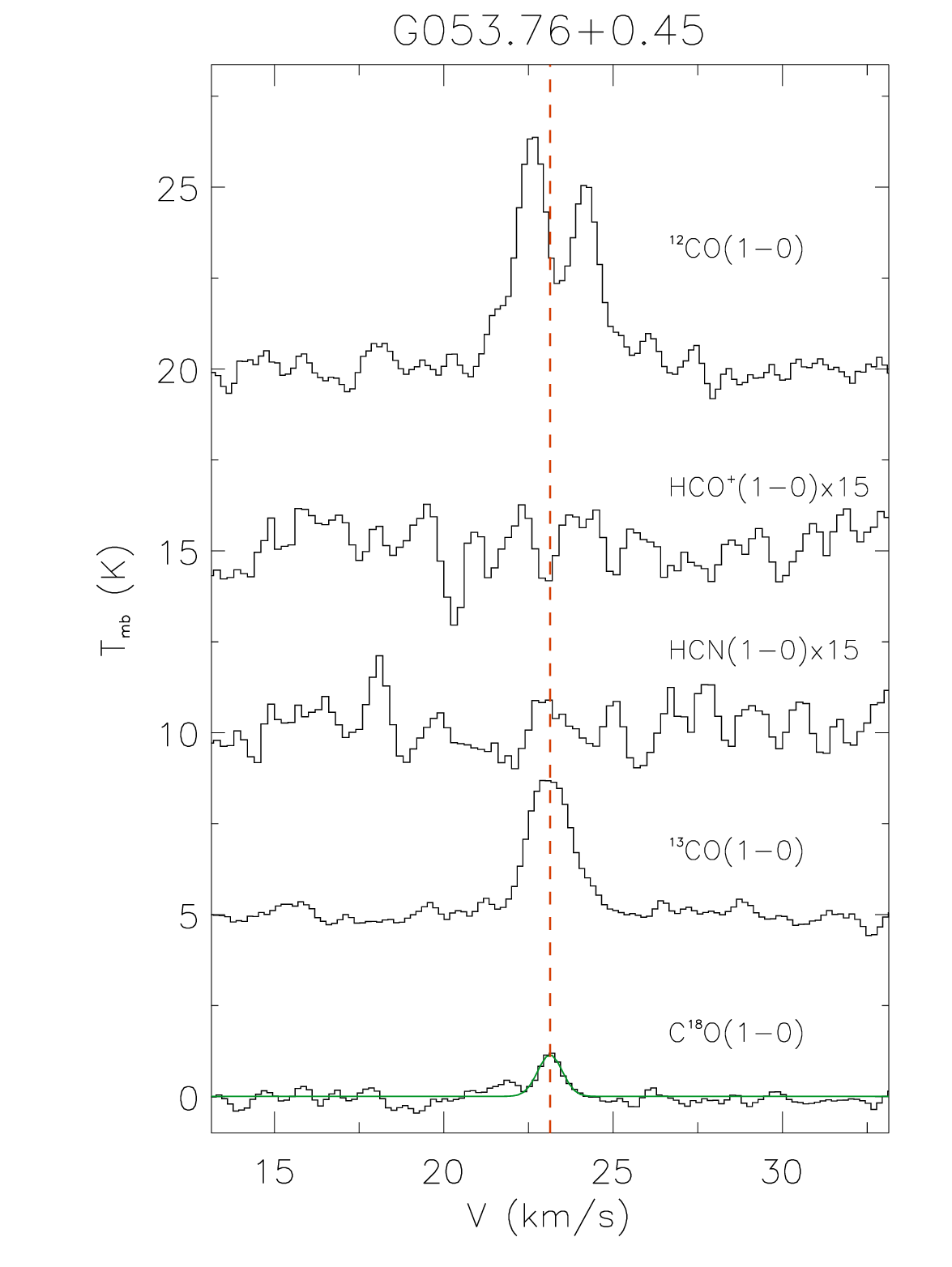}
  \end{minipage}%
  \begin{minipage}[t]{0.325\linewidth}
  \centering
   \includegraphics[width=55mm]{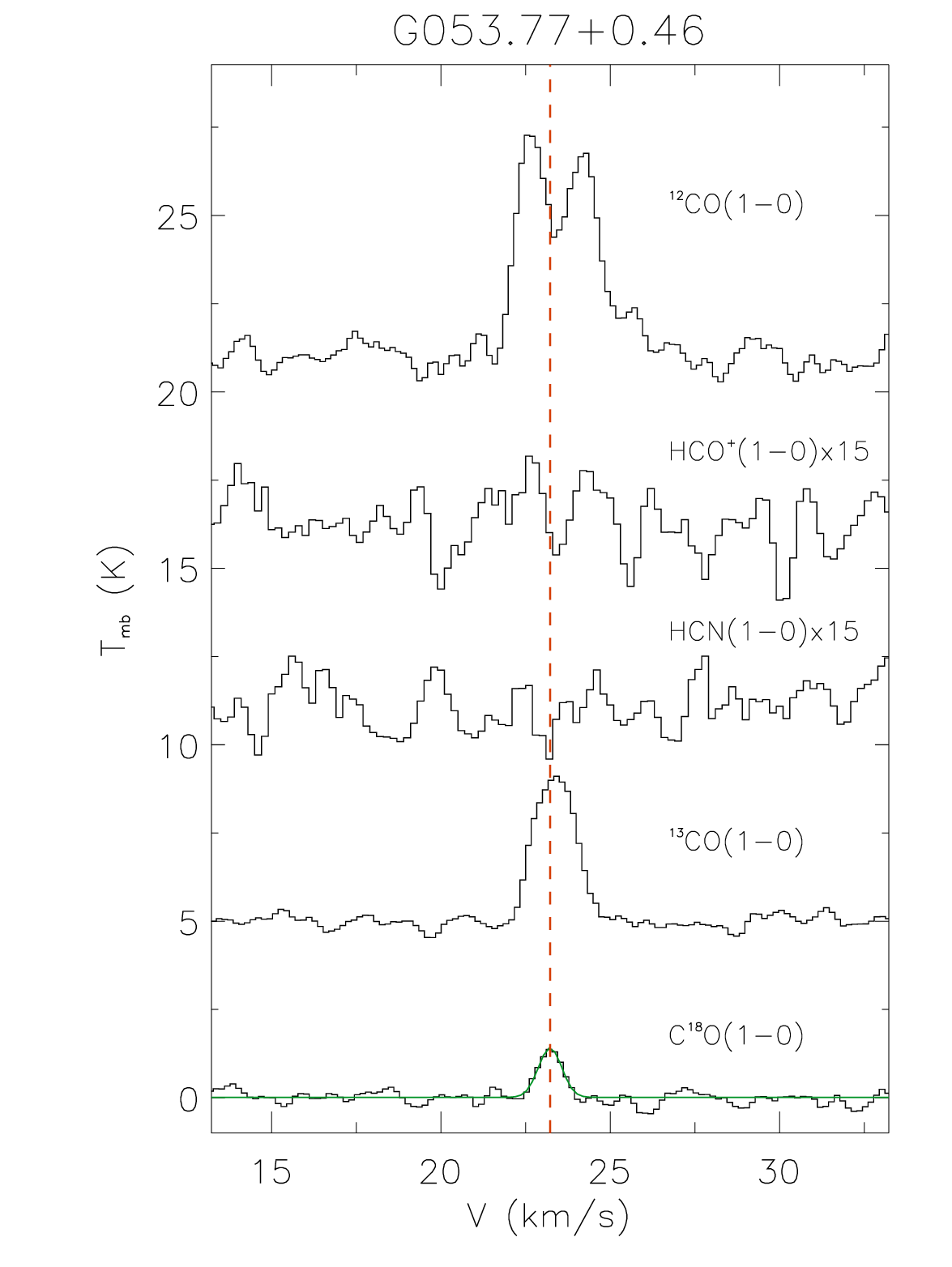}
  \end{minipage}%  
\quad
  \begin{minipage}[t]{0.325\linewidth}
  \centering
   \includegraphics[width=55mm]{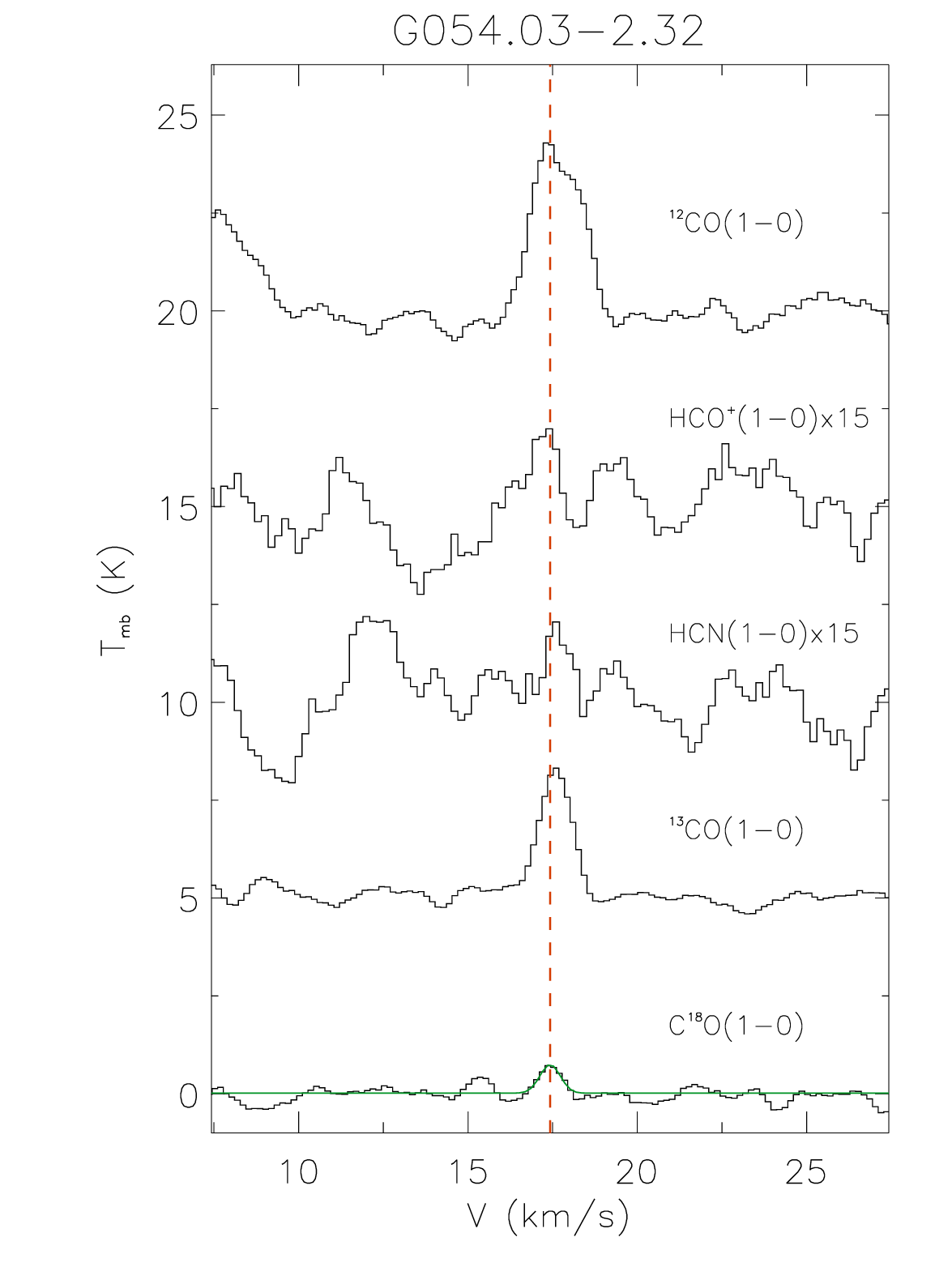}
  \end{minipage}%
  \begin{minipage}[t]{0.325\linewidth}
  \centering
   \includegraphics[width=55mm]{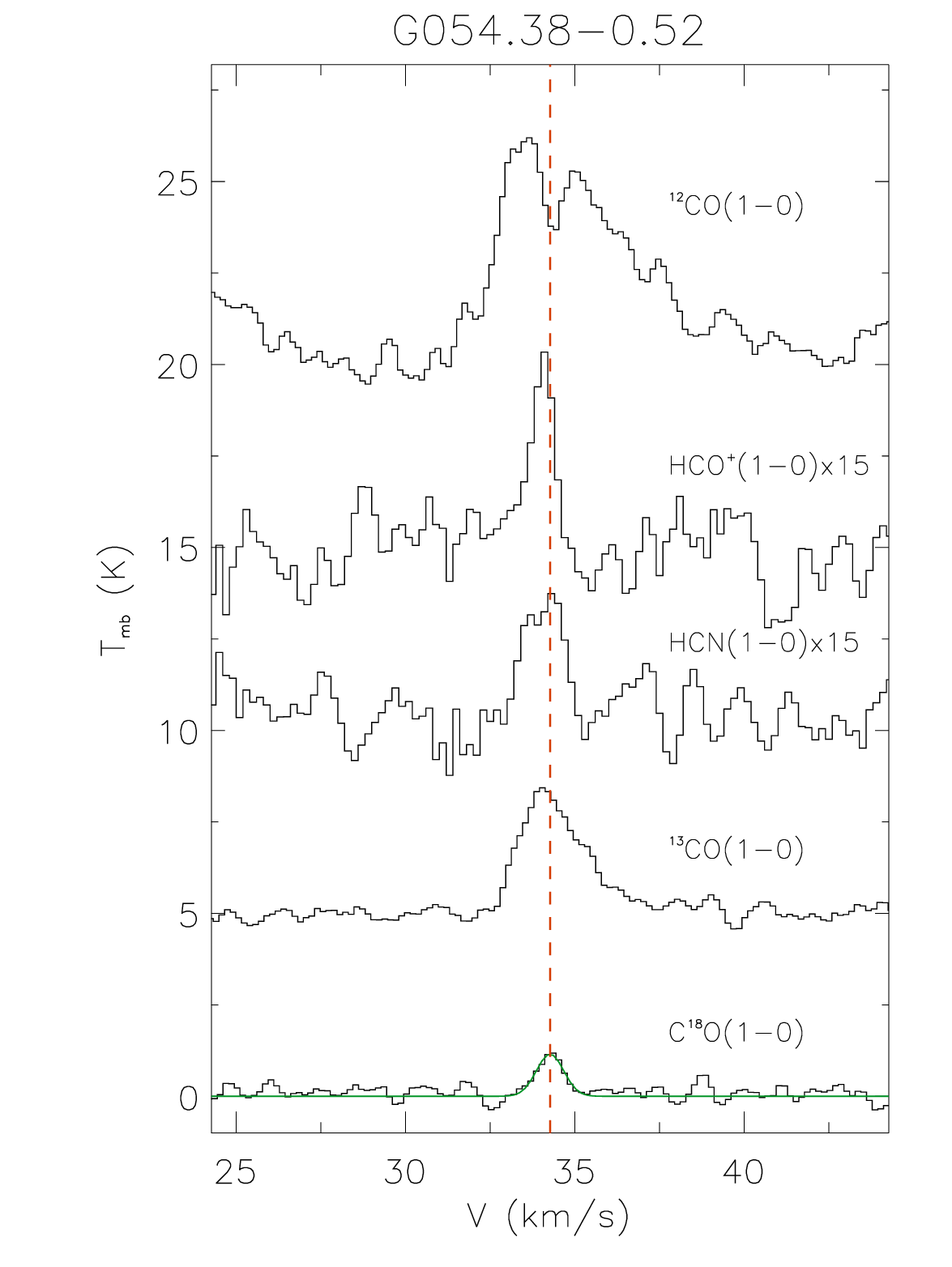}
  \end{minipage}%
  \begin{minipage}[t]{0.325\linewidth}
  \centering
   \includegraphics[width=55mm]{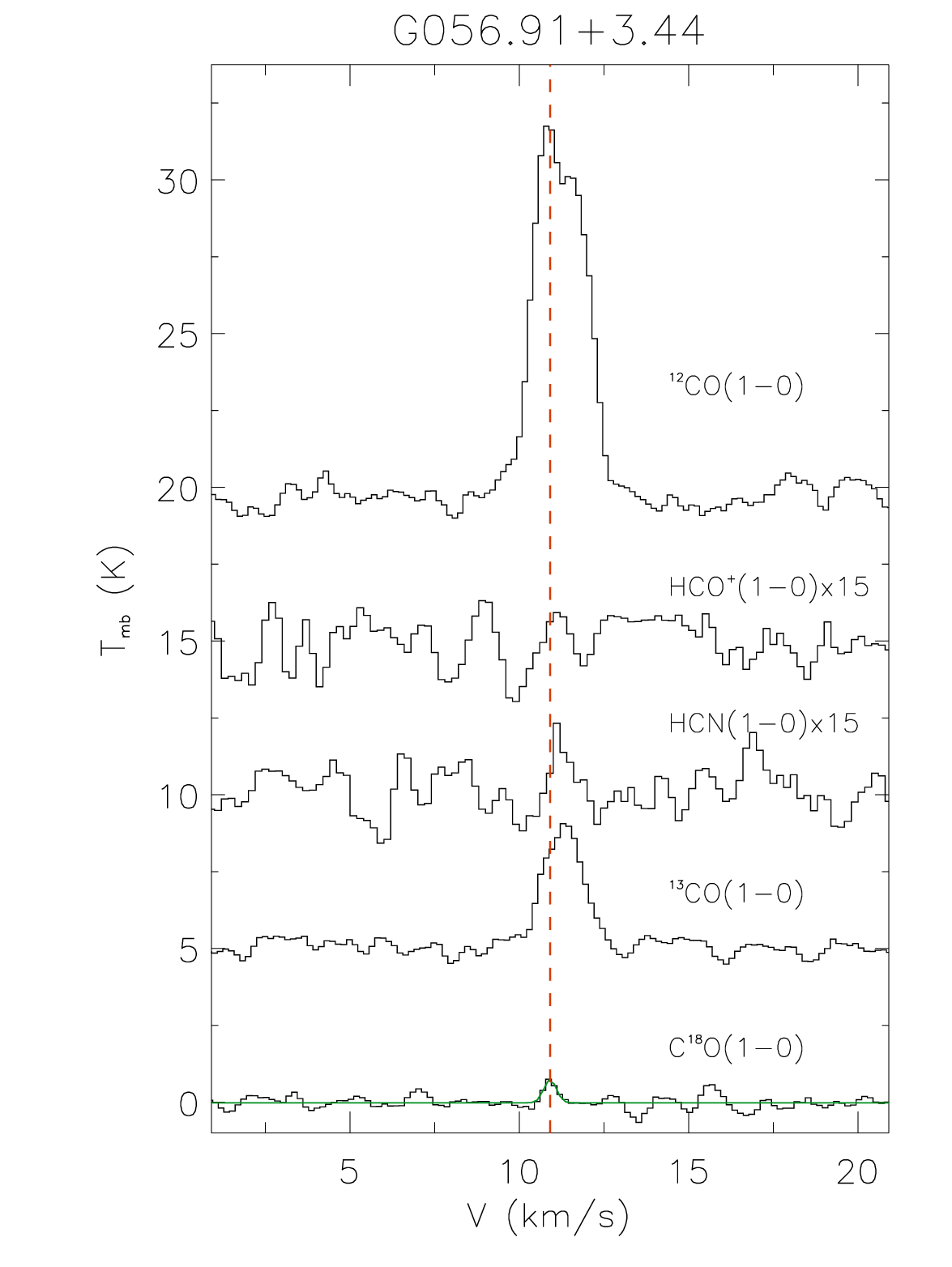}
  \end{minipage}%
\quad
  \begin{minipage}[t]{0.325\linewidth}
  \centering
   \includegraphics[width=55mm]{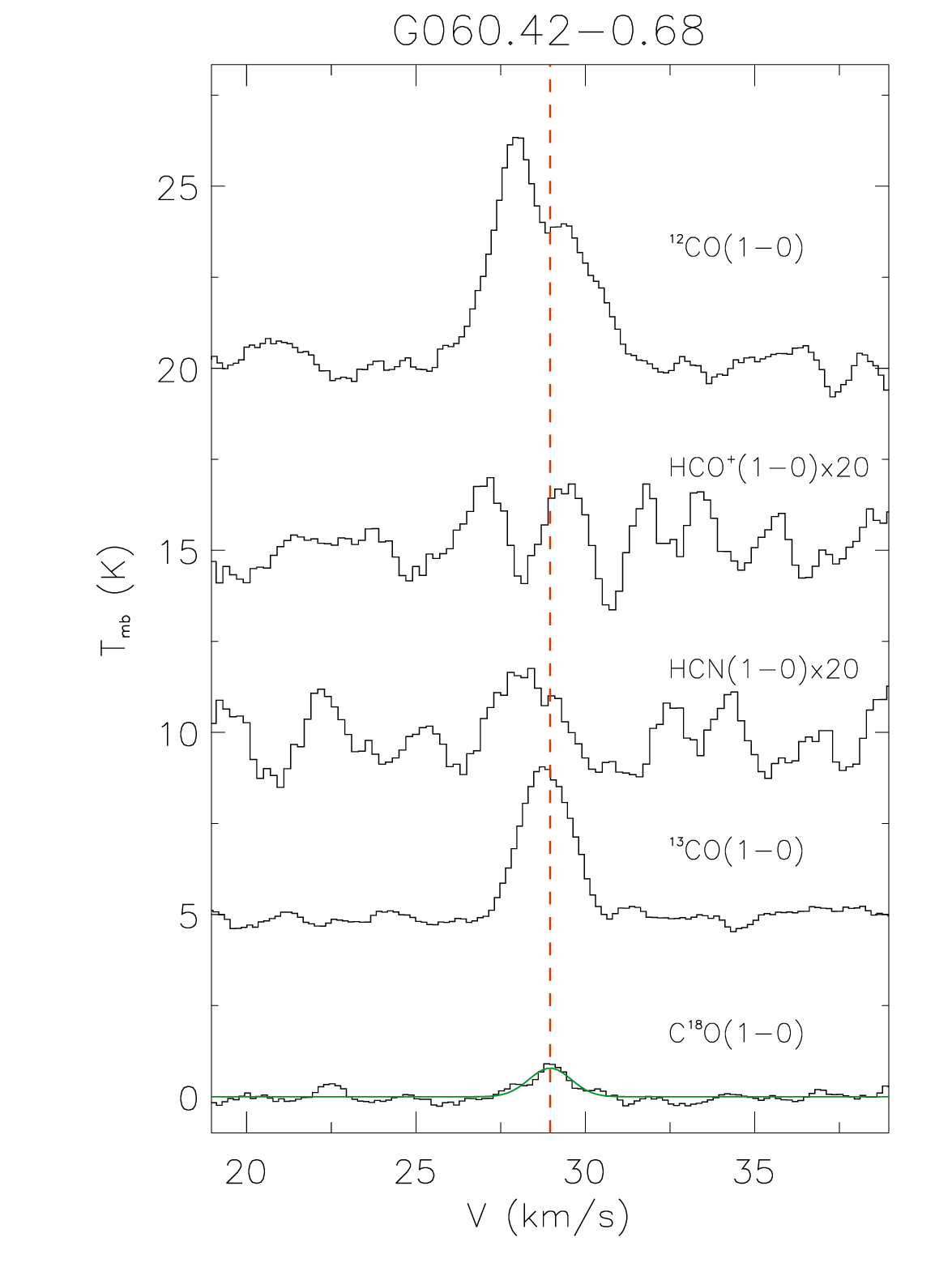}
  \end{minipage}%
  \begin{minipage}[t]{0.325\linewidth}
  \centering
   \includegraphics[width=55mm]{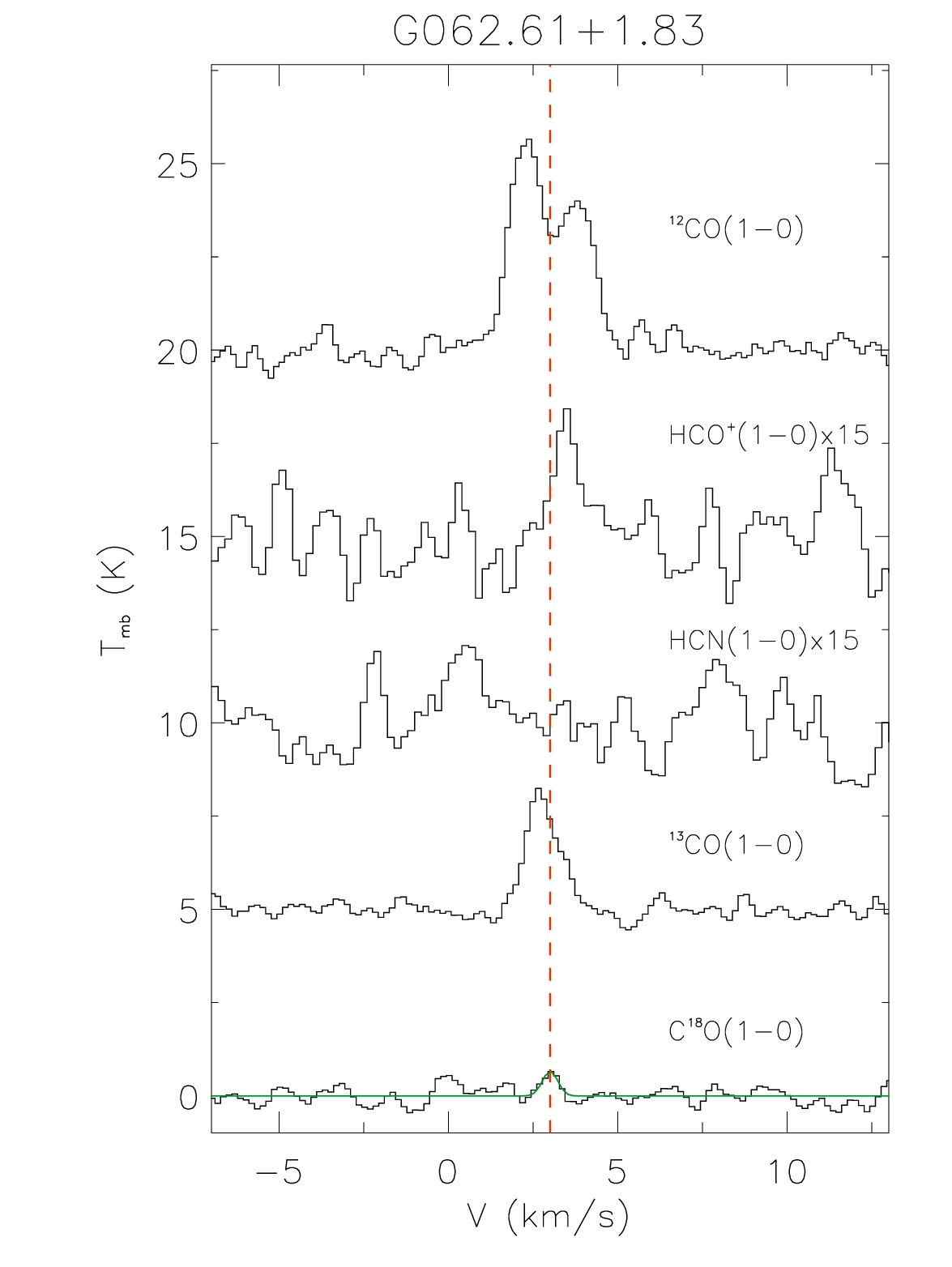}
  \end{minipage}%
  \begin{minipage}[t]{0.325\linewidth}
  \centering
   \includegraphics[width=55mm]{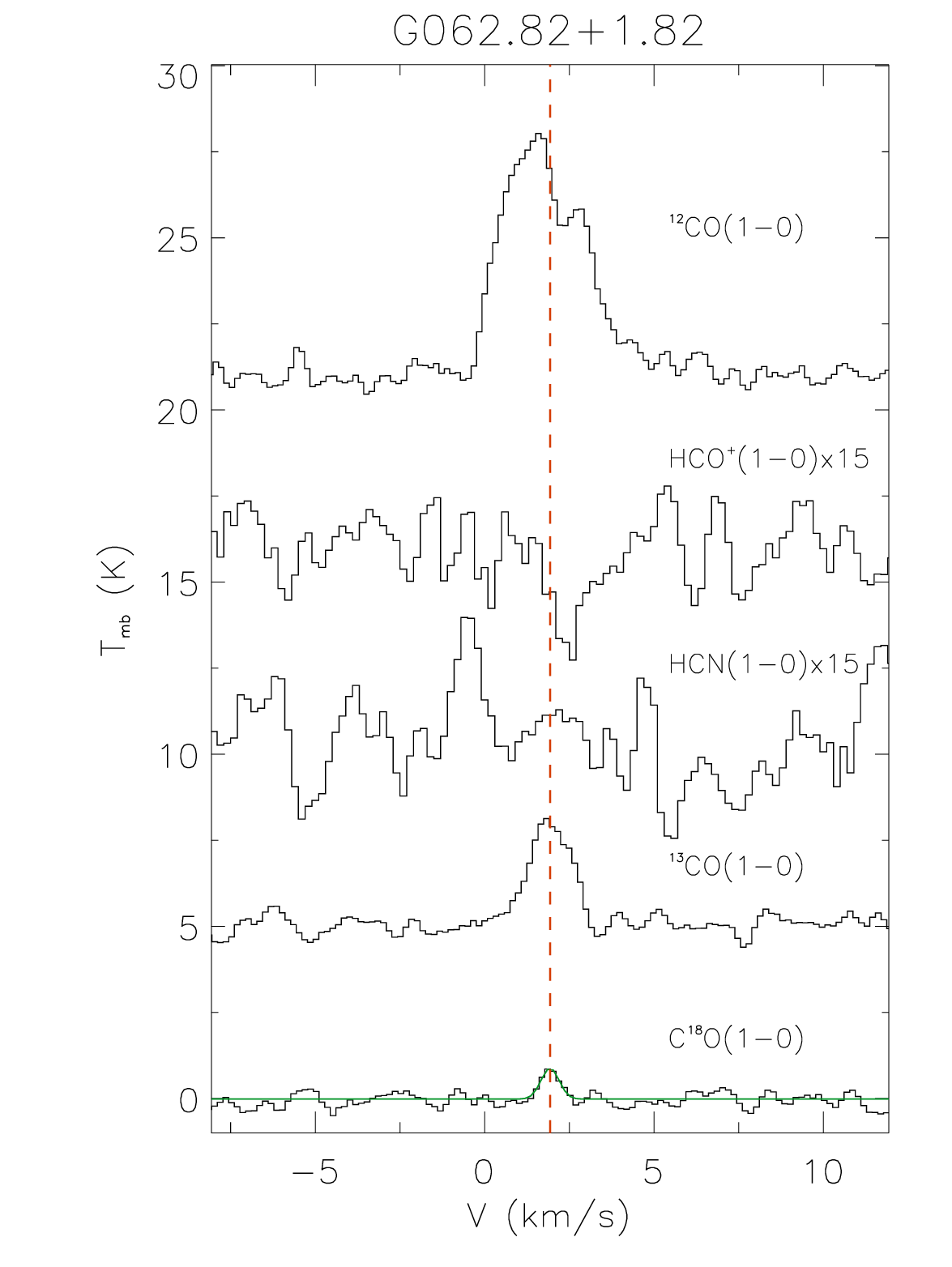}
  \end{minipage}%
  \caption{{\small Line profiles of 133 sources we selected. The lines from bottom to top are C$^{18}$O (1-0), $^{13}$CO (1-0), HCN (1-0) (14 sources lack HCN data), HCO$^+$ (1-0) and $^{12}$CO (1-0), respectively. The dashed red line indicates the central radial velocity of C$^{18}$O (1-0) estimated by Gaussian fitting. For infall candidates, HCO$^+$ (1-0) and HCN (1-0) lines are also Gaussian fitted.}}
  \label{Fig:fig6}
\end{figure} 

\begin{figure}[h]
\ContinuedFloat
  \begin{minipage}[t]{0.325\linewidth}
  \centering
   \includegraphics[width=55mm]{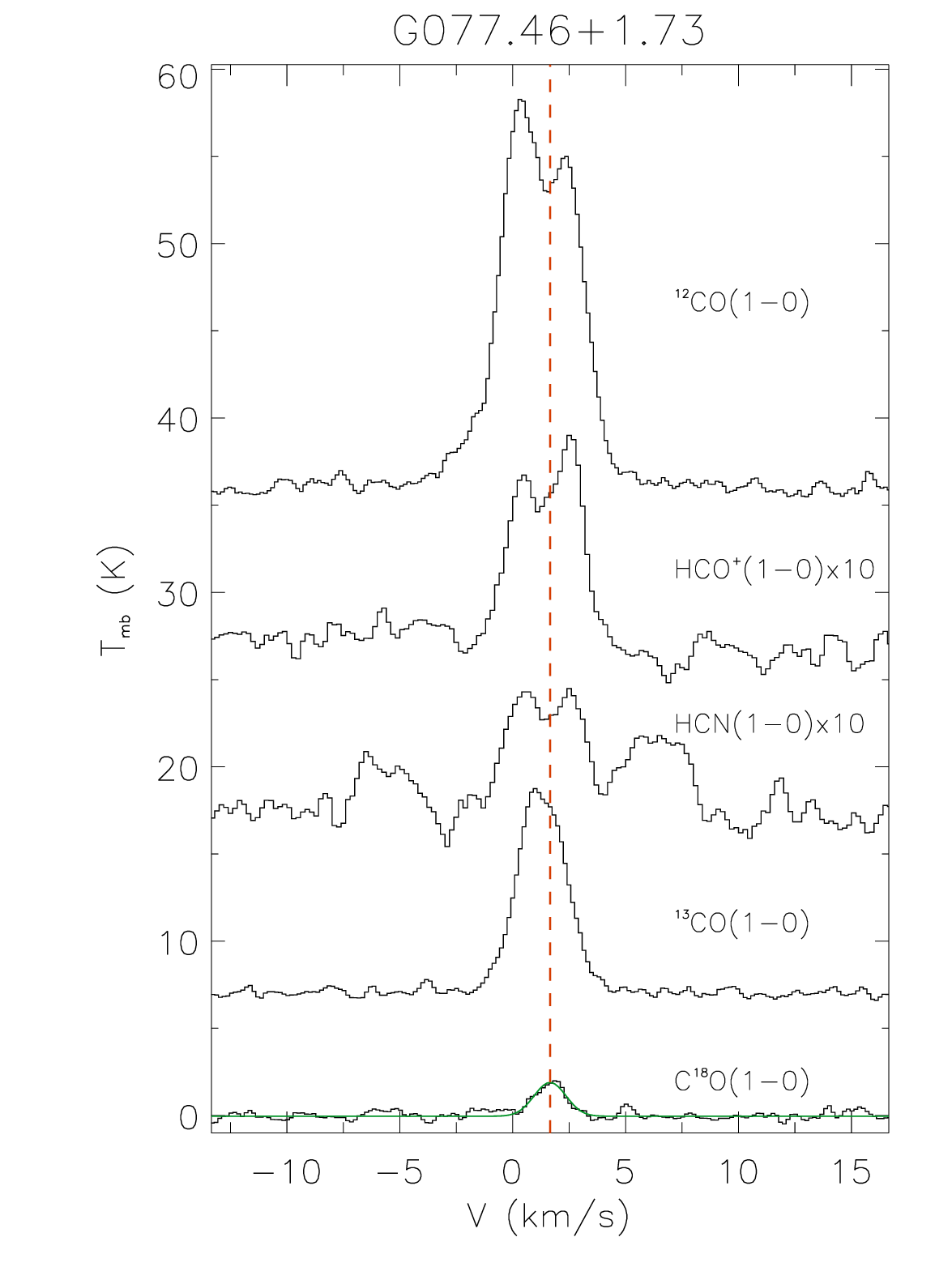}
  \end{minipage}%
  \begin{minipage}[t]{0.325\textwidth}
  \centering
   \includegraphics[width=55mm]{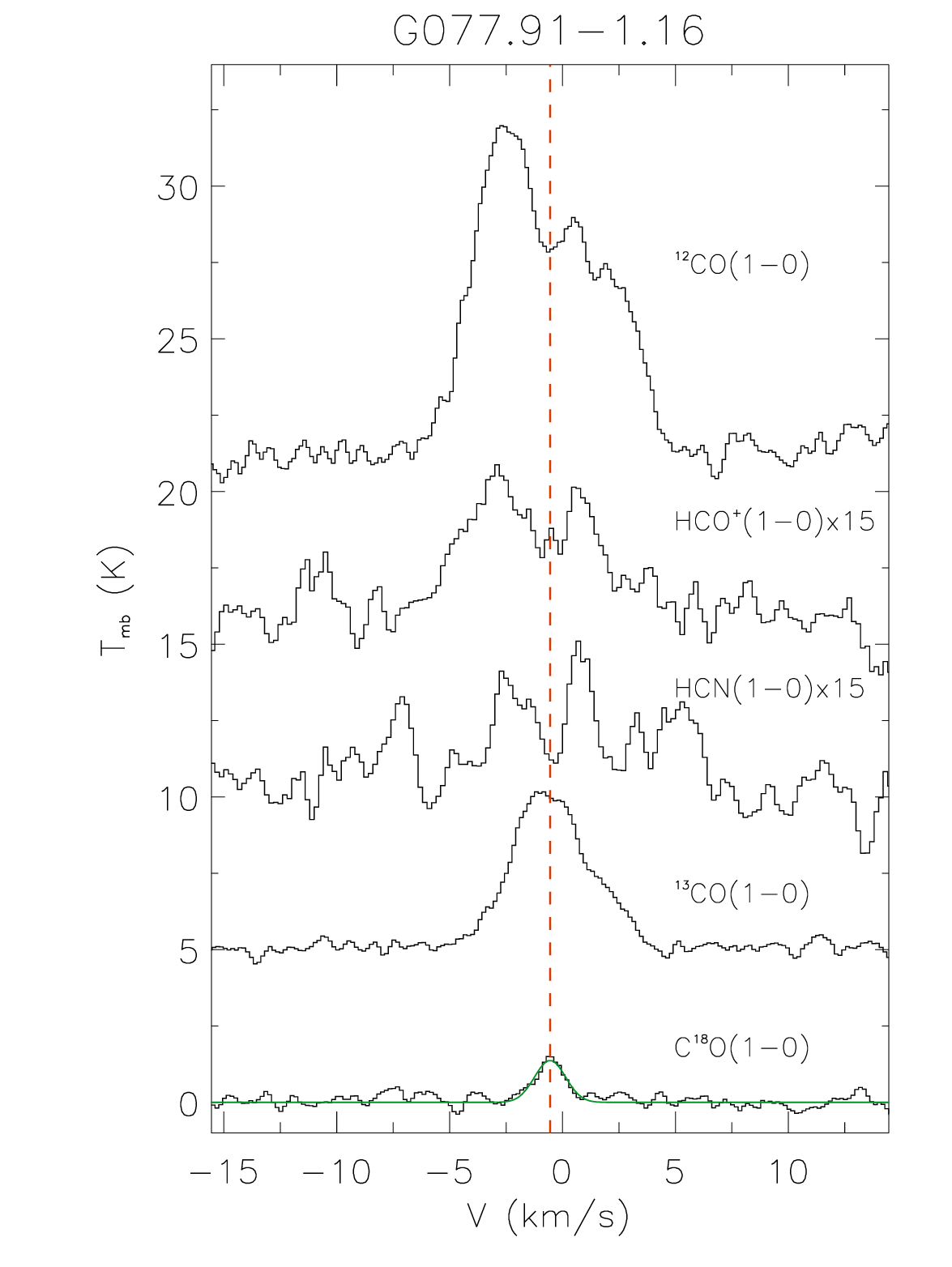}
  \end{minipage}%
  \begin{minipage}[t]{0.325\linewidth}
  \centering
   \includegraphics[width=55mm]{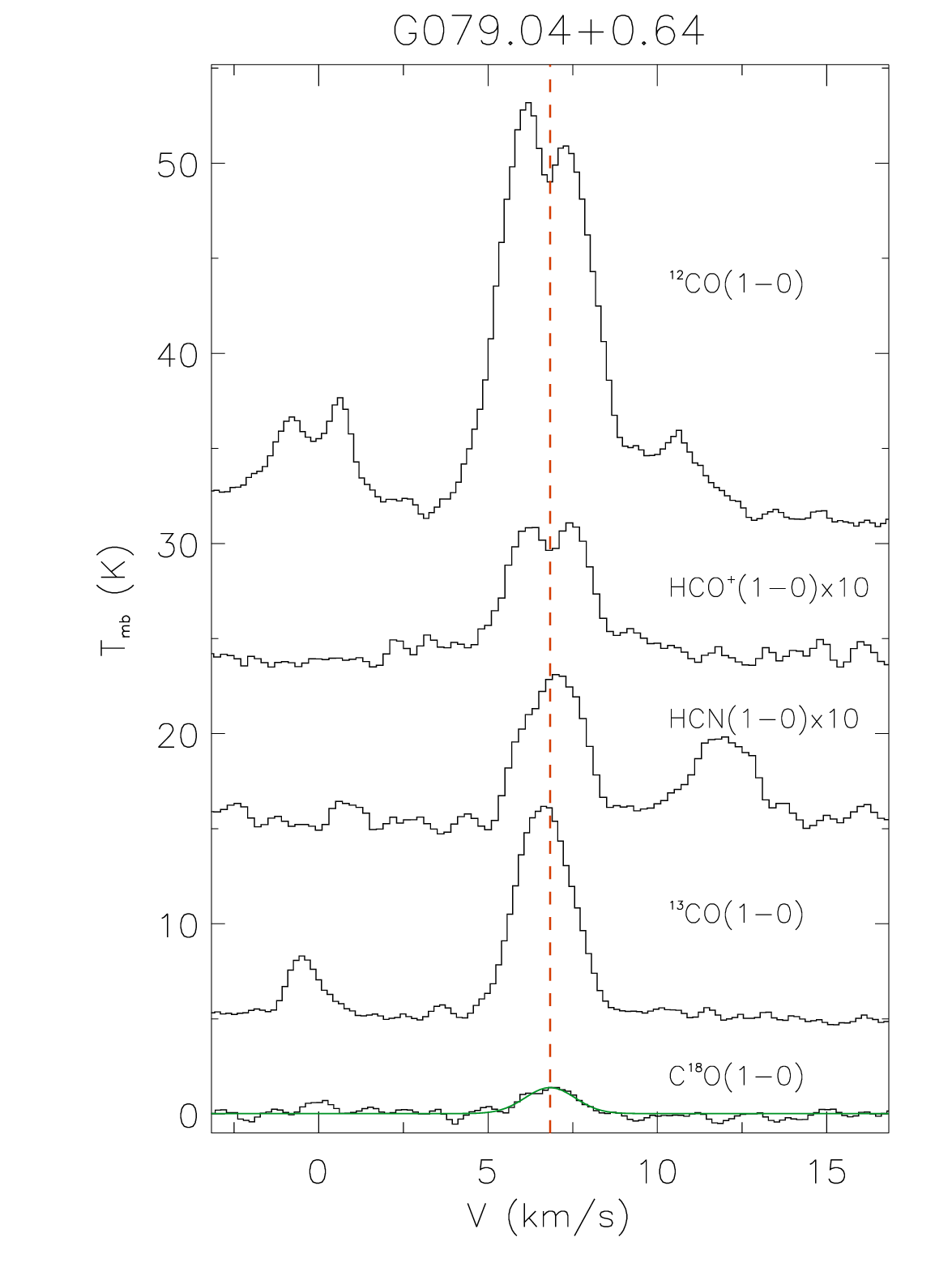}
  \end{minipage}%  
\quad
  \begin{minipage}[t]{0.325\linewidth}
  \centering
   \includegraphics[width=55mm]{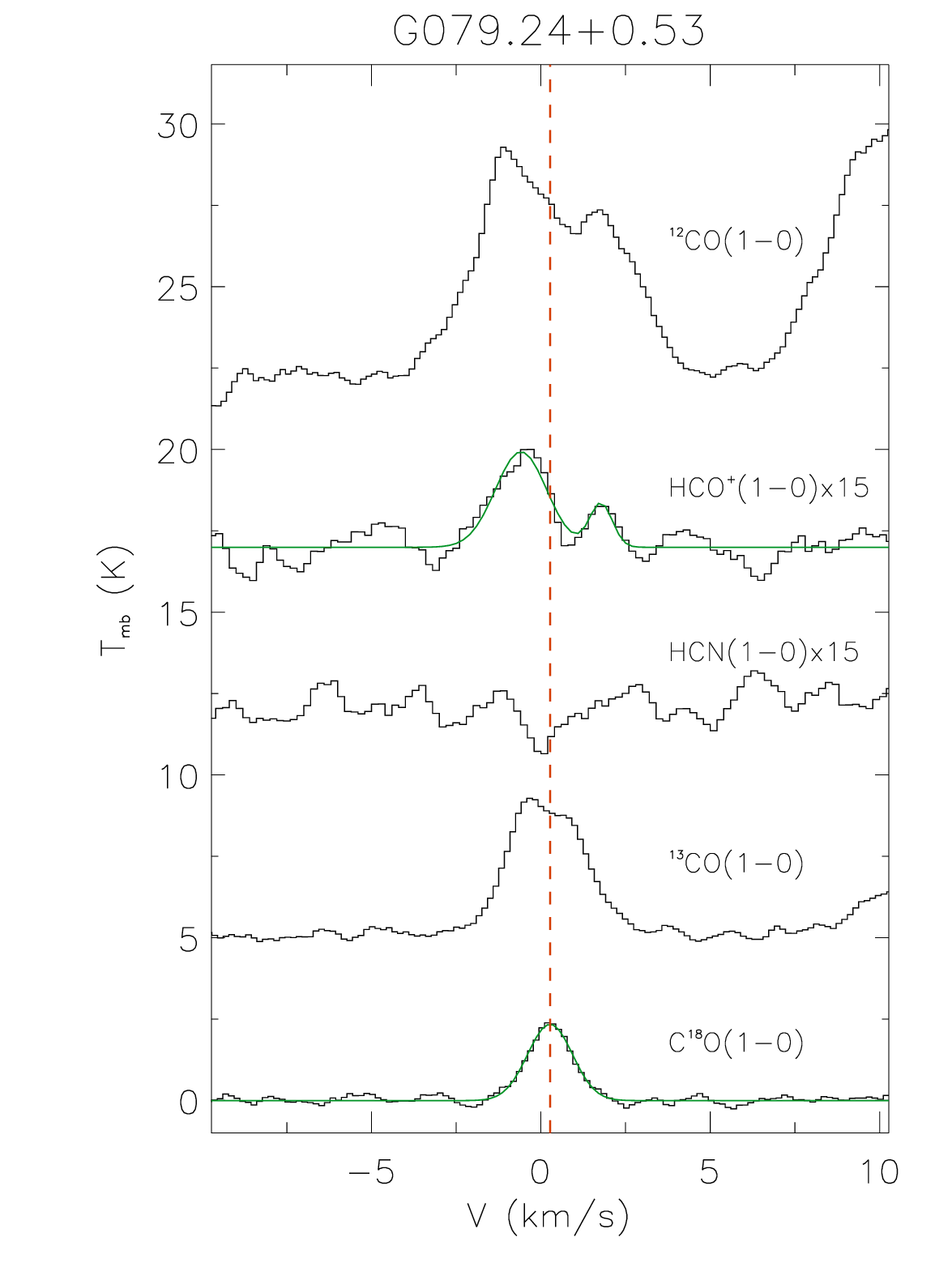}
  \end{minipage}%
  \begin{minipage}[t]{0.325\linewidth}
  \centering
   \includegraphics[width=55mm]{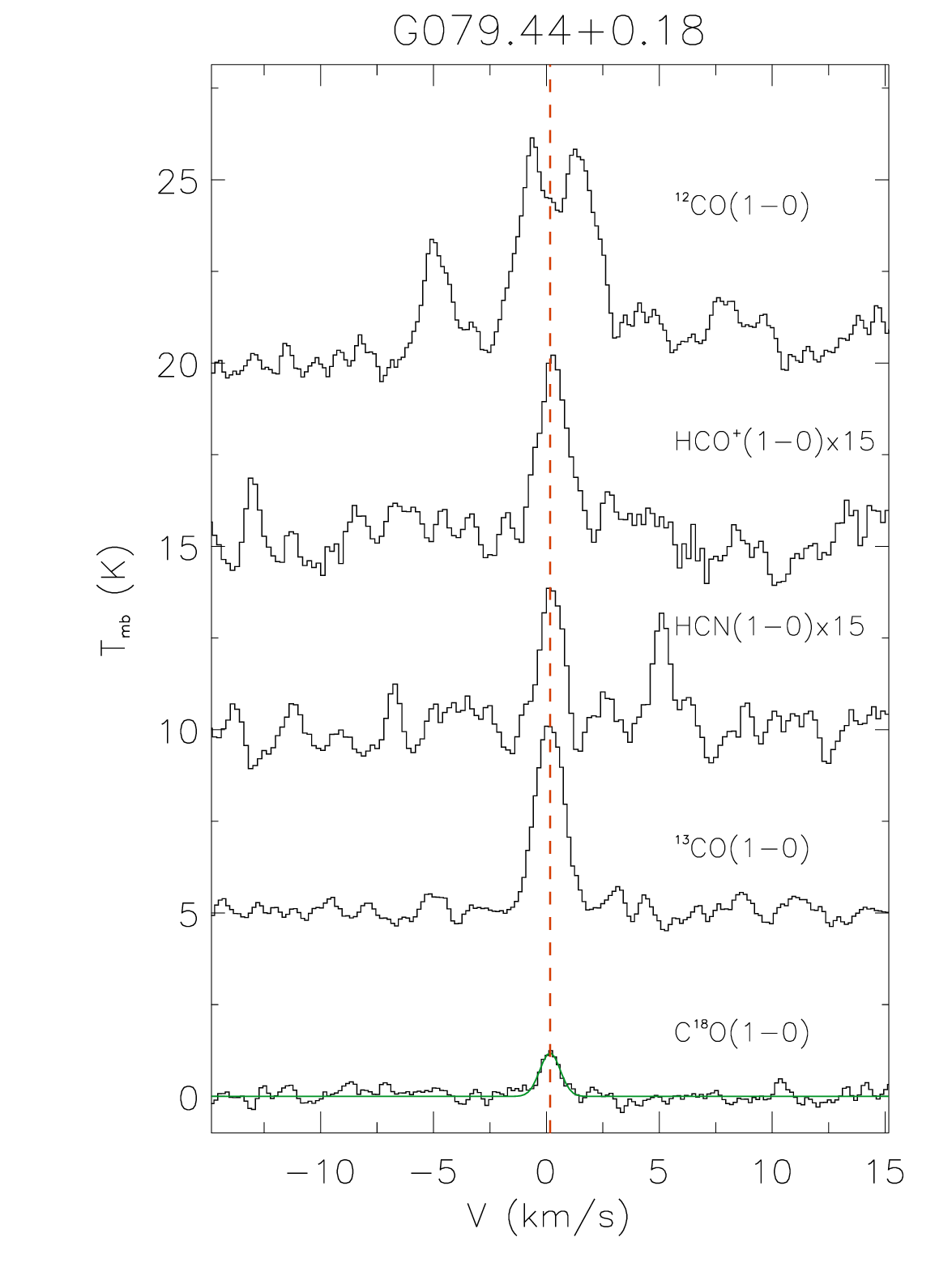}
  \end{minipage}%
  \begin{minipage}[t]{0.325\linewidth}
  \centering
   \includegraphics[width=55mm]{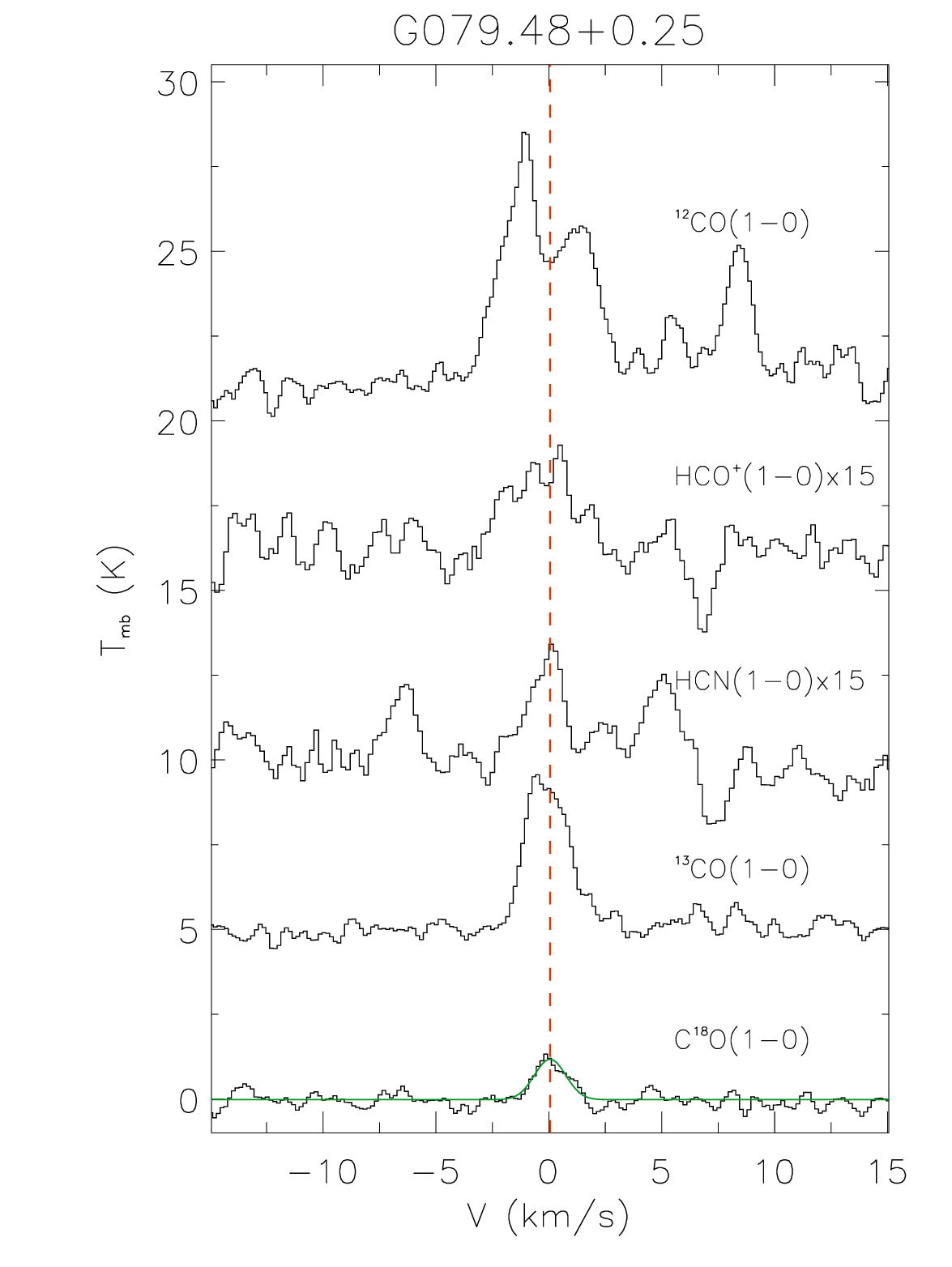}
  \end{minipage}%
\quad
  \begin{minipage}[t]{0.325\linewidth}
  \centering
   \includegraphics[width=55mm]{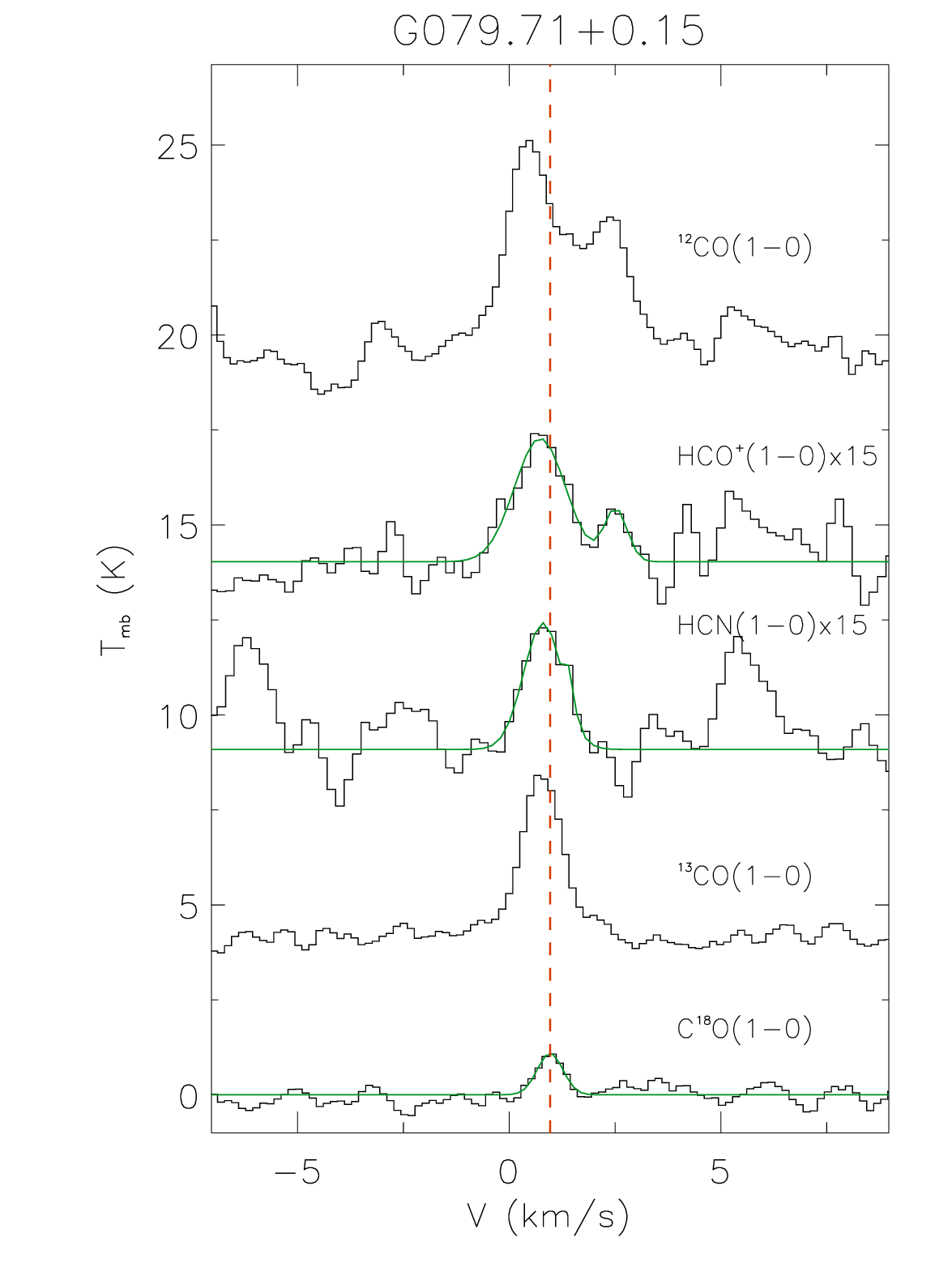}
  \end{minipage}%
  \begin{minipage}[t]{0.325\linewidth}
  \centering
   \includegraphics[width=55mm]{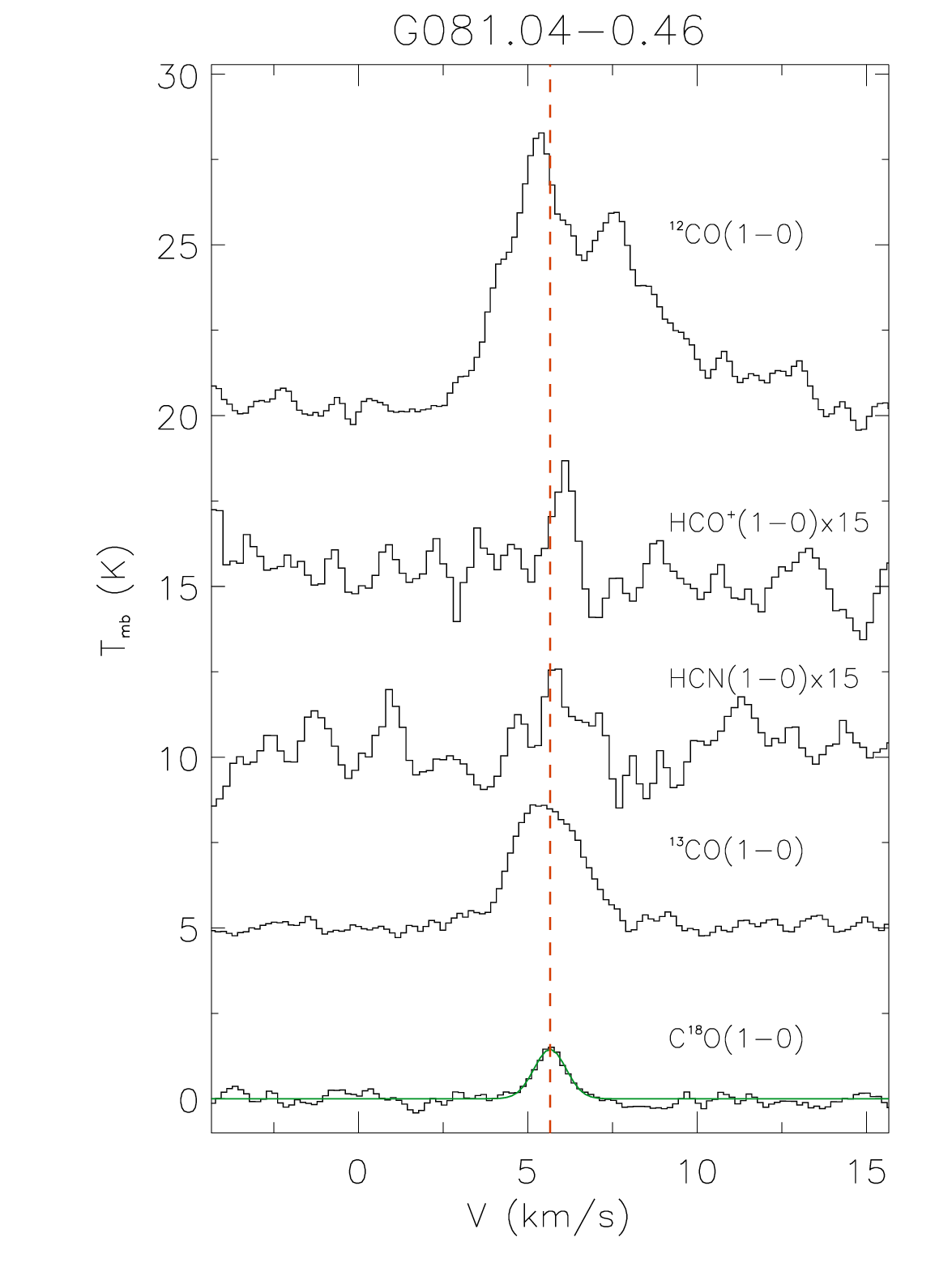}
  \end{minipage}%
  \begin{minipage}[t]{0.325\linewidth}
  \centering
   \includegraphics[width=55mm]{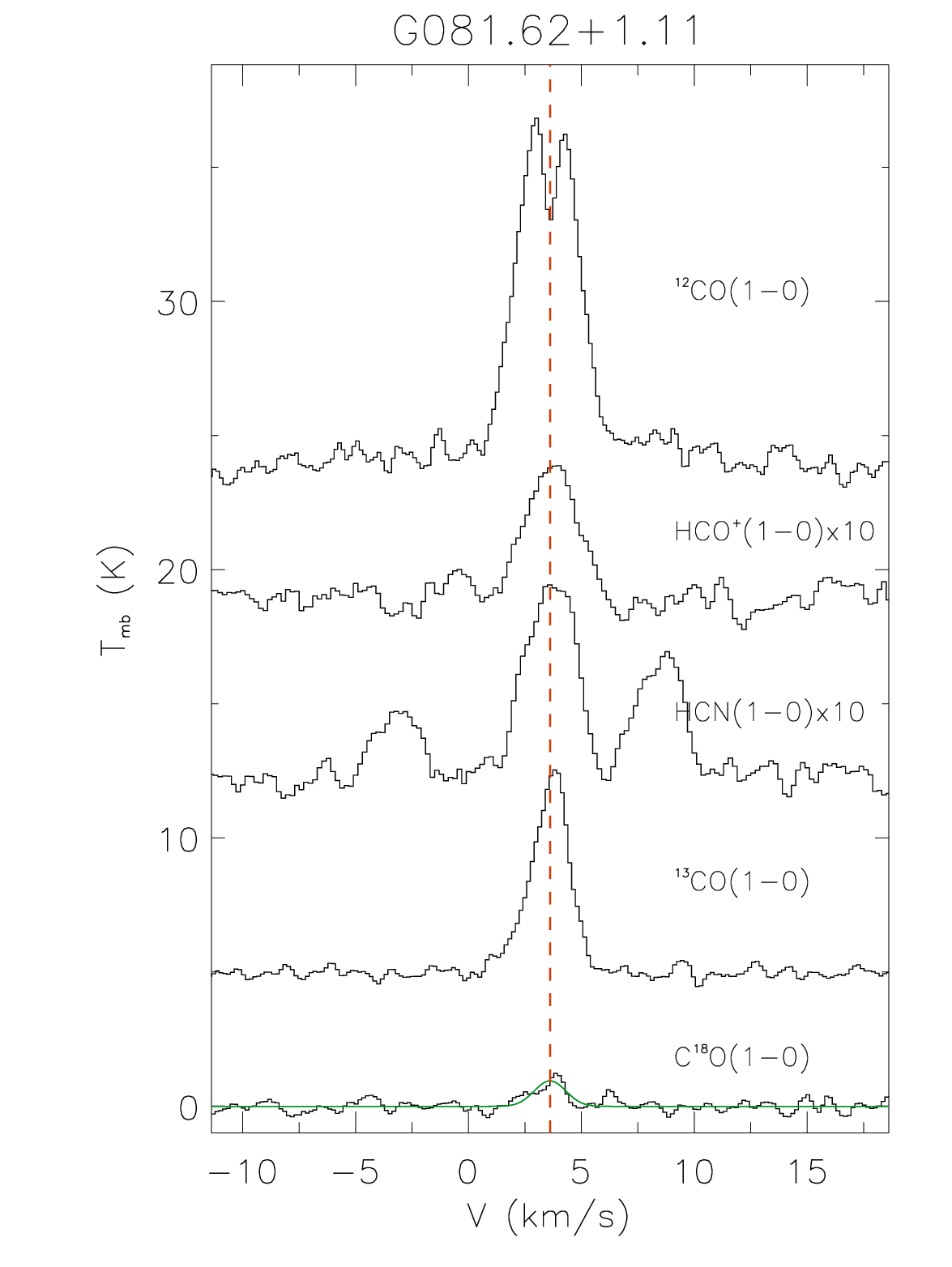}
  \end{minipage}%
  \caption{{\small Line profiles of 133 sources we selected. The lines from bottom to top are C$^{18}$O (1-0), $^{13}$CO (1-0), HCN (1-0) (14 sources lack HCN data), HCO$^+$ (1-0) and $^{12}$CO (1-0), respectively. The dashed red line indicates the central radial velocity of C$^{18}$O (1-0) estimated by Gaussian fitting. For infall candidates, HCO$^+$ (1-0) and HCN (1-0) lines are also Gaussian fitted.}}
  \label{Fig:fig6}
\end{figure} 

\begin{figure}[h]
\ContinuedFloat
  \begin{minipage}[t]{0.325\linewidth}
  \centering
   \includegraphics[width=55mm]{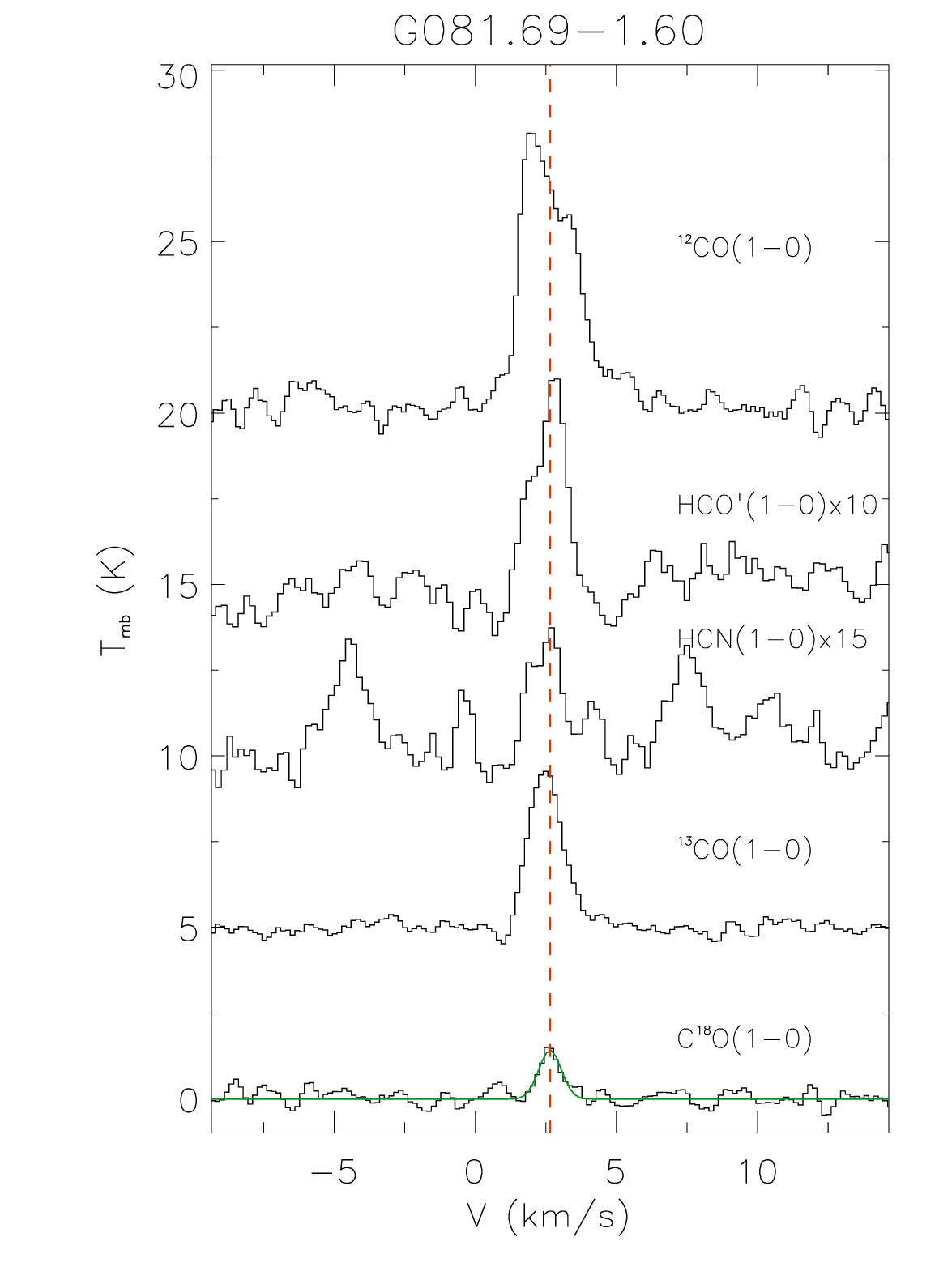}
  \end{minipage}%
  \begin{minipage}[t]{0.325\textwidth}
  \centering
   \includegraphics[width=55mm]{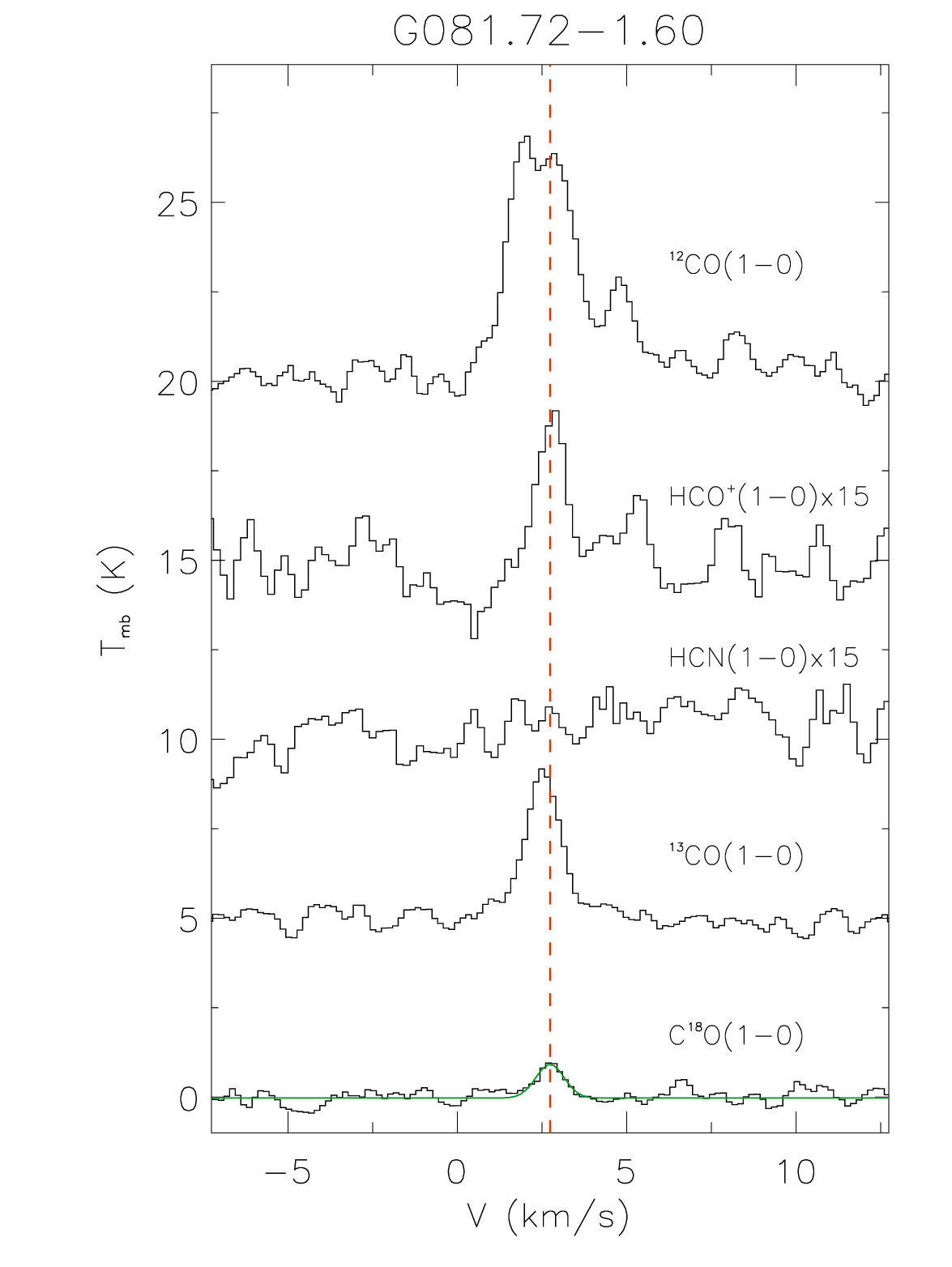}
  \end{minipage}%
  \begin{minipage}[t]{0.325\linewidth}
  \centering
   \includegraphics[width=55mm]{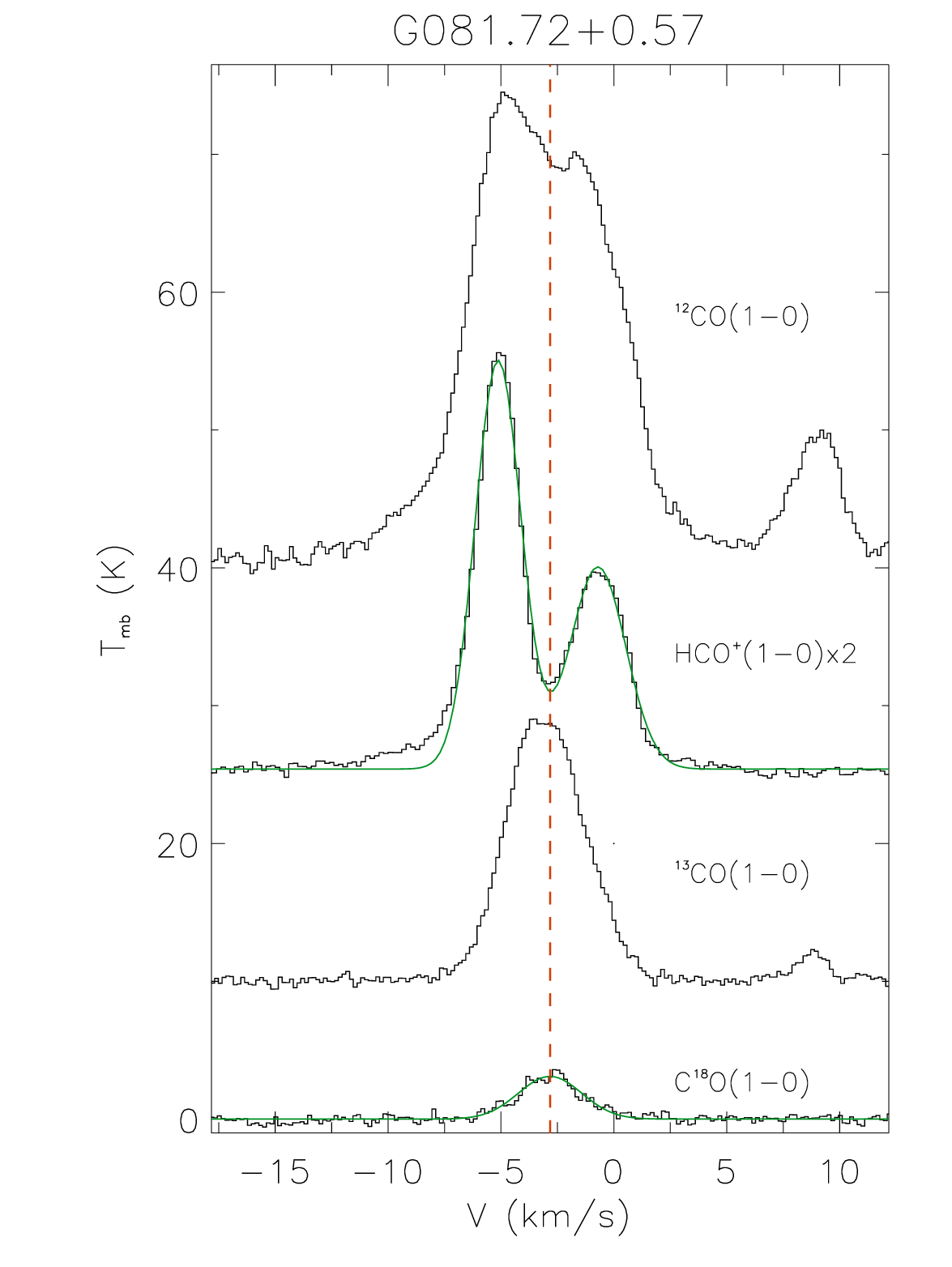}
  \end{minipage}%  
\quad
  \begin{minipage}[t]{0.325\linewidth}
  \centering
   \includegraphics[width=55mm]{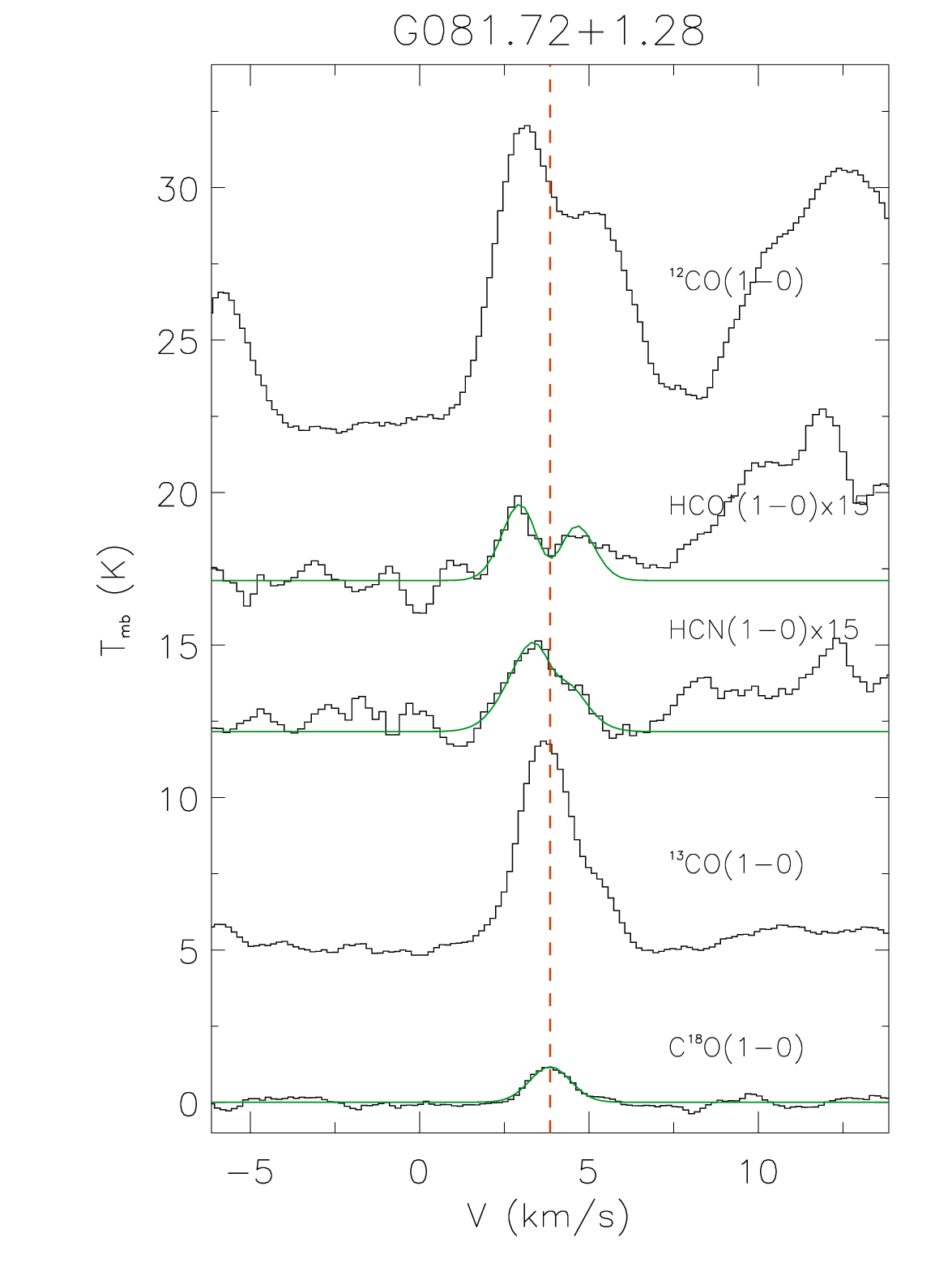}
  \end{minipage}%
  \begin{minipage}[t]{0.325\linewidth}
  \centering
   \includegraphics[width=55mm]{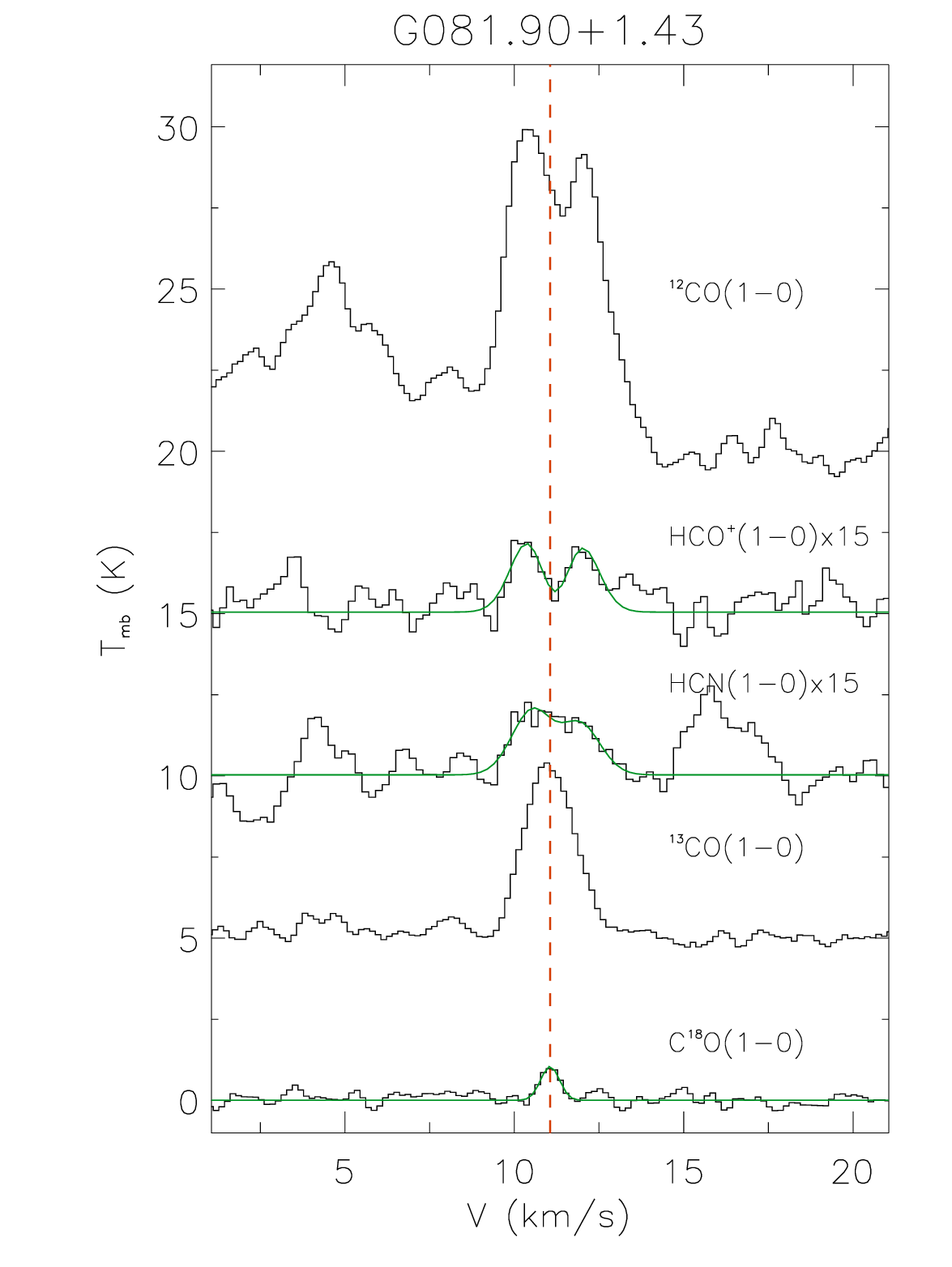}
  \end{minipage}%
  \begin{minipage}[t]{0.325\linewidth}
  \centering
   \includegraphics[width=55mm]{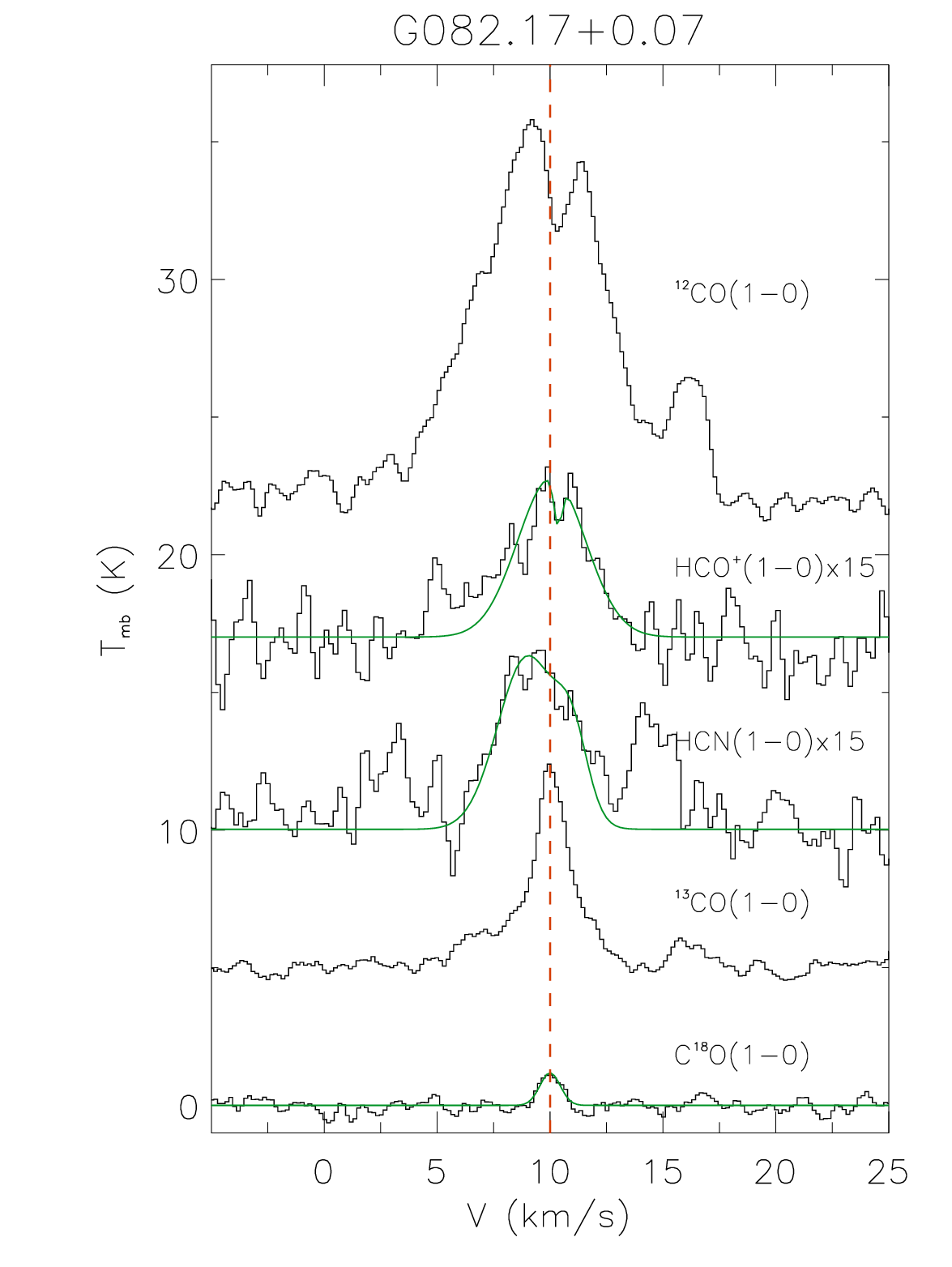}
  \end{minipage}%
\quad
  \begin{minipage}[t]{0.325\linewidth}
  \centering
   \includegraphics[width=55mm]{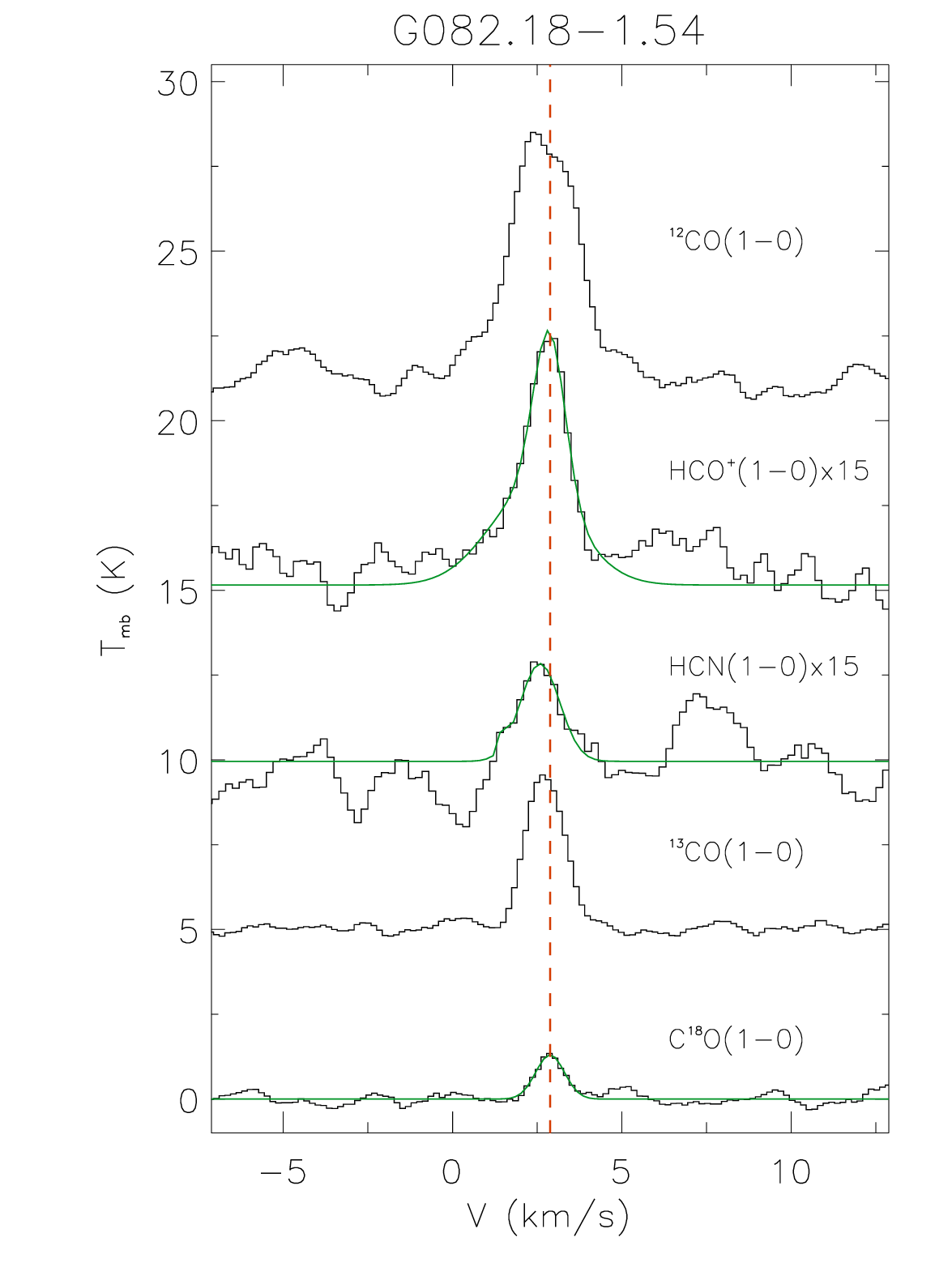}
  \end{minipage}%
  \begin{minipage}[t]{0.325\linewidth}
  \centering
   \includegraphics[width=55mm]{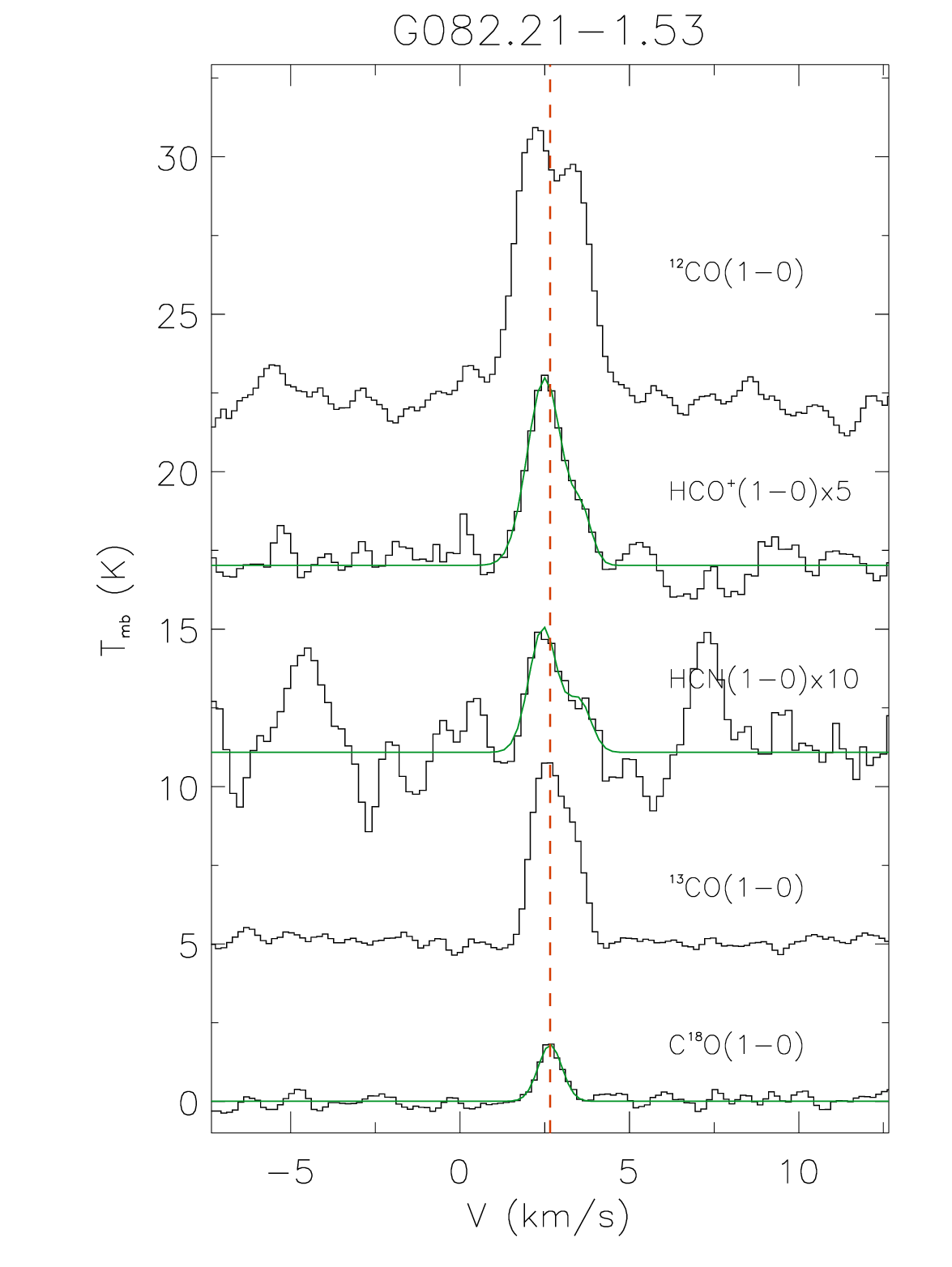}
  \end{minipage}%
  \begin{minipage}[t]{0.325\linewidth}
  \centering
   \includegraphics[width=55mm]{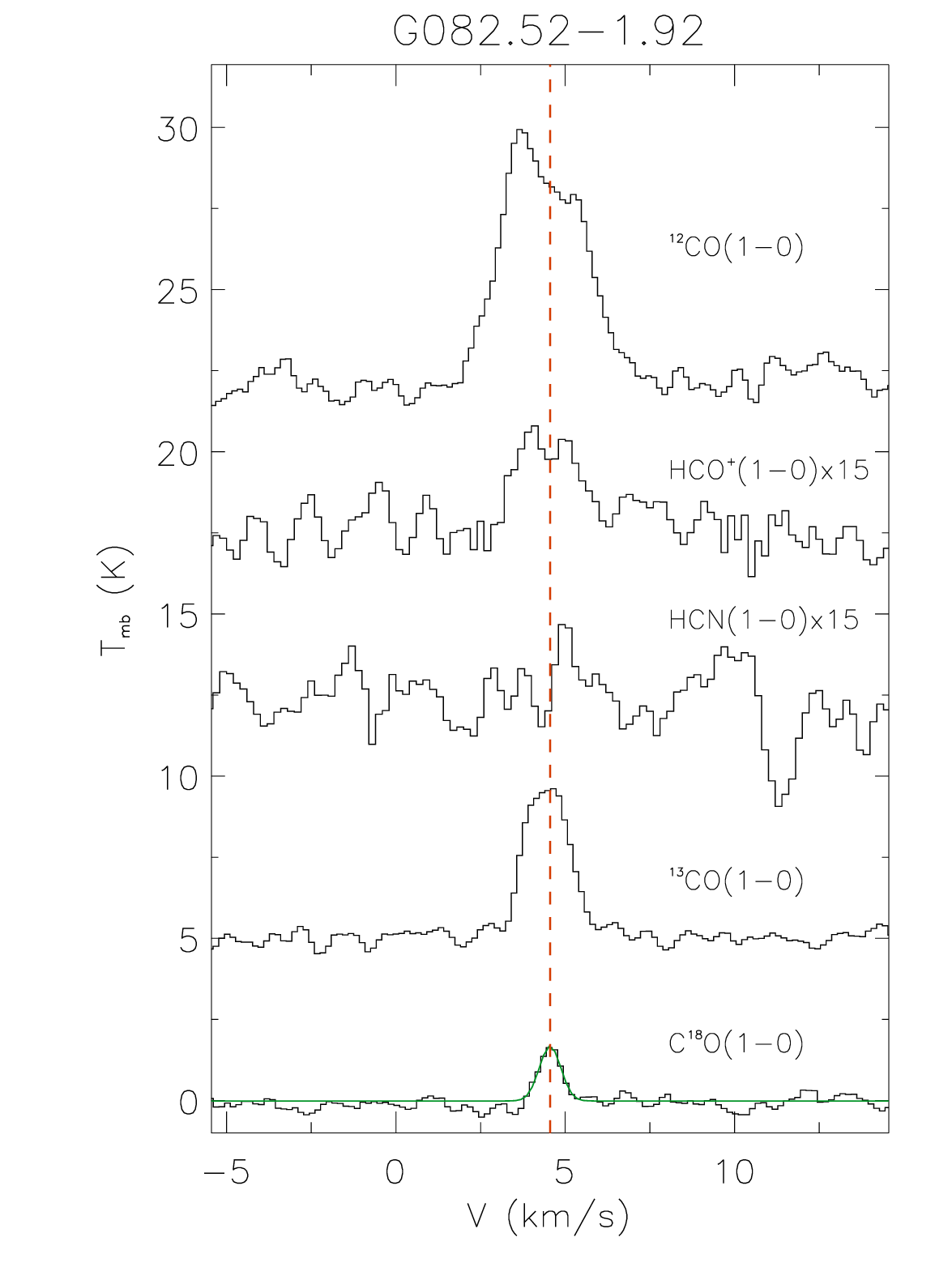}
  \end{minipage}%
  \caption{{\small Line profiles of 133 sources we selected. The lines from bottom to top are C$^{18}$O (1-0), $^{13}$CO (1-0), HCN (1-0) (14 sources lack HCN data), HCO$^+$ (1-0) and $^{12}$CO (1-0), respectively. The dashed red line indicates the central radial velocity of C$^{18}$O (1-0) estimated by Gaussian fitting. For infall candidates, HCO$^+$ (1-0) and HCN (1-0) lines are also Gaussian fitted.}}
  \label{Fig:fig6}
\end{figure} 

\begin{figure}[h]
\ContinuedFloat
  \begin{minipage}[t]{0.325\linewidth}
  \centering
   \includegraphics[width=55mm]{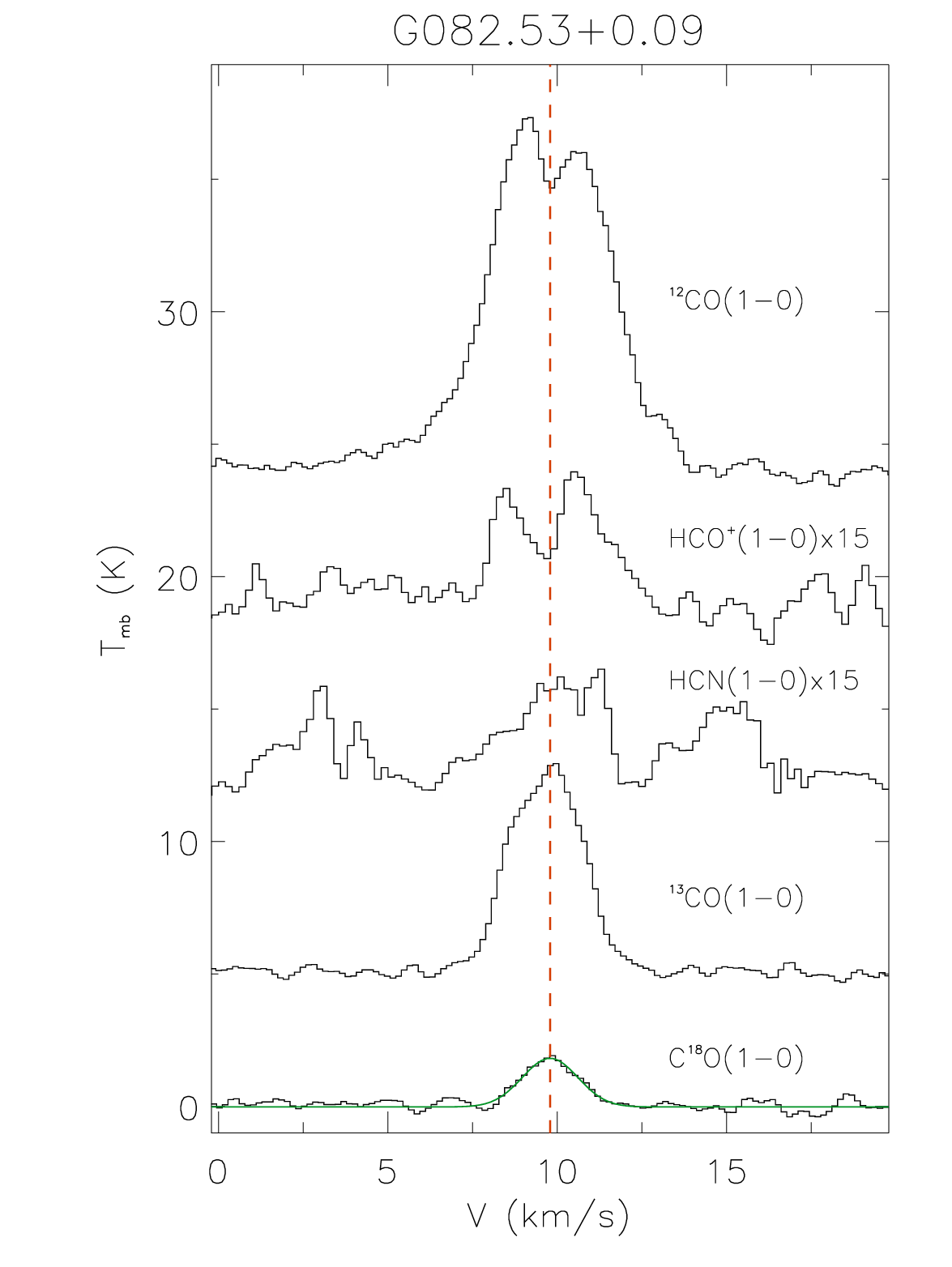}
  \end{minipage}%
  \begin{minipage}[t]{0.325\textwidth}
  \centering
   \includegraphics[width=55mm]{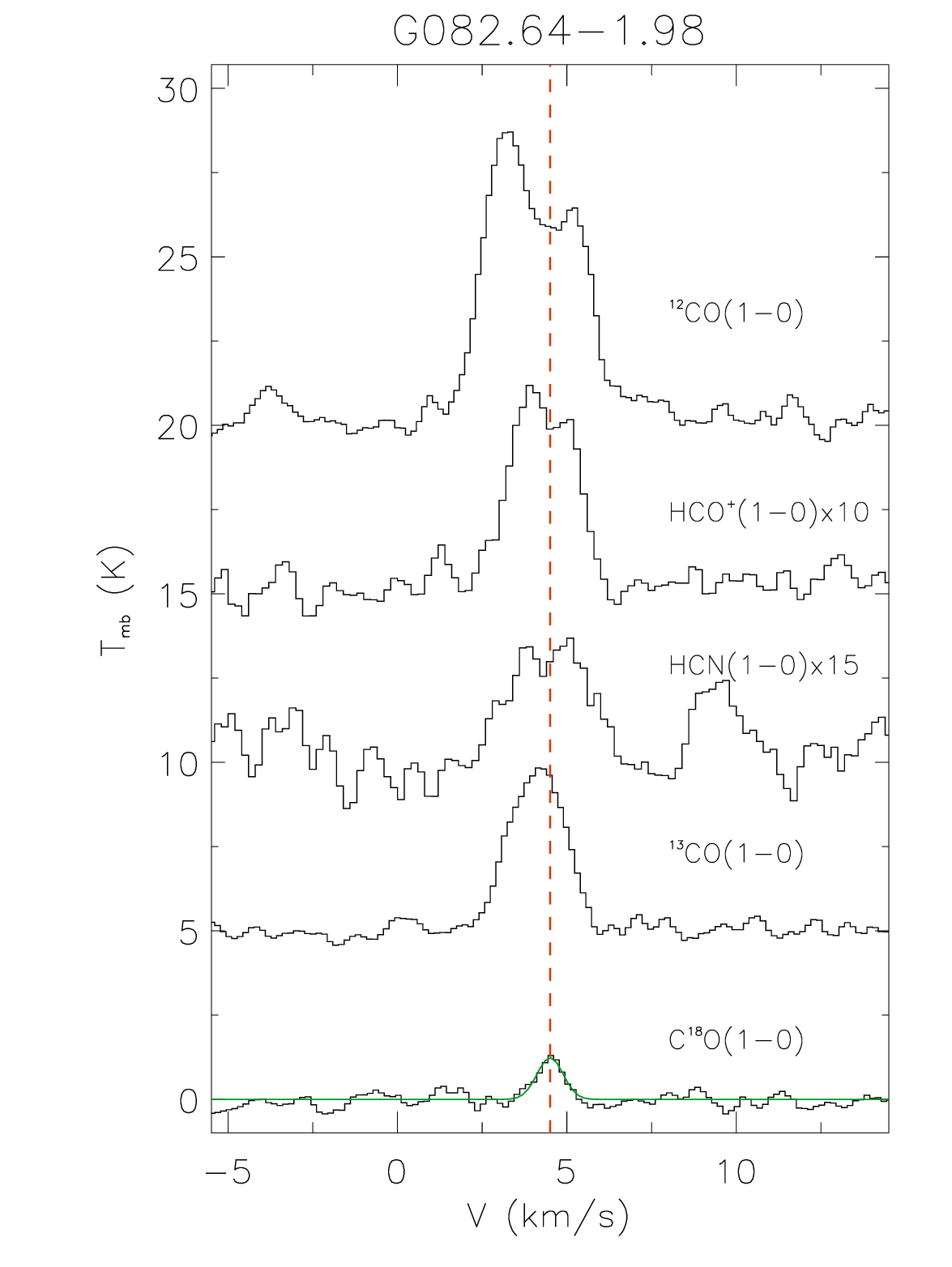}
  \end{minipage}%
  \begin{minipage}[t]{0.325\linewidth}
  \centering
   \includegraphics[width=55mm]{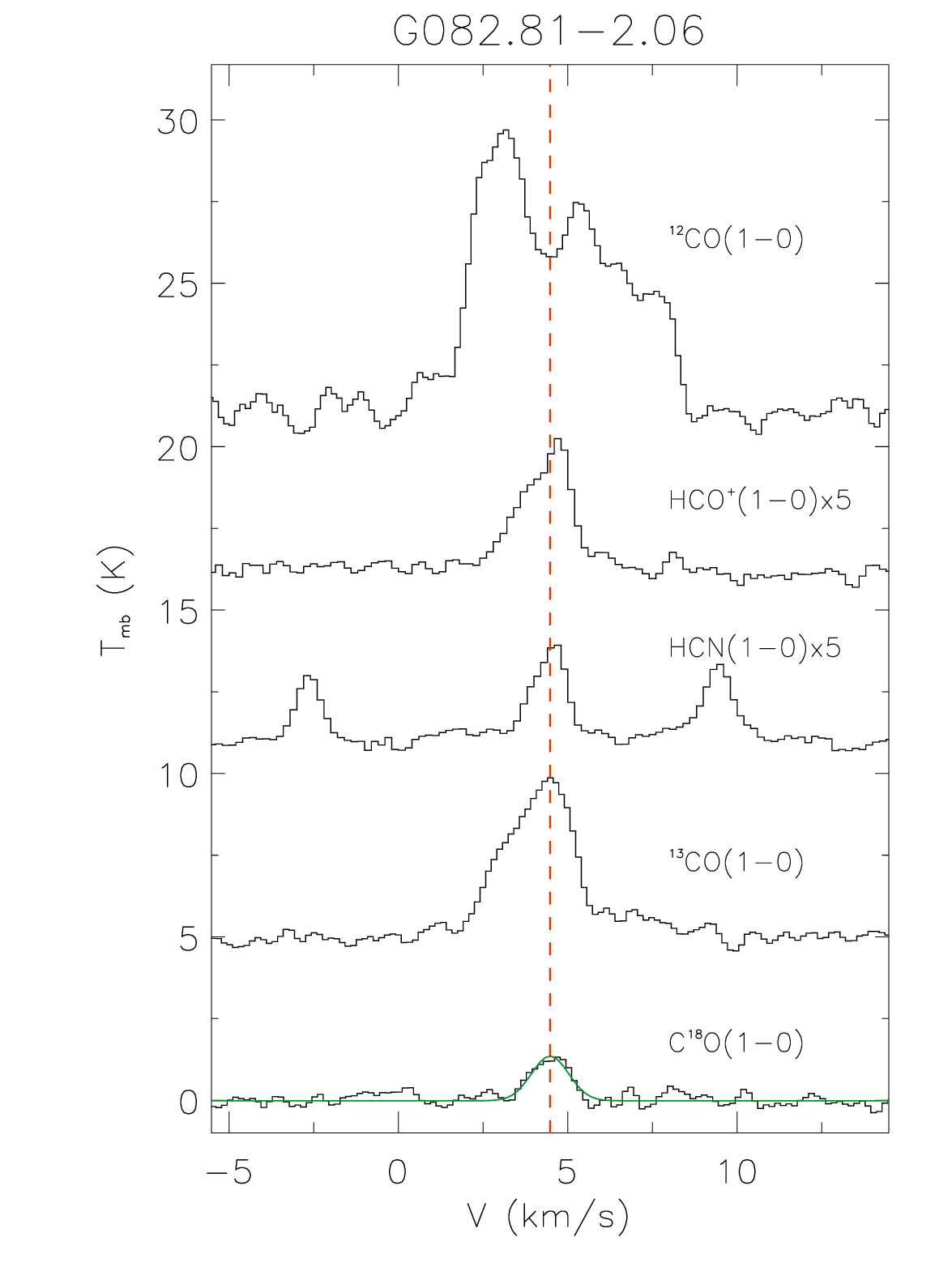}
  \end{minipage}%  
\quad
  \begin{minipage}[t]{0.325\linewidth}
  \centering
   \includegraphics[width=55mm]{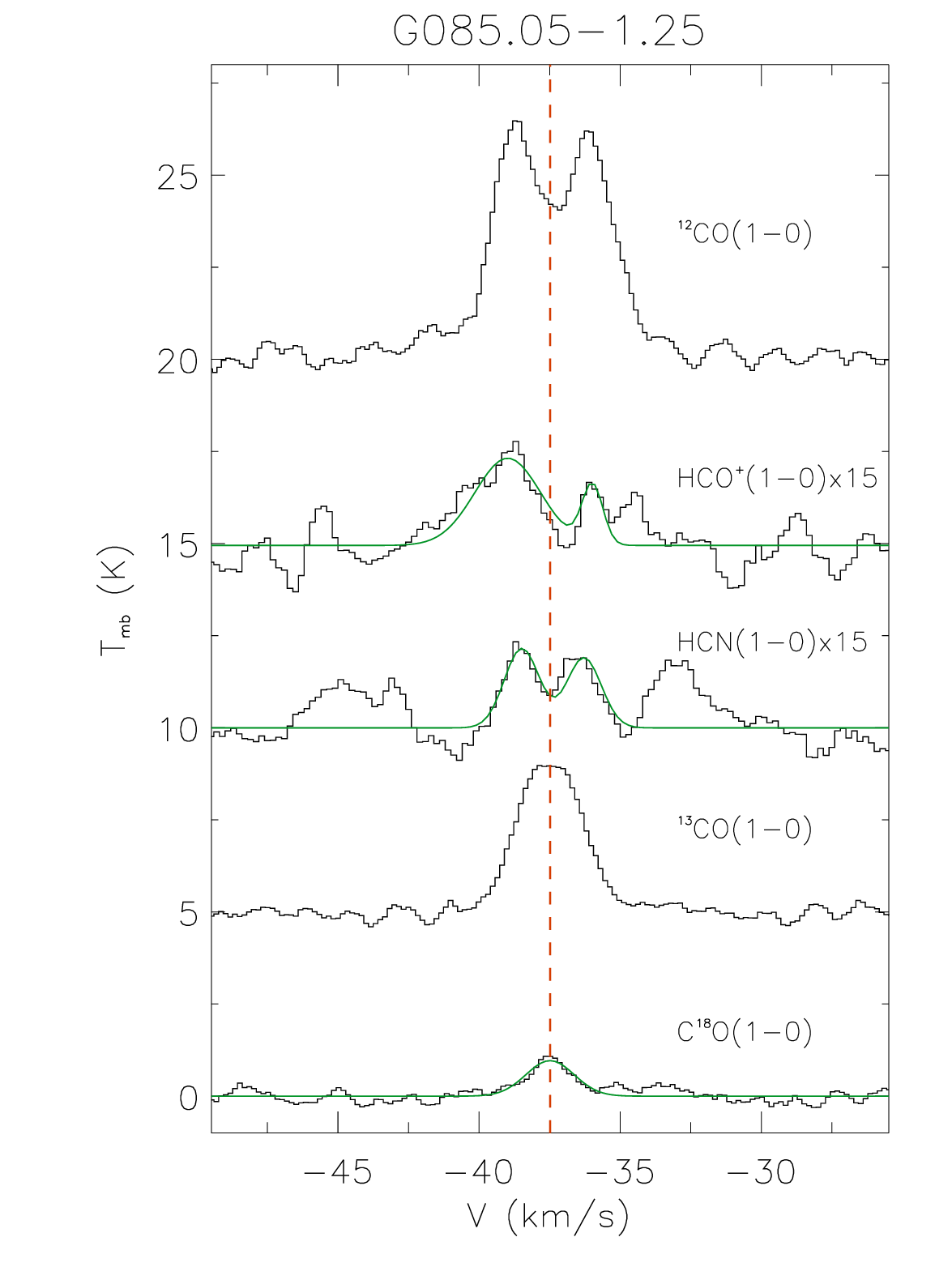}
  \end{minipage}%
  \begin{minipage}[t]{0.325\linewidth}
  \centering
   \includegraphics[width=55mm]{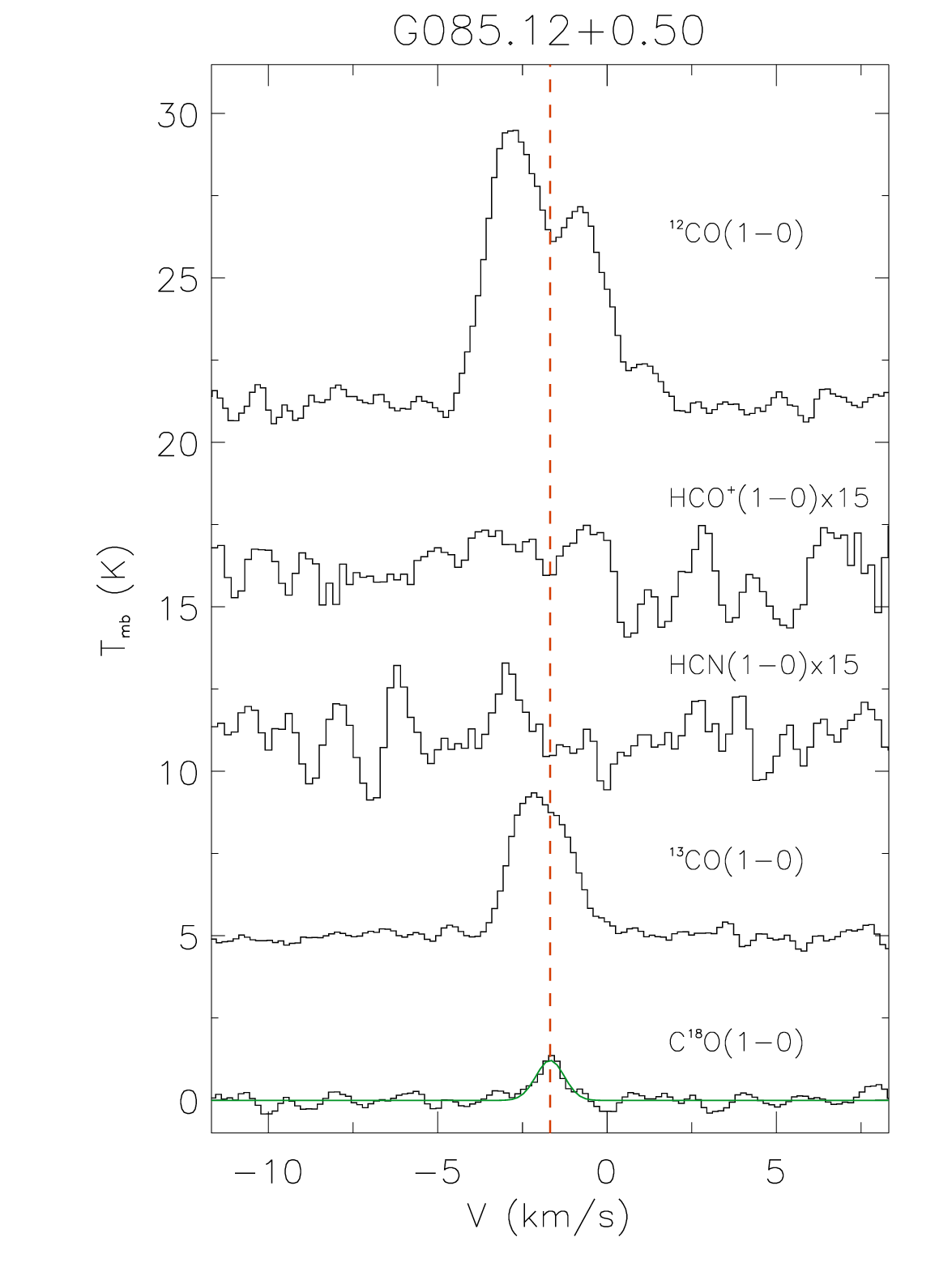}
  \end{minipage}%
  \begin{minipage}[t]{0.325\linewidth}
  \centering
   \includegraphics[width=55mm]{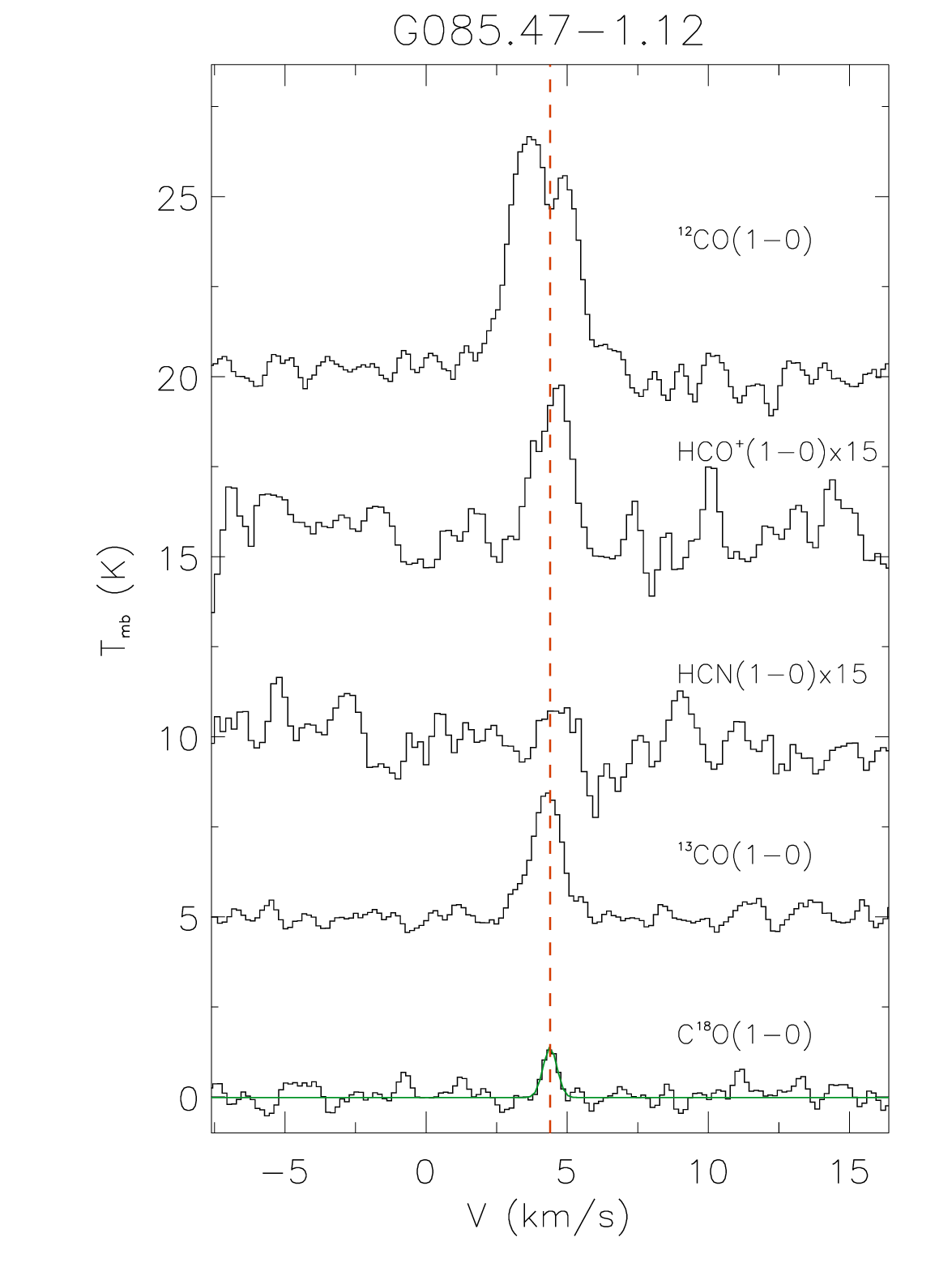}
  \end{minipage}%
\quad
  \begin{minipage}[t]{0.325\linewidth}
  \centering
   \includegraphics[width=55mm]{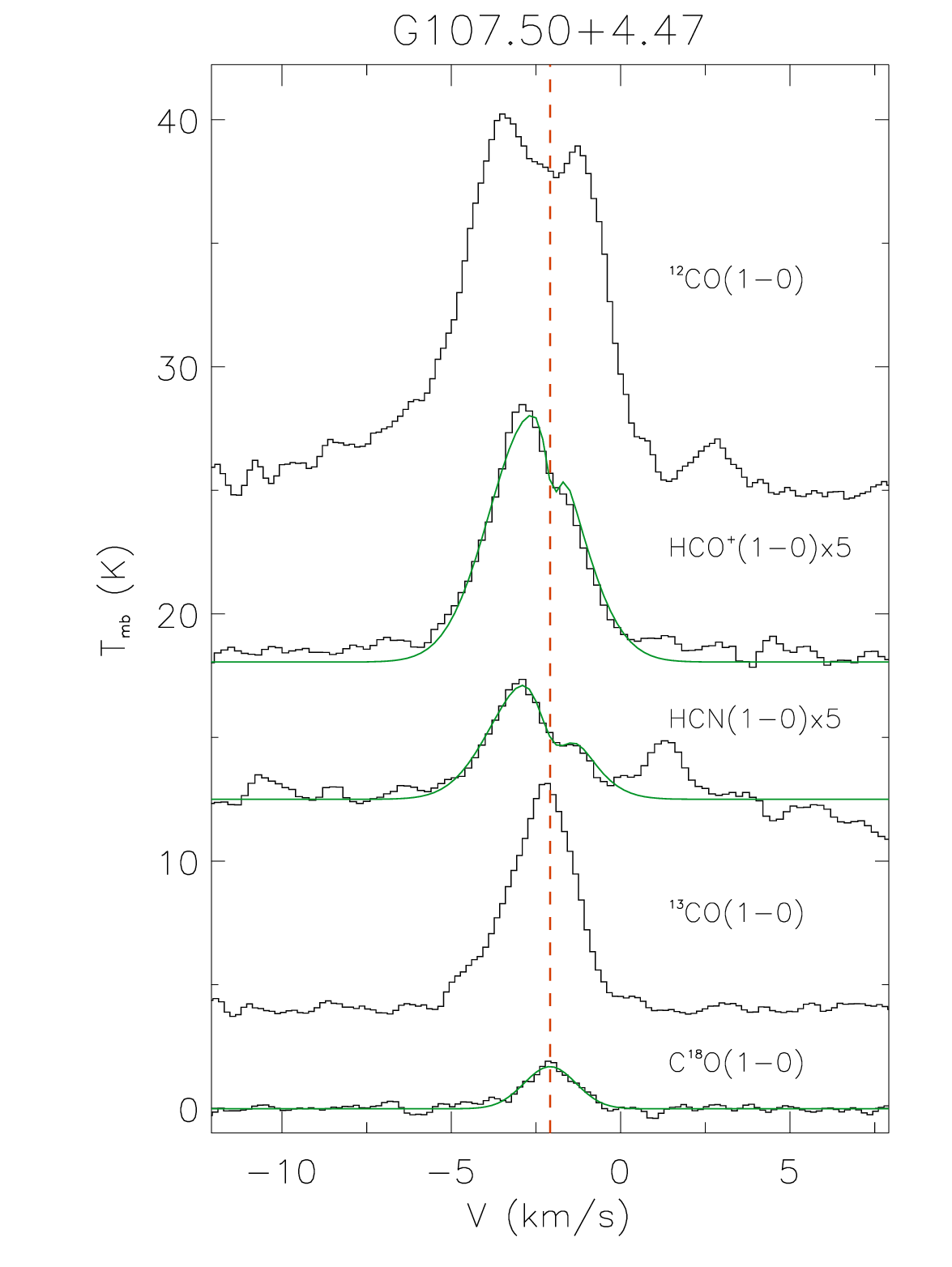}
  \end{minipage}%
  \begin{minipage}[t]{0.325\linewidth}
  \centering
   \includegraphics[width=55mm]{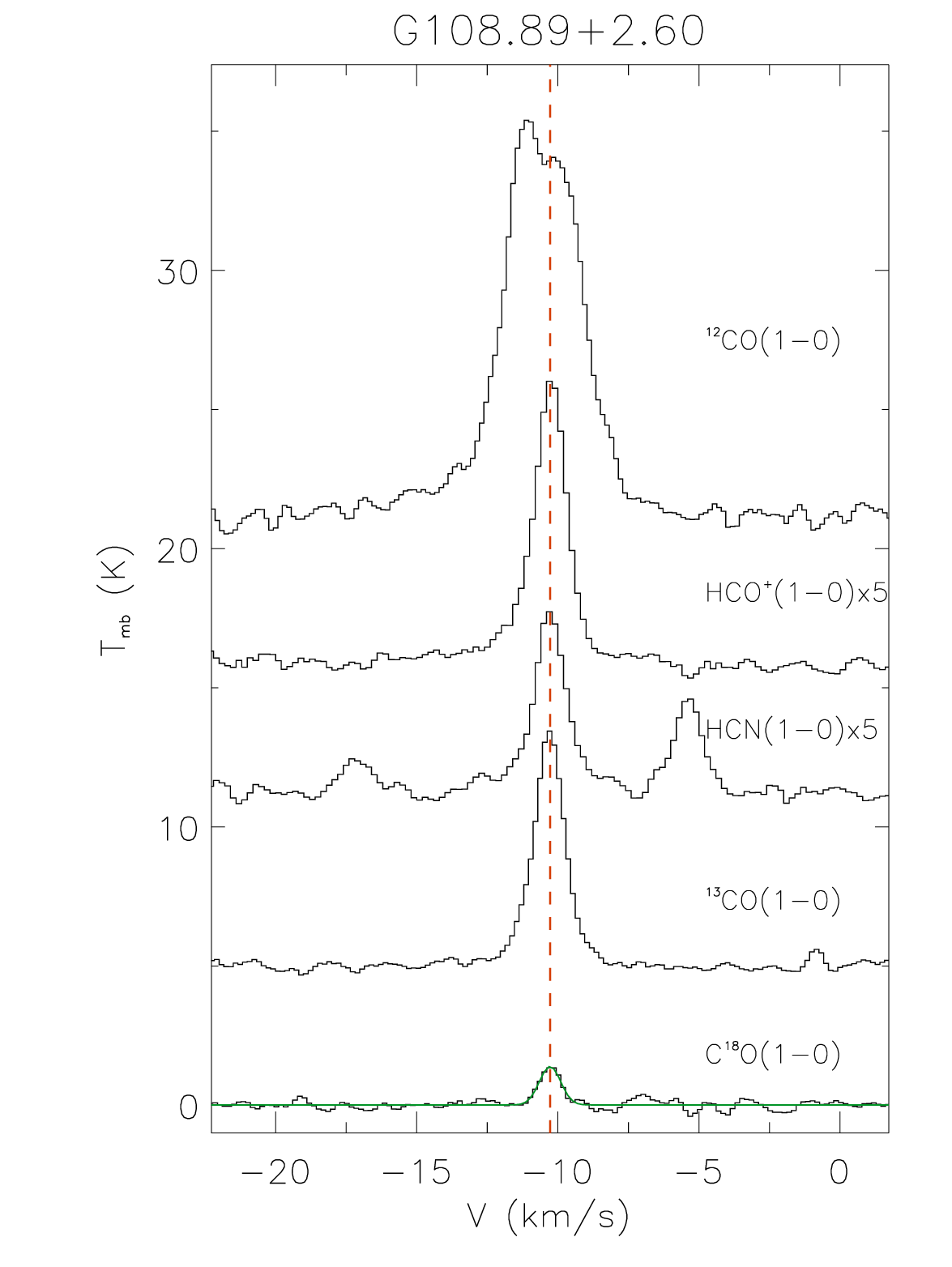}
  \end{minipage}%
  \begin{minipage}[t]{0.325\linewidth}
  \centering
   \includegraphics[width=55mm]{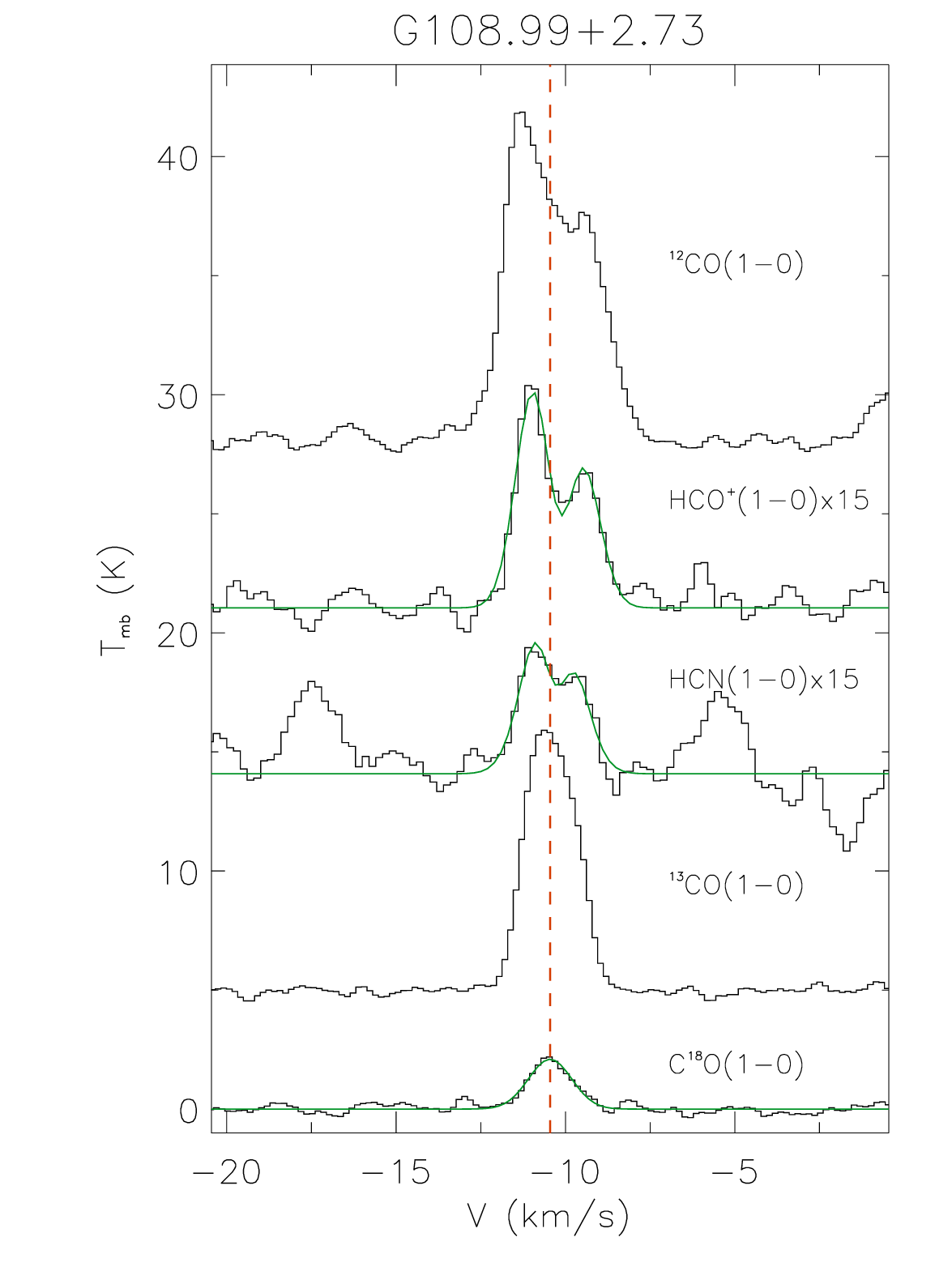}
  \end{minipage}%
  \caption{{\small Line profiles of 133 sources we selected. The lines from bottom to top are C$^{18}$O (1-0), $^{13}$CO (1-0), HCN (1-0) (14 sources lack HCN data), HCO$^+$ (1-0) and $^{12}$CO (1-0), respectively. The dashed red line indicates the central radial velocity of C$^{18}$O (1-0) estimated by Gaussian fitting. For infall candidates, HCO$^+$ (1-0) and HCN (1-0) lines are also Gaussian fitted.}}
  \label{Fig:fig6}
\end{figure} 

\begin{figure}[h]
\ContinuedFloat
  \begin{minipage}[t]{0.325\linewidth}
  \centering
   \includegraphics[width=55mm]{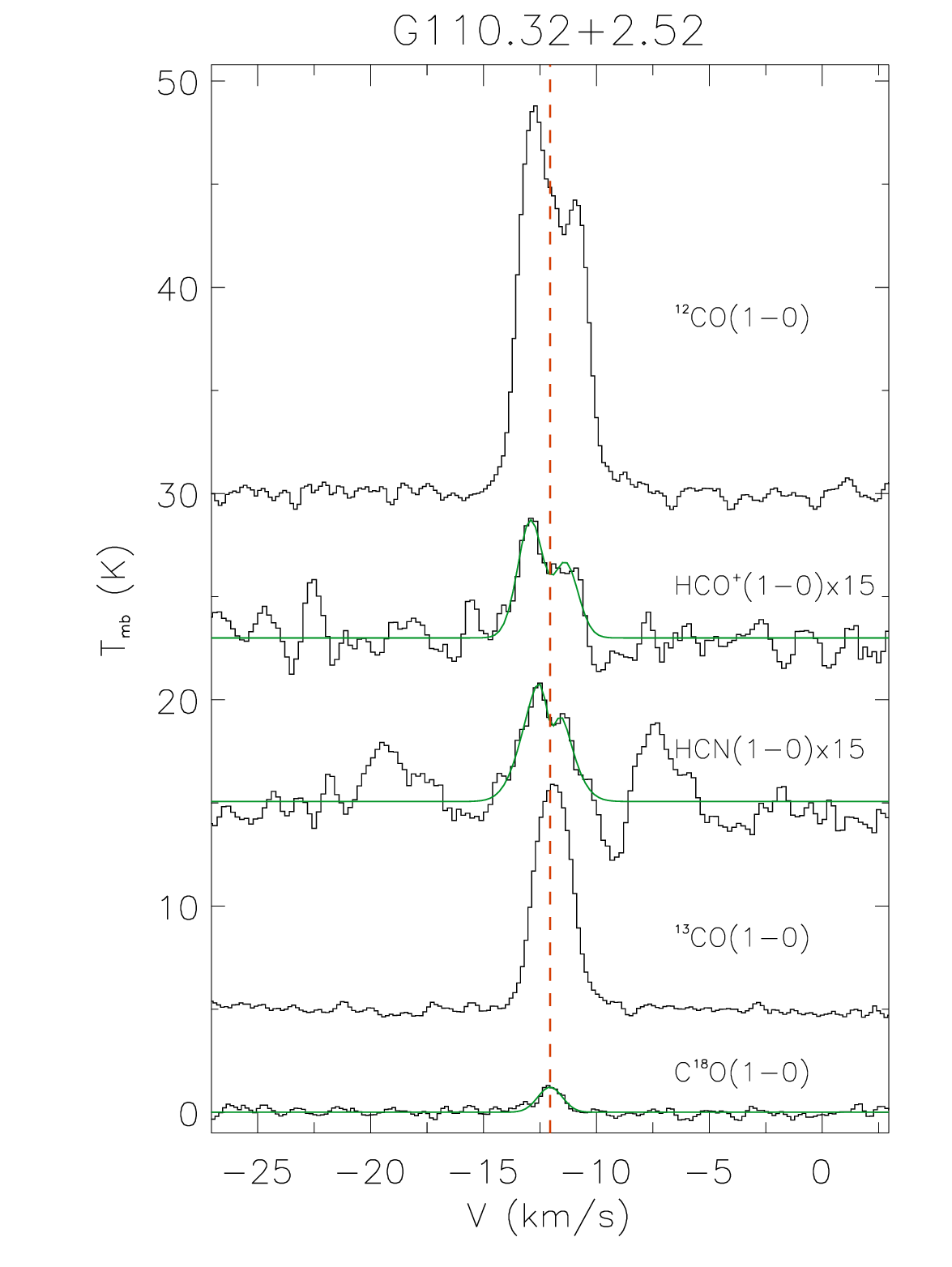}
  \end{minipage}%
  \begin{minipage}[t]{0.325\textwidth}
  \centering
   \includegraphics[width=55mm]{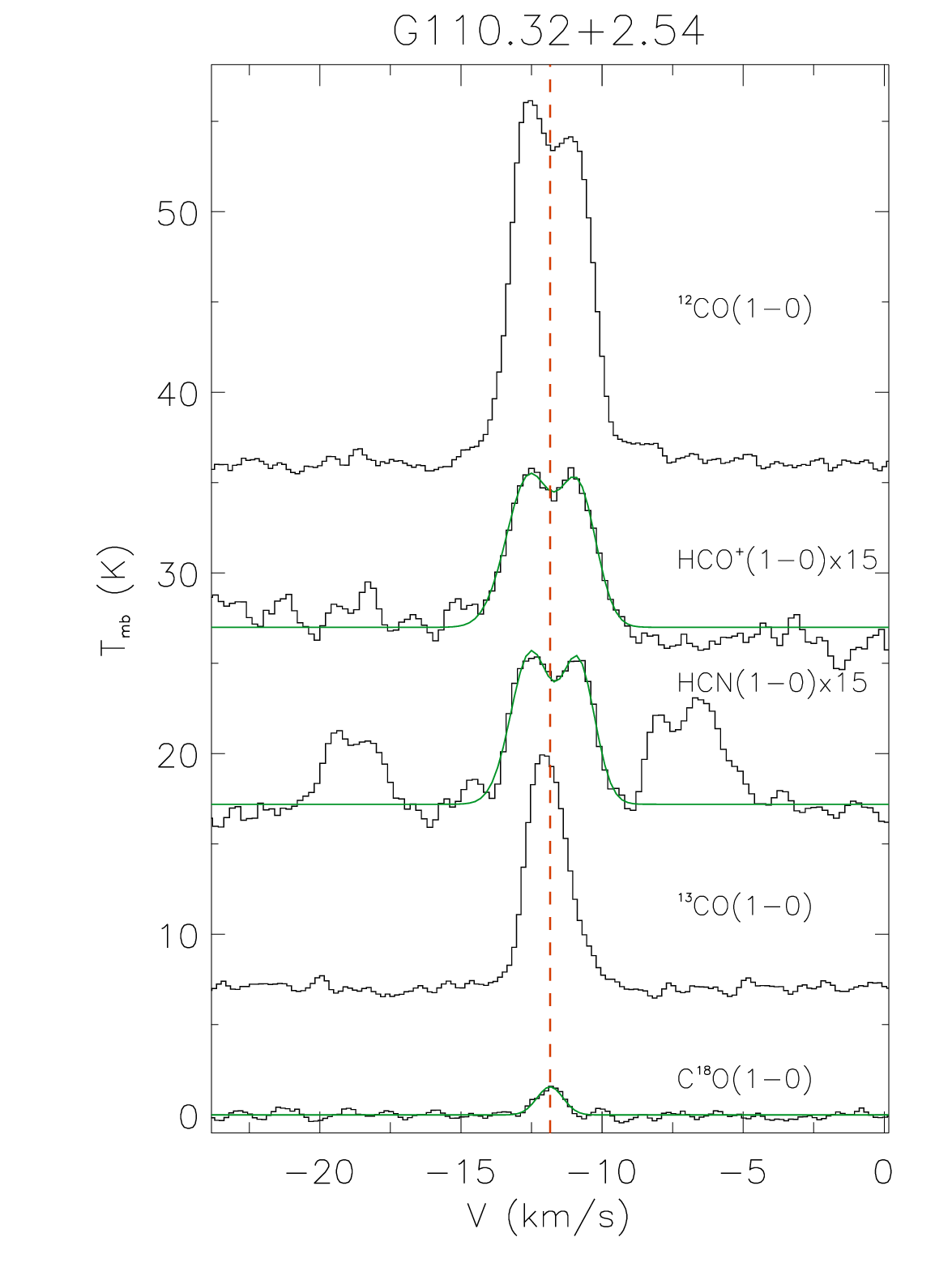}
  \end{minipage}%
  \begin{minipage}[t]{0.325\linewidth}
  \centering
   \includegraphics[width=55mm]{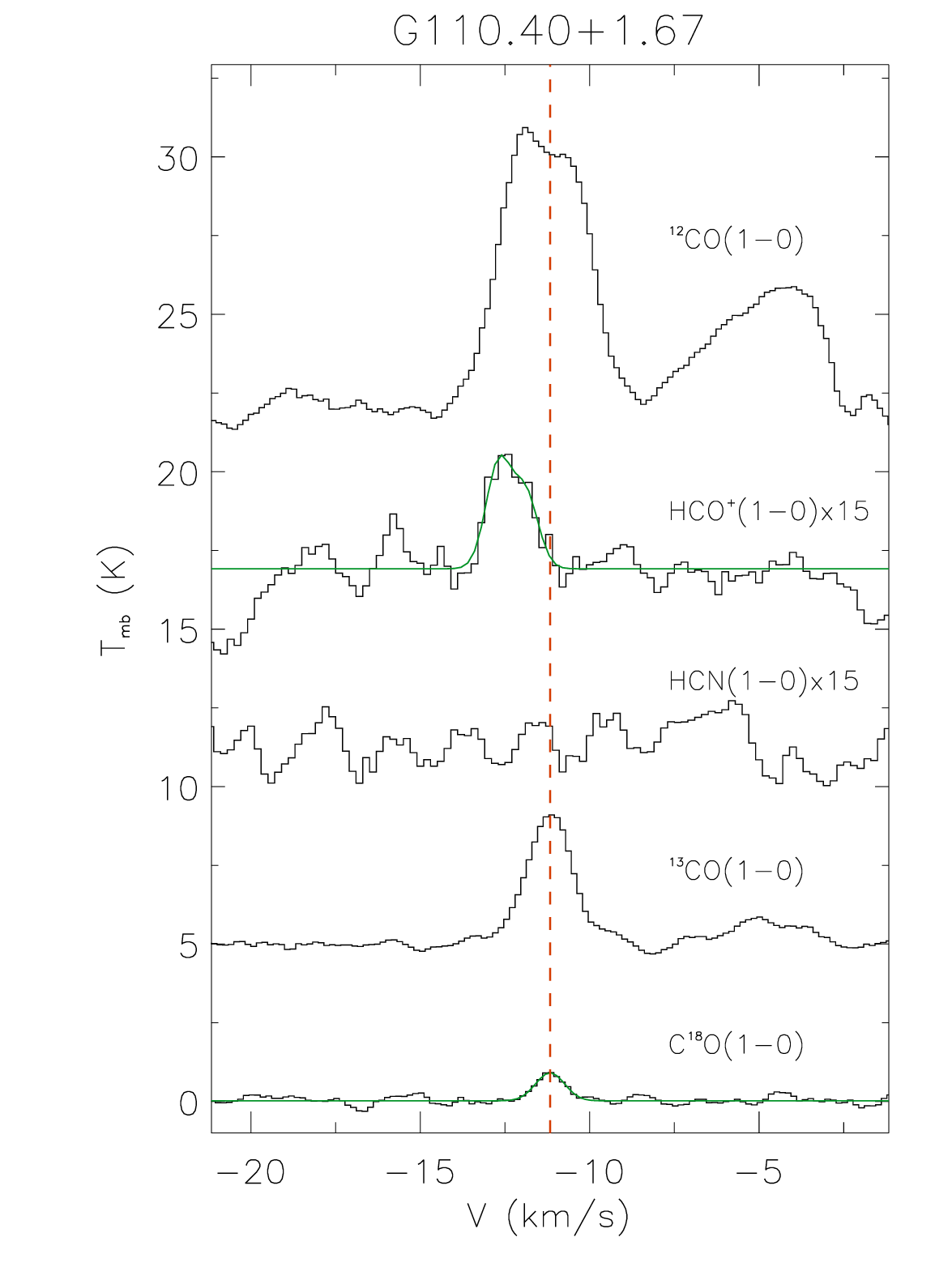}
  \end{minipage}%  
\quad
  \begin{minipage}[t]{0.325\linewidth}
  \centering
   \includegraphics[width=55mm]{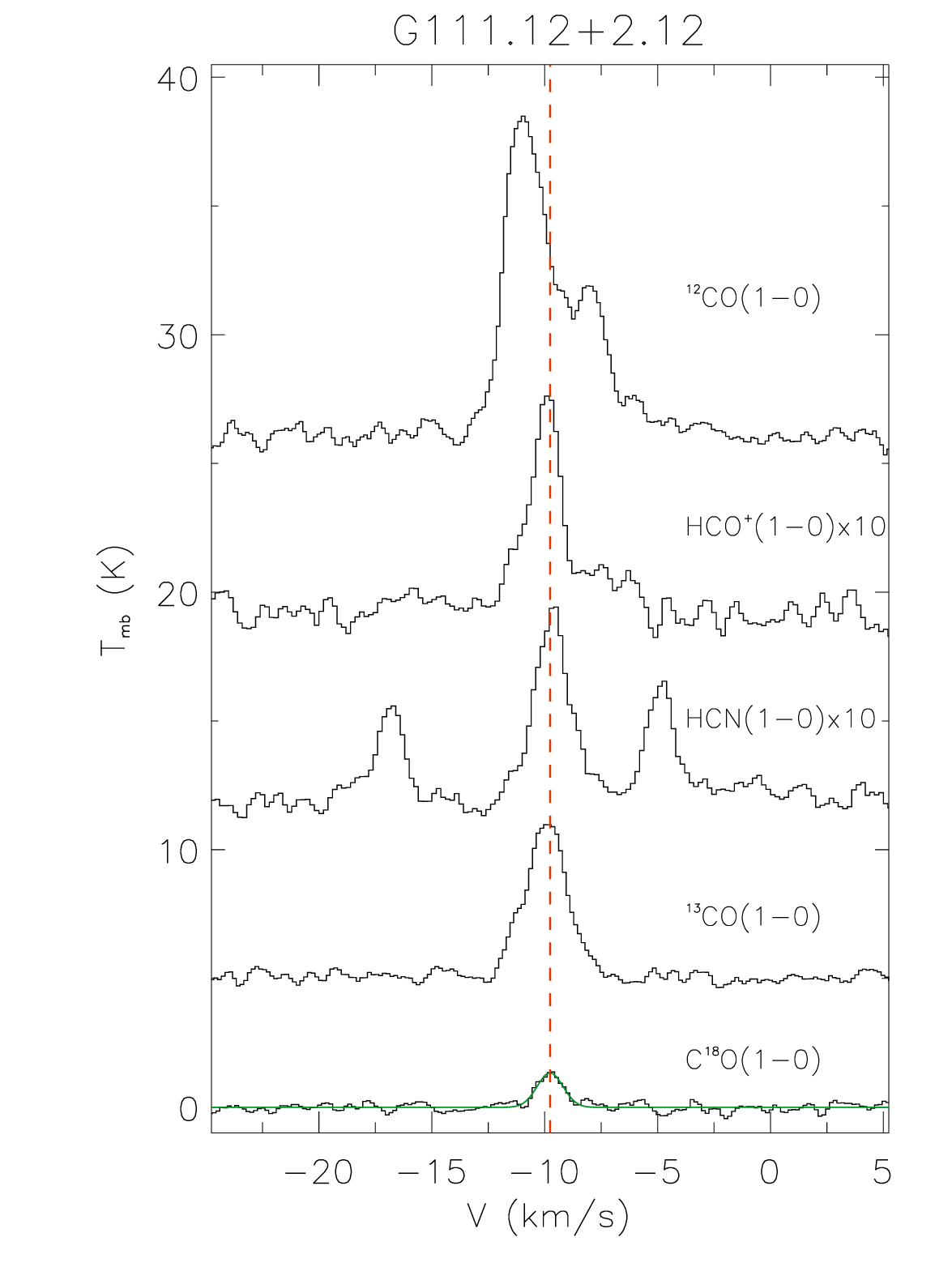}
  \end{minipage}%
  \begin{minipage}[t]{0.325\linewidth}
  \centering
   \includegraphics[width=55mm]{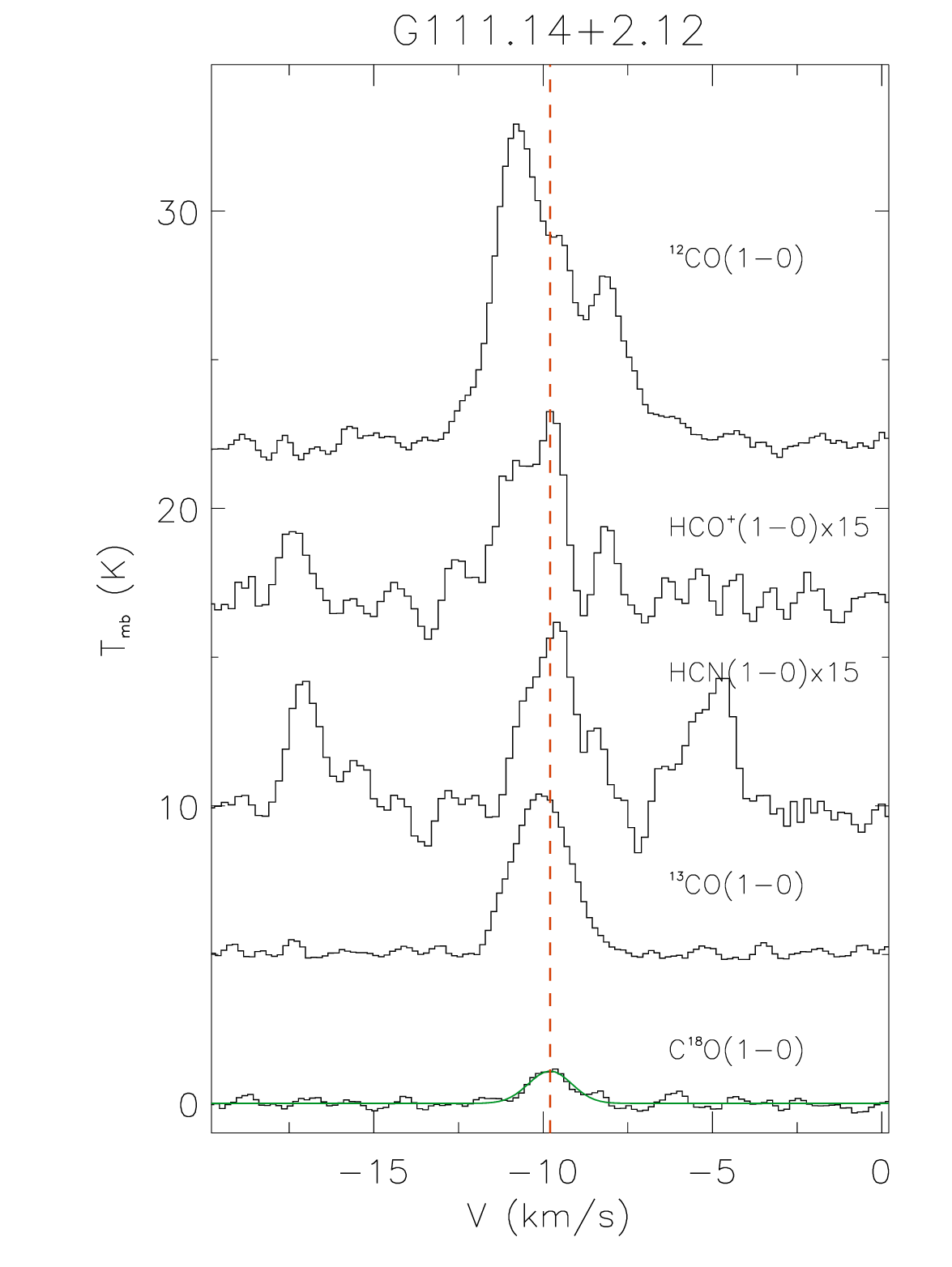}
  \end{minipage}%
  \begin{minipage}[t]{0.325\linewidth}
  \centering
   \includegraphics[width=55mm]{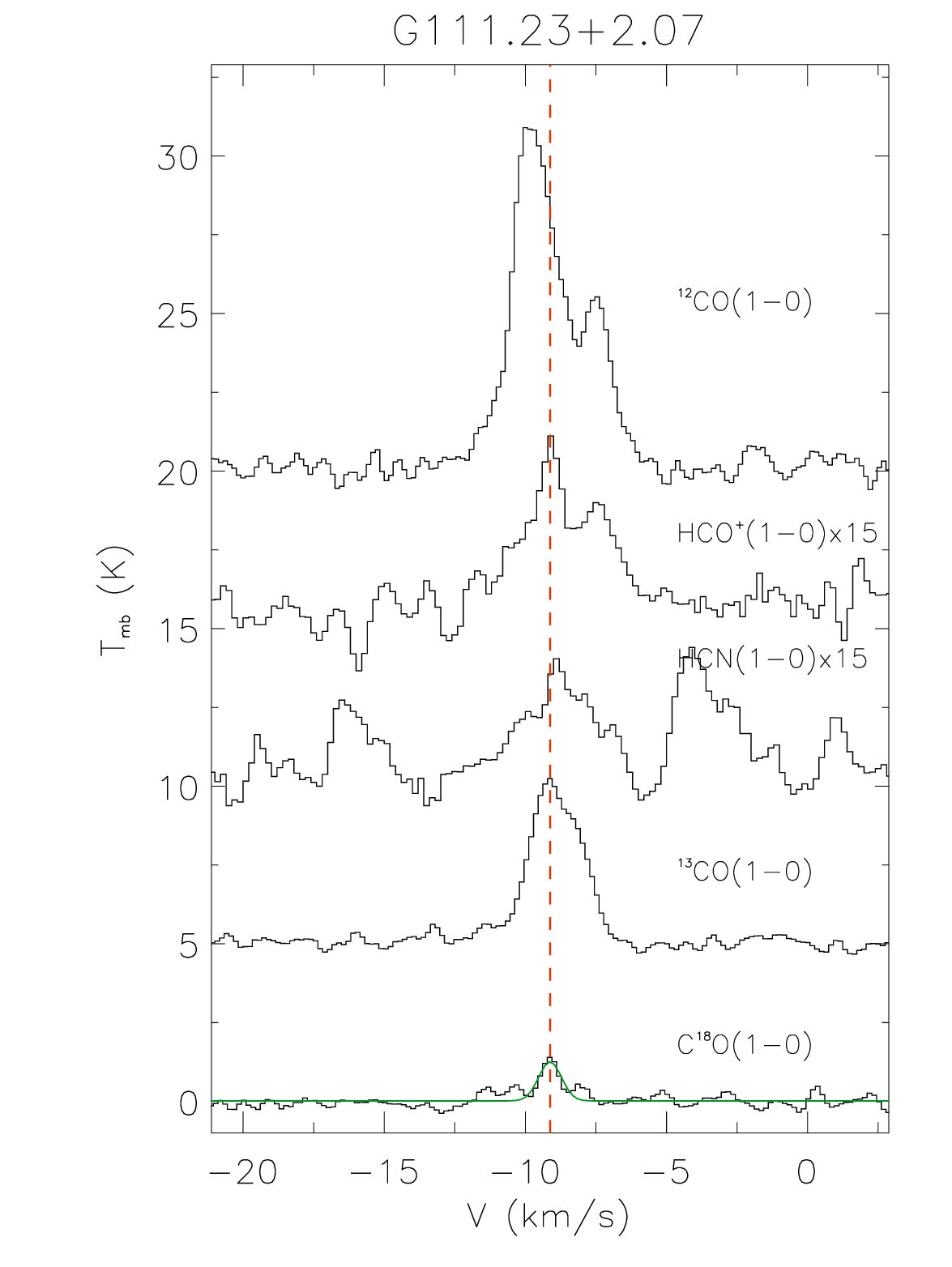}
  \end{minipage}%
\quad
  \begin{minipage}[t]{0.325\linewidth}
  \centering
   \includegraphics[width=55mm]{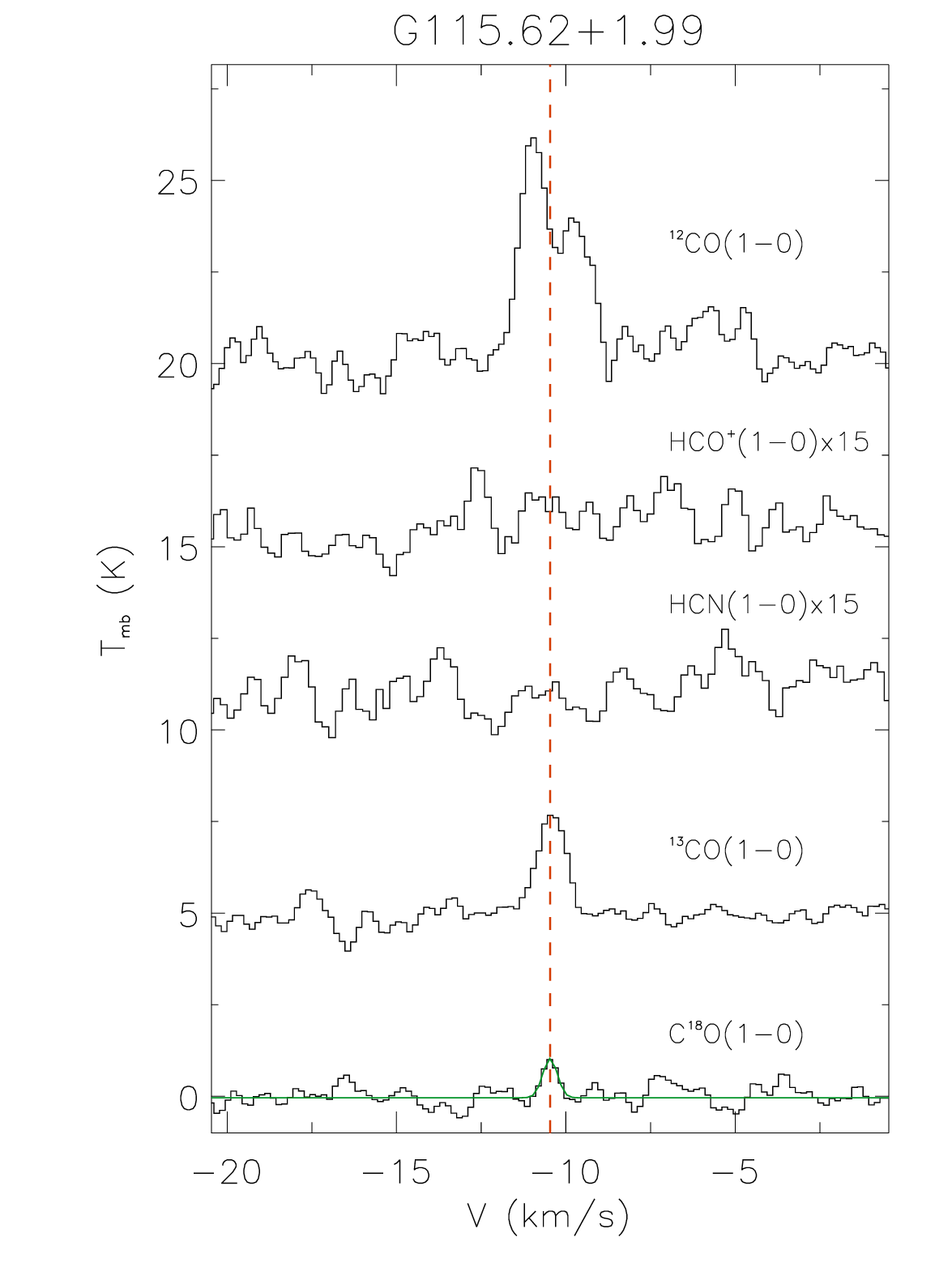}
  \end{minipage}%
  \begin{minipage}[t]{0.325\linewidth}
  \centering
   \includegraphics[width=55mm]{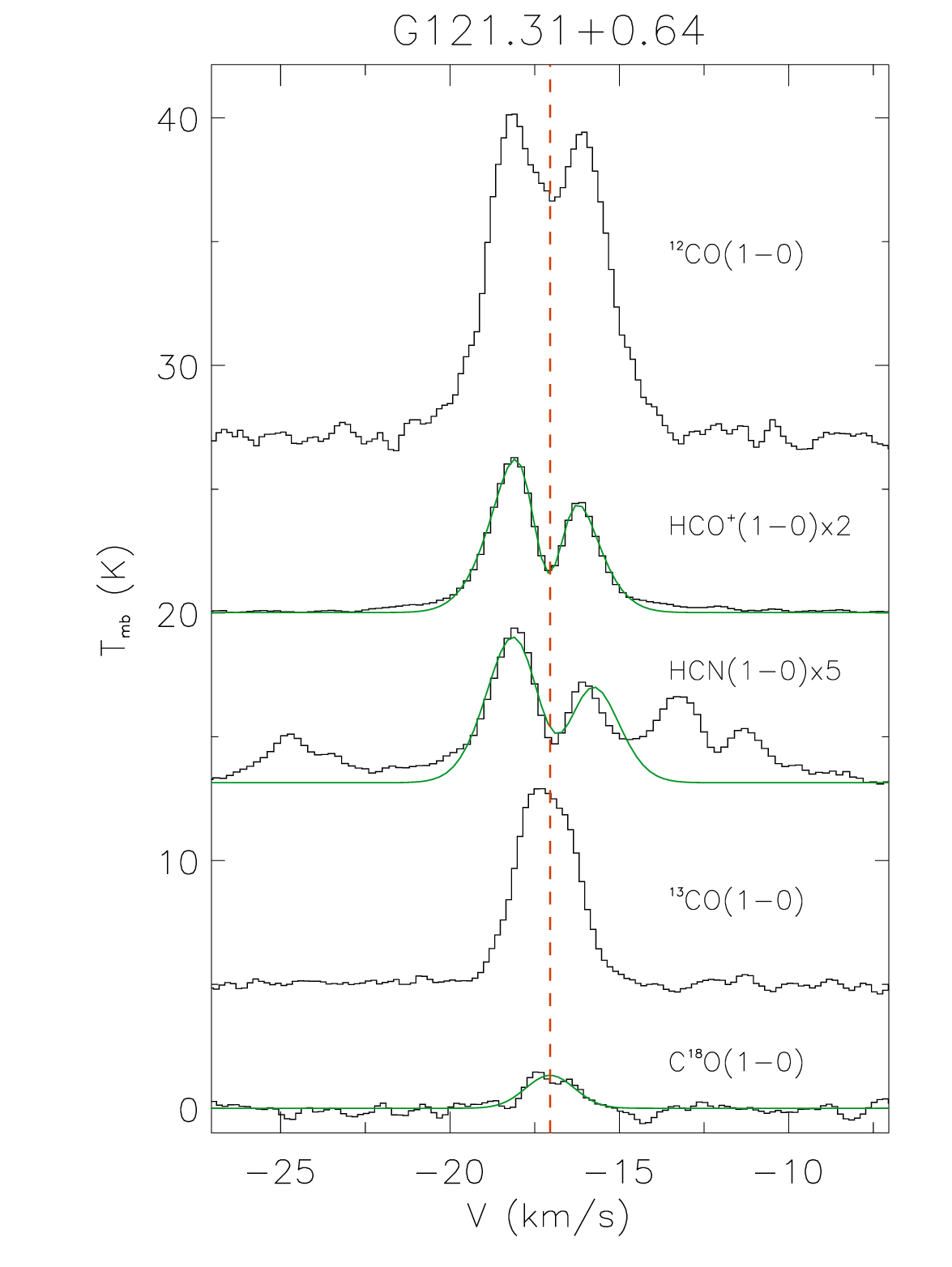}
  \end{minipage}%
  \begin{minipage}[t]{0.325\linewidth}
  \centering
   \includegraphics[width=55mm]{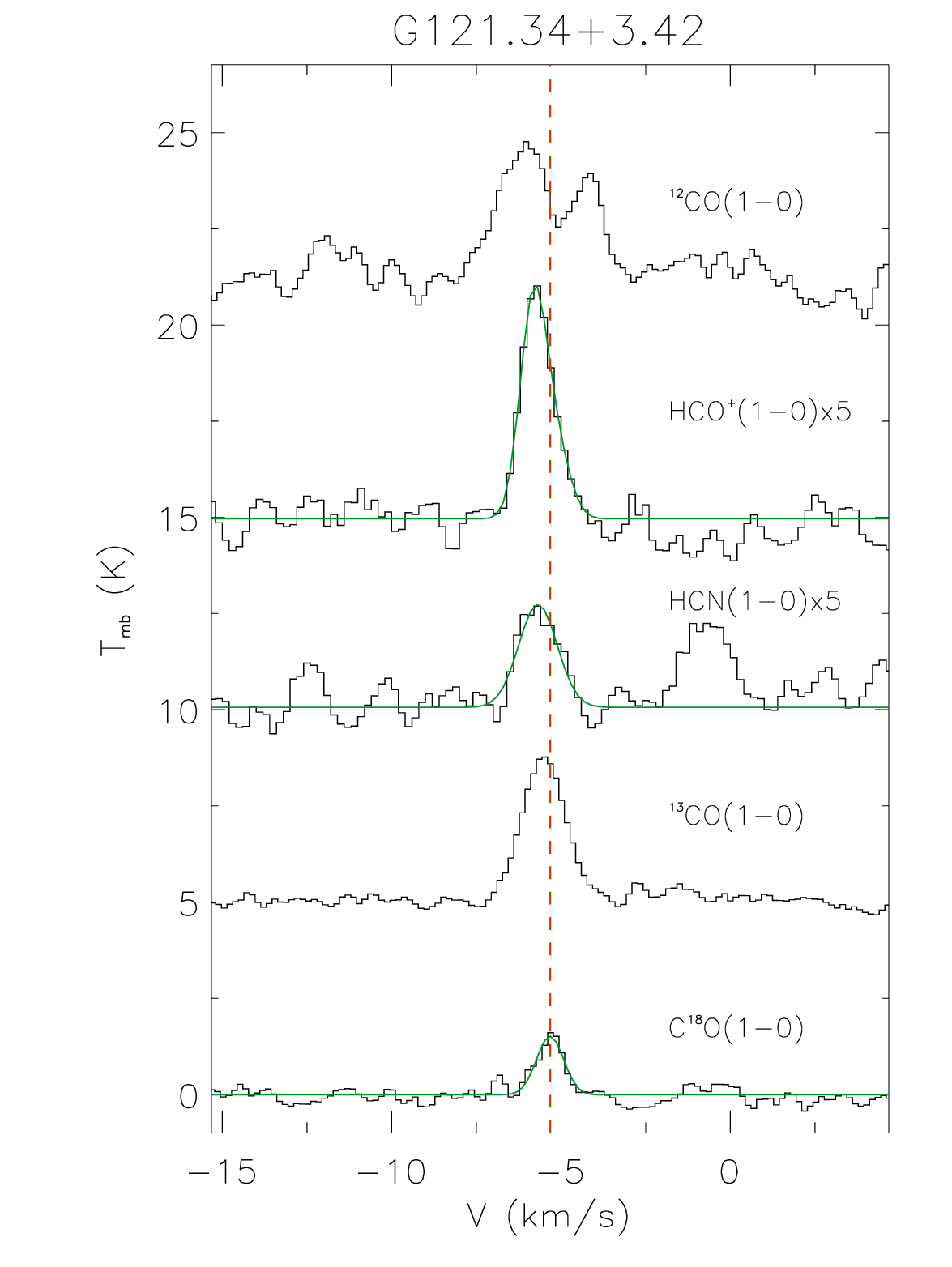}
  \end{minipage}%
  \caption{{\small Line profiles of 133 sources we selected. The lines from bottom to top are C$^{18}$O (1-0), $^{13}$CO (1-0), HCN (1-0) (14 sources lack HCN data), HCO$^+$ (1-0) and $^{12}$CO (1-0), respectively. The dashed red line indicates the central radial velocity of C$^{18}$O (1-0) estimated by Gaussian fitting. For infall candidates, HCO$^+$ (1-0) and HCN (1-0) lines are also Gaussian fitted.}}
  \label{Fig:fig6}
\end{figure} 

\begin{figure}[h]
\ContinuedFloat
  \begin{minipage}[t]{0.325\linewidth}
  \centering
   \includegraphics[width=55mm]{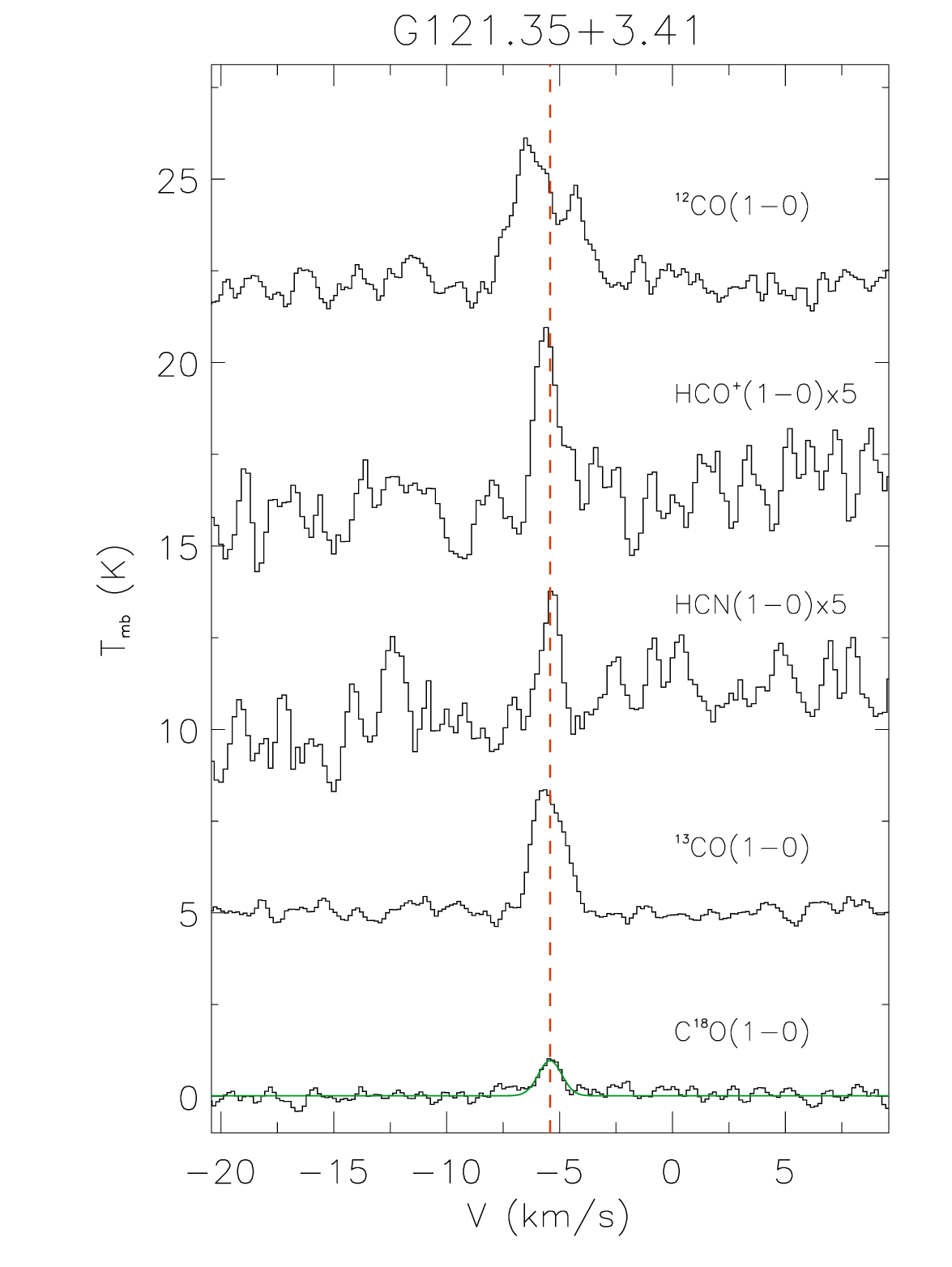}
  \end{minipage}%
  \begin{minipage}[t]{0.325\textwidth}
  \centering
   \includegraphics[width=55mm]{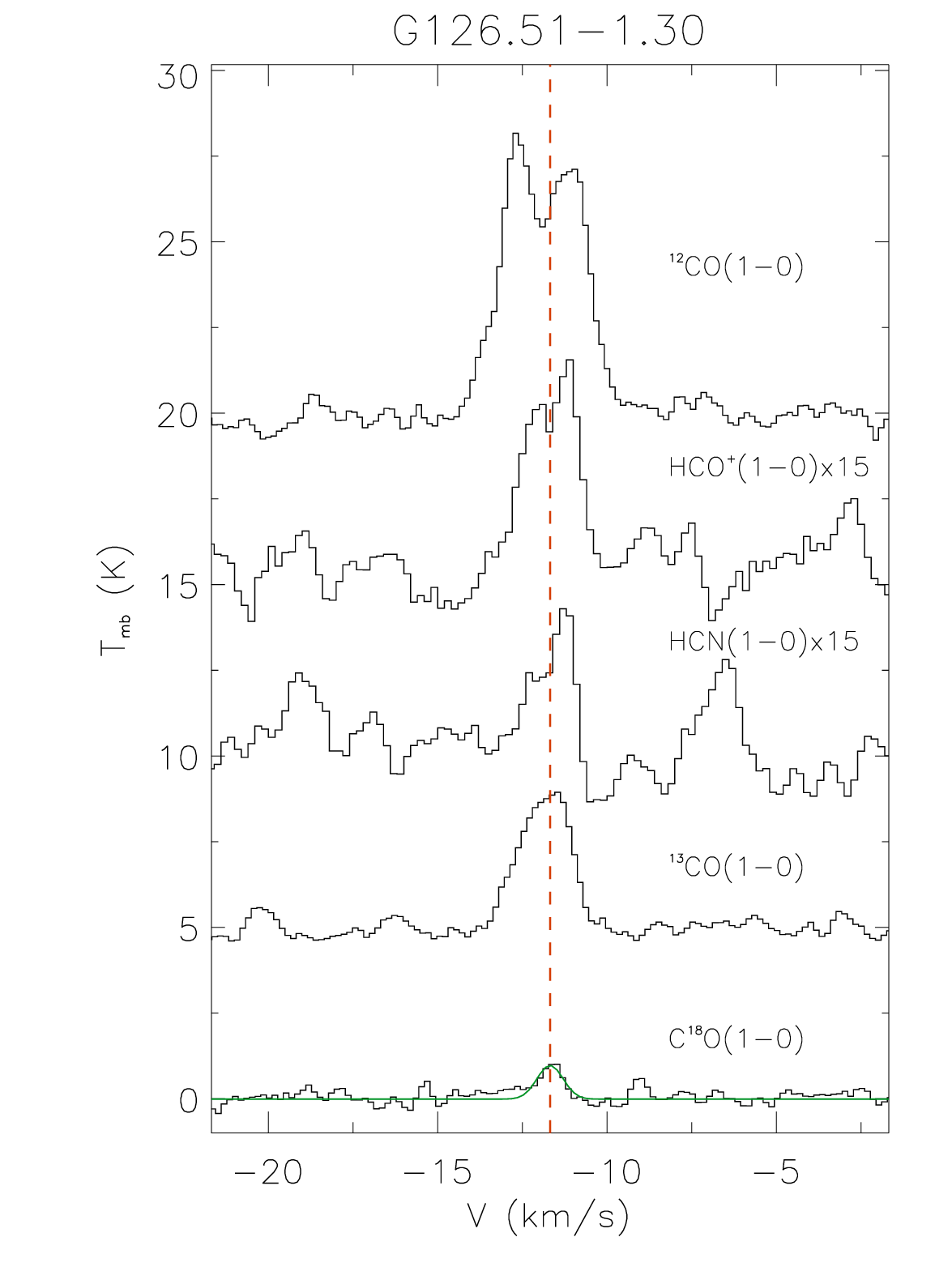}
  \end{minipage}%
  \begin{minipage}[t]{0.325\linewidth}
  \centering
   \includegraphics[width=55mm]{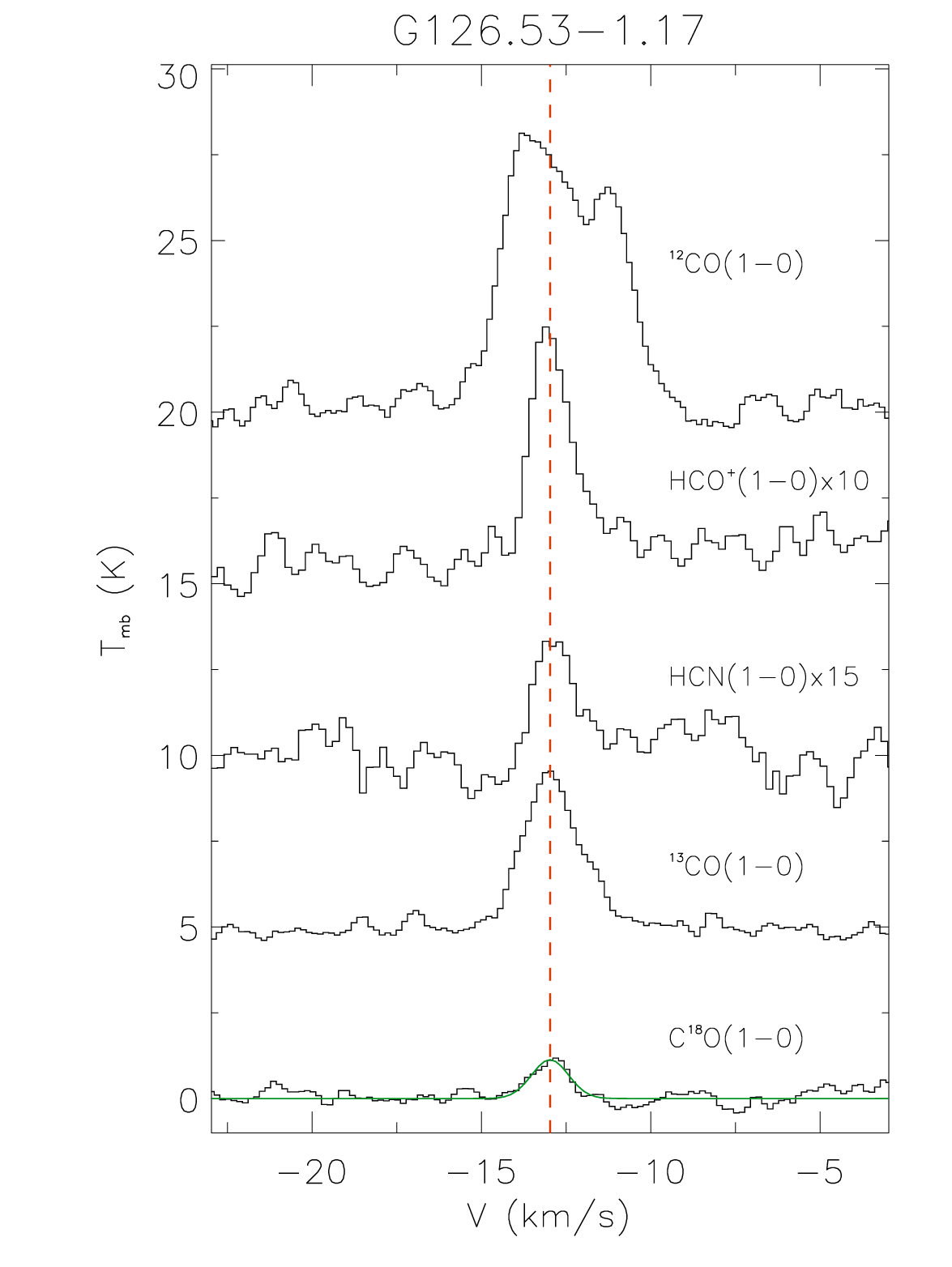}
  \end{minipage}%  
\quad
  \begin{minipage}[t]{0.325\linewidth}
  \centering
   \includegraphics[width=55mm]{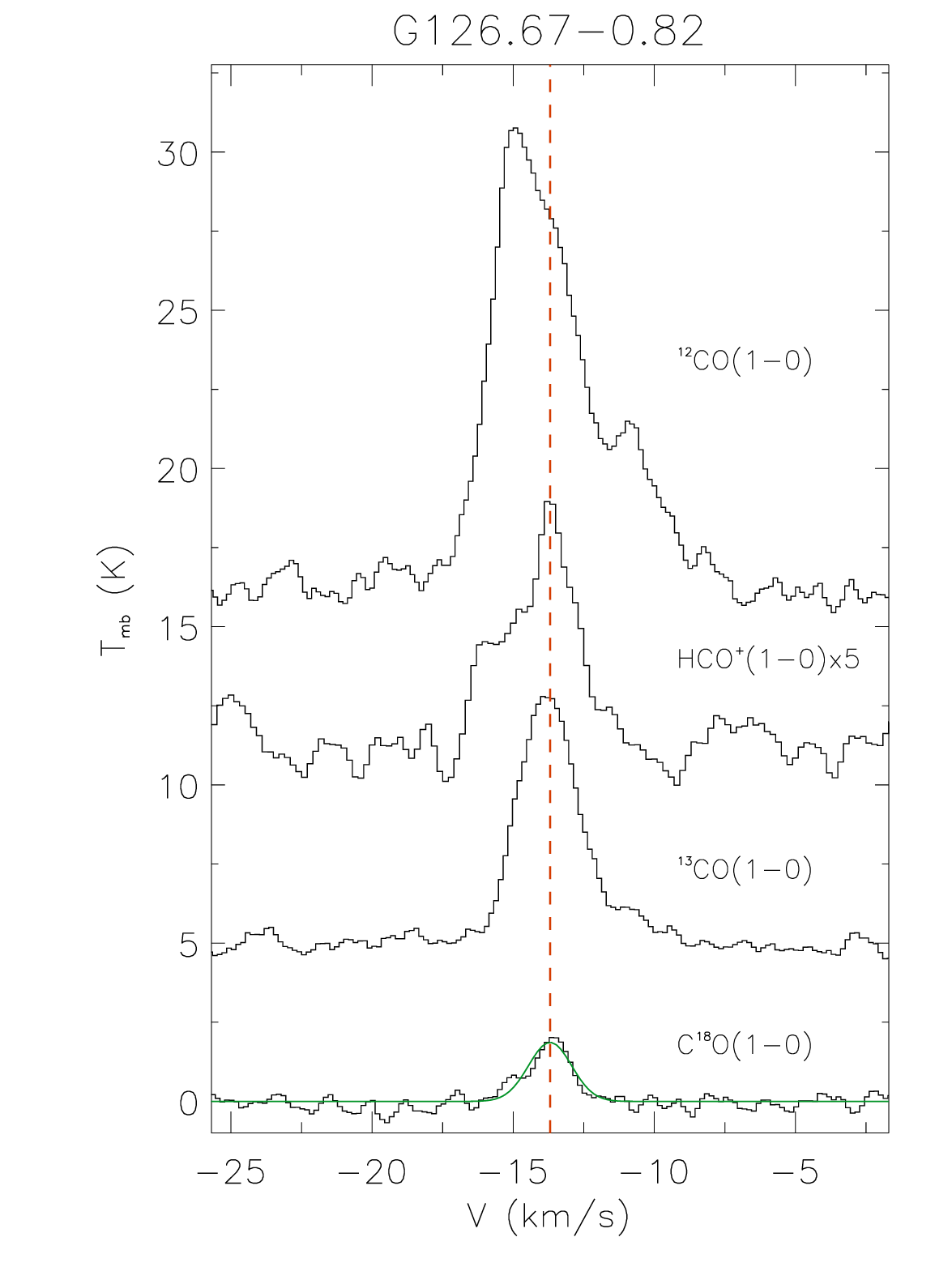}
  \end{minipage}%
  \begin{minipage}[t]{0.325\linewidth}
  \centering
   \includegraphics[width=55mm]{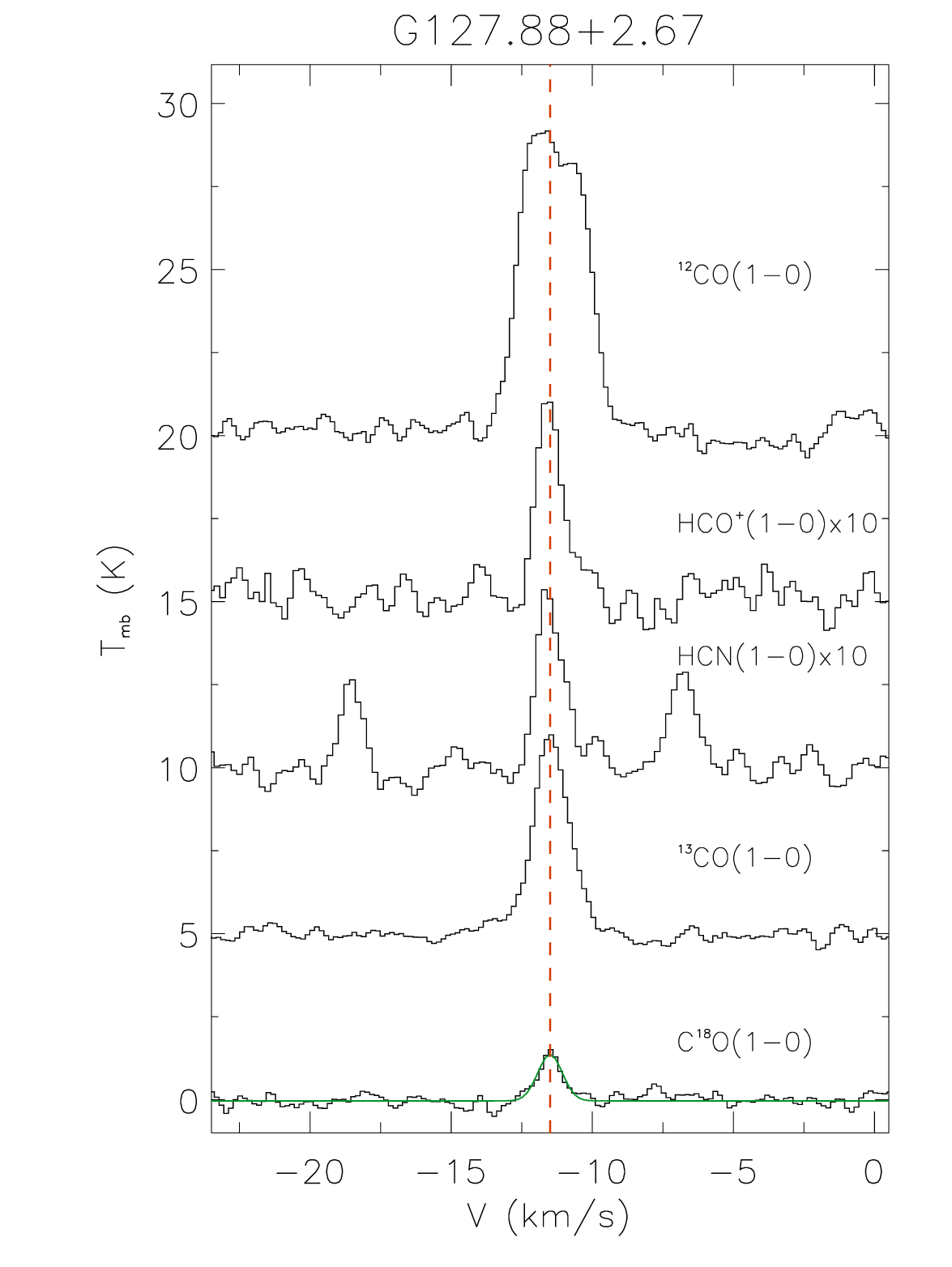}
  \end{minipage}%
  \begin{minipage}[t]{0.325\linewidth}
  \centering
   \includegraphics[width=55mm]{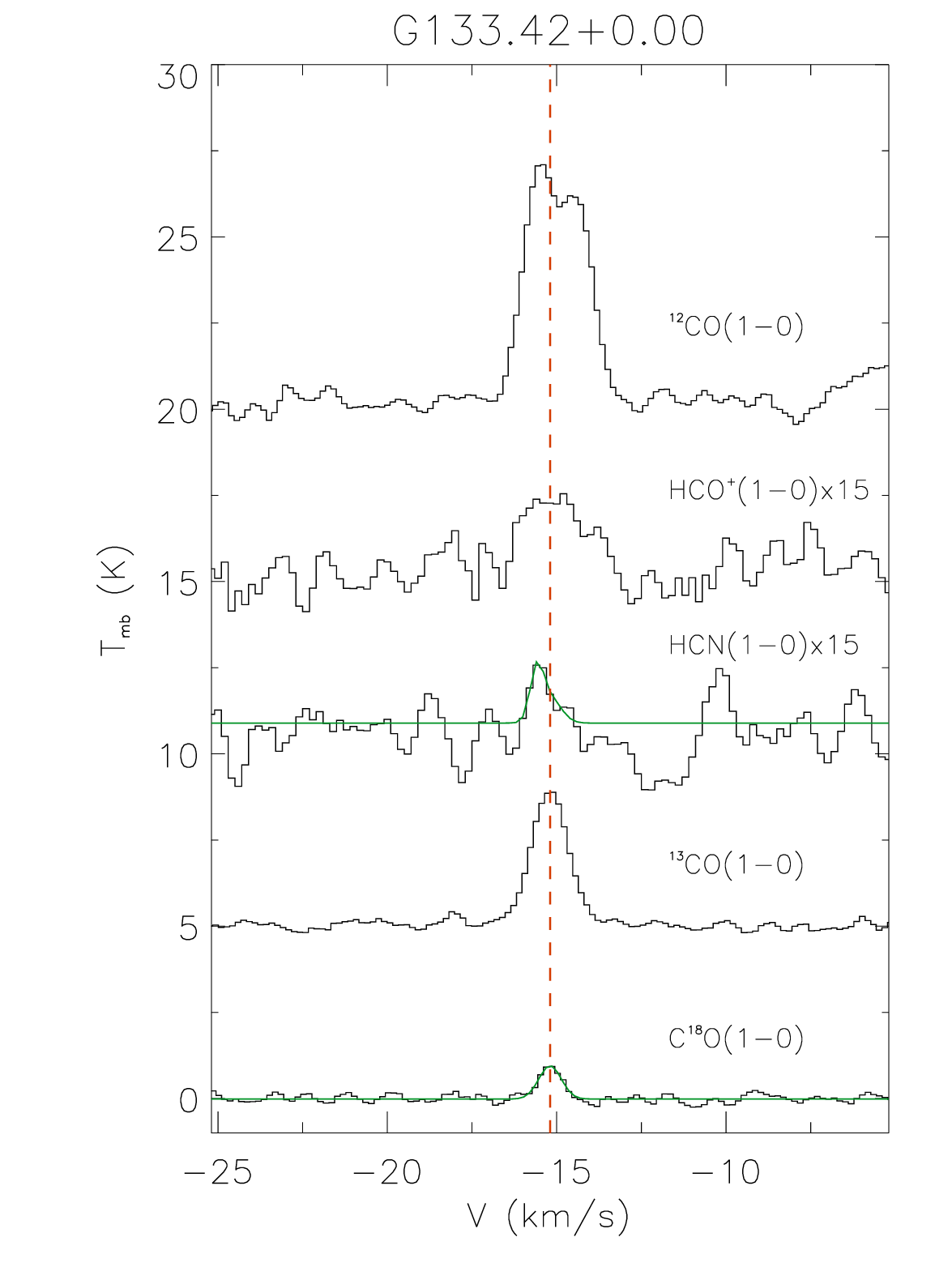}
  \end{minipage}%
\quad
  \begin{minipage}[t]{0.325\linewidth}
  \centering
   \includegraphics[width=55mm]{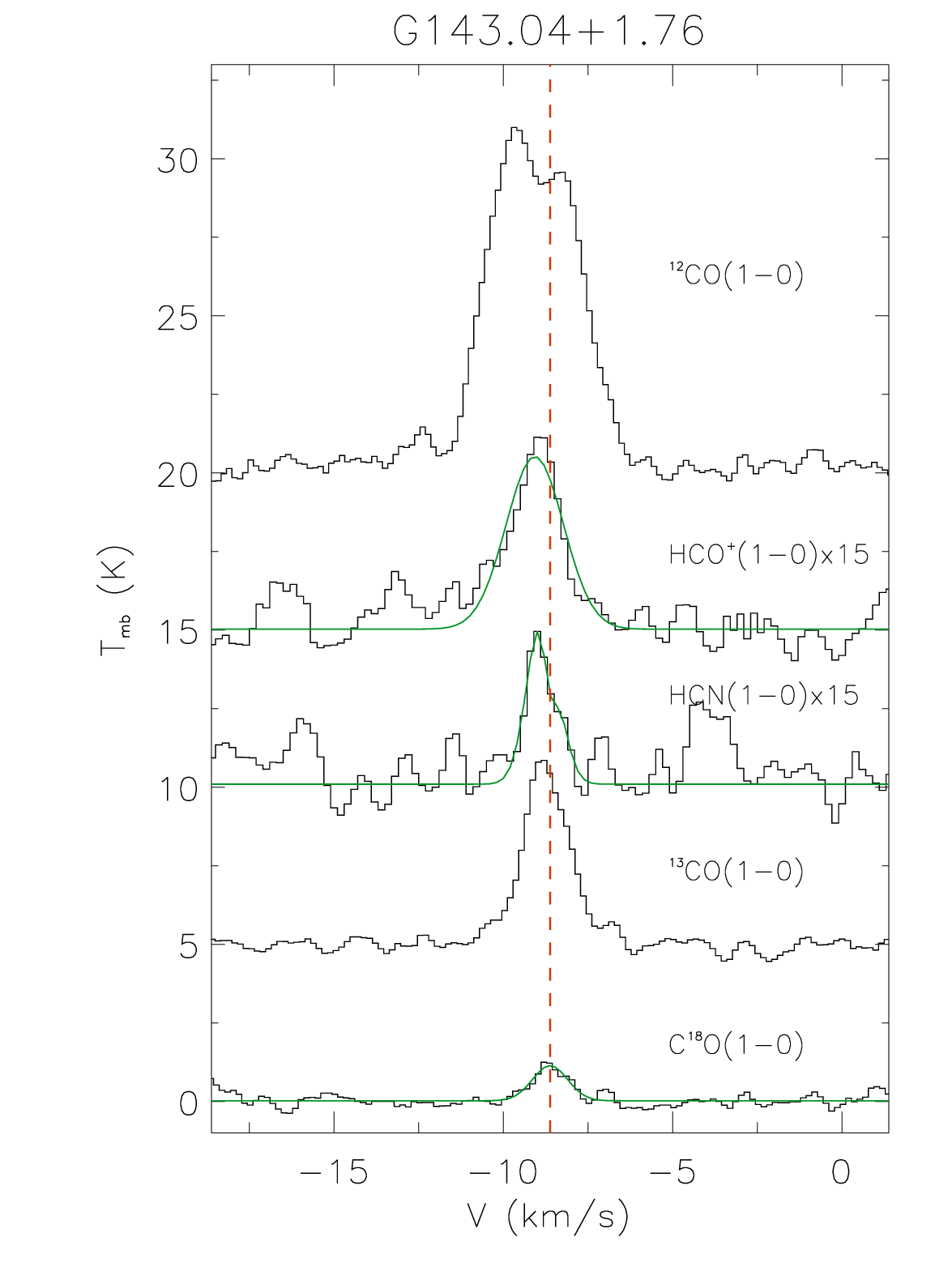}
  \end{minipage}%
  \begin{minipage}[t]{0.325\linewidth}
  \centering
   \includegraphics[width=55mm]{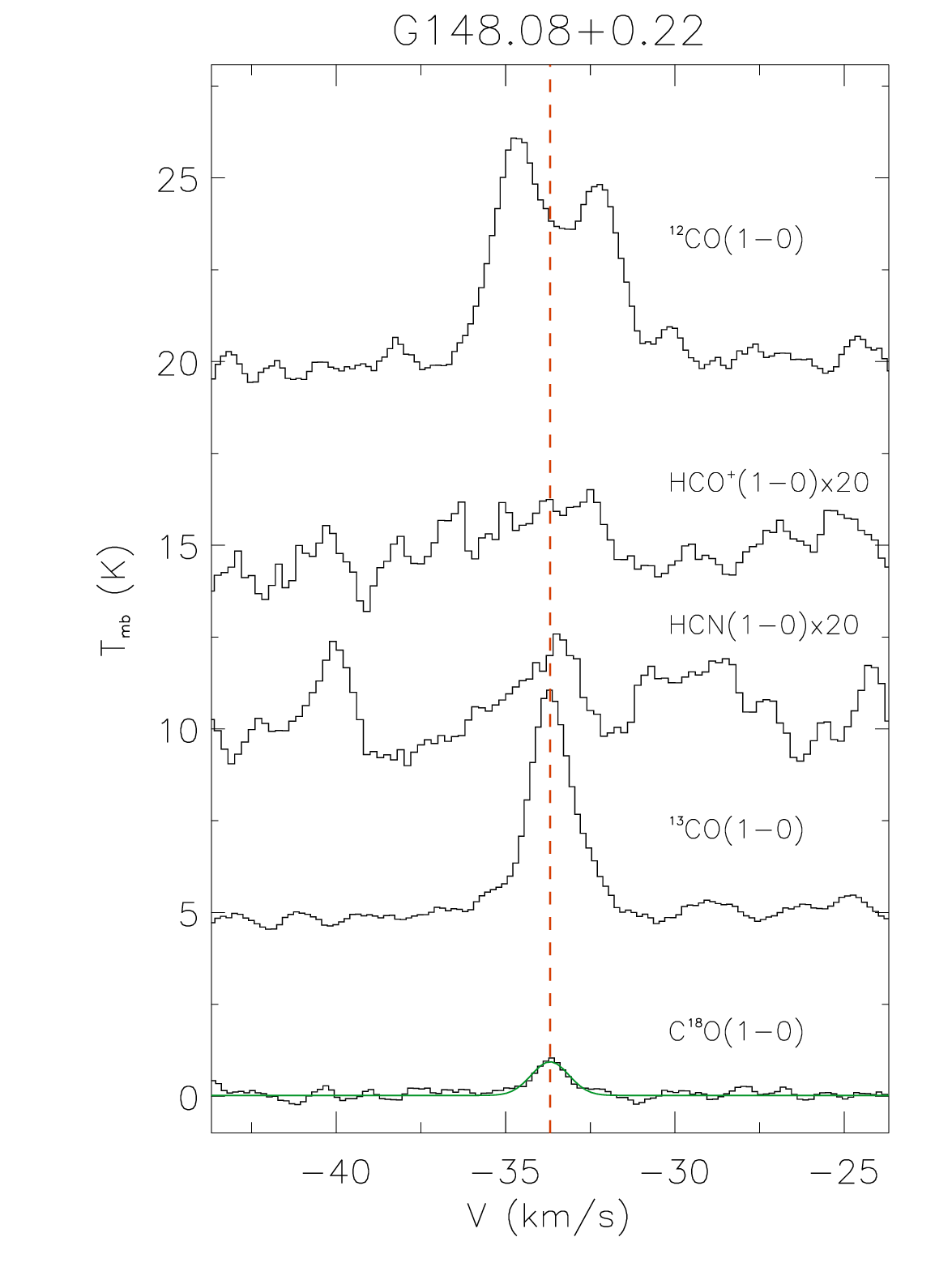}
  \end{minipage}%
  \begin{minipage}[t]{0.325\linewidth}
  \centering
   \includegraphics[width=55mm]{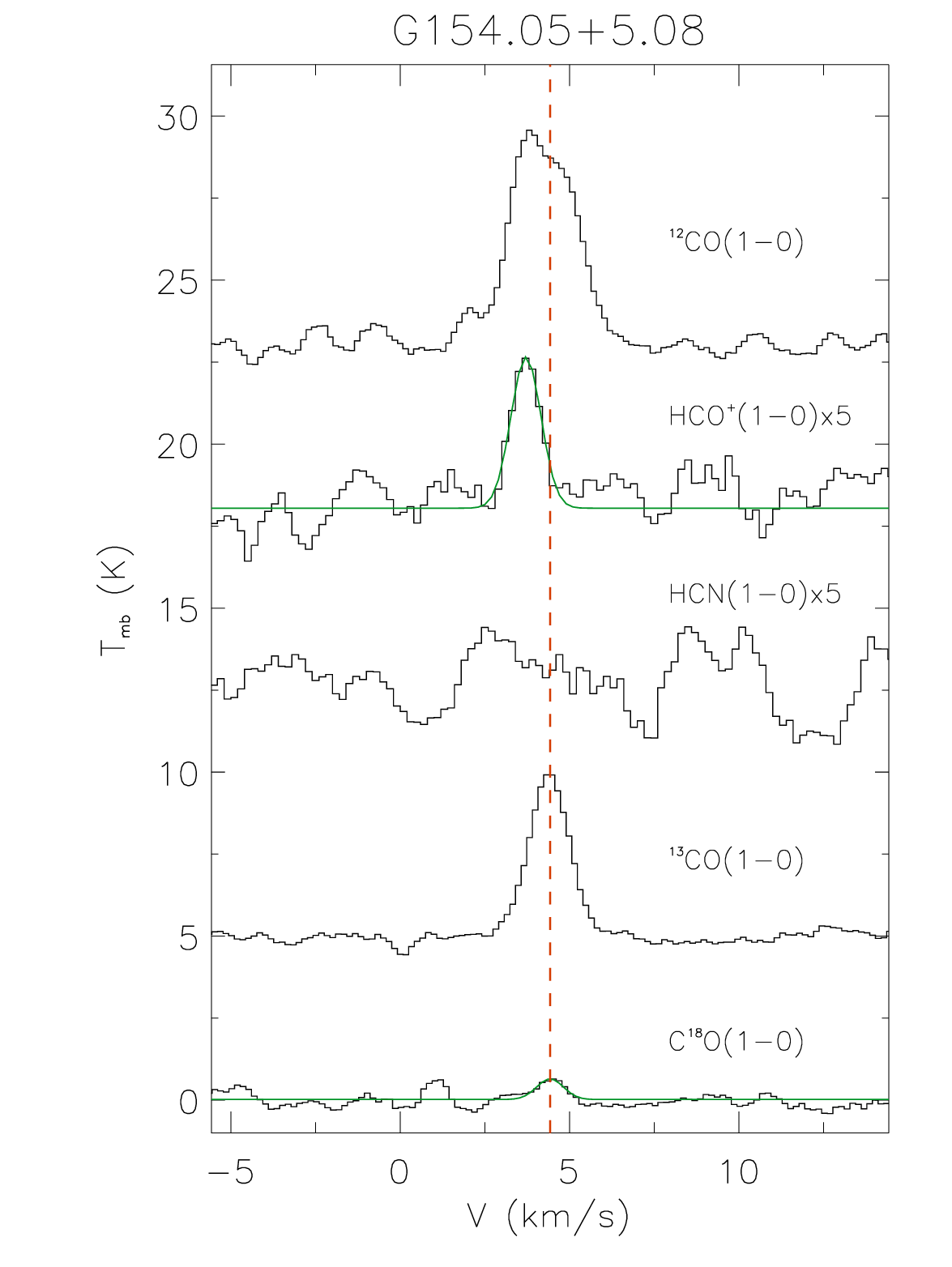}
  \end{minipage}%
  \caption{{\small Line profiles of 133 sources we selected. The lines from bottom to top are C$^{18}$O (1-0), $^{13}$CO (1-0), HCN (1-0) (14 sources lack HCN data), HCO$^+$ (1-0) and $^{12}$CO (1-0), respectively. The dashed red line indicates the central radial velocity of C$^{18}$O (1-0) estimated by Gaussian fitting. For infall candidates, HCO$^+$ (1-0) and HCN (1-0) lines are also Gaussian fitted.}}
  \label{Fig:fig6}
\end{figure} 

\begin{figure}[h]
\ContinuedFloat
  \begin{minipage}[t]{0.325\linewidth}
  \centering
   \includegraphics[width=55mm]{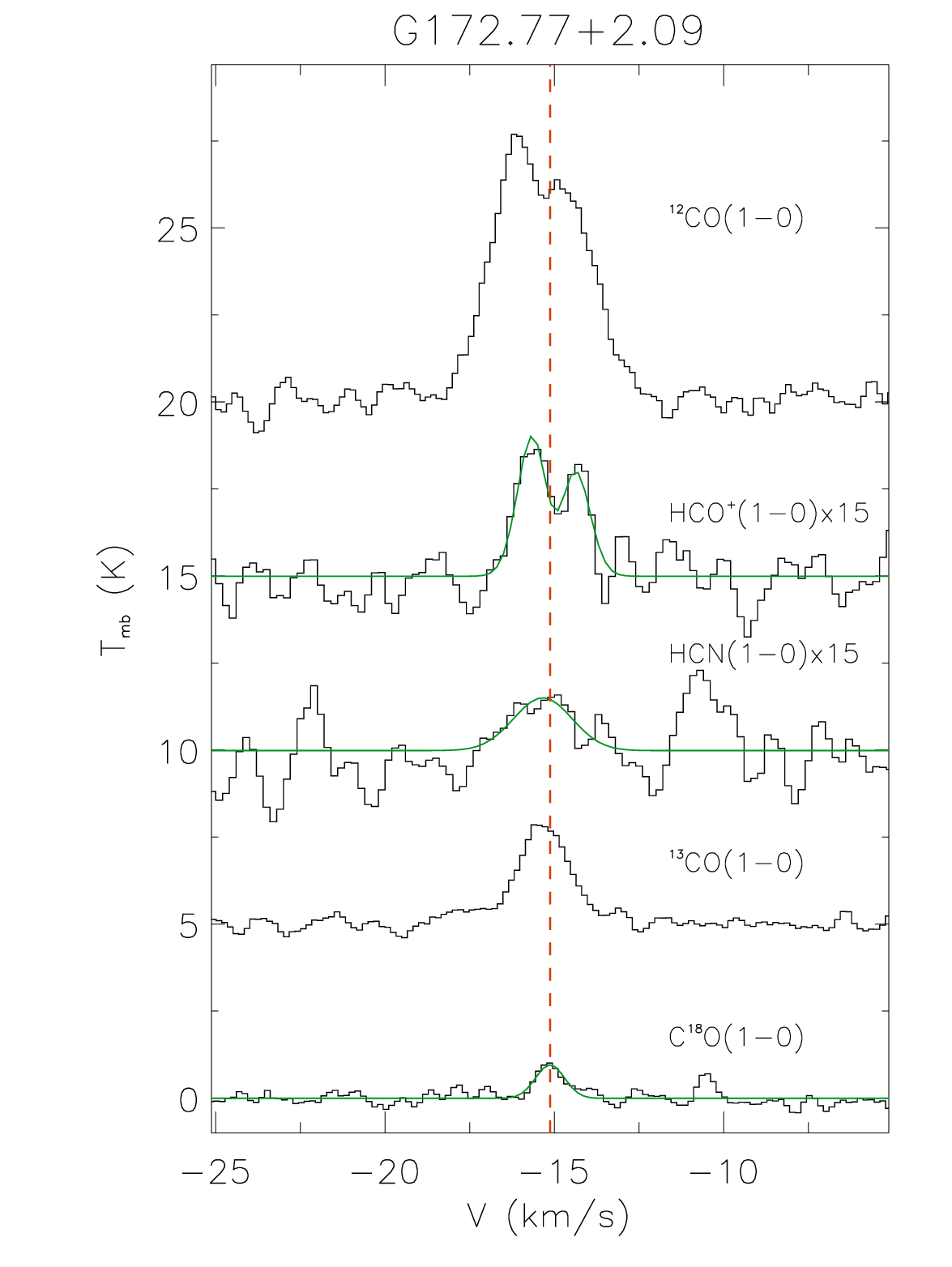}
  \end{minipage}%
  \begin{minipage}[t]{0.325\textwidth}
  \centering
   \includegraphics[width=55mm]{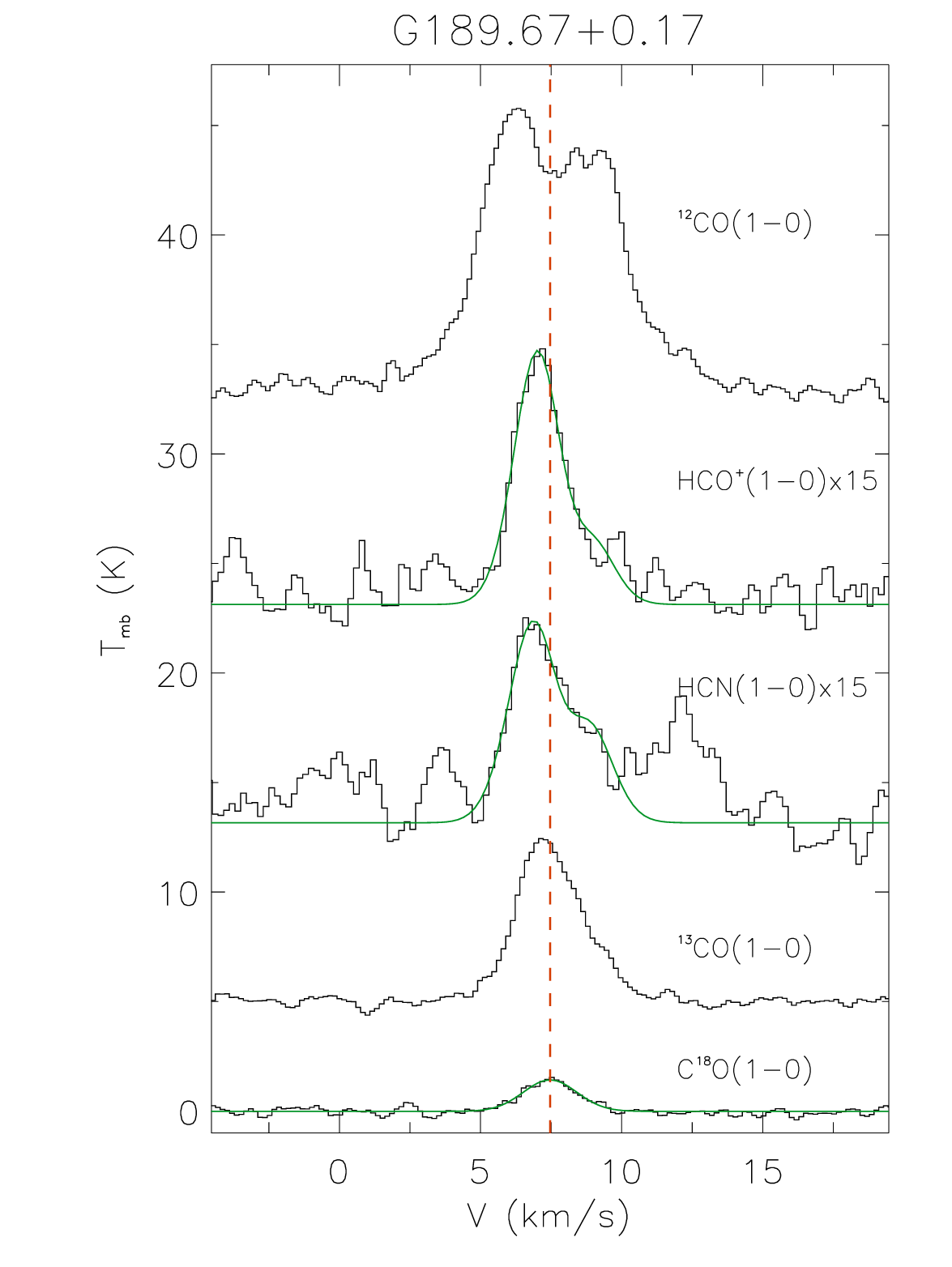}
  \end{minipage}%
  \begin{minipage}[t]{0.325\linewidth}
  \centering
   \includegraphics[width=55mm]{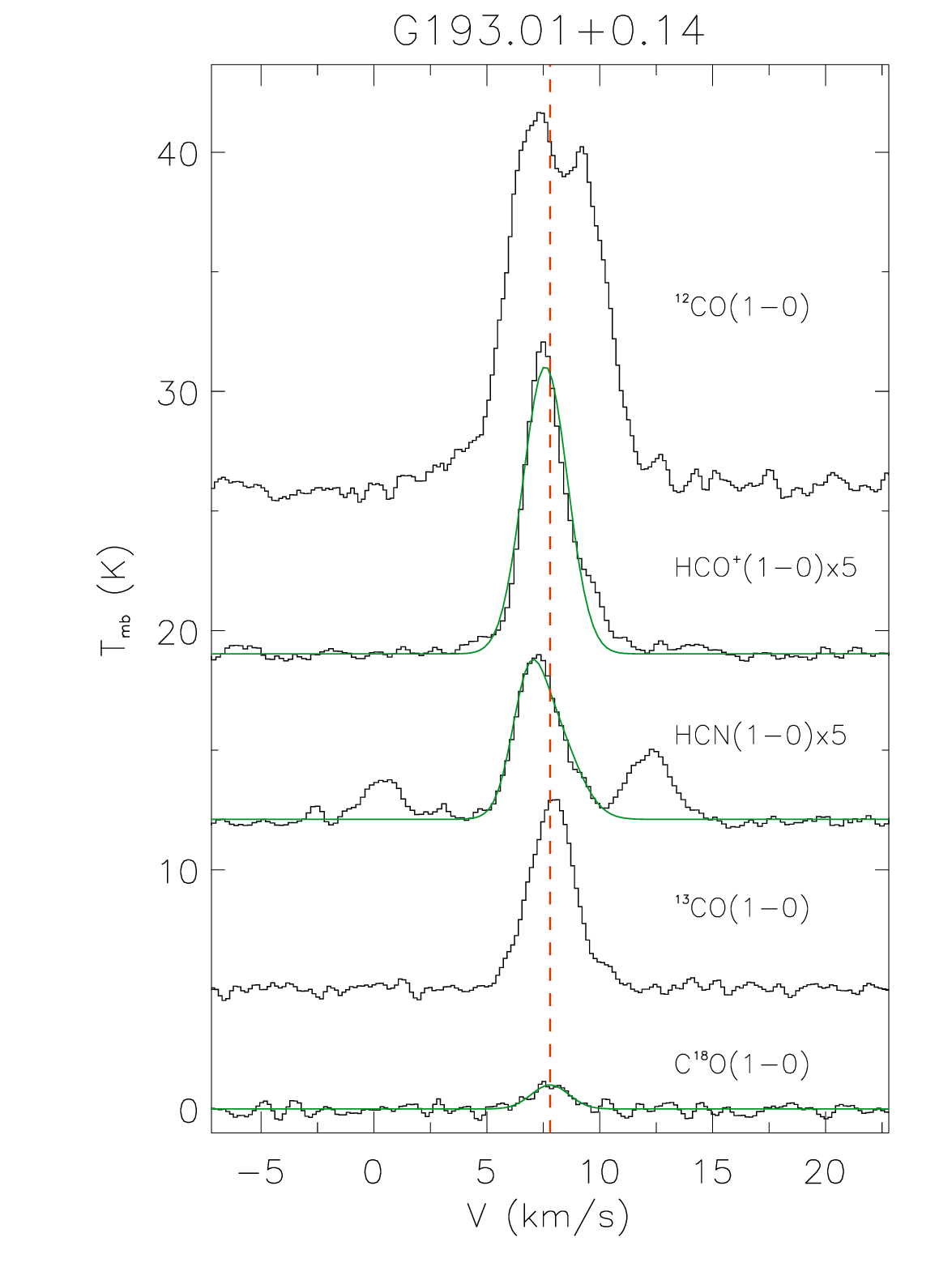}
  \end{minipage}%  
\quad
  \begin{minipage}[t]{0.325\linewidth}
  \centering
   \includegraphics[width=55mm]{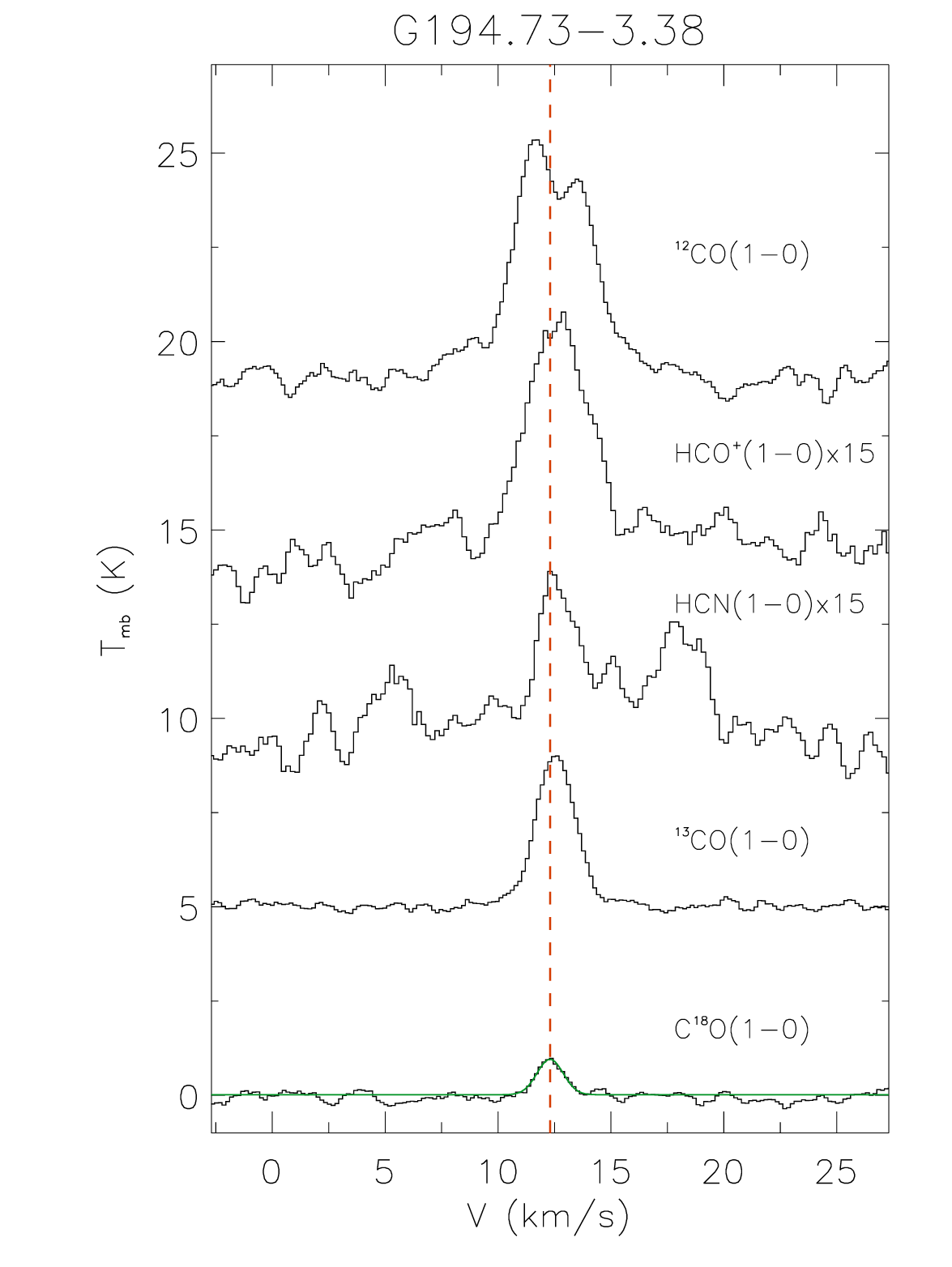}
  \end{minipage}%
  \begin{minipage}[t]{0.325\linewidth}
  \centering
   \includegraphics[width=55mm]{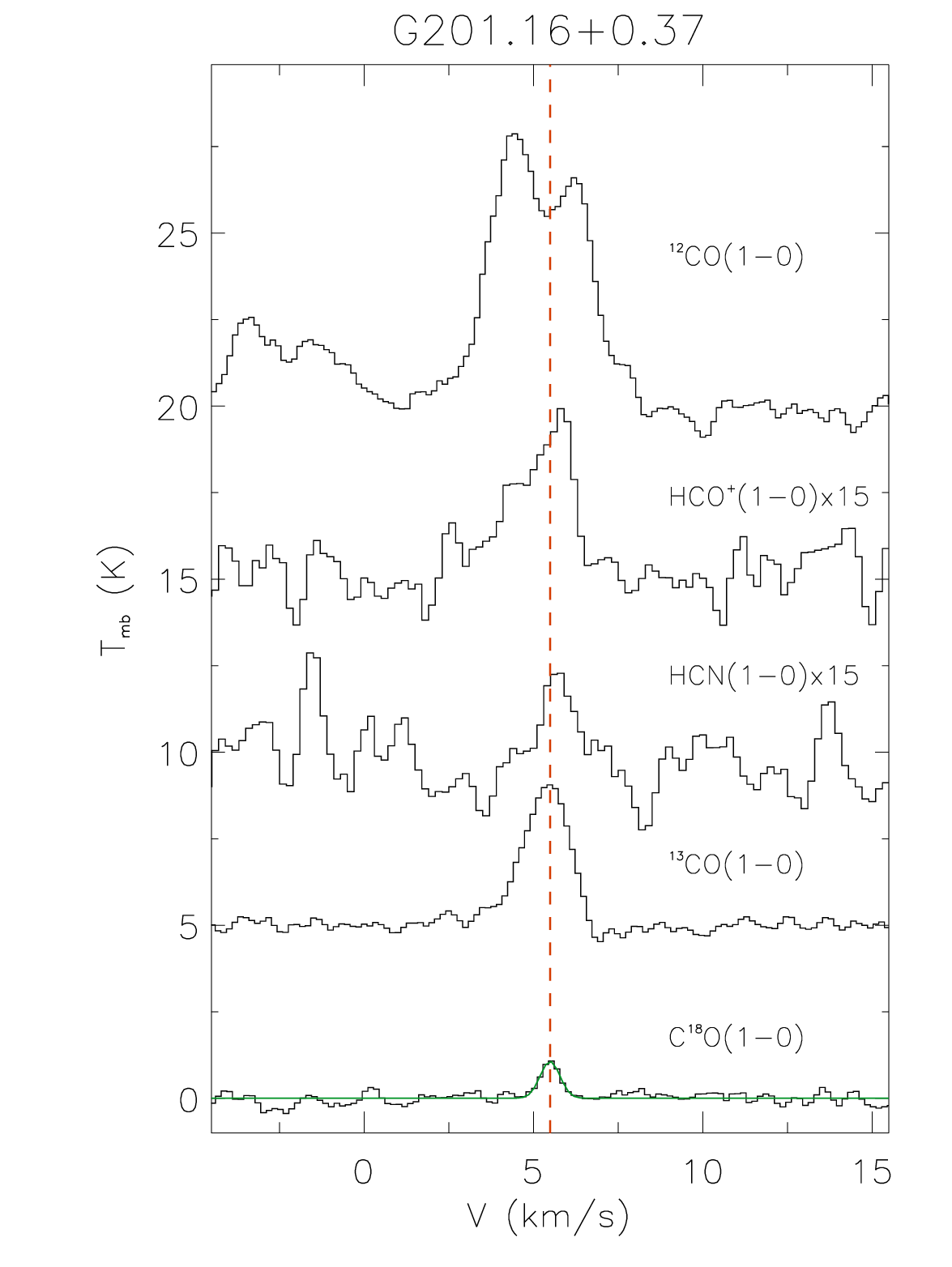}
  \end{minipage}%
  \begin{minipage}[t]{0.325\linewidth}
  \centering
   \includegraphics[width=55mm]{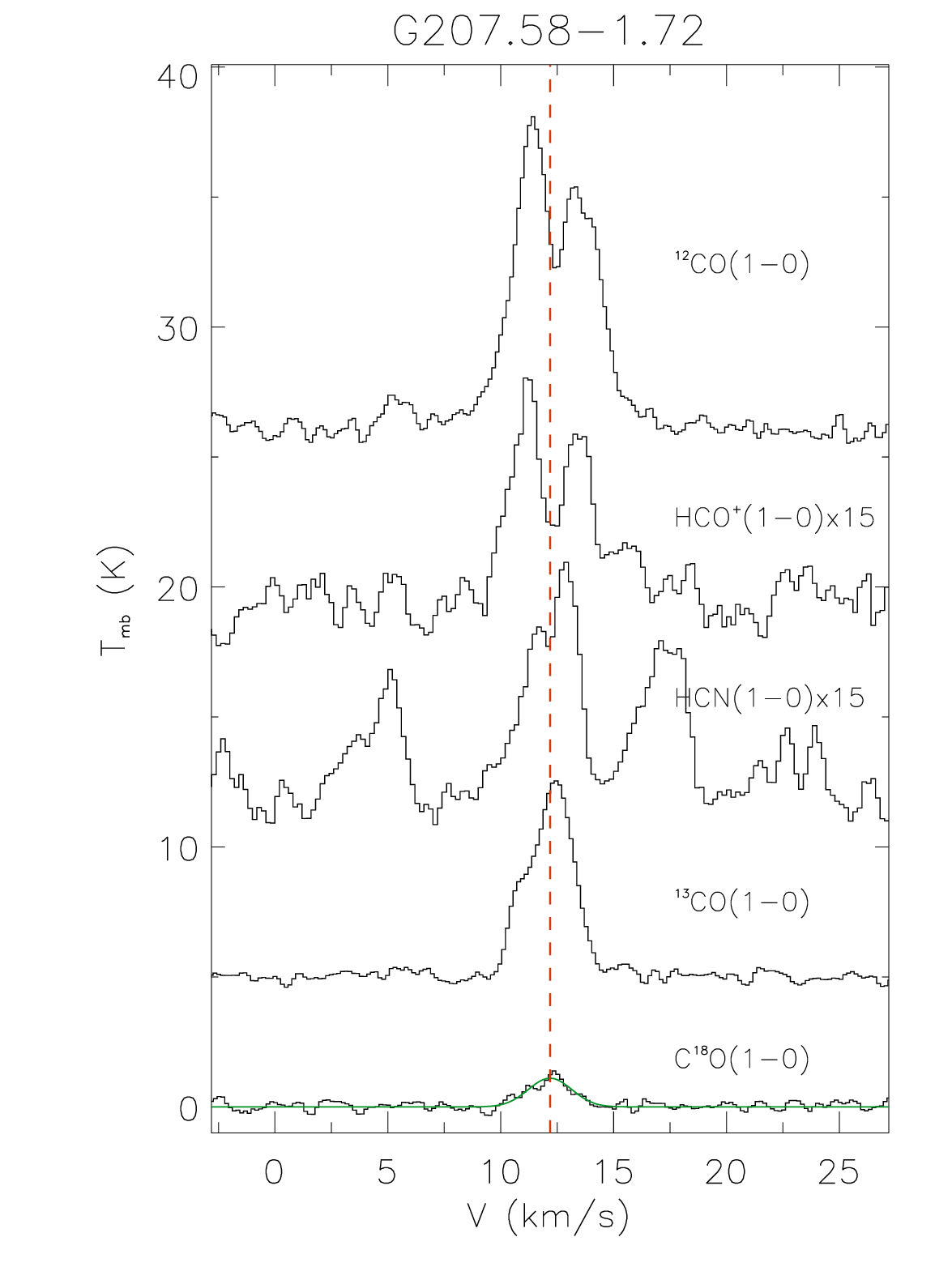}
  \end{minipage}%
\quad
  \begin{minipage}[t]{0.325\linewidth}
  \centering
   \includegraphics[width=55mm]{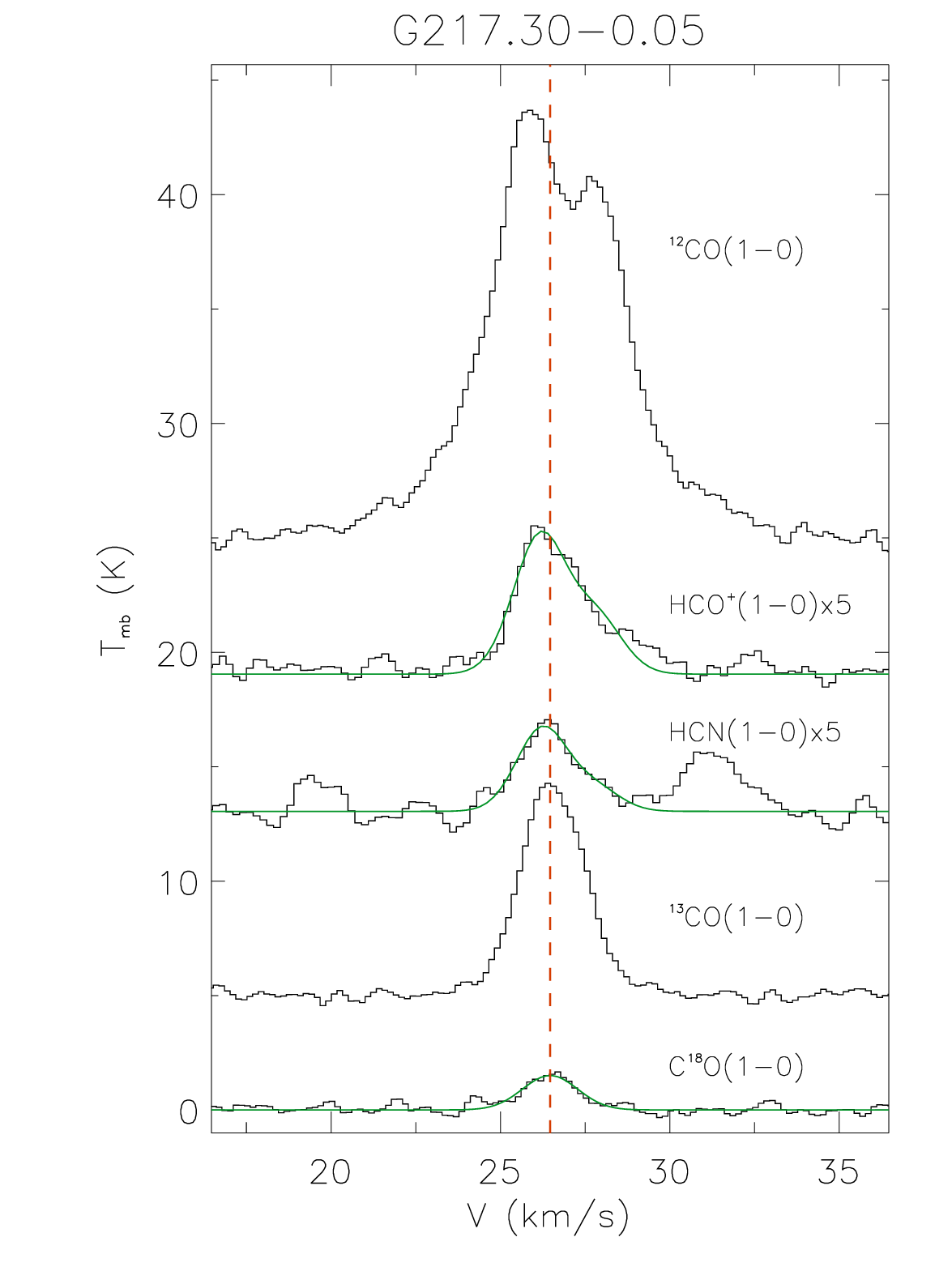}
  \end{minipage}%
  \caption{{\small Line profiles of 133 sources we selected. The lines from bottom to top are C$^{18}$O (1-0), $^{13}$CO (1-0), HCN (1-0) (14 sources lack HCN data), HCO$^+$ (1-0) and $^{12}$CO (1-0), respectively. The dashed red line indicates the central radial velocity of C$^{18}$O (1-0) estimated by Gaussian fitting. For infall candidates, HCO$^+$ (1-0) and HCN (1-0) lines are also Gaussian fitted.}}
  \label{Fig:fig6}
\end{figure} 

\clearpage

\section{The derived parameters of sources}

%________________________________________ Table 1: Basic parameters

\begin{longtable}{ccccccc}
  \caption[]{ The Derived Clump Parameters.}\label{Tab:tabB-1} \\
%Please Capitalize the First Letter of Each Notional Word in table's caption
  \hline\noalign{\smallskip}
Source & RA & Dec & V$_{LSR}$ & Distance & T$_{ex}$($^{12}$CO) & log($N($H$_2))$ \\
Name & (J2000) & (J2000) & (km s$^{-1}$) & (kpc) & (K) & (cm$^{-2}$) \\
  \hline\noalign{\smallskip}
  \endhead\noalign{\smallskip}
% $^*$
G002.97+4.22 & 17:36:36.4 & -24:11:31 & 19.7 & 4.88  & 9.5  & 21.20 \\
G012.34-0.09$^\dagger$ &  18:12:53.7 & -18:17:01 & 34.2  & 3.35  & 14.6  & 22.25 \\
G012.72+0.69$^\dagger$ &  18:10:46.9 & -17:34:15 & 17.3  & 1.94  & 21.8  & 22.19 \\
G012.77-0.19$^\dagger$ &  18:14:08.2 & -17:57:04 & 35.6  & 3.38  & 33.0  & 23.13 \\
G012.82-0.19$^\dagger$ &  18:14:13.3 & -17:54:53 & 34.9  & 3.33  & 38.5  & 23.61 \\
G012.87-0.20$^\dagger$ &  18:14:22.1 & -17:52:02 & 35.5  & 3.36  & 25.5  & 23.01 \\
G012.88-0.24$^\dagger$ &  18:14:32.4 & -17:52:48 & 35.8  & 3.38  & 22.3  & 23.24 \\
G012.96-0.23$^\dagger$ &  18:14:39.6 & -17:48:36 & 35.2  & 3.33  & 17.1  & 23.07 \\
G013.79-0.23$^\dagger$ &  18:16:19.5 & -17:04:39 & 38.0  & 3.39  & 17.9  & 22.63 \\
G013.97-0.15$^\dagger$ &  18:16:22.0 & -16:53:02 & 40.2  & 3.50  & 19.9  & 22.65 \\
G014.00-0.13$^\dagger$ &  18:16:22.3 & -16:50:48 & 39.7  & 3.47  & 22.3  & 22.94 \\
G014.02-0.19$^\dagger$ &  18:16:38.2 & -16:51:09 & 40.0  & 3.48  & 26.0  & 22.84 \\
G014.11-0.16$^\dagger$ &  18:16:40.7 & -16:45:48 & 39.1  & 3.42  & 16.8  & 22.44 \\
G014.24-0.50$^\dagger$ &  18:18:12.0 & -16:48:29 & 19.5  & 2.00  & 23.3  & 22.71 \\
G014.25-0.17$^\dagger$ &  18:17:01.3 & -16:38:48 & 38.3  & 3.36  & 20.9  & 22.85 \\
G014.26-0.17$^\dagger$ &  18:17:02.3 & -16:38:21 & 38.5  & 3.37  & 19.3  & 22.80 \\
G014.27-0.51$^\dagger$ &  18:18:16.8 & -16:47:24 & 19.8  & 2.02  & 19.9  & 22.60 \\
G014.42+3.96$^\dagger$ &  18:02:20.4 & -14:29:53 & 29.7  & 2.78  & 9.5  & 22.08 \\
G015.56+2.03 &  18:11:33.7 & -14:26:32 & 28.9  & 2.61  & 10.0  & 22.01 \\
G016.26+1.00$^\dagger$ & 18:16:41.4 & -14:19:16 & 19.5  & 1.84  & 14.9  & 22.04 \\
G017.09+0.82$^\dagger$ &  18:18:57.2 & -13:40:14 & 22.4  & 2.01  & 25.0  & 22.59 \\
G025.82-0.18$^\dagger$ &  18:39:04.7 & -06:24:21 & 93.7  & 5.01  & 17.4  & 22.75 \\
G026.32-0.07$^\dagger$ &  18:39:36.8 & -05:54:42 & 100.3  & 5.28  & 15.0  & 22.44 \\
G028.20-0.07$^\dagger$ &  18:43:01.8 & -04:14:27 & 97.1  & 5.19  & 15.6  & 22.71 \\
G028.62+4.31 &  18:28:14.3 & -01:51:23 & 7.7  & 0.52  & 7.8  & 21.83 \\
G028.67+3.68$^\dagger$ & 18:30:32.1 & -02:06:24 & 7.2  & 0.48  & 10.1  & 22.12 \\
G028.67+3.70$^\dagger$ & 18:30:29.5 & -02:05:30 & 7.2  & 0.48  & 9.4  & 22.18 \\
G028.97+3.35$^\dagger$ & 18:32:17.2 & -01:59:11 & 7.3  & 0.48  & 12.4  & 22.37 \\
G028.97+3.36$^\dagger$ & 18:32:14.5 & -01:59:24 & 7.3  & 0.48  & 11.7  & 22.40 \\
G028.97+3.54$^\dagger$ & 18:31:35.4 & -01:54:21 & 7.2  & 0.48  & 12.9  & 22.15 \\
G029.03+4.57$^\dagger$ & 18:28:02.3 & -01:22:20 & 7.3  & 0.48  & 8.1  & 21.99 \\
G029.06+4.58 &  18:28:03.3 & -01:20:46 & 7.4  & 0.49  & 8.7  & 21.64 \\
G029.18+3.99 &  18:30:23.2 & -01:30:26 & 8.1  & 0.54  & 11.1  & 22.00 \\
G029.60-0.63$^\dagger$ &  18:47:36.8 & -03:15:15 & 77.2  & 4.31  & 13.9  & 22.56 \\
G029.62+3.19 & 18:34:02.4 & -01:28:57 & 8.0  & 0.53  & 7.7  & 21.41 \\
G030.17+3.69 & 18:33:15.2 & -00:46:22 & 9.1  & 0.61  & 10.4  & 21.80 \\
G030.76+2.74$^\dagger$ &  18:37:42.8 & -00:40:56 & 8.1  & 0.53  & 8.5  & 21.68 \\
G031.41+5.24 &  18:30:00.2 &  01:02:12 & 8.3  & 0.55  & 13.8  & 21.99 \\
G031.84+2.57$^\dagger$ &  18:40:17.0 &  00:12:16 & 8.6  & 0.56  & 8.4  & 21.96 \\
G033.42+0.00$^\dagger$ & 18:52:20.3 & 00:26:20 & 10.6  & 0.69  & 8.8  & 21.60 \\
G033.55+0.29$^\dagger$ &  18:51:31.6 &  00:40:59 & 57.6  & 3.39  & 7.0  & 21.83 \\
G034.25+0.16$^\dagger$ &  18:53:16.7 &  01:14:43 & 58.1  & 3.42  & 19.1  & 23.28 \\
G034.38+0.22$^\dagger$ & 18:53:18.9 & 01:23:26 & 57.5  & 3.39  & 11.1  & 22.53 \\
G034.40+0.21$^\dagger$ & 18:53:22.5 & 01:24:06 & 57.3  & 3.38  & 10.6  & 22.32 \\
G035.28+1.31$^\dagger$ & 18:51:04.0 & 02:41:21 & 13.1  & 0.85  & 7.5  & 21.40 \\
G036.02-1.36 &  19:01:55.3 &  02:07:55 & 31.8  & 1.99  & 10.0  & 21.93 \\
G036.33-1.27 &  19:02:11.4 &  02:26:39 & 30.4  & 1.91  & 7.7  & 21.74 \\
G036.91-0.43$^\dagger$ &  19:00:14.7 &  03:20:25 & 80.2  & 4.79  & 10.9  & 22.17 \\
G036.92+0.71$^\dagger$ & 18:56:12.3 & 03:52:36 & 13.0  & 0.84  & 7.7  & 21.70 \\
G036.94+1.09$^\dagger$ &  18:54:52.0 &  04:03:58 & 13.2  & 0.85  & 7.4  & 21.60 \\
G037.05-0.03$^\dagger$ &  18:59:04.7 &  03:38:57 & 81.3  & 4.89  & 13.2  & 22.51 \\
G037.50+3.05 &  18:48:53.1 &  05:27:14 & 15.5  & 1.00  & 9.9  & 21.39 \\
G038.75-0.97$^\dagger$ &  19:05:31.7 &  04:43:59 & 13.2  & 0.85  & 9.4  & 21.31 \\
G039.28-0.19$^\dagger$ &  19:03:44.7 &  05:33:45 & 70.5  & 4.35  & 11.2  & 22.35 \\
G039.33-1.02$^\dagger$ &  19:06:48.7 &  05:13:28 & 12.8  & 0.82  & 8.8  & 21.75 \\
G039.45-1.17$^\dagger$ &  19:07:32.0 &  05:15:46 & 12.7  & 0.82  & 8.7  & 21.58 \\
G039.98+1.67$^\dagger$ &  18:58:23.0 &  07:02:05 & 27.8  & 1.76  & 10.1  & 21.72 \\
G040.01-0.94$^\dagger$ &  19:07:45.7 &  05:51:43 & 13.5  & 0.87  & 9.5  & 21.84 \\
G044.32-0.79$^\dagger$ &  19:15:16.2 &  09:45:03 & 62.1  & 4.28  & 10.2  & 22.06 \\
G045.45+0.05$^\dagger$ &  19:14:22.9 &  11:08:44 & 59.3  & 4.19  & 20.9  & 22.74 \\
G048.79+0.02$^\dagger$ &  19:20:54.2 &  14:05:04 & 50.1  & 3.79  & 12.6  & 22.14 \\
G049.07-0.33$^\dagger$ & 19:22:42.6 & 14:09:44 & 60.4  & 5.46  & 19.5  & 22.48 \\
G049.25-1.38 &  19:26:52.8 &  13:49:35 & 5.7  & 0.39  & 9.4  & 21.77 \\
G049.27-1.41 &  19:27:00.2 &  13:49:45 & 5.5  & 0.37  & 8.4  & 21.54 \\
G049.28-1.40 &  19:27:00.4 &  13:50:52 & 5.4  & 0.37  & 8.9  & 21.62 \\
G049.32-1.41 &  19:27:07.1 &  13:52:49 & 5.3  & 0.36  & 8.5  & 21.47 \\
G052.99-0.03$^\dagger$ & 19:29:22.4 & 17:45:30 & 22.4  & 1.68  & 13.2  & 21.74 \\
G053.08-0.24$^\dagger$ &  19:30:19.6 &  17:44:19 & 23.6  & 1.78  & 9.5  & 21.81 \\
G053.10+0.11$^\dagger$ &  19:29:04.2 &  17:55:16 & 22.2  & 1.67  & 15.2  & 21.95 \\
G053.11+0.09$^\dagger$ &  19:29:08.9 &  17:55:14 & 22.0  & 1.65  & 16.4  & 22.03 \\
G053.12+0.08$^\dagger$ &  19:29:12.7 &  17:55:52 & 22.3  & 1.68  & 14.2  & 22.38 \\
G053.14+0.07$^\dagger$ & 19:29:16.6 & 17:56:30 & 22.2  & 1.67  & 15.7  & 22.54 \\
G053.14+0.09$^\dagger$ &  19:29:12.9 &  17:56:59 & 22.3  & 1.68  & 13.3  & 22.18 \\
G053.76+0.45$^\dagger$ &  19:29:08.0 &  18:39:46 & 23.1  & 1.77  & 9.8  & 21.77 \\
G053.77+0.46$^\dagger$ &  19:29:07.1 &  18:40:26 & 23.4  & 1.79  & 9.9  & 21.87 \\
G054.03-2.32$^\dagger$ & 19:39:54.4 & 17:33:25 & 17.6  & 1.33  & 8.3  & 21.53 \\
G054.38-0.52$^\dagger$ & 19:33:58.9 & 18:44:41 & 34.3  & 2.87  & 10.0  & 21.83 \\
G056.91+3.44 & 19:24:15.8 & 22:51:13 & 11.3  & 0.92  & 15.7  & 21.42 \\
G060.42-0.68$^\dagger$ &  19:47:24.4 &  23:54:30 & 29.0  & 3.57  & 9.8  & 21.69 \\
G062.61+1.83 &  19:42:36.4 &  27:03:53 & 2.7  & 0.29  & 9.2  & 21.28 \\
G062.82+1.82 &  19:43:07.4 &  27:14:55 & 2.0  & 0.23  & 10.8  & 21.55 \\
G077.46+1.73 &  20:20:45.4 &  39:36:47 & 1.7  & 0.63  & 26.1  & 22.52 \\
G077.91-1.16 & 20:34:11.8 & 38:17:42 & -0.5  & 0.15  & 14.3  & 22.19 \\
G079.04+0.64$^\dagger$ &  20:30:10.6 &  40:16:29 & 6.6  & 1.59  & 25.7  & 22.38 \\
G079.24+0.53$^\dagger$ &  20:31:15.2 &  40:22:21 & 0.3  & 0.39  & 10.4  & 22.41 \\
G079.44+0.18 &  20:33:21.3 &  40:19:33 & 0.1  & 0.35  & 9.0  & 21.89 \\
G079.48+0.25 &  20:33:12.3 &  40:23:57 & -0.1  & 0.30  & 10.0  & 22.09 \\
G079.71+0.15$^\dagger$ &  20:34:19.9 &  40:31:12 & 1.0  & 0.66  & 9.1  & 21.69 \\
G081.04-0.46$^\dagger$ &  20:41:09.8 &  41:12:47 & 5.7  & 1.30  & 11.1  & 22.02 \\
G081.62+1.11$^\dagger$ &  20:36:21.1 &  42:37:18 & 3.8  & 1.22  & 16.7  & 22.06 \\
G081.69-1.60 &  20:48:03.4 &  41:00:48 & 2.7  & 1.21  & 11.9  & 21.94 \\
G081.72+0.57$^\dagger$ & 20:39:00.7 & 42:23:07 & -3.1  & 2.56  & 38.0  & 23.15 \\
G081.72+1.28 &  20:35:56.7 &  42:48:48 & 3.7  & 1.20  & 13.4  & 22.05 \\
G081.72-1.60$^\dagger$ & 20:48:10.1 & 41:02:21 & 2.5  & 1.20  & 10.2  & 21.75 \\
G081.90+1.43 &  20:35:51.9 &  43:02:36 & 11.0  & 1.18  & 13.7  & 21.69 \\
G082.17+0.07$^\dagger$ & 20:42:39.2 & 42:25:23 & 10.1  & 1.14  & 17.6  & 22.00 \\
G082.18-1.54 &  20:49:27.8 &  41:25:54 & 2.8  & 1.13  & 11.4  & 21.93 \\
G082.21-1.53$^\dagger$ &  20:49:30.8 &  41:27:23 & 2.7  & 1.13  & 12.7  & 22.01 \\
G082.52-1.92 &  20:52:08.3 &  41:27:03 & 4.5  & 1.09  & 11.3  & 21.94 \\
G082.53+0.09$^\dagger$ &  20:43:46.7 &  42:43:36 & 9.9  & 1.08  & 16.7  & 22.44 \\
G082.64-1.98 &  20:52:50.3 &  41:30:15 & 4.1  & 1.07  & 12.3  & 21.87 \\
G082.81-2.06 & 20:53:43.0 & 41:35:03 & 4.3  & 1.04  & 13.0  & 22.08 \\
G085.05-1.25$^\dagger$ &  20:58:15.6 &  43:48:54 & -37.5  & 5.78  & 9.8  & 22.05 \\
G085.12+0.50$^\dagger$ &  20:51:00.6 &  44:59:39 & -1.7  & 1.43  & 12.2  & 21.87 \\
G085.47-1.12 &  20:59:16.8 &  44:13:06 & 4.4  & 0.66  & 10.5  & 21.74 \\
G107.50+4.47$^\dagger$ &  22:28:33.3 &  62:58:29 & -2.1  & 0.00  & 18.8  & 22.40 \\
G108.89+2.60$^\dagger$ & 22:46:59.0 & 62:01:49 & -10.3  & 0.78  & 17.8  & 21.99 \\
G108.99+2.73$^\dagger$ &  22:47:13.0 &  62:11:41 & -10.5  & 0.79  & 17.4  & 22.41 \\
G110.32+2.52 &  22:58:15.9 &  62:35:29 & -12.1  & 0.90  & 22.8  & 22.15 \\
G110.32+2.54 &  22:58:12.2 &  62:36:24 & -11.9  & 0.88  & 24.4  & 22.22 \\
G110.40+1.67 &  23:01:58.6 &  61:50:44 & -11.2  & 0.82  & 12.6  & 21.77 \\
G111.12+2.12$^\dagger$ & 23:06:01.2 & 62:33:12 & -9.9  & 0.68  & 16.0  & 22.13 \\
G111.14+2.12$^\dagger$ & 23:06:14.8 & 62:33:20 & -10.0  & 0.69  & 13.8  & 22.05 \\
G111.23+2.07 &  23:07:07.2 &  62:33:12 & -9.1  & 0.61  & 14.6  & 21.88 \\
G115.62+1.99 &  23:44:10.6 &  63:54:14 & -10.5  & 0.65  & 9.9  & 21.56 \\
G121.31+0.64$^\dagger$ & 00:36:53.6 & 63:28:03 & -17.3  & 1.09  & 16.6  & 22.24 \\
G121.34+3.42 &  00:35:40.1 &  66:14:23 & -5.3  & 0.18  & 7.2  & 21.99 \\
G121.35+3.41$^\dagger$ &  00:35:45.3 &  66:13:55 & -5.4  & 0.19  & 7.4  & 21.85 \\
G126.51-1.30$^\dagger$ &  01:21:20.8 &  61:21:50 & -11.7  & 0.63  & 11.3  & 21.73 \\
G126.53-1.17 &  01:21:40.5 &  61:29:06 & -12.9  & 0.72  & 11.9  & 21.98 \\
G126.67-0.82$^\dagger$ & 01:23:09.2 & 61:49:29 & -13.6  & 0.77  & 18.2  & 22.43 \\
G127.88+2.67$^\dagger$ & 01:38:41.2 & 65:05:29 & -11.5  & 0.60  & 13.0  & 21.98 \\
G133.42+0.00$^\dagger$ &  02:19:51.8 &  61:03:26 & -15.2  & 0.86  & 10.6  & 21.68 \\
G143.04+1.76$^\dagger$ &  03:33:54.9 &  58:08:19 & -8.8  & 0.42  & 14.6  & 21.98 \\
G148.08+0.22$^\dagger$ &  03:55:43.7 &  53:50:07 & -33.7  & 2.96  & 9.6  & 21.83 \\
G154.05+5.08 & 04:47:13.1 & 53:03:57 & 4.3  & 0.00  & 10.3  & 21.49 \\
G172.77+2.09$^\dagger$ &  05:35:50.9 &  36:10:41 & -15.3  & 5.95  & 10.9  & 21.78 \\
G189.67+0.17$^\dagger$ & 06:07:41.5 & 20:39:44 & 7.3  & 2.58  & 16.8  & 22.39 \\
G193.01+0.14$^\dagger$ & 06:14:25.2 & 17:43:20 & 7.9  & 1.91  & 20.4  & 22.22 \\
G194.73-3.38$^\dagger$ &  06:05:01.0 &  14:31:04 & 12.3  & 2.58  & 9.9  & 21.90 \\
G201.16+0.37 &  06:31:04.4 &  10:38:12 & 5.4  & 0.79  & 11.4  & 21.66 \\
G207.58-1.72$^\dagger$ &  06:35:29.7 &  03:58:31 & 12.5  & 1.32  & 15.0  & 22.26 \\
G217.30-0.05$^\dagger$ &  06:59:14.3 & -03:54:35 & 26.6  & 2.34  & 22.1  & 22.41 \\
  \hline\noalign{\smallskip}
\end{longtable}
%\tabnote{$^{\rm *}$This is footnote shows what footnote symbols to use.}
\tablecomments{0.95\textwidth}{Columns are (from left to right) the source name, its equatorial coordinate, its local standard of rest (LSR) velocity, its heliocentric distance, its excitation temperature, and its H$_2$ column density. $^\dagger$ Source are associated with Class 0/I YSOs.}
             % $^*$ Source $^{13}$CO has self-absorption and C$^{18}$O is used to estimate $N($H$_2)$.}

%________________________________________ Table 2: Observation parameters

\begin{longtable}{cccccccc}
\caption[]{ The Derived Line Parameters and Profiles of Sample.}\label{Tab:tabB-2} \\
%%Please Capitalize the First Letter of Each Notional Word in table's caption
  \hline\noalign{\smallskip}
Source & V$_{thick}$ & V$_{thick}$ & V$_{thin}$ & $\Delta$ V & $\delta$ V & $\delta$ V & Profile \\
Name & HCO$^+$ (1-0) & HCN (1-0) & C$^{18}$O (1-0) & C$^{18}$O (1-0) & HCO$^+$ (1-0) & HCN (1-0) &  \\
  & (km s$^{-1}$) & (km s$^{-1}$) & (km s$^{-1}$) & (km s$^{-1}$) &  &  &  \\
  \hline\noalign{\smallskip}
  \endhead\noalign{\smallskip}
% $^*$
G002.97+4.22 & - & 19.25(0.19) & 19.67(0.02) & 0.24(0.01) & - & -1.75(0.79) & N,B \\
G012.34-0.09 & - & - & 34.34(0.05) & 2.51(0.11) & - & - & N,N \\
G012.72+0.69 & 16.70(0.09) & 17.26(0.26) & 17.28(0.04) & 1.66(0.08) & -0.35(0.05) & -0.01(0.15) & B,N $^{\alpha}$ \\
G012.77-0.19 & 33.42(0.20) & 35.16(0.20) & 35.78(0.02) & 3.90(0.05) & -0.61(0.05) & -0.16(0.05) & B,N \\
G012.82-0.19 & 34.69(0.04) & 34.68(0.06) & 34.94(0.01) & 4.52(0.02) & -0.06(0.01) & -0.06(0.01) & N,N \\
G012.87-0.20 & 34.90(0.37) & 34.32(0.20) & 35.42(0.03) & 4.41(0.06) & -0.12(0.08) & -0.25(0.05) & N,B \\
G012.88-0.24 & 38.17(0.03) & No data & 35.90(0.01) & 3.94(0.03) & 0.58(0.01) & No data & R \\
G012.96-0.23 & 32.20(0.14) & No data & 35.23(0.02) & 4.87(0.05) & -0.62(0.03) & No data & B $^{\alpha,\gamma}$ \\
G013.79-0.23 & 38.81(0.06) & 38.59(0.11) & 38.01(0.02) & 2.30(0.05) & 0.35(0.03) & 0.25(0.05) & R,R \\
G013.97-0.15 & 37.97(0.12) & 37.82(0.13) & 39.66(0.03) & 3.18(0.07) & -0.53(0.04) & -0.58(0.04) & B,B $^{\alpha}$ \\
G014.00-0.13 & 42.26(0.03) & No data & 39.66(0.02) & 2.85(0.04) & 0.91(0.01) & No data & R \\
G014.02-0.19 & 38.80(0.05) & 38.97(0.20) & 40.02(0.02) & 2.58(0.04) & -0.47(0.02) & -0.41(0.08) & B,B $^{\alpha}$ \\
G014.11-0.16 & 37.50(0.31) & 38.70(0.16) & 38.87(0.04) & 2.84(0.09) & -0.48(0.11) & -0.06(0.06) & B,N \\
G014.24-0.50 & - & No data & 19.65(0.01) & 1.49(0.02) & - & No data & N \\
G014.25-0.17 & 35.73(0.13) & No data & 38.14(0.02) & 3.09(0.05) & -0.78(0.04) & No data & B $^{\beta,\gamma}$ \\
G014.26-0.17 & 35.90(0.20) & 36.86(0.47) & 38.38(0.02) & 2.92(0.05) & -0.85(0.07) & -0.52(0.16) & B,B \\
G014.27-0.51 & 20.59(0.14) & 20.18(0.11) & 19.79(0.02) & 1.81(0.04) & 0.44(0.07) & 0.22(0.06) & R,N \\
G014.42+3.96 & 29.46(0.21) & 29.70(0.29) & 29.56(0.03) & 1.68(0.08) & -0.06(0.13) & 0.08(0.17) & N,N \\
G015.56+2.03 & 29.05(0.24) & - & 28.94(0.02) & 1.13(0.05) & 0.10(0.21) & - & N,N \\
G016.26+1.00 & 19.52(0.18) & - & 19.52(0.02) & 0.98(0.04) & 0.00(0.18) & - & N,N \\
G017.09+0.82 & 20.60(0.19) & 20.98(0.12) & 22.30(0.02) & 2.01(0.05) & -0.85(0.09) & -0.66(0.06) & B,B \\
G025.82-0.18 & 92.50(0.02) & No data & 93.69(0.02) & 3.26(0.06) & -0.37(0.01) & No data & B $^{\alpha,\beta,\gamma}$ \\
G026.32-0.07 & 99.50(0.07) & 99.30(0.11) & 100.54(0.02) & 2.10(0.06) & -0.50(0.03) & -0.59(0.05) & B,B $^{\alpha,\gamma}$ \\
G028.20-0.07 & 96.30(0.16) & No data & 97.31(0.03) & 3.94(0.08) & -0.26(0.04) & No data & B $^{\alpha,\beta,\gamma}$ \\
G028.62+4.31 & - & - & 7.67(0.01) & 0.62(0.03) & - & - & N,N \\
G028.67+3.68 & - & - & 7.22(0.01) & 0.98(0.03) & - & - & N,N \\
G028.67+3.70 & - & - & 7.18(0.01) & 1.03(0.03) & - & - & N,N \\
G028.97+3.35 & 4.14(0.34) & - & 7.36(0.01) & 1.15(0.03) & -2.80(0.29) & - & B,N \\
G028.97+3.36 & 5.72(0.10) & 5.71(0.07) & 7.28(0.01) & 1.25(0.03) & -1.25(0.08) & -1.26(0.06) & B,B \\
G028.97+3.54 & 5.00(0.12) & 5.63(0.15) & 7.14(0.03) & 1.66(0.07) & -1.29(0.07) & -0.91(0.09) & B,B \\
G029.03+4.57 & - & - & 7.33(0.01) & 0.78(0.03) & - & - & N,N \\
G029.06+4.58 & - & - & 7.47(0.01) & 0.48(0.03) & - & - & N,N \\
G029.18+3.99 & - & - & 8.08(0.03) & 1.25(0.06) & - & - & N,N \\
G029.60-0.63 & 74.60(0.02) & No data & 77.11(0.02) & 2.55(0.05) & -0.98(0.01) & No data & B $^{\alpha,\beta,\gamma}$ \\
G029.62+3.19 & 8.76(0.07) & - & 8.06(0.01) & 0.35(0.03) & 2.00(0.19) & - & R,N \\
G030.17+3.69 & 8.67(0.52) & 8.27(0.12) & 9.06(0.02) & 0.84(0.05) & -0.46(0.62) & -0.94(0.14) & B,B \\
G030.76+2.74 & 8.13(0.12) & - & 7.97(0.02) & 0.76(0.06) & 0.21(0.15) & - & N,N \\
G031.41+5.24 & 7.28(0.10) & 8.09(0.11) & 8.28(0.02) & 0.82(0.04) & -1.22(0.13) & -0.23(0.13) & B.N \\
G031.84+2.57 & 9.81(0.14) & - & 8.48(0.03) & 1.18(0.06) & 1.13(0.11) & - & R,N \\
G033.42+0.00 & 10.04(0.07) & No data & 10.60(0.02) & 0.55(0.04) & -1.02(0.12) & No data & B $^{\alpha,\gamma}$ \\
G033.55+0.29 & - & - & 57.52(0.03) & 1.00(0.06) & - & - & N,N \\
G034.25+0.16 & 36.32(0.02) & 55.67(0.09) & 58.34(0.02) & 6.21(0.05) & -3.55(0.00) & -0.43(0.01) & B,B $^{\alpha,\gamma}$ \\
G034.38+0.22 & 60.34(0.13) & 54.29(0.11) & 57.54(0.03) & 2.82(0.06) & 0.99(0.04) & -1.15(0.04) & R,B \\
G034.40+0.21 & 59.45(0.33) & No data & 57.55(0.05) & 3.05(0.11) & 0.62(0.11) & No data & R \\
G035.28+1.31 & 13.11(0.26) & - & 13.12(0.02) & 0.40(0.04) & -0.02(0.66) & - & N,N \\
G036.02-1.36 & 31.20(0.03) & 31.43(0.14) & 31.74(0.02) & 0.90(0.04) & -0.60(0.03) & -0.34(0.16) & B,B \\
G036.33-1.27 & 30.79(0.08) & - & 30.58(0.02) & 0.82(0.06) & 0.26(0.10) & - & R,N \\
G036.91-0.43 & 77.78(0.23) & 81.17(0.06) & 80.23(0.02) & 1.55(0.06) & -1.58(0.15) & 0.61(0.04) & B,R \\
G036.92+0.71 & - & - & 12.92(0.02) & 0.64(0.04) & - & - & N,N \\
G036.94+1.09 & - & - & 13.22(0.02) & 0.49(0.04) & - & - & N,N \\
G037.05-0.03 & 80.65(0.07) & No data & 81.40(0.03) & 2.60(0.06) & -0.29(0.03) & No data & B $^{\alpha,\gamma}$ \\
G037.50+3.05 & 15.49(0.05) & - & 15.50(0.01) & 0.31(0.03) & -0.03(0.17) & - & N,N \\
G038.75-0.97 & - & - & 13.26(0.02) & 0.33(0.04) & - & - & N,N \\
G039.28-0.19 & 71.50(0.35) & 69.93(0.10) & 70.84(0.04) & 2.70(0.09) & 0.24(0.13) & -0.34(0.04) & N,B $^{\gamma}$ \\
G039.33-1.02 & 12.80(0.11) & - & 12.78(0.02) & 0.61(0.04) & 0.03(0.18) & - & N,N \\
G039.45-1.17 & 12.50(0.39) & - & 12.71(0.02) & 0.58(0.05) & -0.36(0.68) & - & B,N \\
G039.98+1.67 & 28.02(0.13) & 27.80(0.14) & 27.63(0.02) & 0.74(0.05) & 0.53(0.17) & 0.23(0.19) & R,N \\
G040.01-0.94 & - & - & 13.59(0.01) & 0.64(0.03) & - & - & N,N \\
G044.32-0.79 & 60.66(0.34) & 62.70(0.13) & 61.90(0.03) & 1.53(0.07) & -0.81(0.22) & 0.52(0.09) & B,R \\
G045.45+0.05 & 60.11(0.07) & 59.45(0.59) & 59.55(0.05) & 4.99(0.12) & 0.11(0.01) & -0.02(0.12) & N,N \\
G048.79+0.02 & 50.70(0.18) & 50.08(0.12) & 50.10(0.02) & 1.05(0.04) & 0.57(0.17) & -0.02(0.11) & R,N \\
G049.07-0.33 & 59.38(0.20) & - & 60.74(0.03) & 2.52(0.08) & -0.54(0.08) & - & B,N $^{\alpha,\beta,\gamma}$ \\
G049.25-1.38 & - & 4.34(0.06) & 5.67(0.02) & 0.81(0.05) & - & -1.64(0.07) & N,B \\
G049.27-1.41 & - & - & 5.51(0.02) & 0.44(0.04) & - & - & N,N \\
G049.28-1.40 & - & - & 5.50(0.01) & 0.44(0.03) & - & - & N,N \\
G049.32-1.41 & - & - & 5.50(0.02) & 0.42(0.04) & - & - & N,N \\
G052.99-0.03 & 22.57(0.14) & 22.19(0.21) & 22.36(0.01) & 0.54(0.03) & 0.39(0.25) & -0.31(0.39) & R,B \\
G053.08-0.24 & 23.35(0.23) & - & 24.09(0.02) & 0.75(0.04) & -0.99(0.30) & - & B,N \\
G053.10+0.11 & 22.00(0.06) & 21.84(0.06) & 22.18(0.02) & 0.85(0.04) & -0.21(0.07) & -0.40(0.07) & N,B $^{\alpha,\gamma}$ \\
G053.11+0.09 & 21.80(0.13) & 22.36(0.19) & 22.14(0.02) & 1.12(0.06) & -0.30(0.12) & 0.20(0.17) & B,N \\
G053.12+0.08 & 21.50(0.12) & 21.50(0.20) & 22.29(0.01) & 1.37(0.03) & -0.58(0.09) & -0.58(0.15) & B,B $^{\gamma}$ \\
G053.14+0.07 & 21.60(0.09) & No data & 22.04(0.02) & 2.22(0.05) & -0.20(0.04) & No data & N \\
G053.14+0.09 & 21.54(0.08) & 21.66(0.13) & 22.64(0.02) & 1.32(0.05) & -0.83(0.06) & -0.74(0.10) & B,B $^{\gamma}$ \\
G053.76+0.45 & - & - & 23.14(0.02) & 0.71(0.04) & - & - & N,N \\
G053.77+0.46 & - & - & 23.22(0.02) & 0.77(0.04) & - & - & N,N \\
G054.03-2.32 & 17.37(0.09) & 17.76(0.06) & 17.49(0.02) & 0.53(0.05) & -0.23(0.17) & 0.51(0.11) & N,R \\
G054.38-0.52 & 34.09(0.06) & 34.29(0.15) & 34.27(0.02) & 0.87(0.05) & -0.21(0.07) & 0.02(0.17) & N,N \\
G056.91+3.44 & - & 11.12(0.02) & 10.88(0.02) & 0.35(0.04) & - & 0.69(0.06) & N,R \\
G060.42-0.68 & - & - & 28.99(0.01) & 0.53(0.03) & - & - & N,N \\
G062.61+1.83 & 3.48(0.09) & - & 3.06(0.01) & 0.27(0.03) & 1.56(0.32) & - & R,N \\
G062.82+1.82 & - & - & 1.92(0.02) & 0.53(0.05) & - & - & N,N \\
G077.46+1.73 & 2.54(0.07) & 2.51(0.15) & 1.66(0.02) & 1.62(0.04) & 0.54(0.04) & 0.52(0.09) & R,R \\
G077.91-1.16 & -2.86(0.18) & 0.71(0.07) & -0.55(0.03) & 1.55(0.06) & -1.49(0.11) & 0.81(0.05) & B,R \\
G079.04+0.64 & 7.40(0.09) & 6.95(0.03) & 6.83(0.03) & 1.71(0.06) & 0.33(0.05) & 0.07(0.02) & R,N \\
G079.24+0.53 & -0.30(0.13) & - & 0.27(0.01) & 1.43(0.03) & -0.40(0.09) & - & B,N $^{\gamma}$ \\
G079.44+0.18 & 0.31(0.07) & 0.12(0.06) & 0.16(0.02) & 0.97(0.05) & 0.15(0.07) & -0.04(0.06) & N,N \\
G079.48+0.25 & 0.52(0.20) & 0.11(0.12) & 0.05(0.03) & 1.61(0.07) & 0.29(0.12) & 0.04(0.07) & R,N \\
G079.71+0.15 & 0.60(0.11) & 0.79(0.09) & 0.97(0.02) & 0.65(0.04) & -0.57(0.17) & -0.28(0.14) & B,B $^{\gamma}$ \\
G081.04-0.46 & 6.08(0.07) & 5.87(0.06) & 5.66(0.02) & 1.04(0.04) & 0.40(0.07) & 0.20(0.06) & R,N \\
G081.62+1.11 & 3.97(0.06) & 3.64(0.04) & 3.65(0.03) & 1.50(0.08) & 0.21(0.04) & -0.01(0.02) & N,N \\
G081.69-1.60 & 2.87(0.05) & 2.71(0.11) & 2.64(0.02) & 0.82(0.04) & 0.28(0.06) & 0.09(0.13) & R,N \\
G081.72+0.57 & -5.10(0.03) & No data & -2.83(0.02) & 3.31(0.04) & -0.69(0.01) & No data & B $^{\gamma}$ \\
G081.72+1.28 & 2.90(0.06) & 3.50(0.15) & 3.84(0.03) & 1.31(0.06) & -0.72(0.05) & -0.26(0.11) & B,B \\
G081.72-1.60 & 2.90(0.07) & - & 2.75(0.03) & 0.88(0.06) & 0.17(0.08) & - & N,N \\
G081.90+1.43 & 10.00(0.13) & 10.39(0.20) & 11.06(0.02) & 0.60(0.04) & -1.77(0.22) & -1.12(0.34) & B,B \\
G082.17+0.07 & 9.70(0.55) & 8.49(0.14) & 10.00(0.02) & 1.05(0.06) & -0.29(0.52) & -1.44(0.13) & B,B \\
G082.18-1.54 & 2.78(0.07) & 2.56(0.13) & 2.87(0.01) & 0.74(0.03) & -0.12(0.09) & -0.42(0.18) & N,B \\
G082.21-1.53 & 2.50(0.05) & 2.30(0.14) & 2.65(0.01) & 0.73(0.03) & -0.21(0.07) & -0.48(0.19) & N,B \\
G082.52-1.92 & 4.11(0.20) & 4.91(0.05) & 4.55(0.01) & 0.69(0.03) & -0.64(0.29) & 0.52(0.07) & B,R \\
G082.53+0.09 & 10.47(0.10) & 11.30(0.14) & 9.79(0.02) & 1.91(0.05) & 0.36(0.05) & 0.79(0.07) & R,R \\
G082.64-1.98 & 3.85(0.09) & 5.13(0.14) & 4.52(0.02) & 0.83(0.05) & -0.81(0.11) & 0.73(0.17) & B,R \\
G082.81-2.06 & 4.68(0.04) & 4.67(0.03) & 4.48(0.02) & 1.27(0.05) & 0.16(0.03) & 0.15(0.03) & N,N \\
G085.05-1.25 & -38.90(0.22) & -38.48(0.13) & -37.49(0.04) & 1.73(0.08) & -0.82(0.13) & -0.57(0.07) & B,B \\
G085.12+0.50 & - & - & -1.67(0.02) & 0.78(0.04) & - & - & N,N \\
G085.47-1.12 & 4.81(0.09) & - & 4.39(0.01) & 0.52(0.03) & 0.81(0.17) & - & R,N \\
G107.50+4.47 & -2.90(0.02) & -2.94(0.04) & -2.08(0.02) & 1.76(0.05) & -0.47(0.01) & -0.49(0.02) & B,B \\
G108.89+2.60 & -10.27(0.01) & -10.31(0.02) & -10.27(0.02) & 0.83(0.04) & 0.00(0.02) & -0.05(0.02) & N,N \\
G108.99+2.73 & -11.10(0.05) & -11.08(0.10) & -10.46(0.02) & 1.47(0.04) & -0.44(0.03) & -0.42(0.07) & B,B $^{\gamma}$ \\
G110.32+2.52 & -12.90(0.13) & -12.48(0.20) & -12.04(0.02) & 1.18(0.06) & -0.73(0.11) & -0.37(0.17) & B,B \\
G110.32+2.54 & -12.50(0.16) & -12.27(0.12) & -11.83(0.02) & 1.05(0.04) & -0.64(0.15) & -0.42(0.11) & B,B $^{\gamma}$ \\
G110.40+1.67 & -12.42(0.13) & - & -11.17(0.02) & 0.80(0.05) & -1.56(0.16) & - & B,N \\
G111.12+2.12 & -9.87(0.04) & -9.67(0.03) & -9.76(0.02) & 1.30(0.06) & -0.08(0.03) & 0.07(0.03) & N,N \\
G111.14+2.12 & -9.76(0.09) & -9.59(0.20) & -9.80(0.03) & 1.43(0.07) & 0.03(0.06) & 0.15(0.14) & N,N \\
G111.23+2.07 & -8.92(0.15) & -8.91(0.16) & -9.14(0.02) & 0.68(0.04) & 0.32(0.23) & 0.34(0.24) & R,R \\
G115.62+1.99 & - & - & -10.46(0.01) & 0.44(0.03) & - & - & N,N \\
G121.31+0.64 & -18.10(0.01) & -18.06(0.01) & -17.05(0.03) & 1.68(0.06) & -0.63(0.01) & -0.60(0.01) & B,B \\
G121.34+3.42 & -5.71(0.04) & -5.69(0.09) & -5.31(0.02) & 0.85(0.04) & -0.47(0.05) & -0.45(0.10) & B,B \\
G121.35+3.41 & -5.61(0.09) & -5.43(0.08) & -5.41(0.03) & 1.10(0.07) & -0.18(0.08) & -0.02(0.08) & N,N \\
G126.51-1.30 & -11.09(0.06) & -11.28(0.22) & -11.64(0.02) & 0.74(0.05) & 0.74(0.08) & 0.49(0.30) & R,R \\
G126.53-1.17 & -13.05(0.05) & -13.08(0.10) & -12.97(0.03) & 1.24(0.06) & -0.06(0.04) & -0.09(0.08) & N,N \\
G126.67-0.82 & -13.83(0.04) & No data & -13.68(0.02) & 1.74(0.05) & -0.09(0.03) & No data & N \\
G127.88+2.67 & -11.54(0.04) & -11.68(0.04) & -11.49(0.02) & 0.89(0.04) & -0.06(0.04) & -0.21(0.05) & N,N \\
G133.42+0.00 & -15.19(0.21) & -15.61(0.09) & -15.19(0.02) & 0.71(0.05) & 0.00(0.29) & -0.59(0.13) & N,B \\
G143.04+1.76 & -9.02(0.08) & -8.97(0.07) & -8.62(0.03) & 1.17(0.06) & -0.34(0.07) & -0.30(0.06) & B,B \\
G148.08+0.22 & - & -33.12(0.26) & -33.68(0.02) & 0.95(0.06) & - & 0.59(0.27) & N,R \\
G154.05+5.08 & 3.74(0.06) & - & 4.58(0.02) & 0.44(0.04) & -1.91(0.13) & - & B,N \\
G172.77+2.09 & -15.70(0.09) & -15.13(0.30) & -15.15(0.03) & 0.89(0.06) & -0.62(0.10) & 0.02(0.34) & B,N \\
G189.67+0.17 & 7.24(0.08) & 6.64(0.20) & 7.46(0.03) & 2.22(0.07) & -0.10(0.04) & -0.37(0.09) & N,B \\
G193.01+0.14 & 7.48(0.01) & 7.27(0.02) & 7.80(0.04) & 1.92(0.09) & -0.17(0.01) & -0.28(0.01) & N,B $^{\gamma}$ \\
G194.73-3.38 & 12.65(0.10) & 12.40(0.18) & 12.29(0.03) & 1.24(0.07) & 0.29(0.08) & 0.09(0.14) & R,N \\
G201.16+0.37 & 5.84(0.13) & 5.70(0.12) & 5.49(0.02) & 0.55(0.04) & 0.64(0.24) & 0.38(0.21) & R,R \\
G207.58-1.72 & 11.11(0.20) & 12.87(0.08) & 12.19(0.04) & 2.24(0.09) & -0.48(0.09) & 0.30(0.03) & B,R \\
G217.30-0.05 & 26.01(0.05) & 26.42(0.07) & 26.47(0.03) & 1.85(0.06) & -0.25(0.03) & -0.03(0.04) & B,N $^{\gamma}$ \\
  \hline\noalign{\smallskip}
\end{longtable}
%\footnotesize{$^*$ Indicates infall candidates with higher reliability.}\\
\tablecomments{1.1\textwidth}{Columns are (from left to right) the source name, peak velocity of HCO$^+$  (1-0), peak velocity of HCN (1-0), peak velocity of C$^{18}$O (1-0), FWHM of C$^{18}$O (1-0), asymmetry of HCO$^+$ (1-0),asymmetry of HCN (1-0), and profile of HCO$^+$ (1-0) and HCN (1-0). The values in parentheses give the uncertainties. The HCO$^+$ (1-0) and HCN (1-0) profiles are evaluated: B denotes blue profile, R denotes red profile, and N denotes neither blue or red or no emission. $^{\alpha,\beta,\gamma}$ indicates a source which are associated with the ATLASGAL compact sources, the ATLASGAL cold high-mass clumps, and the BGPS sources.}
%              $^*$ Indicates infall candidates with higher reliability.} 

\label{lastpage}

\end{document}